\documentclass[11pt]{article}

\usepackage[top=1in,bottom=1in,left=1in,right=1in]{geometry} 
\usepackage[doublespacing]{setspace}
\usepackage{amsfonts, amssymb, amsmath, amsthm}
\usepackage{graphicx, hyperref, subcaption, xcolor, booktabs, enumitem}
\usepackage[sans]{dsfont}
\usepackage[scr=rsfs,cal=boondox]{mathalpha}
\hypersetup{pdfborder = {0 0 0},colorlinks=true,allcolors=blue}
\usepackage[numbers]{natbib}

\newtheorem{thm}{Theorem}
\newtheorem{assumption}{Assumption}

\allowdisplaybreaks[1]

\DeclareMathOperator*{\argmin}{arg\,min}
\DeclareMathOperator{\T}{T}
\DeclareMathOperator{\CI}{I}
\DeclareMathOperator{\perim}{perim}
\DeclareMathOperator{\supp}{supp}
\newcommand{\dif}{\mathop{}\!\mathrm{d}}
\newcommand{\Haus}{\mathfrak{H}}

\renewcommand{\P}{\mathbb{P}}
\newcommand{\E}{\mathbb{E}}
\newcommand{\Var}{\mathbb{V}}
\newcommand{\Indicator}{\mathds{1}}
\newcommand{\En}{\E_n}
\renewcommand{\d}{\mathcal{d}}

\newcommand{\bA}{\mathbf{A}}
\newcommand{\bb}{\mathbf{b}}
\newcommand{\be}{\mathbf{e}}
\newcommand{\bv}{\mathbf{v}}
\newcommand{\bO}{\mathbf{O}}
\newcommand{\br}{\mathbf{r}}
\newcommand{\bz}{\mathbf{z}}

\newcommand{\bu}{\mathbf{u}}
\newcommand{\bU}{\mathbf{U}}
\newcommand{\bV}{\mathbf{V}}
\newcommand{\bx}{\mathbf{x}}
\newcommand{\bX}{\mathbf{X}}
\newcommand{\by}{\mathbf{y}}
\newcommand{\bZ}{\mathbf{Z}}

\newcommand{\bnu}{\boldsymbol{\nu}}
\newcommand{\bgamma}{\boldsymbol{\gamma}}

\newcommand{\bPsi}{\boldsymbol{\Psi}}
\newcommand{\bUpsilon}{\boldsymbol{\Upsilon}}

\newcommand{\A}{\mathcal{A}}
\newcommand{\B}{\mathcal{B}}
\newcommand{\I}{\mathcal{I}}
\newcommand{\q}{\mathfrak{q}}
\newcommand{\U}{\mathcal{U}}
\newcommand{\X}{\mathcal{X}}

\begin{document}
\pagenumbering{roman}

\title{Estimation and Inference in Boundary Discontinuity Designs: Distance-Based Methods\thanks{We thank Alberto Abadie, Boris Hanin, Kosuke Imai, Xinwei Ma, Victor Panaretos, J\"org Stoye, and Jeff Wooldridge for comments and discussions. The co-Editor, Associate Editor, and two reviewers provided excellent comments and suggestions. Cattaneo and Titiunik gratefully acknowledge financial support from the National Science Foundation (SES-2019432, DMS-2210561, SES-2241575 and SES-2342226). Cattaneo gratefully acknowledges financial support from the National Institute for Food and Agriculture (NIFA) through grant 2024-67023-42704, and the Data-Driven Social Science initiative at Princeton University.}\bigskip}

\author{Matias D. Cattaneo\thanks{Department of Operations Research and Financial Engineering, Princeton University.} \and
	    Rocio Titiunik\thanks{Department of Politics, Princeton University.} \and
	    Ruiqi (Rae) Yu\thanks{Department of Operations Research and Financial Engineering, Princeton University.} 
	    }
\maketitle

\begin{abstract}
    \onehalfspacing
    We study nonparametric distance-based (isotropic) local polynomial methods for estimating the boundary average treatment effect curve, a causal functional that captures treatment effect heterogeneity in boundary discontinuity designs. We establish identification, estimation, and inference results both pointwise and uniformly along the treatment assignment boundary. We show that the geometric regularity of the boundary, a one-dimensional manifold, plays a central role in determining feasible convergence rates and valid inference procedures. Our theoretical contributions are threefold. First, we derive uniform lower and upper bounds on the convergence rate of the misspecification bias of isotropic local polynomial estimators. Second, we obtain uniform distributional approximations that justify boundary-robust inference. Third, we establish minimax lower bounds for a broad class of nonparametric isotropic regression estimators. These results yield practical guidance for empirical implementation, including new bandwidth selection rules that adapt to local irregularities of the treatment-assignment boundary. We illustrate the proposed methods using simulation evidence and an empirical application, and provide companion general-purpose software. 
\end{abstract}


\thispagestyle{empty}

\clearpage
\tableofcontents
\thispagestyle{empty}

\clearpage
\pagenumbering{arabic}

\section{Introduction}

Discontinuities in treatment assignment are widely used in observational studies to investigate causal effects. In regression discontinuity (RD) designs, each unit receives a univariate score (or running variable), and treatment is assigned according to a threshold rule: units with scores equal to or above a known cutoff are assigned to treatment, while units with scores below the cutoff remain in the control condition. Local treatment effects at the cutoff are identifiable under continuity \citep{Hahn-Todd-vanderKlaauw_2001_ECMA} or local randomization \citep{Cattaneo-Frandsen-Titiunik_2015_JCI} assumptions, and estimation and inference are typically conducted using local polynomial least squares regression methods. See \citet{Cattaneo-Titiunik_2022_ARE} for an overview of the methodological RD literature and \citet{Cattaneo-Idrobo-Titiunik_2020_CUP} for a practical introduction.

Boundary discontinuity (BD) designs generalize canonical RD settings to contexts in which units receive a bivariate score $\bX_i=(X_{1i},X_{2i})^\top$ with support $\X\subseteq\mathbb{R}^2$, and treatment assignment is determined by a threshold rule based on a one-dimensional boundary curve $\B$ that partitions the support of the score. A prototypical example is the geographic RD design and its variants \citep{Keele-Titiunik_2015_PA,Keele-Titiunik-Zubizarreta_2015_JRSSA,Keele-Titiunik_2016_PSRM,Keele-etal_2017_AIE,Galiani-McEwan-Quistorff_2017_AIE,Rischard-Branson-Miratrix-Bornn_2021_JASA,Diaz-Zubizarreta_2023_AOAS}, where a geographic boundary separates units into treatment and control areas according to their location. These designs are also referred to as Multi-Score RD Designs \citep{Papay-Willett-Murnane_2011_JoE,Reardon-Robinson_2012_JREE,Wong-Steiner-Cook_2013_JEBS}. A practical introduction to BD designs is provided in \citet[Section~5]{Cattaneo-Idrobo-Titiunik_2024_CUP}, while \citet{Cattaneo-Titiunik-Yu_2026_BookCh} offers an overview of the empirical literature.

Using standard causal inference notation \citep{Hernan-Robins_2020_Book}, let $Y_i(0)$ and $Y_i(1)$ denote the potential outcomes for unit $i=1,2,\ldots,n$ under control and treatment assignments, respectively. The key building block in BD designs is the \textit{Boundary Average Treatment Effect Curve} (BATEC):
\begin{align*}
    \tau(\bx) = \E[Y_i(1) - Y_i(0) | \bX_i = \bx], \qquad \bx\in\B,
\end{align*}
a functional causal parameter that characterizes heterogeneous treatment effects along the assignment boundary. Identification, estimation, and inference for the BATEC can proceed through two distinct methodological approaches:
\begin{itemize}
    \item \textit{Location-Based Methods}. This approach directly exploits the bivariate score $\bX_i$, thereby extending canonical RD methods to multidimensional settings. Early identification and estimation results are provided by \citet{Papay-Willett-Murnane_2011_JoE}, \citet{Reardon-Robinson_2012_JREE}, \citet{Wong-Steiner-Cook_2013_JEBS}, and \citet{Keele-Titiunik_2015_PA}, while \citet{Cattaneo-Titiunik-Yu_2026_JASA} develop modern pointwise and uniform estimation and inference procedures.
    
    \item \textit{Distance-Based Methods}. This approach reduces the bivariate score $\bX_i$ to a univariate distance-based running variable defined for each $\bx \in \B$, thereby enabling the use of standard univariate RD methods. \citet[Section~4]{Reardon-Robinson_2012_JREE} refer to this strategy as distance-based RD, and \citet{Keele-Titiunik_2015_PA} discuss its use in geographic RD applications. \citet{Velez-Newman_2019_AJPS} provide an empirical illustration employing distance-based polynomial regression to study treatment-effect heterogeneity in a geographic RD design.
\end{itemize}

The main goal of this paper is to study the identification, estimation, and inference properties of distance-based methods for the BATEC in BD designs, and to leverage the resulting theoretical insights to provide practical guidance for empirical implementation. The defining feature of these methods is the use of univariate local polynomial regression, where the running variable is the distance between $\bX_i$ and each point $\bx \in \B$. From a nonparametric smoothing perspective, this approach is closely related to isotropic regression techniques studied in statistics and machine learning. By contrast, location-based methods employ local polynomial regressions that directly incorporate the bivariate score into the polynomial approximation.

In practice, researchers implementing BD designs frequently transform the multidimensional score into a univariate measure of distance to the assignment boundary and conduct estimation and inference using this distance as the running variable. Evidence compiled in \citet[Table~1]{Cattaneo-Titiunik-Yu_2026_BookCh} documents more than $80$ recent empirical studies in economics and related quantitative disciplines adopting this strategy. When the goal is to study treatment-effect heterogeneity along the boundary, distance-based polynomial regression replaces the full bivariate location information in $\bX_i$ with a scalar distance defined relative to each boundary point $\bx \in \B$. In other applications, the objective is instead to pool treatment effects along the boundary, in which case researchers often use the distance to the closest point on $\B$. This alternative strategy gives rise to pooling-based methods, which are not the focus of this paper. Pooling-based methods discard location information and therefore identify a scalar (weighted) \textit{Boundary Average Treatment Effect} (BATE) causal parameter. \citet{Cattaneo-Titiunik-Yu_2026_BookCh} provide an introductory discussion, while ongoing work \citep{Cattaneo-Titiunik-Yu_2026_BDD-Pooling} develops formal identification, estimation, and inference results for pooling-based methods.

We focus on distance-based methods. Beyond their empirical prevalence, distance-based methods offer important practical advantages. In many applications, precise location information may be unavailable due to confidentiality restrictions, data aggregation, or measurement limitations, whereas distance-to-boundary measures can still be constructed or released. In such settings, distance-based procedures may provide the only feasible way to estimate the heterogeneous treatment effects summarized by the boundary average treatment effect curve $\tau(\bx)$. Moreover, by reducing the dimensionality of the running variable, these methods simplify implementation, facilitate interpretation, and enhance comparability with standard univariate RD analyses. These considerations underscore the importance of formally understanding the statistical properties of distance-based estimators in BD designs.

\subsection{Contributions}

To outline our main contributions, we first formalize the distance-based approach. The core ingredient is the scalar signed distance-based score for each unit $i=1,\ldots,n$,
\begin{align*}
    D_i(\bx) = \big(\Indicator(\bX_i \in \A_1) - \Indicator(\bX_i \in \A_0)\big) \cdot \d(\bX_i, \bx),
    \qquad
    \bx \in \B,
\end{align*}
where $\d(\cdot,\cdot)$ denotes a distance function, $\X = \A_0 \cup \A_1$ with $\A_0$ and $\A_1$ denoting the disjoint (connected) control and treatment regions, respectively, and $\B = \mathtt{bd}(\A_0) \cap \mathtt{bd}(\A_1)$ with $\mathtt{bd}(\A_t)$ denoting the boundary of the set $\A_t$ for $t\in\{0,1\}$. Without loss of generality, we assume that the assignment boundary belongs to the treatment region, that is, $\B\subset\A_1$ and $\B\cap\A_0=\emptyset$. A canonical example is the Euclidean distance $\d(\bX_i,\bx)=|\bX_i-\bx|=\sqrt{(X_{1i}-x_1)^2+(X_{2i}-x_2)^2}$ for $\bx=(x_1,x_2)^\top\in\B$, although alternative distance measures may be appropriate in spatial applications \citep{Banerjee_2005_Biometrics}. For each $\bx\in\B$, the distance-based setup is analogous to a standard univariate RD design in which $D_i(\bx)\in\mathbb{R}$ is the running variable, the cutoff is $c=0$, and $D_i(\bx)\ge0$ ($<0$) corresponds to treatment (control) assignment.

The observed outcome is $Y_i=\Indicator(D_i(\bx)\in\I_0)Y_i(0)+\Indicator(D_i(\bx)\in\I_1)Y_i(1)$, where $\I_0=(-\infty,0)$ and $\I_1=[0,\infty)$. Following standard practice in the RD literature, the distance-based local polynomial treatment effect estimator for each $\bx\in\B$ is
\begin{align*}
    \widehat{\vartheta}(\bx)
    = \be_0^{\top} \widehat{\bgamma}_1(\bx) - \be_0^{\top} \widehat{\bgamma}_0(\bx),
    \qquad \bx \in \B,
\end{align*}
where, for $t \in \{0,1\}$,
\begin{align}\label{eq: locpoly estimator}
    \widehat{\bgamma}_{t}(\bx)
    = \argmin_{\bgamma \in \mathbb{R}^{p+1}}
      \frac{1}{n} \sum_{i=1}^n \big(Y_i - \br_p(D_i(\bx)/h)^{\top} \bgamma \big)^2 K_h(D_i(\bx)) \Indicator(D_i(\bx)\in\I_t),
\end{align}
with $\be_0$ the conformable intercept selector, $\br_p(u)=(1,u,u^2,\ldots,u^p)^\top$ the usual polynomial basis, and $K_h(u)=K(u/h)/h^2$ a distance-based kernel with bandwidth $h$. The kernel down-weights observations as their distance to $\bx\in\B$ increases, while the bandwidth controls the degree of localization. The normalization by $h^2$ reflects the bivariate score dimension; multiplying all weights by a common positive constant does not affect the weighted least squares estimator. In this setup, observations contribute isotropically according to their univariate distance $\d(\bX_i,\bx)$ from the boundary point $\bx$.

The estimator $\widehat{\vartheta}(\bx)$ is thus the difference of two univariate (isotropic) local polynomial regression estimators evaluated along the one-dimensional manifold $\B$. Importantly, this estimator does not directly target $\tau(\bx)$. Rather, it targets the difference $\theta_{1,\bx}(0)-\theta_{0,\bx}(0)$, where
\begin{align}\label{eq: induced distance-based cond exp function}
    \theta_{t,\bx}(r)
    = \E\big[Y_i    \big| D_i(\bx) = r \big]
    = \E\big[Y_i(t) \big| \d(\bX_i,\bx) = |r|, \bX_i \in \A_t\big],
    \qquad r\in\I_t,
\end{align}
are univariate conditional expectations induced by transforming the bivariate score through the distance function. Our objective is therefore to characterize the conditions under which valid pointwise and uniform identification, estimation, and inference for the BATEC $\tau(\bx)$ can be conducted using the distance-based estimator $\widehat{\vartheta}(\bx)$ and its associated estimand $\theta_{1,\bx}(0)-\theta_{0,\bx}(0)$.

We begin by studying identification of $\tau(\bx)$ through distance-based approximations. The induced conditional expectations $\theta_{t,\bx}(r)$ generally differ from the bivariate regression functions $\mu_t(\bx)=\E[Y_i(t)\mid\bX_i=\bx]$. Consequently, without restrictions on the data-generating process, the geometry of the assignment boundary, and the distance function, $\lim_{r\to0}\theta_{t,\bx}(r)$ need not coincide with $\mu_t(\bx)$. Theorem \ref{thm: Identification} provides sufficient conditions for these parameters to agree, including restrictions ensuring that $\B$ is a rectifiable curve \citep{federer2014geometric}. This result contributes to the multidimensional RD literature by establishing a new identification theory for distance-based methods \citep[cf.,][]{Hahn-Todd-vanderKlaauw_2001_ECMA,Papay-Willett-Murnane_2011_JoE,Reardon-Robinson_2012_JREE,Wong-Steiner-Cook_2013_JEBS,Cattaneo-Keele-Titiunik-VazquezBare_2016_JOP}.

The geometry of the boundary influences not only identification but also bias approximation. We establish two complementary results. First, Theorem \ref{thm: Approximation Bias: Uniform Guarantee} shows that near irregular points of the assignment boundary, such as kinks that may arise even when $\B$ is a well-behaved one-dimensional manifold, the $p$th-order estimator $\widehat{\vartheta}(\bx)$ exhibits an irreducible bias of order $h$, regardless of the polynomial order or the smoothness of the underlying regression functions $\mu_0(\bx)$ and $\mu_1(\bx)$. This phenomenon arises because the induced regression function $\theta_{t,\bx}(r)$ is at most Lipschitz continuous uniformly in neighborhoods of such irregular points. Second, Theorem \ref{thm: Approximation Bias: Smooth Boundary} shows that when $\B$ is smooth, the usual approximation bias of order $h^{p+1}$ from the nonparametric smoothing literature is recovered \citep{Hardle-etal_2004_Book}. Hence, the bias of the distance-based estimator varies along the boundary according to its local geometric features.

These findings imply that standard univariate RD procedures can exhibit unexpected behavior when applied to BD designs through distance-based running variables: geometric irregularities of the assignment boundary may induce substantially larger misspecification bias. This issue is particularly relevant in geographic RD applications, where boundaries often contain kinks or other irregularities. For instance, the influential work of \cite{mandelbrot1967long,mandelbrot1983fractal} has argued that geographic borders (and other shapes in nature) are fractals; see \cite{avnir1998geometry}, and references therein, for a recent discussion on this ongoing debate among mathematicians and philosophers. From a practical standpoint, our geometric analysis yields concrete foundational guidance for bandwidth choice, point estimation, and statistical inference.

We next develop estimation and inference theory for $\tau(\bx)$ based on $\widehat{\vartheta}(\bx)$. Theorem \ref{thm: Convergence Rates} establishes pointwise and uniform convergence rates, while Theorem \ref{thm: Statistical Inference} provides valid uncertainty quantification both pointwise and uniformly over $\B$. These results are formulated in terms of a generic misspecification bias component to accommodate the different interactions between the distance function and the boundary geometry. Under appropriate regularity conditions and when $\B$ is sufficiently smooth, the estimator achieves the optimal nonparametric convergence rate pointwise and uniformly \citep{tsybakov2008introduction}. We also construct confidence intervals for $\tau(\bx)$ and confidence bands for the entire curve $(\tau(\bx):\bx\in\B)$, leveraging a new strong approximation result for empirical processes developed in the supplemental appendix.

We also establish a minimax optimality result that highlights the fundamental limitations of distance-based methods when the assignment boundary is not smooth. Theorem \ref{thm: minimax for distance based} derives a minimax lower bound convergence rate for isotropic nonparametric estimators of a smooth bivariate regression function evaluated along a one-dimensional rectifiable manifold. We show that, irrespective of the smoothness of the underlying bivariate regression function, isotropic univariate estimators can achieve at most the uniform convergence rate $n^{-1/4}$ (up to polylogarithmic factors). This rate coincides with the optimal minimax rate for Lipschitz continuous bivariate regression functions over the same domain, thereby demonstrating that the convergence rate attained by $\widehat{\vartheta}(\bx)$ is unimprovable without imposing stronger geometric restrictions on the boundary or exploiting additional structural features of the data-generating process.

Our theoretical results characterize identification, estimation, and inference for the BATEC, both pointwise and uniformly over $\B$, when this causal parameter is estimated using univariate distance-based (isotropic) local polynomial regression methods. As discussed in Section \ref{sec: Estimation and Inference}, these findings also inform empirical practice by motivating new ``regularized'' distance-based estimation and inference procedures when additional information about the geometry of the assignment boundary $\B$ is available. Section \ref{sec: Numerical Results} presents numerical evidence based on the dataset of \cite{LondonoVelezRodriguezSanchez_2020_AEJ}, illustrating the finite-sample performance of the proposed methods.

The supplemental appendix further extends the distance-based framework to policy-relevant functionals of the BATEC: the \textit{Weighted Boundary Average Treatment Effect} (WBATE) and \textit{Largest Boundary Average Treatment Effect} (LBATE), as well as to fuzzy (imperfect treatment compliance) BD designs, covering fuzzy BATEC, fuzzy WBATE, and fuzzy LBATE estimands.
Section \ref{sec: Extensions} summarizes these extensions and also discusses adjustments for pre-intervention covariates aimed at either improving efficiency or heterogeneity analysis.

Concrete guidance for empirical implementation based on the results developed in this paper and in our ongoing companion work \citep{Cattaneo-Titiunik-Yu_2026_JASA,Cattaneo-Titiunik-Yu_2026_BDD-Pooling} is given in Section \ref{sec: Recommendations for Practice}. All our methodological results are implemented in our companion \texttt{R} package \texttt{rd2d} (\url{https://rdpackages.github.io/rd2d}); see \cite{Cattaneo-Titiunik-Yu_2025_rd2d} for further details.

From a broader theoretical perspective, this paper establishes a connection between the econometric methodology for BD designs and the statistical theory of nonparametric smoothing over submanifolds, thereby contributing to two distinct yet closely related literatures in econometrics, statistics, machine learning, and data science. Methodologically, we provide the first foundational results for the analysis and interpretation of BD designs based on distance-based estimators, an approach widely used in applied work but lacking formal theoretical justification. From a nonparametric perspective, we develop pointwise and uniform estimation and inference results for isotropic local polynomial regression on submanifolds, thereby advancing the theory of smoothing methods in low-dimensional geometric settings.

Our results are complementary to concurrent work by \cite{Chen-Gao_2026_thinsets}, who study estimation and inference for integral functionals on submanifolds using series (sieve) methods. In contrast, we focus on estimation based on one-dimensional distance transformations evaluated pointwise along the submanifold, analyze isotropic local polynomial smoothing procedures, and uncover phenomena related to uniform misspecification bias. In addition, we establish a novel minimax lower bound convergence rate for isotropic nonparametric regression estimators in this setting.

\subsection{Notation}

We employ standard concepts and notation from empirical process theory \citep{van-der-Vaart-Wellner_1996_Book,Gine-Nickl_2016_Book} and geometric measure theory \citep{simon1984lectures,federer2014geometric}. For a random variable $\bV_i$, we write $\En[g(\bV_i)] = n^{-1} \sum_{i = 1}^n g(\bV_i)$. For a vector $\bv \in \mathbb{R}^k$, the Euclidean norm is $\|\bv\| = (\sum_{j = 1}^k \bv_j^2)^{1/2}$. For a matrix $\bA$, $\lambda_{\min}(\bA)$ denotes the smallest eigenvalue. $C^k(\mathcal{X},\mathcal{Y})$ denotes the class of $k$-times continuously differentiable functions from $\mathcal{X}$ to $\mathcal{Y}$, and $C^k(\mathcal{X})$ is a shorthand for $C^k(\mathcal{X},\mathbb{R})$. For a Borel set $\mathcal{S} \subseteq \X$, the De Giorgi perimeter of $\mathcal{S}$ is $\perim(\mathcal{S}) = \sup_{g \in \mathscr{D}_2(\X)} \int_{\mathbb{R}^2} \Indicator(\bx \in \mathcal{S}) \operatorname{div} g(\bx) d \bx / \lVert g \rVert_{\infty}$, where $\operatorname{div}$ is the divergence operator, and $\mathscr{D}_2(\X)$ denotes the space of $C^{\infty}$ functions with values in $\mathbb{R}^2$ and with compact support included in $\X$. When $\mathcal{S}$ is connected, and the boundary $\mathtt{bd}(\mathcal{S})$ is a smooth simple closed curve, $\perim(\mathcal{S})$ simplifies to the curve length of $\mathtt{bd}(\mathcal{S})$. A curve $\mathcal{B} \subseteq \mathbb{R}^2$ is a \emph{rectifiable curve} if there exists a Lipschitz map $\gamma:[0,1] \mapsto \mathbb{R}^2$ such that $\mathcal{B}=\gamma([0,1])$. For a function $f:\mathbb{R}^2 \mapsto\mathbb{R}$, $\supp(f)$ denotes the closure of the set $\{\bx \in \mathbb{R}^2: f(\bx) \neq 0\}$. For real sequences, $a_n=o(b_n)$ means $\limsup_{n\to\infty} |a_n/b_n|=0$, and $a_n \lesssim b_n$ means there exist constants $C<\infty$ and $N<\infty$ such that $|a_n| \leq C |b_n|$ for all $n>N$. For sequences of random variables, $a_n=o_{\P}(b_n)$ means $|a_n/b_n|\to_{\P}0$, and $a_n \lesssim_\P b_n$ means $\limsup_{M \to \infty} \limsup_{n \to \infty} \P[|a_n/b_n| \geq M] = 0$. Throughout, $\bx$ denotes a generic point in $\X$ and, when $\bx\in\B$, a generic boundary point; $\bb$ and $\bb_j$ denote selected or discretized evaluation points on the boundary. Finally, $\Phi(x)$ denotes the standard Gaussian cumulative distribution function.

\subsection{Organization}

Section \ref{sec: Setup} introduces the setup and assumptions used throughout the paper. Section \ref{sec: Identification and Interpretation} studies identification of $\tau(\bx)$ via distance-based approximations, while Section \ref{sec: Approximation Bias} analyzes the bias properties of the distance-based estimator $\widehat{\vartheta}(\bx)$. Section \ref{sec: Estimation and Inference} develops estimation and inference procedures for $\tau(\bx)$ based on distance-based local polynomial regression. Section \ref{sec: Distance-Based Minimax Convergence Rate} presents a minimax convergence rate result for isotropic (distance-based) regression estimation. Section \ref{sec: Numerical Results} illustrates the finite-sample performance of the distance-based methods. Section \ref{sec: Extensions} discusses extensions, Section \ref{sec: Recommendations for Practice} provides implementation guidance, and Section \ref{sec: Conclusion} concludes.

The supplemental appendix contains generalizations of the main theoretical results, their proofs, and additional findings that may be of independent interest. In particular, it presents a new strong approximation theorem for empirical processes with multiplicative separable structure and bounded polynomial moments, extending recent work by \cite{Cattaneo-Yu_2025_AOS}.

\section{Setup and Assumptions}\label{sec: Setup}

The following assumption collects the core conditions imposed on the underlying data generating process.

\begin{assumption}[Data Generating Process]\label{assump: DGP}
    Let $t\in\{0,1\}$, $p \geq 0$, and $v \geq 2$.
    \begin{enumerate}[label=\normalfont(\roman*),noitemsep,leftmargin=*]
    
        \item $(Y_1(t),\bX_1^\top)^\top,\ldots,(Y_n(t),\bX_n^\top)^\top$ are independent and identically distributed random vectors.
    
        \item The distribution of $\bX_i$ admits a Lebesgue density $f_X(\bx)$ that is continuous and bounded away from zero on its support $\X=[\mathtt{L},\mathtt{U}]^2$, for $-\infty<\mathtt{L}<\mathtt{U}<\infty$.

        \item $\mu_t(\bx)=\E[Y_i(t)\mid \bX_i=\bx]$ has a $(p+1)$ times continuously differentiable extension to an open neighborhood of $\X$.
    
        \item $\sigma_t^2(\bx)=\Var[Y_i(t)\mid \bX_i=\bx]$ is continuous and bounded away from zero on $\X$.
    
        \item $\sup_{\bx\in\X}\E\!\left[|Y_i(t)|^{2+v}\mid \bX_i=\bx\right]<\infty$.

    \end{enumerate}
\end{assumption}

Assumption \ref{assump: DGP} generalizes standard regularity conditions used in univariate RD designs \citep[see, e.g.,][and references therein]{Cattaneo-Titiunik_2022_ARE}. Additional restrictions specific to distance-based methods in BD designs are also required. Let
\begin{align*}
        \bPsi_{t,\bx}
        = \E \Big[\br_p\Big(\frac{D_i(\bx)}{h}\Big) \br_p\Big(\frac{D_i(\bx)}{h}\Big)^{\top} K_h(D_i(\bx)) \Indicator(D_i(\bx) \in \I_t)\Big]
\end{align*}
denote the fixed-$h$ population Gram matrix associated with the distance-based estimator.

\begin{assumption}[Kernel, Distance, and Boundary]\label{assump: Kernel, Distance, and Boundary}
    Let $t \in \{0,1\}$.
    \vspace{-.1in}
    \begin{enumerate}[label=\normalfont(\roman*),noitemsep,leftmargin=*]
        \item $\B\subset\operatorname{int}(\X)$ is a rectifiable curve with positive length.
        \item The distance function $\d:\mathbb{R}^2\times\mathbb{R}^2\to[0,\infty)$ is a metric satisfying $\|\bx_1-\bx_2\|\lesssim \d(\bx_1,\bx_2)\lesssim \|\bx_1-\bx_2\|$ for all $\bx_1,\bx_2\in\X$.
        \item $K:\mathbb{R}\to[0,\infty)$ is either compactly supported and Lipschitz continuous, or $K(u)=\Indicator(u\in[-1,1])$ and the distance balls $\{\bv\in\X:\d(\bv,\bx)\le r\}$, indexed by $(\bx,r)\in\X\times\mathbb{R}_{+}$, form a VC class.
        \item $\liminf_{h\downarrow 0}\inf_{\bx\in\B}\lambda_{\min}(\bPsi_{t,\bx})\gtrsim 1$.
        \item For each $\bx\in\B$ and $t\in\{0,1\}$, the Hausdorff-measure denominator defining $\theta_{t,\bx}(r)$ is positive and finite for all sufficiently small $|r|>0$ with $r\in\I_t$.
    \end{enumerate}
\end{assumption}

Assumption \ref{assump: Kernel, Distance, and Boundary}(i) imposes minimal regularity on the assignment boundary $\B$, which facilitates the evaluation of integrals and the derivation of uniform results over the one-dimensional submanifold. Assumption \ref{assump: Kernel, Distance, and Boundary}(ii) requires the distance function to be a metric equivalent (up to constants) to the Euclidean distance. Assumption \ref{assump: Kernel, Distance, and Boundary}(iii) imposes standard conditions on the univariate kernel function; for the uniform kernel, the additional VC condition is automatic for Euclidean distance balls and other primitive cases covered by Lemma SA-1 in the supplemental appendix. Assumption \ref{assump: Kernel, Distance, and Boundary}(iv) further restricts (implicitly) the geometry of the boundary $\B$ relative to the kernel and distance functions, ruling out highly irregular shapes that would generate regions with too few observations. Assumption \ref{assump: Kernel, Distance, and Boundary}(v) ensures that the induced one-dimensional conditional expectations are well defined near the cutoff. Lemma SA-2 in the supplemental appendix provides primitive conditions for Assumption \ref{assump: Kernel, Distance, and Boundary}(iv) when $\d(\cdot)$ is the Euclidean norm. The conditions imposed by Assumption \ref{assump: Kernel, Distance, and Boundary} are sufficient for uniform (over $\B$) estimation and inference, although some can be weakened for pointwise results; see the supplemental appendix.

\section{Identification and Interpretation}\label{sec: Identification and Interpretation}

For each boundary point $\bx \in \B$ and corresponding signed distance score $D_i(\bx)$, the univariate distance-based local polynomial estimator is $\widehat{\vartheta}(\bx) = \widehat{\theta}_{1,\bx}(0) - \widehat{\theta}_{0,\bx}(0)$, where $\widehat{\theta}_{t,\bx}(0) = \be_0^{\top}\widehat{\bgamma}_t(\bx) = \br_p(0)^{\top}\widehat{\bgamma}_t(\bx)$. Conceptually, the estimator $\widehat{\theta}_{t,\bx}(0)$ targets the estimand $\theta_{t,\bx}(0)$ defined in \eqref{eq: induced distance-based cond exp function}, which is the univariate conditional expectation induced by the distance transformation applied to the bivariate location variable $\bX_i$ for each point $\bx\in\B$.

The following theorem establishes identification of the causal functional parameter $\tau(\bx)$ through the distance-based induced conditional expectations.

\begin{thm}[Identification]\label{thm: Identification}
    Suppose Assumptions \ref{assump: DGP}(i)--(iii) and \ref{assump: Kernel, Distance, and Boundary}(i), (ii), and (v) hold. Then,
    \begin{align*}
        \tau(\bx) = \lim_{r\downarrow0} \theta_{1,\bx}(r) - \lim_{r\uparrow0} \theta_{0,\bx}(r)
    \end{align*}
    for all $\bx\in\B$.
\end{thm}

This identification result is obtained by representing $\theta_{t,\bx}(r)$ as a sequence of integrals over submanifolds near $\bx\in\B$, that is, over level sets of shrinking spheres generated by the distance function $\d(\cdot, \bx)$. Without restrictions on the data generating process, the geometry of the assignment boundary, and the distance function, the limits $\lim_{r\to0}\theta_{t,\bx}(r)$ and $\mu_t(\bx)$ need not coincide.

The identification result in Theorem \ref{thm: Identification} is new to the literature. Related continuity-based identification arguments in RD designs include \cite{Hahn-Todd-vanderKlaauw_2001_ECMA} for one-dimensional score settings; \cite{Papay-Willett-Murnane_2011_JoE}, \cite{Reardon-Robinson_2012_JREE}, and \cite{Keele-Titiunik_2015_PA} for multi-score settings; and \cite{Cattaneo-Keele-Titiunik-VazquezBare_2016_JOP} for multi-cutoff designs. For example, identification of the BATEC using location-based methods takes the form
\begin{align*}
    \tau(\bx) 
    = \E[Y_i(1) - Y_i(0) | \bX_i = \bx]
    = \lim_{\bu\to\bx,\bu\in\A_1} \E[Y_i | \bX_i = \bu] - \lim_{\bu\to\bx,\bu\in\A_0} \E[Y_i | \bX_i = \bu],
\end{align*}
for all $\bx\in\B$. Our result does not cover pooling-based methods, where identification, estimation, and inference rely on the distance to the closest boundary point and therefore aggregate treatment effects along $\B$. See \cite{Cattaneo-Titiunik-Yu_2026_BookCh} for an overview; formal analysis of pooling-based methods is the subject of ongoing work \citep{Cattaneo-Titiunik-Yu_2026_BDD-Pooling}.

The isotropic nonparametric procedures underlying distance-based methods naturally introduce approximation (misspecification) error when targeting $\tau(\bx)$. To study this issue, we interpret the estimator $\widehat{\bgamma}_t(\bx)$ as the sample analogue of the coefficients from the best local mean-square approximation of the conditional expectation $\E[Y_i(t)\mid D_i(\bx)]$ using the polynomial basis $\br_p(D_i(\bx)/h)$:
\begin{align*}
    \bgamma^{\ast}_{t}(\bx)
    = \argmin_{\bgamma \in \mathbb{R}^{p+1}}
      \E \Big[ \big(Y_i - \br_p(D_i(\bx)/h)^{\top} \bgamma \big)^2 K_h(D_i(\bx)) \Indicator(D_i(\bx) \in \I_t) \Big].
\end{align*}
Letting $\theta_{t,\bx}^{\ast}(r) = \br_p(r/h)^{\top}\bgamma^{\ast}_{t}(\bx)$, so that $\theta_{t,\bx}^{\ast}(0) = \be_0^{\top}\bgamma^{\ast}_{t}(\bx) = \br_p(0)^{\top}\bgamma^{\ast}_{t}(\bx)$, we obtain the standard least-squares decomposition
\begin{align}\label{eq: distance error decomposition}
    \widehat{\theta}_{t,\bx}(0) - \theta_{t,\bx}(0)
    = \big[\theta_{t,\bx}^{\ast}(0) - \theta_{t,\bx}(0)\big]
    + \be_0^{\top} \bPsi_{t,\bx}^{-1} \bO_{t,\bx}
    + \big[\be_0^{\top} (\widehat{\bPsi}_{t,\bx}^{-1} - \bPsi_{t,\bx}^{-1}) \bO_{t,\bx}\big],
\end{align}
where
\begin{align*}
    \widehat{\bPsi}_{t,\bx} &= \En \Big[\br_p\Big(\frac{D_i(\bx)}{h}\Big) \br_p\Big(\frac{D_i(\bx)}{h}\Big)^{\top} K_h(D_i(\bx)) \Indicator(D_i(\bx) \in \I_t)\Big],
    \qquad\text{and}\\
    \bO_{t,\bx} &= \En \Big[\br_p\Big(\frac{D_i(\bx)}{h}\Big) K_h(D_i(\bx)) (Y_i - \theta_{t,\bx}^{\ast}(D_i(\bx)))\Indicator(D_i(\bx) \in \I_t)\Big].
\end{align*}

In decomposition \eqref{eq: distance error decomposition}, the first term represents the nonrandom approximation bias, the second term is a stochastic linear component (a sample average of mean-zero random variables), and the third term is a higher-order linearization error arising from estimation of the Gram matrix. Importantly, our analysis does not impose conditional mean-zero restrictions, allowing for general forms of local misspecification.

Theorem \ref{thm: Identification} together with decomposition \eqref{eq: distance error decomposition} implies that
\begin{align*}
    \widehat{\vartheta}(\bx) - \tau(\bx)
    = \mathfrak{B}(\bx) + \mathfrak{L}(\bx) + \mathfrak{Q}(\bx),
    \qquad \bx\in\B,
\end{align*}
where
\begin{align*}
    \mathfrak{B}(\bx)
    &= \theta_{1,\bx}^{\ast}(0) - \theta_{0,\bx}^{\ast}(0) - \tau(\bx),\\
    \mathfrak{L}(\bx)
    &= \be_0^{\top} \bPsi_{1,\bx}^{-1} \bO_{1,\bx}
    - \be_0^{\top} \bPsi_{0,\bx}^{-1} \bO_{0,\bx},\\
    \mathfrak{Q}(\bx)
    &= \be_0^{\top} (\widehat{\bPsi}_{1,\bx}^{-1} - \bPsi_{1,\bx}^{-1}) \bO_{1,\bx}
    - \be_0^{\top} (\widehat{\bPsi}_{0,\bx}^{-1} - \bPsi_{0,\bx}^{-1}) \bO_{0,\bx}.
\end{align*}
Here, $\mathfrak{B}(\bx)$ denotes the nonrandom approximation bias, $\mathfrak{L}(\bx)$ is an unconditional mean-zero linear statistic, and $\mathfrak{Q}(\bx)$ captures higher-order linearization effects.

In conventional local polynomial settings, $\mathfrak{B}(\bx)$ is of order $h^{p+1}$, $\mathfrak{L}(\bx)$ is approximately Gaussian, and $\mathfrak{Q}(\bx)$ is asymptotically negligible. In the present setting, however, these standard results need not hold because estimation is conducted along the one-dimensional boundary $\B$ and relies on isotropic smoothing based on distance to each boundary point. We show that additional geometric and smoothness conditions are required to recover the usual approximation properties, and that these properties may fail in some empirically relevant configurations.

More specifically, the approximation bias $\mathfrak{B}(\bx)$ plays a central role in both pointwise and uniform inference for the BATEC. As emphasized in the RD literature \citep{Calonico-Cattaneo-Titiunik_2014_ECMA} and in kernel-based nonparametric inference more broadly \citep{Calonico-Cattaneo-Farrell_2018_JASA,Calonico-Cattaneo-Farrell_2022_Bernoulli}, valid large-sample inference requires a standardized ``small bias" condition of the form
\begin{align}\label{eq: small bias condition}
    \frac{\mathfrak{B}(\bx)}{\sqrt{\Var[\mathfrak{L}(\bx)]}}
    \lesssim \sqrt{nh^2} \mathfrak{B}(\bx)
    \to 0
\end{align}
for the bandwidth sequence employed. Because bandwidths are typically chosen to minimize mean squared error (MSE), this condition may fail in general, implying that inference cannot ignore first-order misspecification bias.

Robust bias correction addresses this issue by explicitly estimating and removing the leading bias term and adjusting the variance accordingly. A common implementation proceeds in two steps: first, an MSE-optimal bandwidth and corresponding MSE-optimal point estimator are constructed using a polynomial of order $p$; second, inference is conducted using a higher-order polynomial (typically $p+1$) while retaining the bandwidth from the first step. This approach yields valid confidence intervals when $\mathfrak{B}(\bx)\lesssim h^{p+1}$. While such bias behavior is standard under smoothness conditions in canonical nonparametric settings, we show in the next section that there exist configurations in which $\mathfrak{B}(\bx)\lesssim h$ regardless of the polynomial order. Consequently, approximation bias can substantially affect both estimation and robust bias-corrected inference when employing distance-based methods.

\section{Approximation Bias}\label{sec: Approximation Bias}

The smoothness of the induced distance-based conditional expectation function $r\mapsto\theta_{t,\bx}(r)=\E[Y_i(t) | D_i(\bx) = r]$ depends on the smoothness of the conditional expectation $\mu_t(\bx)=\E[Y_i(t) |\bX_i=\bx]$, the distance function $\d(\cdot,\cdot)$, and the geometric regularity of the boundary $\B$. Consequently, the bias of the distance-based local polynomial estimator may be affected by the shape of the boundary $\B$, regardless of the polynomial order $p$ imposed in Assumption \ref{assump: DGP}.

\begin{figure}
    \centering
    \begin{subfigure}[b]{0.45\textwidth}
        \centering
        \includegraphics[width=\linewidth]{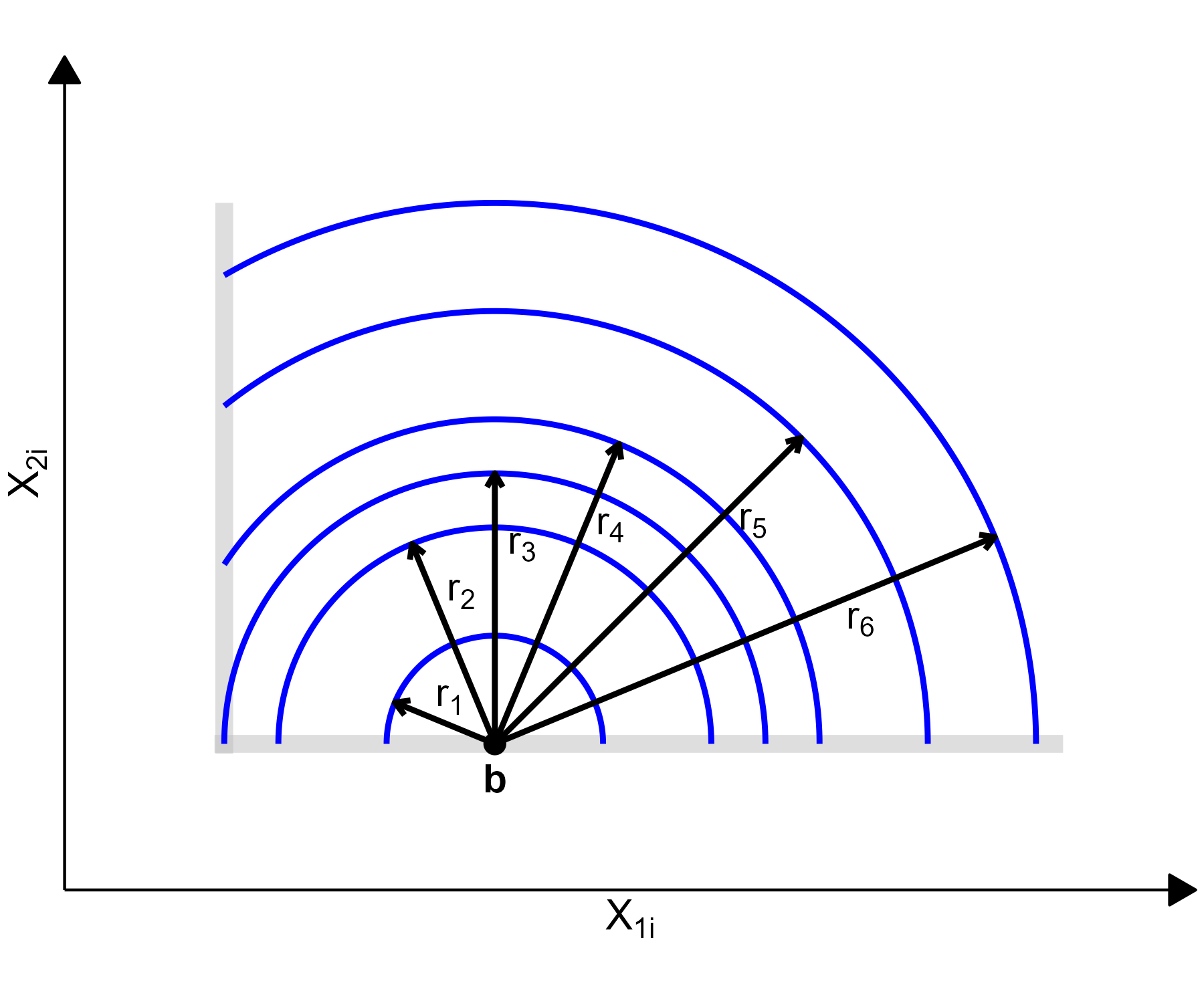}
        \caption{Distance to $\bb\in\B$.}
        \label{fig-bias-arcdist}
    \end{subfigure}
    \quad
    \begin{subfigure}[b]{0.45\textwidth}
        \centering
        \includegraphics[width=\linewidth]{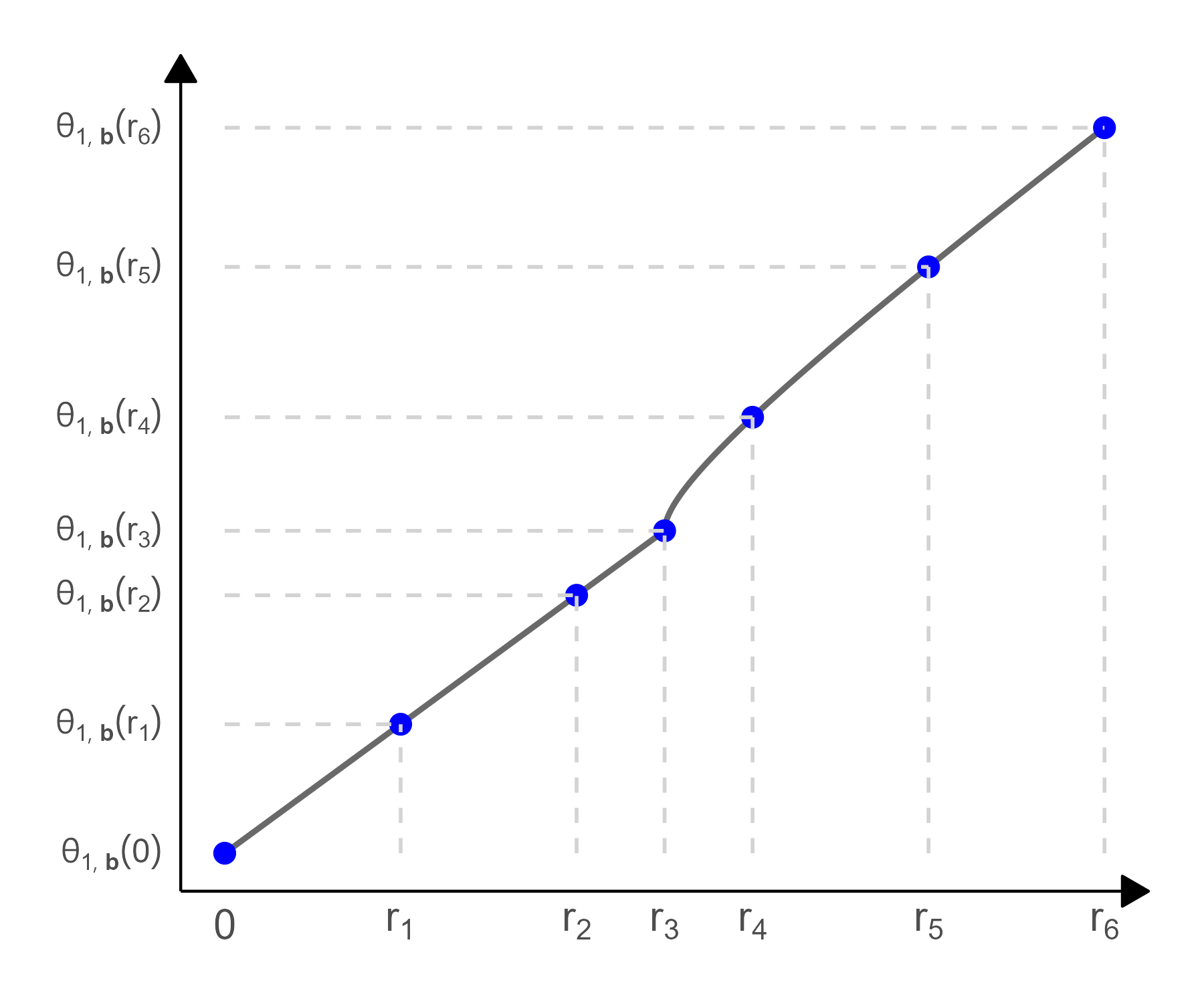}
        \caption{Distance-based Conditional Expectation.}
        \label{fig:fig-bias-inducedE}
    \end{subfigure}
    
    \caption{Lack of smoothness of distance-based conditional expectation near a kink.\\
    \footnotesize Note: Analytic example of $\theta_{1,\bb}(r) = \E[Y(1)|D_i(\bb)=r]$, $r\geq0$, for distance transformation $D_i(\bb)=\d(\bX_i, \bb) = \|\bX_i - \bb\|$ to point $\bb\in\B$ near a kink point on the boundary, based on location $\bX_i=(X_{1i},X_{2i})^\top$. The induced univariate conditional expectation $r\mapsto\theta_{1,\bb}(r)$ is continuous but not differentiable at $r=r_3$.}
    \label{fig:Induced 1d conditional expectation}
\end{figure}

Figure \ref{fig:Induced 1d conditional expectation} illustrates this issue graphically. For a point $\bx\in\B$ sufficiently close to a kink of the boundary, the conditional expectation $\theta_{1,\bx}(r)$ may fail to be differentiable for some $r\ge0$. The problem arises because, given the distance function $\d(\cdot,\cdot)$, a sufficiently small value of $r$ yields a complete arc $\{\bu\in\A_1: \d(\bu,\bx)=r\}$, whereas for larger values of $r$ this arc becomes truncated by the boundary. In the example depicted in Figure \ref{fig:Induced 1d conditional expectation}, the function $\theta_{1,\bx}(r)$ is smooth for $r<r_3$ and for $r>r_3$, but is not differentiable at $r=r_3$. In particular, the left derivative is finite while the right derivative diverges. Details of this analytic example are provided in the supplemental appendix.

Although the smoothness of the boundary $\B$ influences the regularity of $\theta_{t,\bx}(r)$, the locality of the distance-based estimator implies that the approximation error cannot be smaller than that of a local constant estimator, regardless of the polynomial order $p\ge0$ or the smoothness assumptions imposed on $\mu_t(\bx)$. The following theorem formalizes this observation and shows that the resulting bias order cannot be improved by increasing $p$.

\begin{thm}[Approximation Bias: Uniform Guarantee]\label{thm: Approximation Bias: Uniform Guarantee}
    For some $L \geq 1$, let $\mathcal{P}$ be the class of data generating processes satisfying Assumptions \ref{assump: DGP}(i)-(iii) with $\X\subseteq[-L,L]^2$, Assumption \ref{assump: Kernel, Distance, and Boundary}, and the following additional conditions, with (ii) and (iii) imposed for each $t\in\{0,1\}$:
    \begin{enumerate}[label=\normalfont(\roman*),noitemsep,leftmargin=*]
    \item $L^{-1} \leq \inf_{\bx \in \X} f_X(\bx) \leq \sup_{\bx \in \X} f_X(\bx) \leq L$,
    \item $\max_{0 \leq |\bnu| \leq p} \sup_{\bx \in \X}  |\partial^{\bnu} \mu_t(\bx)| + \max_{0 \leq |\bnu| \leq p} \sup_{\bx,\by \in \X} \frac{|\partial^{\bnu} \mu_t(\bx) - \partial^{\bnu}\mu_t(\by)|}{\|\bx - \by\|} \leq L$, and
    \item $\liminf_{h\downarrow0}\inf_{\bx \in \B} \int_{\A_t} K_h((2t-1)\d(\bu,\bx)) d \bu \geq L^{-1}$.
    \end{enumerate}
    For any $p\geq0$, if $nh^2\to\infty$ and $h\to0$, then
    \begin{align*}
         1 \lesssim \liminf_{n\to\infty} \sup_{\P\in\mathcal{P}} \sup_{\bx \in \B}\frac{|\mathfrak{B}(\bx)|}{h}
         \leq \limsup_{n\to\infty} \sup_{\P\in\mathcal{P}} \sup_{\bx \in \B}\frac{|\mathfrak{B}(\bx)|}{h}
         \lesssim 1.
    \end{align*}
\end{thm}

This result precisely characterizes the uniform (over $\B$ and the data generating process) approximation bias of the distance-based local polynomial estimator $\widehat{\vartheta}(\bx)$. The upper bound follows from the fact that, in general, $|\theta_{t,\bx}(0)-\theta_{t,\bx}(r)|\lesssim r$ for $t\in\{0,1\}$. The lower bound is established using the following example. Let $R_K<\infty$ satisfy $\supp(K)\subseteq[-R_K,R_K]$ and set $\bar R_K=\max\{1,R_K\}$. Suppose $\bX_i \sim \mathsf{Uniform}([-2\bar R_K,2\bar R_K]^2)$, $\mu_0(x_1, x_2) = 0$, $\mu_1(x_1,x_2) = x_2$ for all $(x_1,x_2) \in [-2\bar R_K,2\bar R_K]^2$, and $Y_i(0)|\bX_i \sim \mathsf{Normal}(\mu_0(\bX_i),1)$ and $Y_i(1)|\bX_i \sim \mathsf{Normal}(\mu_1(\bX_i),1)$. Let $\d(\cdot,\cdot)$ be the Euclidean distance, and define the treatment and control regions as $\A_1 = \{(x,y) \in [-2\bar R_K,2\bar R_K]^2: x \geq 0, y \geq 0\}$ and $\A_0 = [-2\bar R_K,2\bar R_K]^2 \setminus \A_1$. The resulting assignment boundary $\B = ([0,2\bar R_K]\times\{0\})\cup(\{0\}\times[0,2\bar R_K])$ is L-shaped, with a $90^\circ$ kink at $\bx=(0,0)$, similar to the empirical application considered in Section \ref{sec: Numerical Results}. This data generating process satisfies the conditions of Theorem \ref{thm: Approximation Bias: Uniform Guarantee}. The supplemental appendix establishes the lower bound through a detailed analysis of the induced approximation bias.

As a point of contrast, one might expect that $\mathfrak{B}(\bx)\lesssim h^{p+1}$ pointwise in $\bx\in\B$ for sufficiently small bandwidth $h$, provided that kinks on the boundary are sufficiently separated relative to the bandwidth and other regularity conditions hold. However, Theorem \ref{thm: Approximation Bias: Uniform Guarantee} shows that, regardless of sample size (and hence bandwidth), there always exists a neighborhood of a boundary kink where the misspecification bias of $\widehat{\vartheta}(\bx)$ is at best of order $h$, independently of the polynomial order $p$. This phenomenon arises because nonsmooth changes in the geometry of $\B$ induce nondifferentiability in $\theta_{t,\bx}(r)$, as illustrated in Figure \ref{fig:fig-bias-inducedE}.

When the boundary $\B$ is sufficiently smooth, sharper bias bounds can be obtained.

\begin{thm}[Approximation Bias: Smooth Boundary]\label{thm: Approximation Bias: Smooth Boundary}
    Suppose Assumption \ref{assump: DGP}(i)-(iii) and Assumptions \ref{assump: Kernel, Distance, and Boundary} hold, with $\d(\cdot,\cdot)$ the Euclidean distance. Let $h\to0$.
    
    \begin{enumerate}[label=\normalfont(\roman*),noitemsep,leftmargin=*]
        \item For $\bx \in \B$, and for some $\delta,\varepsilon > 0$, suppose that $\B \cap \{\by: \|\by - \bx\| \leq \varepsilon \} = \gamma([-\delta,\delta])$, where $\gamma: \mathbb{R} \to \mathbb{R}^2$ is a one-to-one function in $C^{\kappa + 2}([-\delta, \delta],\mathbb{R}^2)$. Let $m_{\bx}=\kappa\wedge (p+1)$. Then, for each $t\in\{0,1\}$, $r\mapsto\theta_{t,\bx}(r)$ is $m_{\bx}$-times continuously differentiable on $\I_t$ near zero. Therefore, there exists a positive constant $C_{\bx}$ such that $|\mathfrak{B}(\bx)| \leq C_{\bx} h^{m_{\bx}}$ for all sufficiently small $h$.
        
        \item Suppose $\B = \gamma([0,L])$ where $\gamma$ is a one-to-one function in $C^{\iota+2}([0,L],\mathbb{R}^2)$ for some $L > 0$.  Suppose there exists $\delta, \varepsilon > 0$ such that for all $\bx \in \B^o = \gamma([\delta,L-\delta])$, $r \in [0,\varepsilon]$, and $t = 0,1$, the set $\{\bu \in \mathbb{R}^2: \d(\bu, \bx) = r\}$ intersects $\mathtt{bd}(\mathcal{A}_t)$ at only two points in $\B$. Let $m=\iota \wedge (p+1)$. Then, for each $t\in\{0,1\}$, $r\mapsto\theta_{t,\bx}(r)$ is $m$-times continuously differentiable on $\I_t$ near zero with derivatives uniformly bounded over $\bx\in\B^o$. In particular, the one-sided limits at zero of the derivatives of order $0\leq v\leq m$ exist and are finite. Therefore, there exists a positive constant $C$ such that $\sup_{\bx \in \B^o}|\mathfrak{B}(\bx)| \leq C h^m$ for all sufficiently small $h$.
    \end{enumerate}
\end{thm}

This theorem provides sufficient smoothness conditions on the boundary $\B$ under which the distance-based estimator $\widehat{\vartheta}(\bx)$ attains the usual nonparametric smoothing bias rate, improving upon the minimal guarantee established in Theorem \ref{thm: Approximation Bias: Uniform Guarantee}. Achieving a uniform bias bound requires that the boundary be uniformly smooth in the sense that it admits a global smooth parameterization. This condition is essential because, as demonstrated by Theorem \ref{thm: Approximation Bias: Uniform Guarantee}, even a single kink in an otherwise smooth boundary can deteriorate the convergence rate of the misspecification bias; see Figure \ref{fig:Induced 1d conditional expectation}.

\section{Estimation and Inference}\label{sec: Estimation and Inference}

The results in this section are established for a general misspecification bias $\mathfrak{B}(\bx)$ of the distance-based estimator $\widehat{\vartheta}(\bx)$, which depends on the properties of the kernel and distance functions as well as on the geometry (smoothness) of the assignment boundary $\B$, as demonstrated by Theorems \ref{thm: Approximation Bias: Uniform Guarantee} and \ref{thm: Approximation Bias: Smooth Boundary}. We also discuss the implications of these results for empirical implementation and compare them with standard methods for univariate RD designs.

\subsection{Convergence Rates}\label{sec: Convergence Rates}

Using technical results established in the supplemental appendix, we obtain the following convergence rates for the univariate distance-based local polynomial treatment effect estimator.

\begin{thm}[Convergence Rates]\label{thm: Convergence Rates}
    Suppose Assumptions \ref{assump: DGP} and \ref{assump: Kernel, Distance, and Boundary} hold. If $n^{\frac{1+v}{2+v}} h^2/\log(1/h) \to \infty$ and $h\to0$, then
    \vspace{-.1in}
    \begin{enumerate}[label=\normalfont(\roman*),noitemsep,leftmargin=*]
        \item $|\widehat{\vartheta}(\bx) - \tau(\bx) |
        \lesssim_\P \frac{1}{\sqrt{n h^2}} + \frac{1}{n^{\frac{1+v}{2+v}}h^2} + |\mathfrak{B}(\bx)|$ for $\bx \in \B$, and
        
        \item $\sup_{\bx \in \B} |\widehat{\vartheta}(\bx) - \tau(\bx)|
        \lesssim_\P \sqrt{\frac{\log(1/h)}{ n h^2}} + \frac{\log(1/h)}{n^{\frac{1+v}{2+v}}h^2} + \sup_{\bx \in \B}|\mathfrak{B}(\bx)|$.
    \end{enumerate}
\end{thm}

This theorem establishes pointwise (for each $\bx\in\B$) and uniform (over $\B$) convergence rates for the distance-based treatment effect estimator. In particular, by Theorem \ref{thm: Approximation Bias: Uniform Guarantee}, $\widehat{\vartheta}(\bx)\to_\P\tau(\bx)$ both pointwise and uniformly under the stated bandwidth condition. Furthermore, by Theorem \ref{thm: Approximation Bias: Smooth Boundary}(i), we obtain the pointwise bound $|\mathfrak{B}(\bx)| \lesssim h^{p+1}$ for sufficiently small $h$ when the boundary $\B$ satisfies appropriate smoothness conditions. Similarly, Theorem \ref{thm: Approximation Bias: Smooth Boundary}(ii) shows that the uniform bias rate of $\widehat{\vartheta}(\bx)$ improves as the assignment boundary becomes smoother.

The ``variance" component of the distance-based estimator $\widehat{\vartheta}(\bx)$ is of order $(nh^{2})^{-1}$, even though the procedure is formally a univariate local polynomial estimator. A na\"ive analogy with standard univariate nonparametric regression would instead suggest a rate of order $(nh)^{-1}$. This difference is expected because the target estimand involves differences of bivariate regression functions, and the isotropic distance-based approach cannot circumvent the curse of dimensionality. See also Section \ref{sec: Distance-Based Minimax Convergence Rate} for further discussion.

Combining Theorems \ref{thm: Approximation Bias: Uniform Guarantee} and \ref{thm: Convergence Rates}(ii), and for an appropriate bandwidth choice, the estimator $\widehat{\vartheta}(\bx)$ can attain the uniform convergence rate $(n/\log n)^{-1/4}$. Section \ref{sec: Distance-Based Minimax Convergence Rate} shows that, over the class of rectifiable boundaries, no isotropic nonparametric estimator can achieve a uniform convergence rate faster than $n^{-1/4}$, which coincides with the minimax mean square rate for estimating Lipschitz continuous bivariate regression functions \citep[see, e.g.,][and references therein]{tsybakov2008introduction}. Consequently, the distance-based estimator $\widehat{\vartheta}(\bx)$ is minimax optimal (up to logarithmic factors) for suitable bandwidth sequences.

Finally, approximate pointwise and integrated (over $\B$) MSE criteria can be formulated from the same linearization, with leading variance and squared-bias terms matching the rates in Theorem \ref{thm: Convergence Rates}. Details are provided in the supplemental appendix.

\subsection{Uncertainty Quantification}\label{sec: Uncertainty Quantification}

To develop companion pointwise and uniform inference procedures along the treatment-assignment boundary $\B$, consider the feasible $t$-statistic for a given bandwidth choice and each boundary point $\bx\in\B$:
\begin{align*}
    \widehat{\T}(\bx) = \frac{\widehat{\vartheta}(\bx) - \tau(\bx)}{\sqrt{\widehat{\Xi}_{\bx,\bx}}},
\end{align*}
where, using standard least squares algebra, for all $\bx_1,\bx_2\in\B$ and $t\in\{0,1\}$,
\begin{align*}
    \widehat{\Xi}_{\bx_1,\bx_2}
    = \widehat{\Xi}_{0,\bx_1,\bx_2} + \widehat{\Xi}_{1,\bx_1,\bx_2},
    \qquad 
    \widehat{\Xi}_{t,\bx_1,\bx_2}
    = \frac{1}{nh^2}\be_0^\top\widehat{\bPsi}^{-1}_{t,\bx_1} \widehat{\bUpsilon}_{t,\bx_1,\bx_2} \widehat{\bPsi}^{-1}_{t,\bx_2}\be_0,
\end{align*}
and
\begin{align*}
    \widehat{\bUpsilon}_{t,\bx_1,\bx_2}
    &= h^2\En \Big[\br_p\Big(\frac{D_i(\bx_1)}{h}\Big) K_h(D_i(\bx_1)) \big(Y_i - \br_p(D_i(\bx_1)/h)^{\top}\widehat{\bgamma}_t(\bx_1)\big) \Indicator(D_i(\bx_1)\in\I_t)\\
    &\qquad\qquad \times
      \br_p\Big( \frac{D_i(\bx_2)}{h}\Big)^{\top} K_h(D_i(\bx_2)) \big(Y_i - \br_p(D_i(\bx_2)/h)^{\top}\widehat{\bgamma}_t(\bx_2)\big)
               \Indicator(D_i(\bx_2)\in\I_t)\Big].
\end{align*}
Accordingly, feasible confidence intervals and confidence bands over $\B$ take the form
\begin{align*}
    \widehat{\CI}_\alpha(\bx)
  = \left[\;\widehat{\vartheta}(\bx) - \q_{\alpha} \sqrt{\widehat{\Xi}_{\bx,\bx}} \; , \;
            \widehat{\vartheta}(\bx) + \q_{\alpha} \sqrt{\widehat{\Xi}_{\bx,\bx}}\;\right],
  \qquad \bx\in\B,
\end{align*}
for any $\alpha\in(0,1)$, where $\q_{\alpha}$ denotes the appropriate critical value depending on the inference procedure.

For pointwise inference, we show in the supplemental appendix that $\sup_{t\in\mathbb{R}} |\P[\widehat{\T}(\bx) \leq t] - \Phi(t)| \to 0$ for each $\bx\in\B$, under standard regularity conditions and provided that $nh^{2}\mathfrak{B}^{2}(\bx)\to0$. In this case, $\q_{\alpha}=\Phi^{-1}(1-\alpha/2)$ is asymptotically valid.

Uniform inference over $\B$ is more challenging. The stochastic process $(\widehat{\T}(\bx):\bx\in\B)$ is generally not asymptotically tight and therefore does not converge weakly in the space of uniformly bounded real-valued functions on $\B$ equipped with the supremum norm \citep{van-der-Vaart-Wellner_1996_Book,Gine-Nickl_2016_Book}. Consequently, classical empirical process arguments based on weak convergence cannot be applied directly. Moreover, the geometry of the manifold $\B$ and the reliance on signed distance scores introduce additional technical complications. We address these issues through Gaussian approximations for the supremum of $(\widehat{\T}(\bx):\bx\in\B)$, with a stronger process-level coupling provided in the supplemental appendix. These approximations allow uniform inference based on critical values $\q_{\alpha}$ satisfying
\begin{align*}
    \P \big[\tau(\bx) \in \widehat{\CI}_\alpha(\bx) \; , \text{ for all } \bx \in \B \big]
    = \P \Big[ \sup_{\bx\in\B} \big| \widehat{\T}(\bx) \big| \leq  \q_{\alpha} \Big].
\end{align*}
Our technical analysis combines a process-level strong approximation (Theorem SA-7 in the supplemental appendix), the corresponding confidence-band result (Theorem SA-8), and a direct supremum approximation (Theorem SA-9) with recent results from \cite{Chernozhukov-Chetverikov-Kato_2014a_AoS,Chernozhukov-Chetverikov-Kato_2014b_AoS} and \cite{Chernozhukov-Chetverikov-Kato-Koike_2022_AOS}, thereby improving upon classical coupling approaches such as Yurinskii's coupling \citep[see also][]{Cattaneo-Masini-Underwood_2025_AOS}. Let $\U_n$ denote the $\sigma$-algebra generated by $\big((Y_i,(D_i(\bx):\bx\in\B)):1\le i\le n\big)$.

\begin{thm}[Statistical Inference]\label{thm: Statistical Inference}
    Suppose Assumptions \ref{assump: DGP} and \ref{assump: Kernel, Distance, and Boundary} hold, and let $h\to0$.
    \begin{enumerate}[label=\normalfont(\roman*),noitemsep,leftmargin=*]
        \item For all $\bx \in \B$, if $n^{\frac{v}{2+v}} h^2 \to \infty$ and $n h^2 \mathfrak{B}^2(\bx) \to 0$, then 
        \begin{align*}
            \P \big[\tau(\bx) \in \widehat{\CI}_\alpha(\bx) \big] \to  1 - \alpha,
        \end{align*}
        for $\q_{\alpha} = \Phi^{-1}(1-\alpha/2)$.
        
        \item If $\liminf_{n\to\infty} \frac{\log h}{\log n} > -\infty$, $\frac{n^{\frac{v}{2+v}} h^2}{(\log n)^3} \to \infty$, $n h^2\log n \sup_{\bx \in \B} \mathfrak{B}^2(\bx) \to 0$,
        and $\perim(\{\by \in \A_t: (2t-1)\d(\by,\bx)/h \in \supp(K)\}) \lesssim h$ for all $\bx \in \B$ and $t \in \{0,1\}$, then
        \begin{align*}
            \P \big[\tau(\bx) \in \widehat{\CI}_\alpha(\bx), \text{ for all } \bx \in \B \big] \to 1 - \alpha,
        \end{align*}
        for $\q_{\alpha} = \inf \{c > 0: \P[\sup_{\bx \in \B} |\widehat{Z}_n(\bx)| \geq c | \U_n] \leq \alpha\}$, where $(\widehat{Z}_n: \bx \in \B)$ is a Gaussian process conditional on $\U_n$ satisfying $\E[\widehat{Z}_n(\bx_1)|\U_n]=0$ and
        $\E[\widehat{Z}_n(\bx_1) \widehat{Z}_n(\bx_2) | \U_n]
        = \widehat{\Xi}_{\bx_1,\bx_2}/\sqrt{\widehat{\Xi}_{\bx_1,\bx_1}\widehat{\Xi}_{\bx_2,\bx_2}}$ for all $\bx_1, \bx_2 \in \B$.
    \end{enumerate}
\end{thm}

This theorem establishes asymptotically valid inference procedures based on the distance-based local polynomial treatment-effect estimator $\widehat{\vartheta}(\bx)$. For uniform inference, an additional geometric restriction on the assignment boundary $\B$ is required. Specifically, the De Giorgi perimeter condition can be verified when the boundary of the set $\{\by\in\A_t:(2t-1)\d(\by,\bx)/h\in\supp(K)\}$ has length of order $h$. Because this set is contained in a constant multiple of an $h$-ball centered at $\bx$ when $K$ is compactly supported, the condition holds provided that the local segment $\B\cap\{\by\in\mathbb{R}^2:\d(\by,\bx)\lesssim h\}$ is not excessively irregular. Intuitively, the one-dimensional boundary $\B$ must not be locally ``too long" or highly oscillatory.

\subsection{Discussion and Implementation}

Although the distance-based estimator $\widehat{\vartheta}(\bx)$ resembles a univariate local polynomial procedure based on the scalar running variable $D_i(\bx)$, Theorem \ref{thm: Convergence Rates} shows that its pointwise and uniform variance convergence rates coincide with those of a bivariate nonparametric estimator and are therefore unimprovable. Theorem \ref{thm: Statistical Inference} further shows that inference procedures developed for univariate local polynomial regression can be directly applied in distance-based settings, provided the required side conditions are satisfied. This result follows because $\widehat{\T}(\bx)$ is constructed as a self-normalizing statistic and is therefore \textit{adaptive} to the fact that the scalar running variable $D_i(\bx)$ is induced by the underlying bivariate covariate $\bX_i$. This finding highlights another advantage of pre-asymptotic variance estimation and self-normalization techniques for distributional approximation and inference \citep{Calonico-Cattaneo-Farrell_2018_JASA}. Consequently, standard estimation and inference methods from the univariate RD design literature \citep{Calonico-Cattaneo-Titiunik_2014_ECMA} can be employed in BD designs based on distance-to-boundary running variables, provided that bandwidth choices ensure that the bias induced by the geometry of the assignment boundary $\B$ (documented in Section \ref{sec: Approximation Bias}) is sufficiently small.

For implementation, first consider the case in which $\mathfrak{B}(\bx) \lesssim h^{p+1}$, that is, the assignment boundary $\B$ is smooth (in the sense of Theorem \ref{thm: Approximation Bias: Smooth Boundary}) or the bias is otherwise negligible. Establishing an exact MSE expansion for $\widehat{\vartheta}(\bx)$ is cumbersome due to the additional complexity introduced by the distance transformation, but the relevant convergence rates follow from Theorem \ref{thm: Convergence Rates}. The \textit{incorrect} univariate MSE-optimal bandwidth is $h_{\mathtt{1d},\bx}\asymp n^{-1/(3+2p)}$, whereas the \textit{correct} MSE-optimal bandwidth is $h_{\mathtt{MSE},\bx}\asymp n^{-1/(4+2p)}$. Treating the distance variable as an intrinsically univariate score therefore leads to a smaller-than-optimal bandwidth choice because $n^{-1/(3+2p)}<n^{-1/(4+2p)}$, resulting in undersmoothing relative to the correct MSE-optimal bandwidth. As a consequence, the point estimator exhibits higher variance and lower bias, and associated inference procedures tend to be conservative. A simple remedy is to rescale the incorrect univariate bandwidth by the factor $n^{1/(3+2p)-1/(4+2p)}$, although this adjustment may be unnecessary when empirical bandwidth selection employs pre-asymptotic variance estimation, as implemented in the software package \texttt{rdrobust} (\url{https://rdpackages.github.io/rdrobust/}); see \cite{Calonico-Cattaneo-Farrell_2020_ECTJ} and references therein.

When the assignment boundary $\B$ exhibits kinks or other geometric irregularities, there always exists a neighborhood around each kink where the bias is irreducibly of order $h$, regardless of sample size or polynomial order, as established by Theorem \ref{thm: Approximation Bias: Uniform Guarantee}. In this case, the correct local MSE-optimal bandwidth near each kink is $h\asymp n^{-1/4}$. Away from kink points, the MSE-optimal bandwidth behaves as described above. This lack of smoothness leads to spatially varying optimal bandwidths along $\B$, complicating automatic implementation. A simple global strategy is therefore to set $h=\widehat{\mathtt{C}}\cdot n^{-1/4}$ for all $\bx\in\B$, where $\widehat{\mathtt{C}}$ is a rule-of-thumb constant. Although generically suboptimal, this choice is never larger than the pointwise MSE-optimal bandwidth and therefore yields a more variable (but conservative) estimator and associated inference procedure. A more adaptive alternative is to allow bandwidths to vary with proximity to kink points, for example by defining
\begin{align*}
    \widehat{h}_{\mathtt{kink},\bx}(\B)
    = \min\Big\{ \widehat{h}_{\mathtt{MSE},\bx} ,
                 \max\big\{ \widehat{\mathtt{C}} \cdot n^{-1/4} , \min_{\bb\in\B_\mathtt{kink}} \d(\bx, \bb) \big\}
          \Big\},
\end{align*}
for all $\bx\in\B$, where $\B_{\mathtt{kink}}$ denotes the set of kink points on the boundary.

Given a bandwidth choice, valid statistical inference can be achieved by appropriately controlling the remaining misspecification bias. When the boundary $\B$ is smooth, robust bias-correction methods developed for univariate RD designs remain applicable in the distance-based setting \citep{Calonico-Cattaneo-Titiunik_2014_ECMA,Calonico-Cattaneo-Farrell_2018_JASA,Calonico-Cattaneo-Farrell_2022_Bernoulli}. When $\B$ contains kinks, however, undersmoothing relative to the MSE-optimal bandwidth for $p=0$ becomes necessary because increasing the polynomial order does not guarantee a reduction in misspecification bias, rendering bias-correction techniques ineffective uniformly over $\bx\in\B$. In such cases, bandwidth and polynomial order choices should account for proximity to boundary irregularities.

The companion software package \texttt{rd2d} implements a rule-of-thumb bandwidth selector targeting $h\asymp n^{-1/4}$ as the default choice for all $\bx\in\B$ when the smoothness of $\B$ is unknown. This choice is valid for point estimation regardless of whether the boundary is smooth. For inference, following \cite{Calonico-Cattaneo-Farrell_2018_JASA}, the package employs an undersmoothed bandwidth of order $n^{-1/3}$ to target coverage-error optimality. When $\B$ is known to be smooth, the package instead uses a rule-of-thumb selector targeting $h\asymp n^{-1/(4+2p)}$ for point estimation and implements robust bias-corrected inference based on that bandwidth. If the locations of kink points are known, the adaptive selector $\widehat{h}_{\mathtt{kink},\bx}(\B)$ can also be employed. Section \ref{sec: Numerical Results} illustrates the performance of these approaches and compares them with simple rule-of-thumb bandwidth choices obtained using the \texttt{rdrobust} package with $p=0$. Further implementation details and simulation evidence are provided in \citet{Cattaneo-Titiunik-Yu_2025_rd2d}.

Finally, uniform inference based on Theorem \ref{thm: Statistical Inference}(ii) is implemented by discretizing the boundary at evaluation points $\bb_{1},\ldots,\bb_{M}\in\B$. The conditional Gaussian process $(\widehat{Z}_{n}(\bx):\bx\in\B)$ is then approximated by the $M$-dimensional (conditional) Gaussian vector $\widehat{\bZ}_n = (\widehat{Z}_n(\bb_1),\ldots,\widehat{Z}_n(\bb_M))$ whose covariance matrix has generic element $\E[\widehat{Z}_n(\bb_1) \widehat{Z}_n(\bb_2) | \U_n]$. In finite samples this estimated covariance matrix may fail to be positive definite, in which case simple regularization can be applied \citep[see, e.g.,][]{Cattaneo-Feng-Underwood_2024_JASA}. Critical values $\q_{\alpha}$ are then obtained by simulating the distribution of $\max_{1\le l\le M}|\widehat{Z}_n(\bb_l)|$.

\section{Distance-based Minimax Convergence Rate}\label{sec: Distance-Based Minimax Convergence Rate}

Theorems \ref{thm: Approximation Bias: Uniform Guarantee} and \ref{thm: Approximation Bias: Smooth Boundary} provide precise conditions characterizing the pointwise and uniform (over $\B$) bias and convergence rates of the distance-based local polynomial estimator $\widehat{\vartheta}(\bx)$. In particular, the estimator can attain at best the uniform convergence rate $(n/\log n)^{-1/4}$ for an appropriate bandwidth sequence $h$. A natural follow-up theoretical (and potentially practical) question is whether alternative distance-based estimators could achieve faster rates. This section provides one answer: if the boundary $\B$ is rectifiable, then no isotropic nonparametric estimator of a bivariate regression function can improve upon this rate, up to polylogarithmic factors, regardless of the degree of smoothness imposed on the underlying conditional expectation function.

This section is largely self-contained, as it concerns minimax theory for nonparametric estimation on curves \citep[see, e.g.,][and references therein]{tsybakov2008introduction}. The following theorem establishes a minimax lower bound convergence rate for estimation of compactly supported bivariate regression functions using a class of isotropic nonparametric estimators.

\begin{thm}[Distance-based Minimax Convergence Rate]\label{thm: minimax for distance based}
    For constants $q\geq1$ and $L \geq 1$, let $\mathcal{P}_{\mathtt{NP}}=\mathcal{P}_{\mathtt{NP}}(L,q)$ be the class of (joint) probability laws $\P$ of $(Y_1,\bX_1),\cdots,(Y_n,\bX_n)$ satisfying the following:
    \begin{enumerate}[label=\normalfont(\roman*),noitemsep,leftmargin=*]
        \item $((Y_i,\bX_i):1\leq i \leq n)$ are i.i.d. taking values in $\mathbb{R} \times \mathbb{R}^2$.
        \item $\bX_i$ admits a Lebesgue density $f$ that is continuous on its compact support $\X \subseteq [-L,L]^2$, with $L^{-1} \leq \inf_{\bx \in \X} f(\bx) \leq \sup_{\bx \in \X} f(\bx) \leq L$, and $\B=\mathtt{bd}(\X)$ is a rectifiable curve.
        \item $\mu(\bx) = \E[Y_i|\bX_i = \bx]$ belongs to a H\"older ball of smoothness $q$ on $\X$ with 
        \begin{align*}
            \max_{0 \leq |\bnu| \leq \lfloor q \rfloor} \sup_{\bx \in \X}  |\partial^{\bnu} \mu(\bx)| + \max_{|\bnu| = \lfloor q \rfloor} \sup_{\bx_1,\bx_2 \in \X} \frac{|\partial^{\bnu} \mu(\bx_1) - \partial^{\bnu}\mu(\bx_2)|}{\|\bx_1 - \bx_2\|^{q - \lfloor q \rfloor}} \leq L.
        \end{align*}
        \item $\sigma^2(\bx) = \Var[Y_i|\bX_i = \bx]$ is continuous on $\X$ with $L^{-1} \leq \inf_{\bx \in \X} \sigma^2(\bx) \leq \sup_{\bx \in \X} \sigma^2(\bx) \leq L$.
    \end{enumerate}
    In addition, let $\mathcal{T}$ be the class of all distance-based estimators $T_n(\bU_n(\bx))$ with $\bU_n(\bx) = ((Y_i,\|\bX_i - \bx\|)^\top : 1\leq i \leq n)$ for each  $\bx \in \X$. Then,
    \begin{align*}
        \liminf_{n\to\infty} n^{1/4}
        \inf_{T_n \in \mathcal{T}} \sup_{\P \in \mathcal{P}_{\mathtt{NP}}}
        \E_{\P} \Big[\sup_{\bx \in \B} \big|T_n(\bU_n(\bx)) - \mu(\bx)\big| \Big] \gtrsim 1,
    \end{align*}
    where $\E_{\P}[\cdot]$ denotes an expectation taken under the data generating process $\P$.
\end{thm}

In sharp contrast, under the same assumptions the uniform minimax convergence rate over the unrestricted class of nonparametric estimators is
\begin{align*}
    \liminf_{n\to\infty} \Big(\frac{n}{\log n}\Big)^{\frac{q}{2q+2}}
    \inf_{S_n \in \mathcal{S}} \sup_{\P \in \mathcal{P}_{\mathtt{NP}}}
    \E_{\P} \Big[\sup_{\bx \in \B} \big|S_n(\bx;\bV_n) - \mu(\bx)\big| \Big] \gtrsim 1, 
\end{align*}
where $\mathcal{S}$ denotes the unrestricted class of estimators based on $\bV_n=((Y_i,\bX_i^\top)^\top:1\le i\le n)$. Therefore, Theorem \ref{thm: minimax for distance based} shows that restricting attention to estimators that use only the scalar distance $\|\bX_i-\bx\|$ (rather than the full covariate $\bX_i$) necessarily limits the attainable uniform estimation accuracy along the boundary. In particular, for any such estimator there exists a data-generating process for which the maximal estimation error along $\B$ is of order at least $n^{-1/4}$ when the boundary may contain countably many kinks. Notably, increasing the smoothness of the underlying regression function $\mu(\bx)$ does not improve the achievable uniform convergence rate within the class $\mathcal{T}$, because nonsmooth boundary geometry induces effective nondifferentiability in the induced regression problem.

Finally, these results can be connected to the analysis of BD designs. For expositional clarity, consider estimation of a single regression function rather than a treatment-effect difference. By Theorems \ref{thm: Convergence Rates} and \ref{thm: Approximation Bias: Uniform Guarantee}, the distance-based $p$-th order local polynomial estimator
\begin{align*}
    \widehat{\mu}(\bx) = \be_0^\top \widehat{\bgamma}(\bx),
    \qquad
    \widehat{\bgamma}(\bx)
    = \argmin_{\bgamma \in \mathbb{R}^{p+1}} \En\Big[ \big(Y_i - \br_p(D_i(\bx)/h)^{\top} \bgamma \big)^2 K_h(D_i(\bx)) \Big], \qquad \bx \in \B,
\end{align*}
satisfies
\begin{align*}
    \limsup_{M \to \infty} \limsup_{n \to \infty}\sup_{\P \in \mathcal{P}}
    \P \Big[\Big(\frac{n}{\log n}\Big)^{1/4} \sup_{\bx \in \B} \big|\widehat{\mu}(\bx) - \mu(\bx)\big| \geq  M \Big] = 0.
\end{align*}
Since $\widehat{\mu}(\bx)\in\mathcal{T}$, the distance-based local polynomial estimator is minimax optimal in the sense of Theorem \ref{thm: minimax for distance based}, up to the logarithmic factor $\log^{1/4}n$, which we conjecture to be unimprovable. More precisely, this factor arises from the use of the uniform norm, whereas the lower bound in Theorem \ref{thm: minimax for distance based} is derived using pointwise minimax arguments.

\section{Numerical Results}\label{sec: Numerical Results}

We study the numerical performance of the distance-based methods using the real-world dataset analyzed by \cite{LondonoVelezRodriguezSanchez_2020_AEJ}, which evaluates the Colombian post-secondary education subsidy program \textit{Ser Pilo Paga} (SPP). This anti-poverty policy provided tuition support to undergraduate students admitted to high-quality, government-certified higher education institutions. Eligibility for SPP was determined by a deterministic bivariate cutoff combining academic merit and economic need: students were required to obtain a SABER 11 high school exit exam score in the top $9$ percent and to have a SISBEN wealth index below a region-specific threshold. The resulting treatment assignment rule induces an L-shaped boundary $\B$ with a kink at the lowest joint levels of eligibility in both score dimensions (and two additional kinks at the limits of the score support), while remaining linear elsewhere.

The dataset contains $n=363,096$ complete observations for the first program cohort (2014). Each observation $i=1,\ldots,n$ corresponds to an individual student with bivariate score $\bX_i = (X_{1i},X_{2i})^\top = (\mathtt{SABER11}_i,\mathtt{SISBEN}_i)^\top$, where the SABER11 test score ranges from $-310$ to $172$ and the SISBEN wealth index ranges from $-103.41$ to $127.21$. Without loss of generality, both score components are recentered at their respective eligibility cutoffs and standardized to facilitate the use of a common distance function and scalar bandwidth parameter. Under this normalization, the treatment assignment boundary can be written as $\B=\{(\texttt{SABER11},\texttt{SISBEN}): (\texttt{SABER11}\ge 0,\ \texttt{SISBEN}=0)\}\ \cup\ \{(\texttt{SABER11}=0,\ \texttt{SISBEN}\ge 0)\}$. Because the support of $\X$ is compact, the boundary $\B$ contains three kinks: the interior point $(0,0)$, which is of primary interest in our analysis, and two additional points where $\B$ intersects the boundary of the support of $\X$. Our empirical analysis considers $21$ evenly spaced cutoff points $\{\bb_1,\ldots,\bb_{21}\}\subset\B$, located near the interior kink and away from the boundary of $\X$, in order to highlight the impact of a single kink on the bias of the distance-based estimator. Throughout, we employ the Euclidean distance, $\d(\bX_i,\bx) = \| \bX_i - \bx \|$, to construct the signed score $D_i(\bx)$ for each $\bx\in\B$.

\begin{figure}
    \centering
    \begin{subfigure}[b]{0.45\textwidth}
        \centering
        \includegraphics[width=\linewidth]{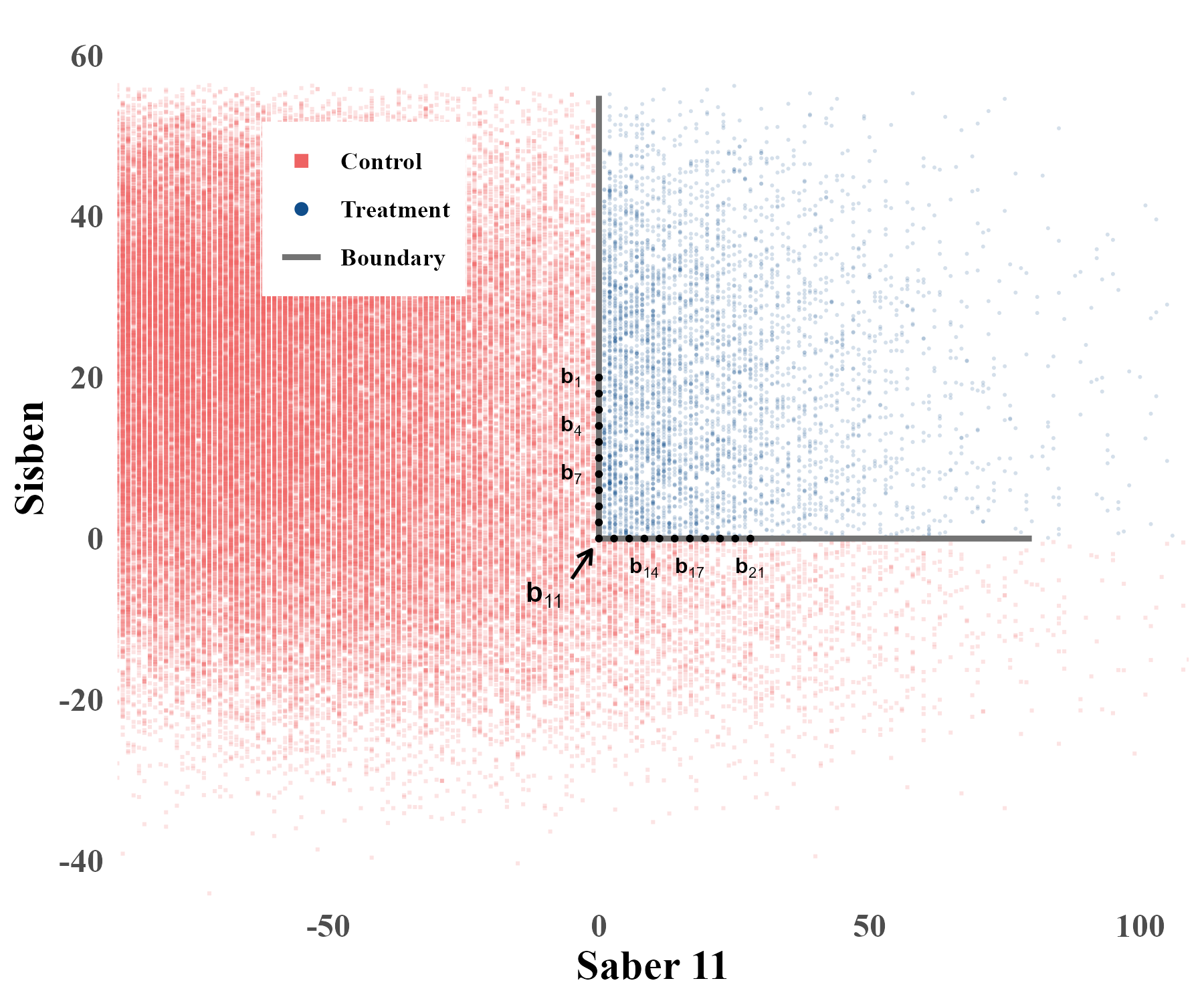}
        \caption{Scatter plot score and grid points.}
    \end{subfigure}
    \quad
    \begin{subfigure}[b]{0.45\textwidth}
        \centering
        \includegraphics[width=\linewidth]{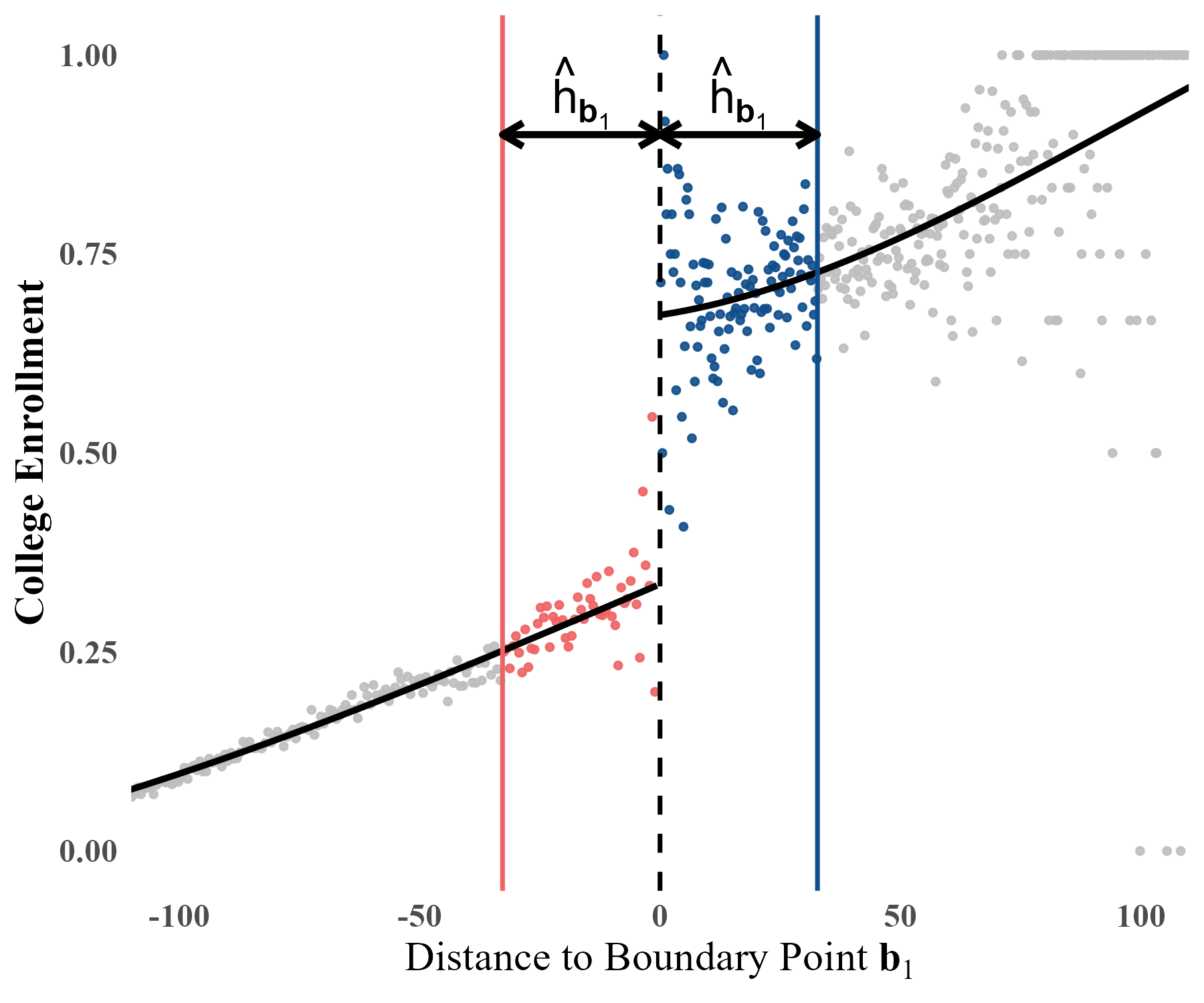}
        \caption{RD Plot based on $D_i(\bb_{1})$.}
    \end{subfigure}
    
    \begin{subfigure}[b]{0.45\textwidth}
        \centering
        \includegraphics[width=\linewidth]{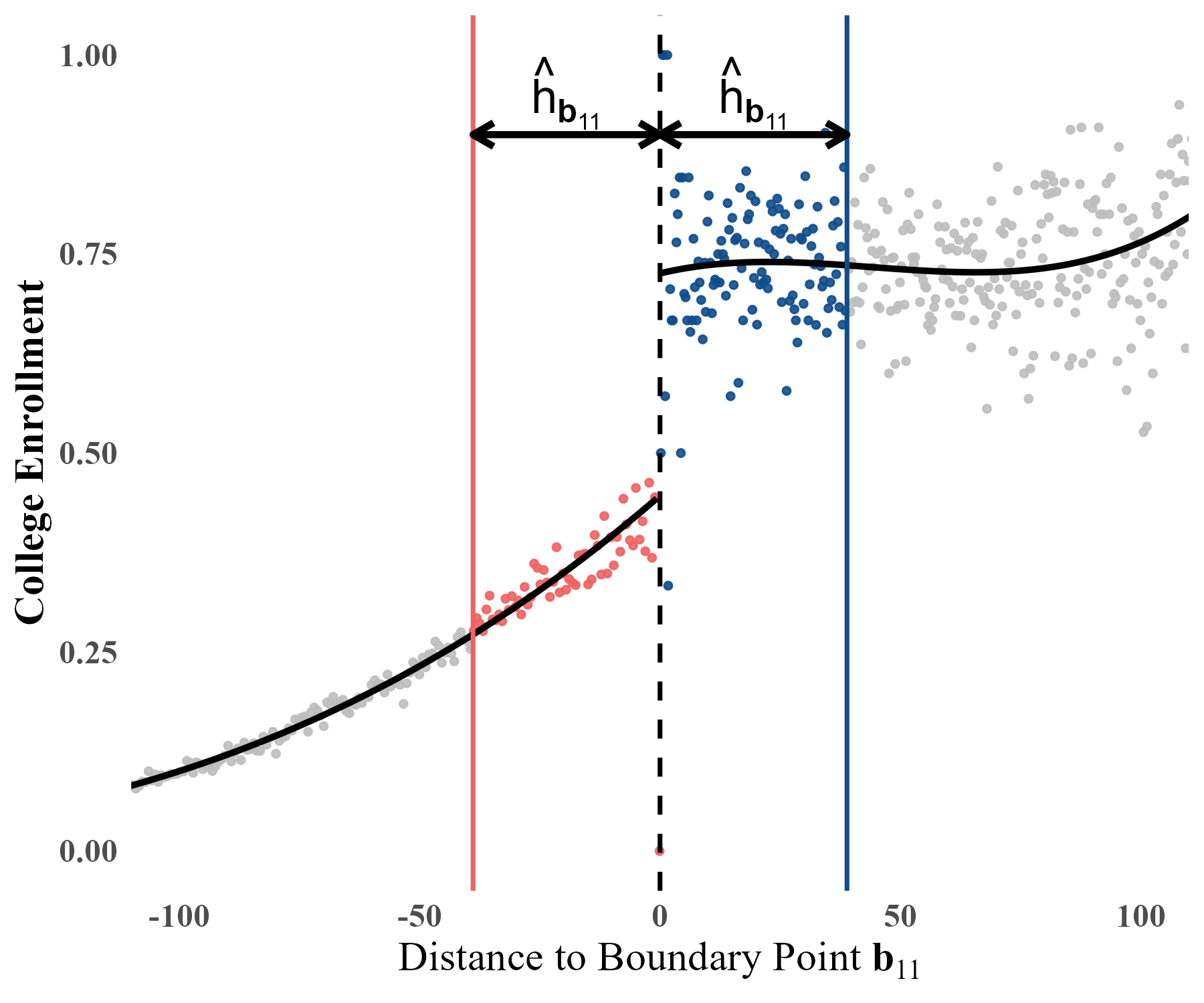}
        \caption{RD Plot based on $D_i(\bb_{11})$.}
    \end{subfigure}
    \quad
    \begin{subfigure}[b]{0.45\textwidth}
        \centering
        \includegraphics[width=\linewidth]{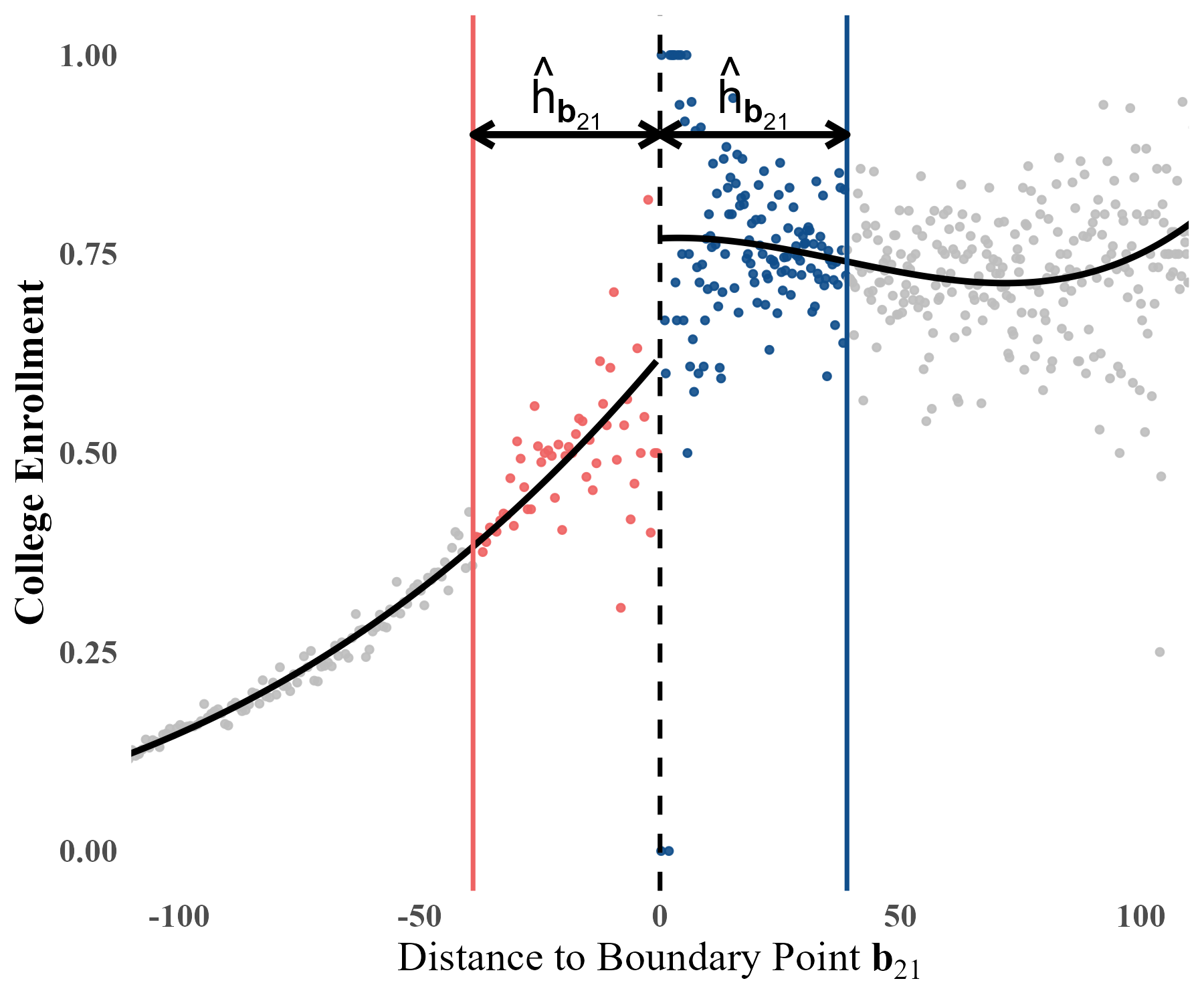}
        \caption{RD Plot based on $D_i(\bb_{21})$.}
    \end{subfigure}

    \caption{Scatter Plot and Selected Distance-Based RD Plots (SPP Data)}
    \label{fig:SPP}
\end{figure}

We analyze both simulated data calibrated to match salient features of the SPP dataset and the SPP dataset itself as an empirical application. All results reported in this section are implemented using our companion software package \texttt{rd2d}, together with the package \texttt{rdrobust} for RD designs with a univariate score. Additional implementation details are provided in the replication materials available at \url{https://rdpackages.github.io/}. Figure \ref{fig:SPP} summarizes key features of the SPP data. Panel (a) displays a scatterplot of the bivariate score, the treatment assignment boundary, and the $21$ evaluation points. Panels (b)--(d) report representative RD plots based on the induced signed distance scores $D_i(\bb_{1})$, $D_i(\bb_{11})$, and $D_i(\bb_{21})$, respectively.

\subsection{Simulation Study}\label{sec: Small Simulation Study}

The score $\bX_i = (X_{1i}, X_{2i})^\top$ is generated from the distribution $(100\cdot\mathsf{Beta}(3,4)-25,\;100\cdot\mathsf{Beta}(3,4)-25)^\top$ with independent components, which approximately reproduces salient features of the SPP dataset. We employ the same L-shaped treatment assignment boundary as in the empirical application, featuring a kink at $\bx=(0,0)^\top$, corresponding to the evaluation grid point $\bb_{11}$. 

For each $t\in\{0,1\}$ and $i=1,\ldots,n$, potential outcomes are generated according to the regression model $Y_{i}(t) = \mu_t(\bX_i) + \sigma_t(\bX_i) u_{t,i}$, where $\bX_i$, $u_{0,i}$ and $u_{1,i}$ are mutually independent, and
\begin{align*}
    \mu_t(\bX_i)
    &= \beta_{t,0} + X_{1i}\beta_{t,11} + X_{2i}\beta_{t,12} + X_{1i}^2\beta_{t,21} + X_{1i}X_{2i}\beta_{t,22} + X_{2i}^2\beta_{t,23},\\
    \sigma_t^2(\bX_i)
    &= \exp\!\big( \delta_{t,0} + X_{1i}\delta_{t,11} + X_{2i}\delta_{t,12} + X_{1i}^2\delta_{t,21} + X_{1i}X_{2i}\delta_{t,22} + X_{2i}^2\delta_{t,23} \big),
\end{align*}
with $u_{t,i}\sim\mathsf{Normal}(0,1)$. The coefficient values are calibrated using estimates obtained from the SPP dataset. We consider four data-generating processes combining linear (i.e., $\beta_{t,21}=\beta_{t,22}=\beta_{t,23}=0$) or quadratic specifications with homoskedastic (i.e., $\delta_{t,11}=\delta_{t,12}=\delta_{t,21}=\delta_{t,22}=\delta_{t,23}=0$) or heteroskedastic disturbances. Table \ref{tab:true-params} reports the corresponding parameter values.

\begin{table}[ht]
    \centering
    \begin{subtable}{0.45\textwidth}
        \centering
        \resizebox{\linewidth}{!}{\begin{tabular}{@{}rrrrr@{}}
\hline\hline
    & \multicolumn{2}{c}{Linear} & \multicolumn{2}{c}{Quadratic} \\ \cmidrule(lr){2-3} \cmidrule(lr){4-5}
    & \multicolumn{1}{c}{$t=0$} & \multicolumn{1}{c}{$t=1$} & \multicolumn{1}{c}{$t=0$} & \multicolumn{1}{c}{$t=1$} \\ \midrule
$\beta_{Y,t,0}$ & $3.35 \times 10^{-1}$ & $6.98 \times 10^{-1}$ & $3.72 \times 10^{-1}$ & $7.44 \times 10^{-1}$ \\
$\beta_{Y,t,11}$ & $2.52 \times 10^{-3}$ & $2.74 \times 10^{-3}$ & $4.23 \times 10^{-3}$ & $2.29 \times 10^{-3}$ \\
$\beta_{Y,t,12}$ & $-1.27 \times 10^{-3}$ & $-6.05 \times 10^{-4}$ & $-2.45 \times 10^{-3}$ & $-5.84 \times 10^{-3}$ \\
$\beta_{Y,t,21}$ & $0$ & $0$ & $1.25 \times 10^{-5}$ & $-1.33 \times 10^{-7}$ \\
$\beta_{Y,t,22}$ & $0$ & $0$ & $-4.92 \times 10^{-6}$ & $2.14 \times 10^{-5}$ \\
$\beta_{Y,t,23}$ & $0$ & $0$ & $3.12 \times 10^{-5}$ & $1.04 \times 10^{-4}$ \\
\hline\hline
\end{tabular}}
        \caption{Population parameters: $\mu_t(\bX_i)$}
    \end{subtable}
    \hspace{.1in}
    \begin{subtable}{0.45\textwidth}
        \centering
        \resizebox{0.79\linewidth}{!}{\begin{tabular}{@{}rrrrr@{}}
\hline\hline
    & \multicolumn{2}{c}{Homoskedastic} & \multicolumn{2}{c}{Heteroskedastic} \\ \cmidrule(lr){2-3} \cmidrule(lr){4-5}
    & \multicolumn{1}{c}{$t=0$} & \multicolumn{1}{c}{$t=1$} & \multicolumn{1}{c}{$t=0$} & \multicolumn{1}{c}{$t=1$} \\ \midrule
$\delta_{t,0}$ & $-2.20$ & $-2.22$ & $-1.57$ & $-5.00$ \\
$\delta_{t,11}$ & $0$ & $0$ & $2.19 \times 10^{-2}$ & $-9.03 \times 10^{-2}$ \\
$\delta_{t,12}$ & $0$ & $0$ & $-5.08 \times 10^{-3}$ & $1.11 \times 10^{-1}$ \\
$\delta_{t,21}$ & $0$ & $0$ & $-1.15 \times 10^{-4}$ & $1.05 \times 10^{-3}$ \\
$\delta_{t,22}$ & $0$ & $0$ & $5.23 \times 10^{-4}$ & $-7.46 \times 10^{-4}$ \\
$\delta_{t,23}$ & $0$ & $0$ & $6.50 \times 10^{-4}$ & $-1.67 \times 10^{-3}$ \\
\hline\hline
\end{tabular}}
        \caption{Population parameters: $\sigma^2_t(\bX_i)$}
    \end{subtable}
    \caption{Parameters calibrated using the SPP dataset.}
    \label{tab:true-params}
\end{table}

The simulation design sets $n=20,000$ and uses $5,000$ Monte Carlo replications. We evaluate performance in terms of point estimation and uncertainty quantification, both pointwise and uniformly over the boundary $\B$. Numerical results for the distance-based local polynomial procedures are reported in Tables \ref{tab:simuls-linhom}--\ref{tab:simuls-quadhet}, corresponding to the linear homoskedastic, linear heteroskedastic, quadratic homoskedastic, and quadratic heteroskedastic specifications, respectively. Each table contains four panels reflecting the bandwidth selection method employed. Panels (a) consider the case of a smooth boundary and use the selector $h=\widehat{h}_{\mathtt{MSE},\bb}$. Panels (b) report results based on the kink-adaptive selector $h=\widehat{h}_{\mathtt{kink},\bb}(\B)$, which exploits knowledge of the kink location. Panels (c) use the rule $h=\widehat{\mathtt{C}}\cdot n^{-1/4}$, which remains valid regardless of whether kinks are present in $\B$. Panels (d) employ the bandwidth $h=h_{\mathtt{1d},\bb}$ from the software package \texttt{rdrobust}, which is designed for univariate RD settings and therefore induces undersmoothing in the present BD design.

\begin{figure}
    \centering
    \begin{subfigure}[b]{0.45\textwidth}
        \centering
        \includegraphics[width=1\linewidth]{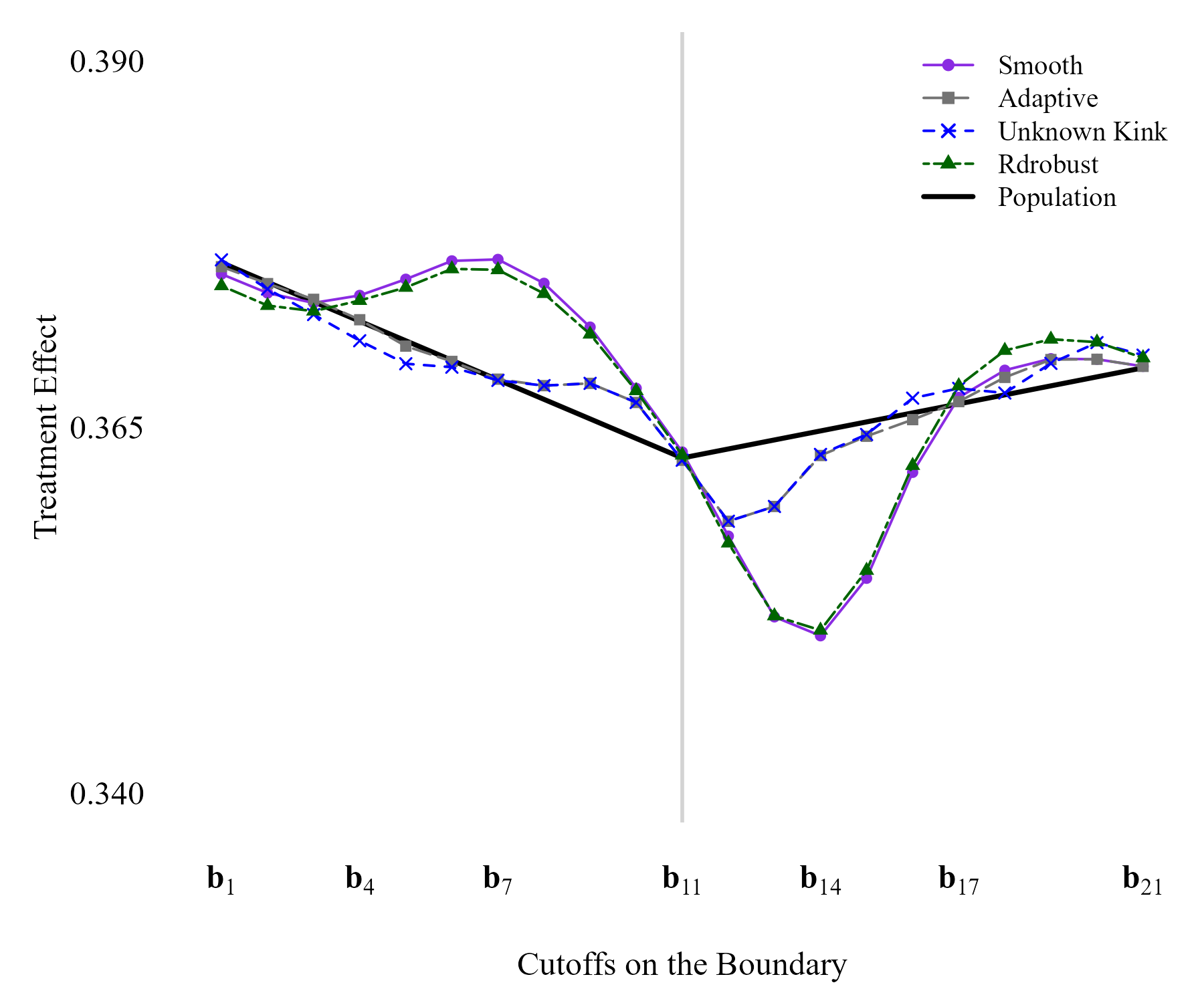}
        \caption{Linear Homoskedastic Model}
    \end{subfigure}
    \quad
    \begin{subfigure}[b]{0.45\textwidth}
        \centering
        \includegraphics[width=1\linewidth]{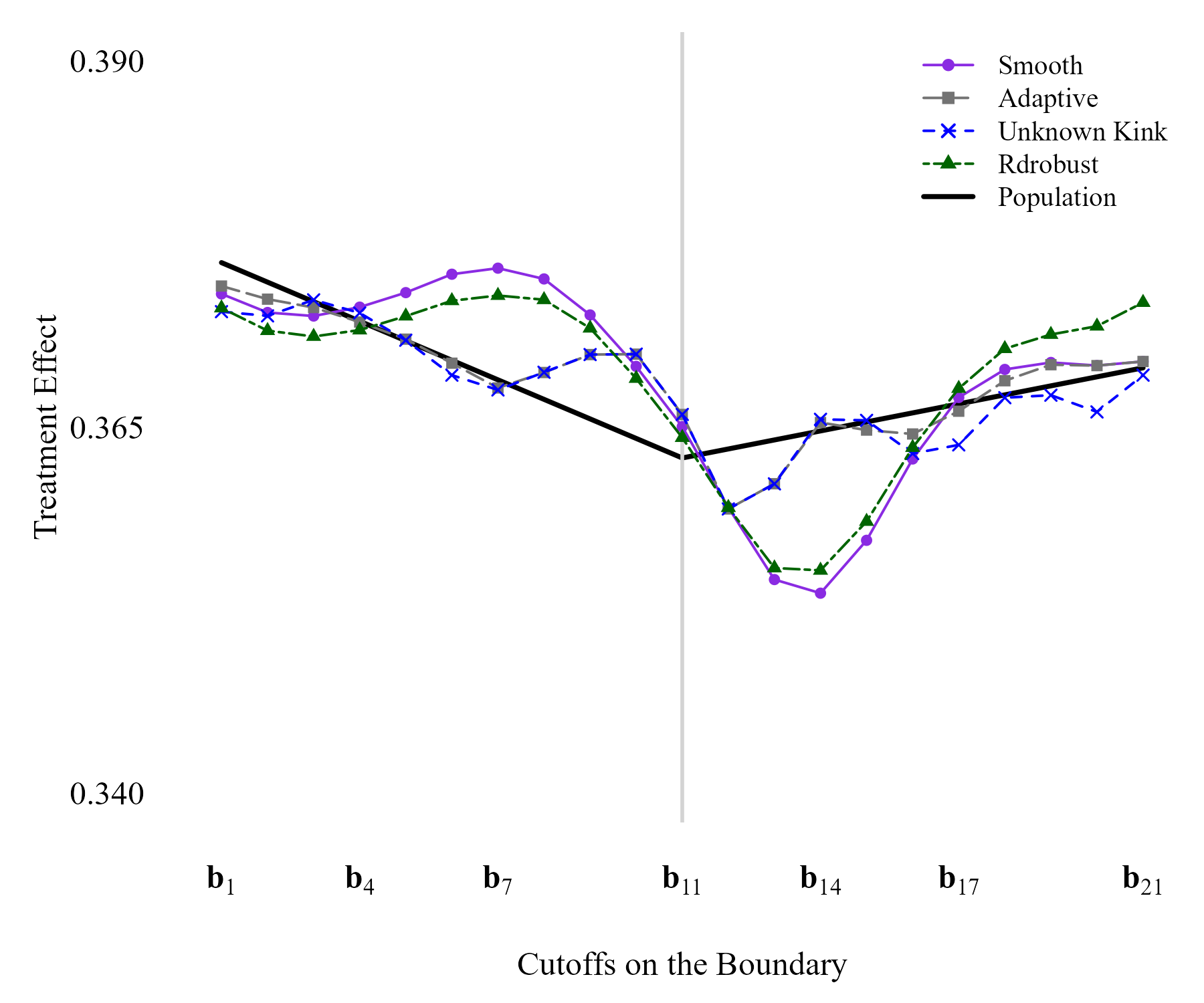}
        \caption{Linear Heteroskedastic Model}
    \end{subfigure}
    
    \begin{subfigure}[b]{0.45\textwidth}
        \centering
        \includegraphics[width=1\linewidth]{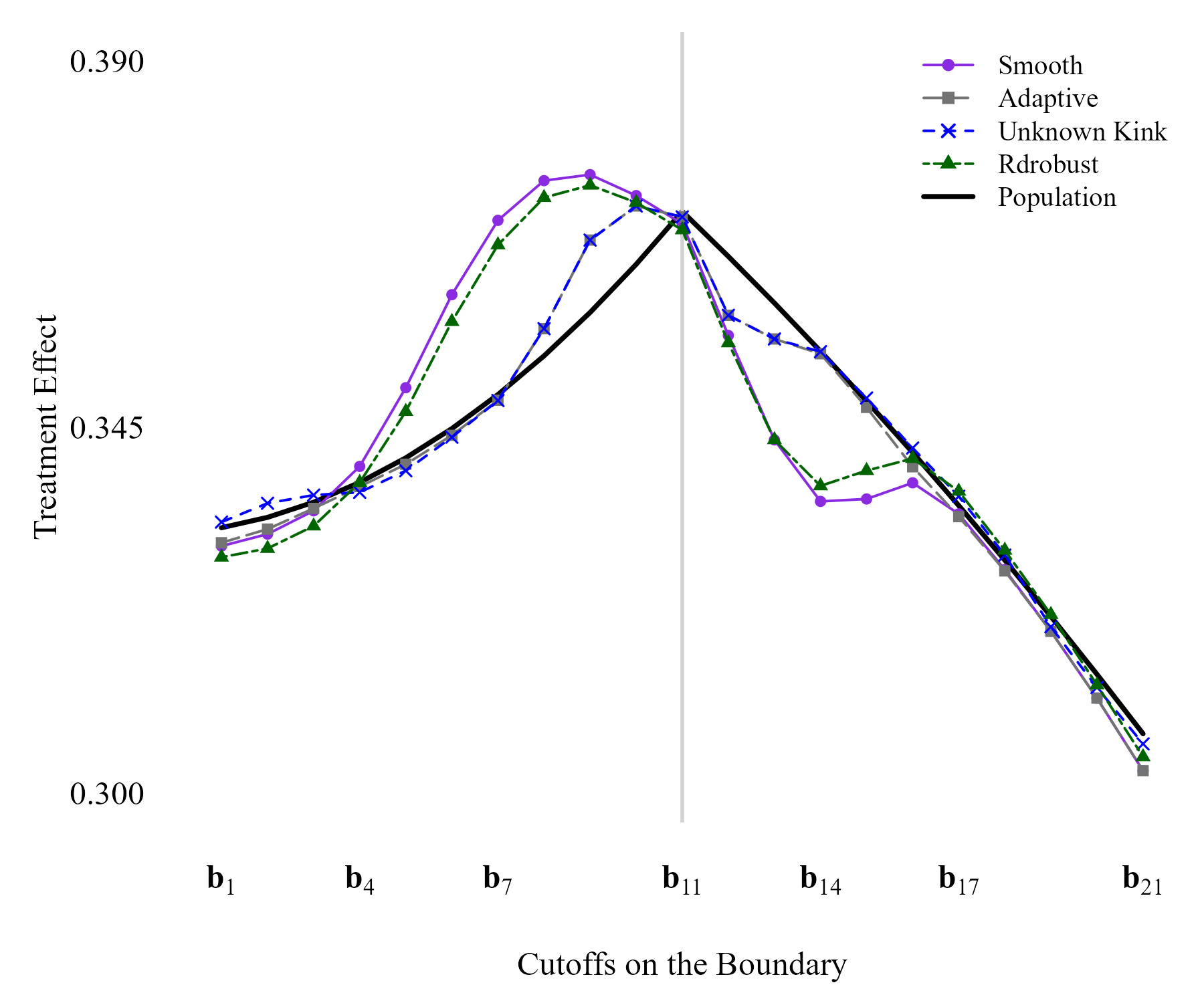}
        \caption{Quadratic Homoskedastic Model}
    \end{subfigure}
    \quad
    \begin{subfigure}[b]{0.45\textwidth}
        \centering
        \includegraphics[width=1\linewidth]{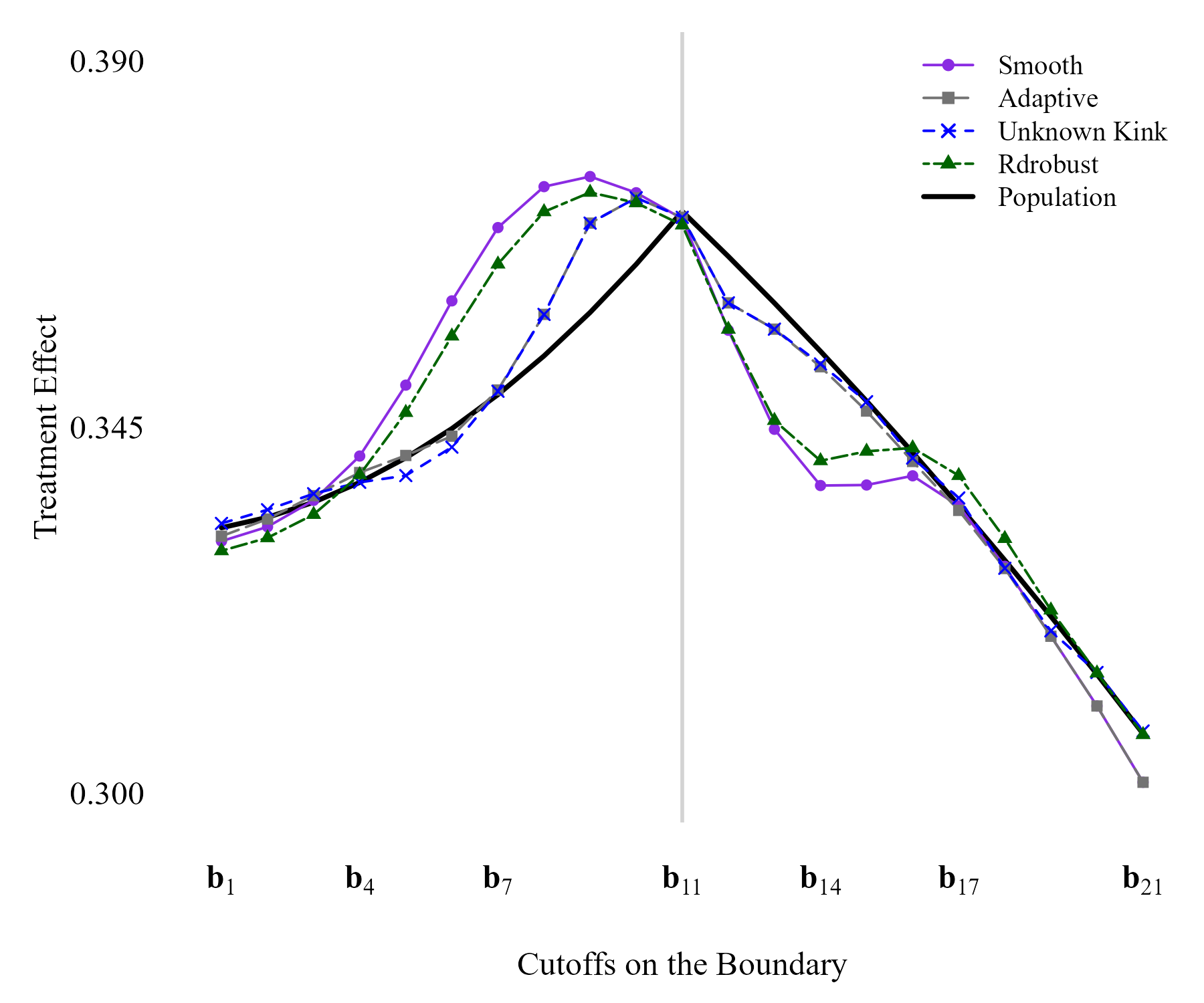}
        \caption{Quadratic Heteroskedastic Model}
    \end{subfigure}
    
    \centering
    \caption{Average of BATEC Estimators (Simulation Results). Notes: (i) \emph{Smooth} denotes using bandwidths $h=\widehat{h}_{\mathtt{MSE},\bx}$ under the assumption that the boundary is smooth; (ii) \emph{Adaptive} denotes using bandwidths $\widehat{h}_{\mathtt{kink},\bx}(\B)$ that adapt to the kink given information about the kink location; (iii) \emph{Unknown Kink} denotes using the bandwidth $h = \widehat{\mathtt{C}} \cdot n^{-1/4}$ under kink rates; (iv) \emph{Rdrobust} denotes the (incorrect) univariate MSE optimal bandwidth $h_{\mathtt{1d},\bx}$ from \texttt{rdrobust}; and (v) \emph{Population} denotes the population BATEC causal parameter, $\tau(\bx)$, calibrated using the SPP dataset (see Table \ref{tab:true-params}).}
    \label{fig:Bias}
\end{figure}

The main findings for bias are summarized in Figure \ref{fig:Bias}, which reports the average value of each point estimator across the $5,000$ replications and the target population regression function for each of the four data generating processes. The results show that distance-based methods that ignore the presence of the kink in $\B$ exhibit a larger bias than procedures that explicitly account for kinks or other irregularities in the assignment boundary. This numerical evidence is consistent with the theoretical bias characterization established in Theorem \ref{thm: Approximation Bias: Uniform Guarantee}. In this calibrated simulation study, however, the relatively large bias has limited consequences for inference because the signal-to-noise ratio is modest. As a result, coverage rates of confidence intervals and confidence bands remain close to their nominal levels across all specifications. These findings suggest that the bias induced by ignoring boundary irregularities can materially affect point estimation even when its impact on uncertainty quantification is muted. Naturally, in settings with stronger signals, failure to account for kinks or other non-smooth features of the assignment boundary may also lead to meaningful distortions in inference.

\renewcommand{\arraystretch}{1.1}

\begin{table}[ht]
    \begin{subtable}{0.5\textwidth}
        \centering
        \resizebox{\linewidth}{!}{\begin{tabular}{lccccrrrrr}
\toprule\toprule
 & $h_0$ & $h_1$ & $N_{\mathrm{Co}}$ & $N_{\mathrm{Tr}}$ & Bias & SD & RMSE & EC & IL \\
\midrule
$\bb_{1}$ & 19.531 & 19.531 & 2120 & 4259 & -0.001 & 0.033 & 0.033 & 0.958 & 0.240 \\
$\bb_{2}$ & 19.133 & 19.133 & 2135 & 4107 & -0.001 & 0.034 & 0.034 & 0.956 & 0.243 \\
$\bb_{3}$ & 18.879 & 18.879 & 2182 & 3953 & -0.000 & 0.035 & 0.035 & 0.955 & 0.245 \\
$\bb_{4}$ & 18.784 & 18.784 & 2267 & 3798 & 0.002 & 0.036 & 0.036 & 0.952 & 0.246 \\
$\bb_{5}$ & 18.826 & 18.826 & 2383 & 3637 & 0.004 & 0.035 & 0.036 & 0.947 & 0.246 \\
$\bb_{6}$ & 18.978 & 18.978 & 2512 & 3469 & 0.007 & 0.035 & 0.035 & 0.953 & 0.245 \\
$\bb_{7}$ & 19.090 & 19.090 & 2617 & 3257 & 0.008 & 0.034 & 0.035 & 0.960 & 0.247 \\
\bottomrule\bottomrule
\end{tabular}}
        \caption{$h=\widehat{h}_{\mathtt{MSE},\bb}$ (smooth $\B$)}
    \end{subtable}
    \hfill
    \begin{subtable}{0.5\textwidth}
        \centering
        \resizebox{\linewidth}{!}{\begin{tabular}{lccccrrrrr}
\toprule\toprule
 & $h_0$ & $h_1$ & $N_{\mathrm{Co}}$ & $N_{\mathrm{Tr}}$ & Bias & SD & RMSE & EC & IL \\
\midrule
$\bb_{1}$ & 18.845 & 18.845 & 2020 & 3988 & -0.000 & 0.034 & 0.034 & 0.958 & 0.247 \\
$\bb_{2}$ & 17.539 & 17.539 & 1876 & 3510 & -0.000 & 0.035 & 0.035 & 0.953 & 0.261 \\
$\bb_{3}$ & 15.909 & 15.909 & 1661 & 2913 & 0.000 & 0.038 & 0.038 & 0.948 & 0.285 \\
$\bb_{4}$ & 13.996 & 13.996 & 1382 & 2256 & 0.000 & 0.043 & 0.043 & 0.950 & 0.322 \\
$\bb_{5}$ & 12.000 & 12.000 & 1081 & 1641 & -0.000 & 0.050 & 0.050 & 0.947 & 0.375 \\
$\bb_{6}$ & 10.000 & 10.000 & 788 & 1113 & -0.000 & 0.060 & 0.060 & 0.947 & 0.451 \\
$\bb_{7}$ & 8.547 & 8.547 & 594 & 781 & 0.000 & 0.071 & 0.071 & 0.942 & 0.532 \\
\bottomrule\bottomrule
\end{tabular}}
        \caption{$h=\widehat{h}_{\mathtt{kink},\bb}(\B)$ (kink adaptive)}
    \end{subtable}

    \vspace{.25in}
    
    \begin{subtable}{0.5\textwidth}
        \centering
        \resizebox{0.983\linewidth}{!}{\begin{tabular}{lccccrrrrr}
\toprule\toprule
 & $h_0$ & $h_1$ & $N_{\mathrm{Co}}$ & $N_{\mathrm{Tr}}$ & Bias & SD & RMSE & EC & IL \\
\midrule
$\bb_{1}$ & 8.557 & 8.557 & 609 & 823 & 0.000 & 0.069 & 0.069 & 0.957 & 0.276 \\
$\bb_{2}$ & 8.382 & 8.382 & 597 & 801 & -0.000 & 0.072 & 0.072 & 0.950 & 0.280 \\
$\bb_{3}$ & 8.271 & 8.271 & 588 & 786 & -0.001 & 0.074 & 0.074 & 0.949 & 0.282 \\
$\bb_{4}$ & 8.229 & 8.229 & 582 & 777 & -0.001 & 0.073 & 0.073 & 0.956 & 0.284 \\
$\bb_{5}$ & 8.248 & 8.248 & 579 & 772 & -0.002 & 0.074 & 0.074 & 0.952 & 0.283 \\
$\bb_{6}$ & 8.314 & 8.314 & 576 & 770 & -0.000 & 0.073 & 0.073 & 0.952 & 0.283 \\
$\bb_{7}$ & 8.364 & 8.364 & 574 & 750 & -0.000 & 0.073 & 0.073 & 0.951 & 0.285 \\
\bottomrule\bottomrule
\end{tabular}}
        \caption{$h=\widehat{\mathtt{C}} \cdot n^{-1/4}$ (unknown kink location)}
    \end{subtable}
    \hfill
    \begin{subtable}{0.5\textwidth}
        \centering
        \resizebox{\linewidth}{!}{\begin{tabular}{lccccrrrrr}
\toprule\toprule
 & $h_0$ & $h_1$ & $N_{\mathrm{Co}}$ & $N_{\mathrm{Tr}}$ & Bias & SD & RMSE & EC & IL \\
\midrule
$\bb_{1}$ & 19.346 & 19.346 & 2103 & 4191 & -0.002 & 0.035 & 0.035 & 0.960 & 0.244 \\
$\bb_{2}$ & 19.187 & 19.187 & 2155 & 4130 & -0.002 & 0.036 & 0.036 & 0.957 & 0.245 \\
$\bb_{3}$ & 19.183 & 19.183 & 2247 & 4066 & -0.001 & 0.037 & 0.037 & 0.958 & 0.244 \\
$\bb_{4}$ & 19.301 & 19.301 & 2372 & 3985 & 0.001 & 0.037 & 0.037 & 0.959 & 0.244 \\
$\bb_{5}$ & 19.499 & 19.499 & 2514 & 3876 & 0.004 & 0.037 & 0.038 & 0.953 & 0.242 \\
$\bb_{6}$ & 19.771 & 19.771 & 2662 & 3744 & 0.006 & 0.037 & 0.037 & 0.953 & 0.241 \\
$\bb_{7}$ & 20.043 & 20.043 & 2791 & 3576 & 0.008 & 0.036 & 0.037 & 0.960 & 0.241 \\
\bottomrule\bottomrule
\end{tabular}}
        \caption{$h=\widehat{h}_{\mathtt{1d},\mathbf{b}}$ ($\mathtt{rdrobust} + \mathtt{rd2d.dist}$)}
    \end{subtable}
    \caption{Linear Homoskedastic Model (Simulation Results)}
    \label{tab:simuls-linhom}
\end{table}

\begin{table}[ht]
    \begin{subtable}{0.5\textwidth}
        \centering
        \resizebox{\linewidth}{!}{\begin{tabular}{lccccrrrrr}
\toprule\toprule
 & $h_0$ & $h_1$ & $N_{\mathrm{Co}}$ & $N_{\mathrm{Tr}}$ & Bias & SD & RMSE & EC & IL \\
\midrule
$\bb_{1}$ & 19.917 & 19.917 & 2174 & 4411 & -0.002 & 0.044 & 0.044 & 0.956 & 0.312 \\
$\bb_{2}$ & 19.329 & 19.329 & 2166 & 4181 & -0.002 & 0.044 & 0.045 & 0.958 & 0.315 \\
$\bb_{3}$ & 18.839 & 18.839 & 2175 & 3939 & -0.001 & 0.045 & 0.045 & 0.957 & 0.318 \\
$\bb_{4}$ & 18.488 & 18.488 & 2211 & 3699 & 0.001 & 0.046 & 0.046 & 0.953 & 0.320 \\
$\bb_{5}$ & 18.264 & 18.264 & 2270 & 3459 & 0.003 & 0.046 & 0.046 & 0.954 & 0.321 \\
$\bb_{6}$ & 18.133 & 18.133 & 2338 & 3215 & 0.006 & 0.046 & 0.046 & 0.958 & 0.323 \\
$\bb_{7}$ & 18.010 & 18.010 & 2394 & 2947 & 0.008 & 0.046 & 0.046 & 0.960 & 0.325 \\
\bottomrule\bottomrule
\end{tabular}}
        \caption{$h=\widehat{h}_{\mathtt{MSE},\bb}$ (smooth $\B$)}
    \end{subtable}
    \hfill
    \begin{subtable}{0.5\textwidth}
        \centering
        \resizebox{\linewidth}{!}{\begin{tabular}{lccccrrrrr}
\toprule\toprule
 & $h_0$ & $h_1$ & $N_{\mathrm{Co}}$ & $N_{\mathrm{Tr}}$ & Bias & SD & RMSE & EC & IL \\
\midrule
$\bb_{1}$ & 19.062 & 19.062 & 2049 & 4073 & -0.002 & 0.045 & 0.045 & 0.954 & 0.324 \\
$\bb_{2}$ & 17.606 & 17.606 & 1887 & 3536 & -0.001 & 0.046 & 0.046 & 0.951 & 0.342 \\
$\bb_{3}$ & 15.908 & 15.908 & 1661 & 2914 & -0.000 & 0.050 & 0.050 & 0.948 & 0.371 \\
$\bb_{4}$ & 13.995 & 13.995 & 1382 & 2256 & -0.000 & 0.055 & 0.055 & 0.945 & 0.414 \\
$\bb_{5}$ & 12.000 & 12.000 & 1082 & 1641 & 0.000 & 0.063 & 0.063 & 0.943 & 0.477 \\
$\bb_{6}$ & 10.000 & 10.000 & 789 & 1112 & -0.000 & 0.075 & 0.075 & 0.944 & 0.571 \\
$\bb_{7}$ & 8.241 & 8.241 & 553 & 727 & -0.001 & 0.090 & 0.090 & 0.937 & 0.688 \\
\bottomrule\bottomrule
\end{tabular}}
        \caption{$h=\widehat{h}_{\mathtt{kink},\bb}(\B)$ (kink adaptive)}
    \end{subtable}

    \vspace{.25in}
    
    \begin{subtable}{0.5\textwidth}
        \centering
        \resizebox{0.983\linewidth}{!}{\begin{tabular}{lccccrrrrr}
\toprule\toprule
 & $h_0$ & $h_1$ & $N_{\mathrm{Co}}$ & $N_{\mathrm{Tr}}$ & Bias & SD & RMSE & EC & IL \\
\midrule
$\bb_{1}$ & 8.726 & 8.726 & 630 & 856 & -0.003 & 0.093 & 0.093 & 0.950 & 0.356 \\
$\bb_{2}$ & 8.468 & 8.468 & 608 & 818 & -0.002 & 0.093 & 0.093 & 0.948 & 0.360 \\
$\bb_{3}$ & 8.253 & 8.253 & 585 & 781 & 0.000 & 0.094 & 0.094 & 0.949 & 0.365 \\
$\bb_{4}$ & 8.100 & 8.100 & 566 & 751 & 0.001 & 0.095 & 0.095 & 0.947 & 0.369 \\
$\bb_{5}$ & 8.002 & 8.002 & 549 & 725 & 0.000 & 0.096 & 0.096 & 0.949 & 0.371 \\
$\bb_{6}$ & 7.944 & 7.944 & 532 & 701 & -0.001 & 0.096 & 0.096 & 0.954 & 0.374 \\
$\bb_{7}$ & 7.890 & 7.890 & 514 & 670 & -0.001 & 0.096 & 0.096 & 0.957 & 0.377 \\
\bottomrule\bottomrule
\end{tabular}}
        \caption{$h=\widehat{\mathtt{C}} \cdot n^{-1/4}$ (unknown kink location)}
    \end{subtable}
    \hfill
    \begin{subtable}{0.5\textwidth}
        \centering
        \resizebox{\linewidth}{!}{\begin{tabular}{lccccrrrrr}
\toprule\toprule
 & $h_0$ & $h_1$ & $N_{\mathrm{Co}}$ & $N_{\mathrm{Tr}}$ & Bias & SD & RMSE & EC & IL \\
\midrule
$\bb_{1}$ & 19.141 & 19.141 & 2074 & 4119 & -0.003 & 0.049 & 0.049 & 0.958 & 0.328 \\
$\bb_{2}$ & 18.748 & 18.748 & 2087 & 3973 & -0.003 & 0.049 & 0.049 & 0.955 & 0.329 \\
$\bb_{3}$ & 18.546 & 18.546 & 2138 & 3846 & -0.002 & 0.050 & 0.050 & 0.959 & 0.328 \\
$\bb_{4}$ & 18.409 & 18.409 & 2206 & 3689 & -0.001 & 0.051 & 0.051 & 0.954 & 0.327 \\
$\bb_{5}$ & 18.352 & 18.352 & 2291 & 3510 & 0.002 & 0.051 & 0.051 & 0.954 & 0.326 \\
$\bb_{6}$ & 18.342 & 18.342 & 2380 & 3306 & 0.004 & 0.050 & 0.051 & 0.964 & 0.325 \\
$\bb_{7}$ & 17.995 & 17.995 & 2387 & 2973 & 0.006 & 0.050 & 0.050 & 0.963 & 0.331 \\
\bottomrule\bottomrule
\end{tabular}}
        \caption{$h=\widehat{h}_{\mathtt{1d},\mathbf{b}}$ ($\mathtt{rdrobust} + \mathtt{rd2d.dist}$)}
    \end{subtable}
    \caption{Linear Heteroskedastic Model (Simulation Results)}
    \label{tab:simuls-linhet}
\end{table}

\begin{table}[ht]
    \begin{subtable}{0.5\textwidth}
        \centering
        \resizebox{\linewidth}{!}{\begin{tabular}{lccccrrrrr}
\toprule\toprule
 & $h_0$ & $h_1$ & $N_{\mathrm{Co}}$ & $N_{\mathrm{Tr}}$ & Bias & SD & RMSE & EC & IL \\
\midrule
$\bb_{1}$ & 17.917 & 17.917 & 1899 & 3638 & -0.002 & 0.036 & 0.036 & 0.948 & 0.260 \\
$\bb_{2}$ & 17.386 & 17.386 & 1869 & 3461 & -0.002 & 0.037 & 0.037 & 0.950 & 0.266 \\
$\bb_{3}$ & 17.186 & 17.186 & 1889 & 3364 & -0.001 & 0.038 & 0.038 & 0.957 & 0.268 \\
$\bb_{4}$ & 17.352 & 17.352 & 1993 & 3331 & 0.002 & 0.038 & 0.039 & 0.957 & 0.265 \\
$\bb_{5}$ & 17.854 & 17.854 & 2186 & 3338 & 0.009 & 0.038 & 0.039 & 0.953 & 0.259 \\
$\bb_{6}$ & 18.432 & 18.432 & 2398 & 3310 & 0.016 & 0.037 & 0.041 & 0.950 & 0.253 \\
$\bb_{7}$ & 18.885 & 18.885 & 2573 & 3201 & 0.021 & 0.036 & 0.042 & 0.958 & 0.250 \\
\bottomrule\bottomrule
\end{tabular}}
        \caption{$h=\widehat{h}_{\mathtt{MSE},\bb}$ (smooth $\B$)}
    \end{subtable}
    \hfill
    \begin{subtable}{0.5\textwidth}
        \centering
        \resizebox{\linewidth}{!}{\begin{tabular}{lccccrrrrr}
\toprule\toprule
 & $h_0$ & $h_1$ & $N_{\mathrm{Co}}$ & $N_{\mathrm{Tr}}$ & Bias & SD & RMSE & EC & IL \\
\midrule
$\bb_{1}$ & 17.644 & 17.644 & 1860 & 3529 & -0.002 & 0.036 & 0.036 & 0.948 & 0.263 \\
$\bb_{2}$ & 16.703 & 16.703 & 1759 & 3205 & -0.001 & 0.038 & 0.038 & 0.948 & 0.274 \\
$\bb_{3}$ & 15.611 & 15.611 & 1617 & 2813 & -0.001 & 0.039 & 0.039 & 0.951 & 0.290 \\
$\bb_{4}$ & 13.979 & 13.979 & 1378 & 2252 & -0.001 & 0.043 & 0.043 & 0.949 & 0.322 \\
$\bb_{5}$ & 12.000 & 12.000 & 1080 & 1642 & -0.001 & 0.050 & 0.050 & 0.945 & 0.375 \\
$\bb_{6}$ & 10.000 & 10.000 & 787 & 1114 & -0.001 & 0.059 & 0.059 & 0.947 & 0.450 \\
$\bb_{7}$ & 8.509 & 8.509 & 589 & 775 & -0.001 & 0.070 & 0.070 & 0.944 & 0.535 \\
\bottomrule\bottomrule
\end{tabular}}
        \caption{$h=\widehat{h}_{\mathtt{kink},\bb}(\B)$ (kink adaptive)}
    \end{subtable}

    \vspace{.25in}
    
    \begin{subtable}{0.5\textwidth}
        \centering
        \resizebox{0.983\linewidth}{!}{\begin{tabular}{lccccrrrrr}
\toprule\toprule
 & $h_0$ & $h_1$ & $N_{\mathrm{Co}}$ & $N_{\mathrm{Tr}}$ & Bias & SD & RMSE & EC & IL \\
\midrule
$\bb_{1}$ & 7.849 & 7.849 & 524 & 693 & 0.001 & 0.078 & 0.078 & 0.945 & 0.300 \\
$\bb_{2}$ & 7.617 & 7.617 & 504 & 662 & 0.002 & 0.079 & 0.079 & 0.949 & 0.307 \\
$\bb_{3}$ & 7.529 & 7.529 & 498 & 651 & 0.001 & 0.079 & 0.079 & 0.948 & 0.310 \\
$\bb_{4}$ & 7.602 & 7.602 & 506 & 663 & -0.001 & 0.080 & 0.080 & 0.946 & 0.307 \\
$\bb_{5}$ & 7.822 & 7.822 & 527 & 696 & -0.002 & 0.078 & 0.078 & 0.945 & 0.300 \\
$\bb_{6}$ & 8.075 & 8.075 & 547 & 729 & -0.001 & 0.075 & 0.075 & 0.954 & 0.292 \\
$\bb_{7}$ & 8.274 & 8.274 & 563 & 736 & -0.001 & 0.073 & 0.073 & 0.954 & 0.288 \\
\bottomrule\bottomrule
\end{tabular}}
        \caption{$h=\widehat{\mathtt{C}} \cdot n^{-1/4}$ (unknown kink location)}
    \end{subtable}
    \hfill
    \begin{subtable}{0.5\textwidth}
        \centering
        \resizebox{\linewidth}{!}{\begin{tabular}{lccccrrrrr}
\toprule\toprule
 & $h_0$ & $h_1$ & $N_{\mathrm{Co}}$ & $N_{\mathrm{Tr}}$ & Bias & SD & RMSE & EC & IL \\
\midrule
$\bb_{1}$ & 17.226 & 17.226 & 1805 & 3381 & -0.004 & 0.038 & 0.038 & 0.951 & 0.271 \\
$\bb_{2}$ & 16.871 & 16.871 & 1794 & 3282 & -0.004 & 0.039 & 0.039 & 0.955 & 0.275 \\
$\bb_{3}$ & 16.770 & 16.770 & 1829 & 3228 & -0.003 & 0.040 & 0.040 & 0.963 & 0.277 \\
$\bb_{4}$ & 16.999 & 16.999 & 1940 & 3229 & -0.000 & 0.041 & 0.041 & 0.961 & 0.275 \\
$\bb_{5}$ & 17.517 & 17.517 & 2124 & 3259 & 0.006 & 0.042 & 0.043 & 0.956 & 0.270 \\
$\bb_{6}$ & 18.251 & 18.251 & 2356 & 3293 & 0.013 & 0.042 & 0.044 & 0.955 & 0.262 \\
$\bb_{7}$ & 18.950 & 18.950 & 2571 & 3265 & 0.018 & 0.040 & 0.044 & 0.956 & 0.256 \\
\bottomrule\bottomrule
\end{tabular}}
        \caption{$h=\widehat{h}_{\mathtt{1d},\mathbf{b}}$ ($\mathtt{rdrobust} + \mathtt{rd2d.dist}$)}
    \end{subtable}
    \caption{Quadratic Homoskedastic Model (Simulation Results)}
    \label{tab:simuls-quadhom}
\end{table}

\begin{table}[ht]
    \begin{subtable}{0.5\textwidth}
        \centering
        \resizebox{\linewidth}{!}{\begin{tabular}{lccccrrrrr}
\toprule\toprule
 & $h_0$ & $h_1$ & $N_{\mathrm{Co}}$ & $N_{\mathrm{Tr}}$ & Bias & SD & RMSE & EC & IL \\
\midrule
$\bb_{1}$ & 18.550 & 18.550 & 1986 & 3881 & -0.002 & 0.042 & 0.042 & 0.949 & 0.302 \\
$\bb_{2}$ & 17.907 & 17.907 & 1948 & 3654 & -0.001 & 0.042 & 0.042 & 0.955 & 0.304 \\
$\bb_{3}$ & 17.562 & 17.562 & 1954 & 3494 & 0.000 & 0.042 & 0.042 & 0.960 & 0.303 \\
$\bb_{4}$ & 17.483 & 17.483 & 2019 & 3373 & 0.003 & 0.043 & 0.043 & 0.960 & 0.299 \\
$\bb_{5}$ & 17.596 & 17.596 & 2134 & 3256 & 0.009 & 0.042 & 0.043 & 0.958 & 0.293 \\
$\bb_{6}$ & 17.768 & 17.768 & 2262 & 3110 & 0.016 & 0.042 & 0.045 & 0.958 & 0.287 \\
$\bb_{7}$ & 17.881 & 17.881 & 2367 & 2912 & 0.021 & 0.041 & 0.046 & 0.951 & 0.283 \\
\bottomrule\bottomrule
\end{tabular}}
        \caption{$h=\widehat{h}_{\mathtt{MSE},\bb}$ (smooth $\B$)}
    \end{subtable}
    \hfill
    \begin{subtable}{0.5\textwidth}
        \centering
        \resizebox{\linewidth}{!}{\begin{tabular}{lccccrrrrr}
\toprule\toprule
 & $h_0$ & $h_1$ & $N_{\mathrm{Co}}$ & $N_{\mathrm{Tr}}$ & Bias & SD & RMSE & EC & IL \\
\midrule
$\bb_{1}$ & 18.127 & 18.127 & 1925 & 3713 & -0.001 & 0.043 & 0.043 & 0.950 & 0.307 \\
$\bb_{2}$ & 16.987 & 16.987 & 1800 & 3309 & -0.000 & 0.044 & 0.043 & 0.950 & 0.317 \\
$\bb_{3}$ & 15.693 & 15.693 & 1630 & 2842 & 0.001 & 0.045 & 0.045 & 0.955 & 0.333 \\
$\bb_{4}$ & 13.973 & 13.973 & 1379 & 2251 & 0.001 & 0.048 & 0.048 & 0.948 & 0.366 \\
$\bb_{5}$ & 12.000 & 12.000 & 1082 & 1642 & 0.000 & 0.055 & 0.055 & 0.945 & 0.418 \\
$\bb_{6}$ & 10.000 & 10.000 & 788 & 1113 & -0.001 & 0.065 & 0.065 & 0.942 & 0.494 \\
$\bb_{7}$ & 8.216 & 8.216 & 549 & 723 & 0.001 & 0.081 & 0.081 & 0.940 & 0.595 \\
\bottomrule\bottomrule
\end{tabular}}
        \caption{$h=\widehat{h}_{\mathtt{kink},\bb}(\B)$ (kink adaptive)}
    \end{subtable}

    \vspace{.25in}
    
    \begin{subtable}{0.5\textwidth}
        \centering
        \resizebox{0.983\linewidth}{!}{\begin{tabular}{lccccrrrrr}
\toprule\toprule
 & $h_0$ & $h_1$ & $N_{\mathrm{Co}}$ & $N_{\mathrm{Tr}}$ & Bias & SD & RMSE & EC & IL \\
\midrule
$\bb_{1}$ & 8.127 & 8.127 & 558 & 743 & 0.001 & 0.090 & 0.090 & 0.948 & 0.347 \\
$\bb_{2}$ & 7.845 & 7.845 & 533 & 703 & 0.001 & 0.090 & 0.090 & 0.946 & 0.350 \\
$\bb_{3}$ & 7.694 & 7.694 & 518 & 680 & 0.001 & 0.088 & 0.088 & 0.953 & 0.349 \\
$\bb_{4}$ & 7.659 & 7.659 & 513 & 673 & 0.000 & 0.088 & 0.088 & 0.947 & 0.345 \\
$\bb_{5}$ & 7.709 & 7.709 & 514 & 674 & -0.002 & 0.087 & 0.087 & 0.952 & 0.338 \\
$\bb_{6}$ & 7.785 & 7.785 & 513 & 674 & -0.002 & 0.086 & 0.086 & 0.951 & 0.332 \\
$\bb_{7}$ & 7.834 & 7.834 & 507 & 661 & 0.000 & 0.086 & 0.086 & 0.947 & 0.328 \\
\bottomrule\bottomrule
\end{tabular}}
        \caption{$h=\widehat{\mathtt{C}} \cdot n^{-1/4}$ (unknown kink location)}
    \end{subtable}
    \hfill
    \begin{subtable}{0.5\textwidth}
        \centering
        \resizebox{\linewidth}{!}{\begin{tabular}{lccccrrrrr}
\toprule\toprule
 & $h_0$ & $h_1$ & $N_{\mathrm{Co}}$ & $N_{\mathrm{Tr}}$ & Bias & SD & RMSE & EC & IL \\
\midrule
$\bb_{1}$ & 17.834 & 17.834 & 1890 & 3614 & -0.003 & 0.045 & 0.045 & 0.951 & 0.315 \\
$\bb_{2}$ & 17.392 & 17.392 & 1874 & 3473 & -0.003 & 0.045 & 0.045 & 0.953 & 0.315 \\
$\bb_{3}$ & 17.123 & 17.123 & 1890 & 3353 & -0.002 & 0.046 & 0.046 & 0.959 & 0.314 \\
$\bb_{4}$ & 17.070 & 17.070 & 1956 & 3254 & 0.001 & 0.047 & 0.047 & 0.964 & 0.311 \\
$\bb_{5}$ & 17.206 & 17.206 & 2067 & 3162 & 0.006 & 0.048 & 0.048 & 0.959 & 0.306 \\
$\bb_{6}$ & 17.450 & 17.450 & 2199 & 3053 & 0.011 & 0.048 & 0.049 & 0.958 & 0.299 \\
$\bb_{7}$ & 17.514 & 17.514 & 2288 & 2849 & 0.016 & 0.047 & 0.050 & 0.949 & 0.296 \\
\bottomrule\bottomrule
\end{tabular}}
        \caption{$h=\widehat{h}_{\mathtt{1d},\mathbf{b}}$ ($\mathtt{rdrobust} + \mathtt{rd2d.dist}$)}
    \end{subtable}
    \caption{Quadratic Heteroskedastic Model (Simulation Results)}
    \label{tab:simuls-quadhet}
\end{table}

\subsection{Empirical Application}\label{sec: Empirical Application}

\begin{figure}
    \centering
    \begin{subfigure}[b]{0.45\textwidth}
        \centering
        \includegraphics[width=1\linewidth]{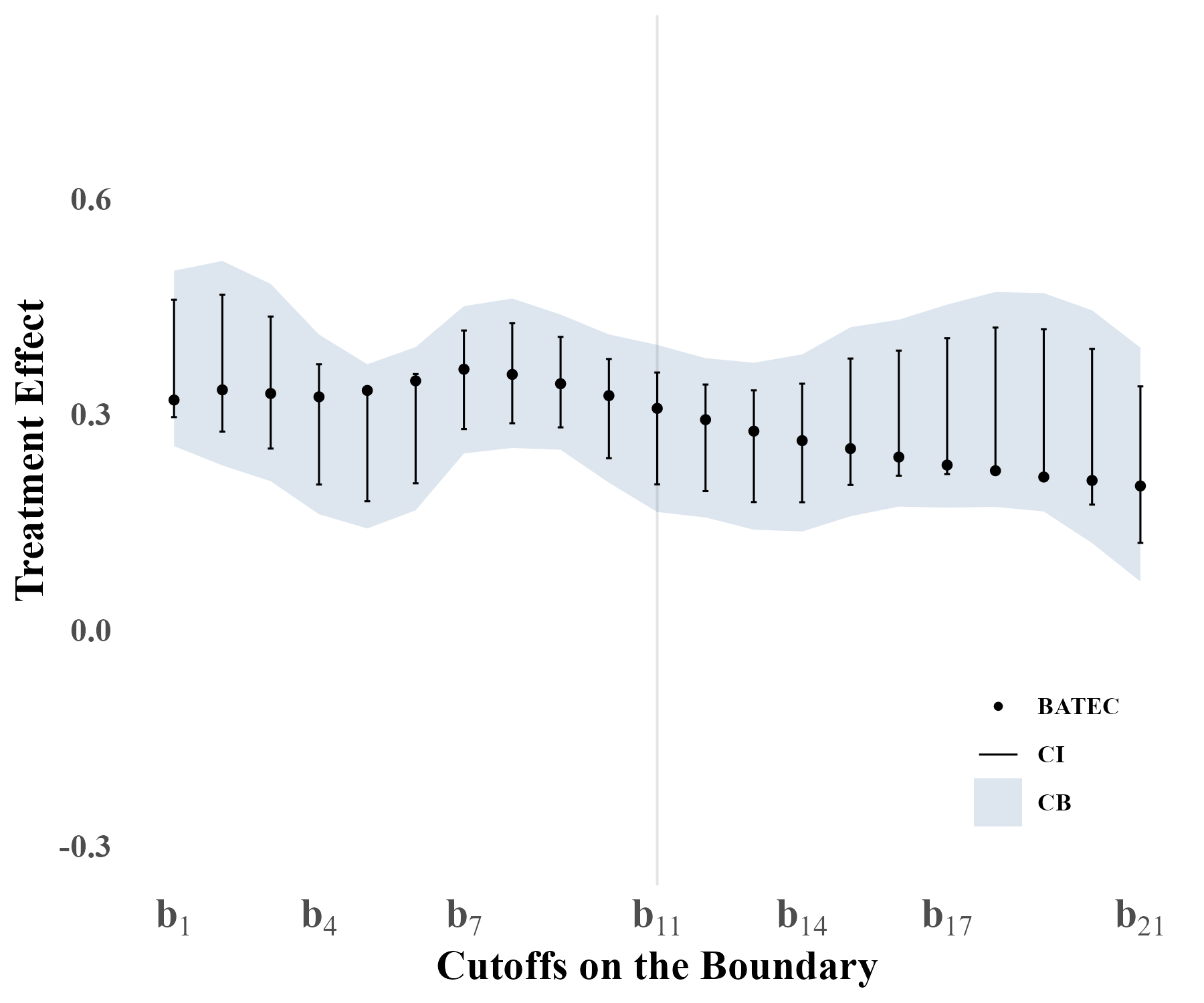}
        \caption{$h=\widehat{h}_{\mathtt{MSE},\bb}$ (smooth $\B$)}
    \end{subfigure}
    \quad
    \begin{subfigure}[b]{0.45\textwidth}
        \centering
        \includegraphics[width=1\linewidth]{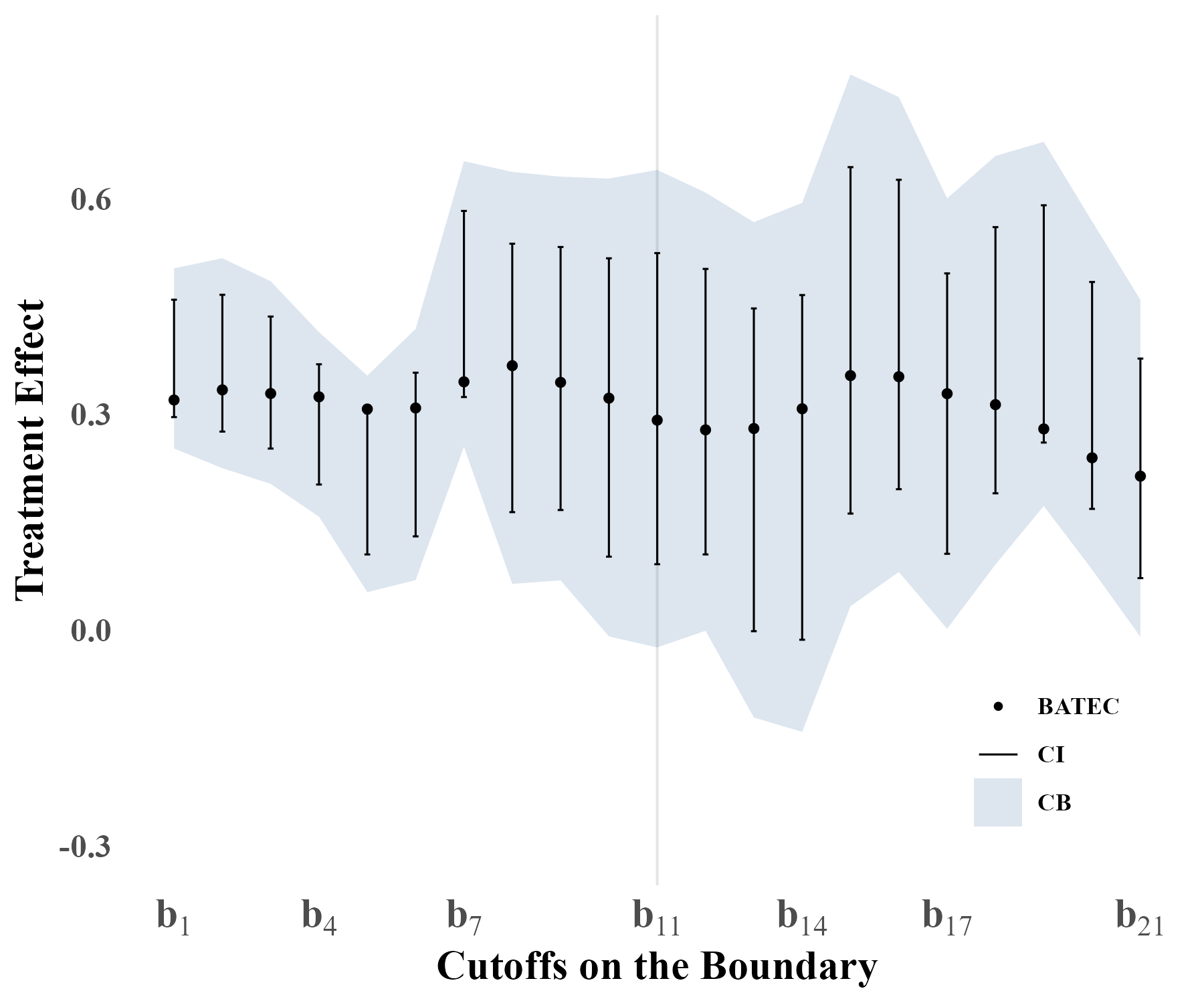}
        \caption{$h=\widehat{h}_{\mathtt{kink},\bb}(\B)$ (kink adaptive)}
    \end{subfigure}
    
    \begin{subfigure}[b]{0.45\textwidth}
        \centering
        \includegraphics[width=1\linewidth]{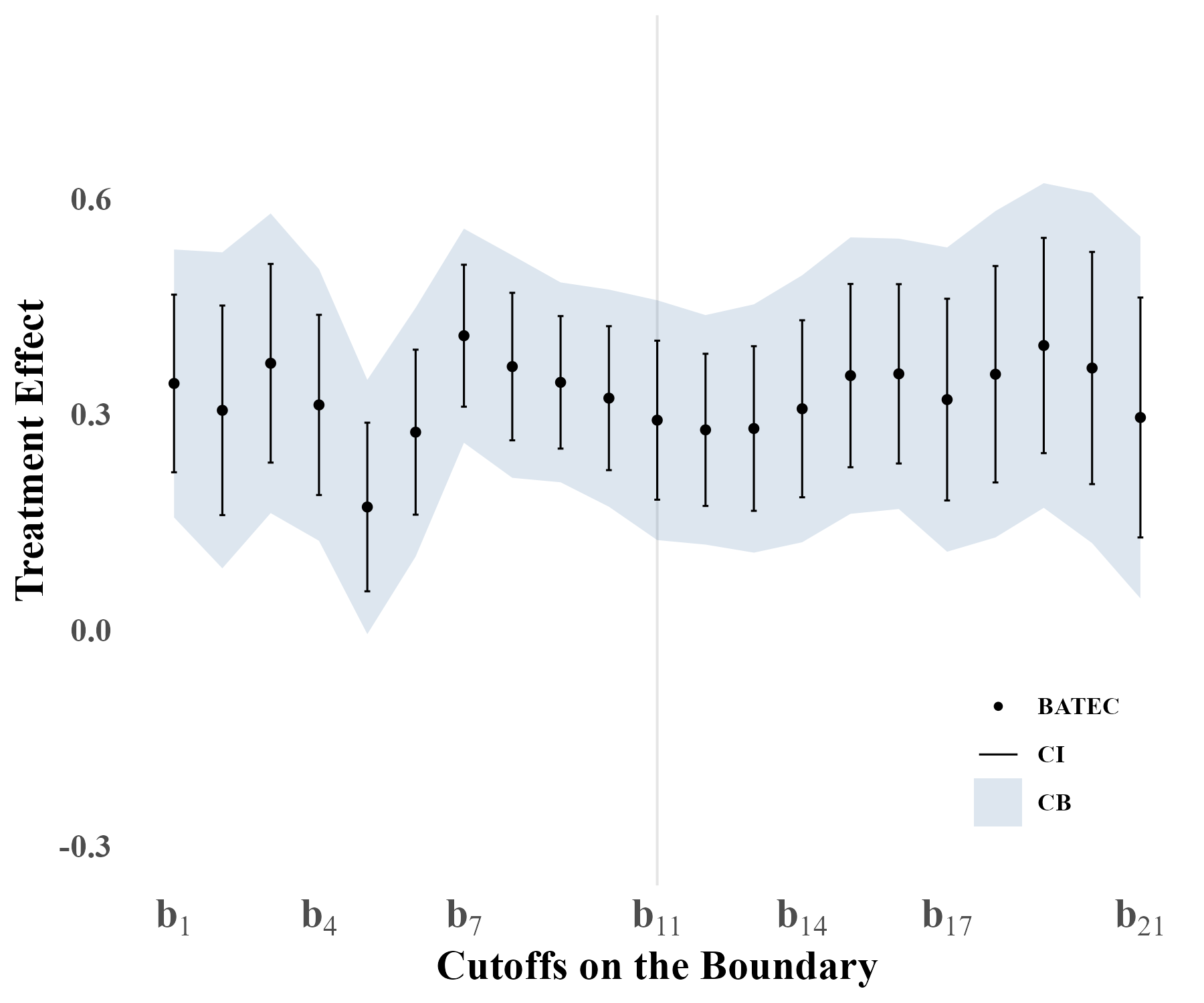}
        \caption{$h=\widehat{\mathtt{C}} \cdot n^{-1/4}$ (unknown kink location)}
    \end{subfigure}
    \quad
    \begin{subfigure}[b]{0.45\textwidth}
        \centering
        \includegraphics[width=1\linewidth]{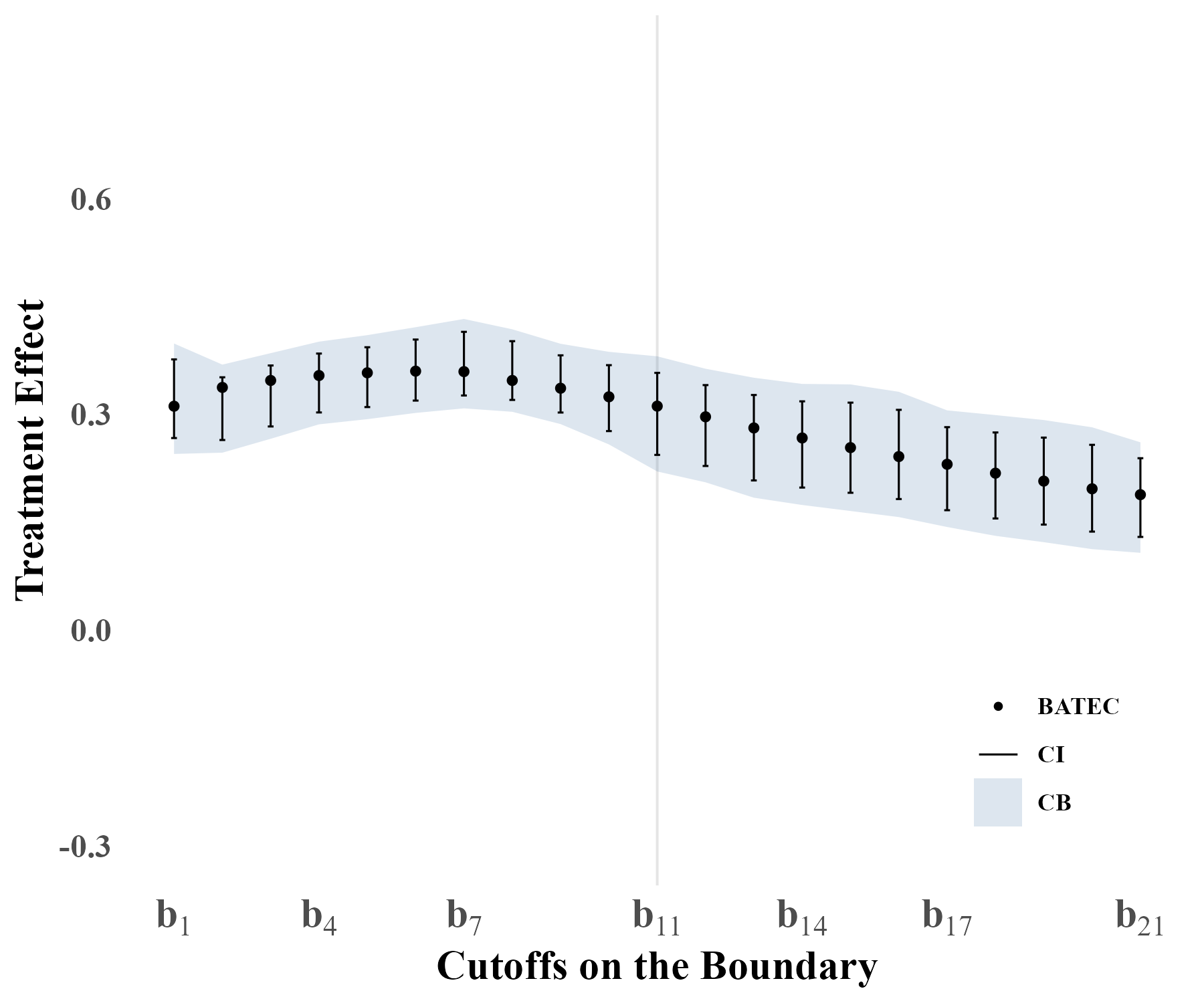}
        \caption{$h=\widehat{h}_{\mathtt{1d},\mathbf{b}}$ ($\mathtt{rdrobust} + \mathtt{rd2d.dist}$)}
    \end{subfigure}
    
    \centering
    \caption{BATEC Estimation and Inference (SPP Empirical Application)}
    \label{fig:empapp}
\end{figure}

Table \ref{tab:empapp} and Figure \ref{fig:empapp} report pointwise and uniform estimation and inference for the BATEC using the distance-based procedures and the SPP dataset. The empirical findings are broadly consistent across the alternative methods proposed in this paper and align closely with the results obtained using location-based approaches in \cite{Cattaneo-Titiunik-Yu_2026_JASA}. Overall, the evidence reveals heterogeneity in treatment effects along the assignment boundary: estimated effects are larger for students who are poorer and have lower academic achievement (evaluation points $\bb_{1}$--$\bb_{11}$), and smaller for students who are relatively wealthier and have higher academic achievement (evaluation points $\bb_{11}$--$\bb_{21}$).

\begin{table}[ht]
    \begin{subtable}{0.5\textwidth}
        \centering
        \resizebox{0.8\linewidth}{!}{\begin{tabular}{@{}crrrrrc@{}}
\toprule\toprule
\multicolumn{1}{c}{$\bb\in\B$} & \multicolumn{1}{c}{$h$} & \multicolumn{1}{c}{$N_{\mathrm{Co}}$} & \multicolumn{1}{c}{$N_{\mathrm{Tr}}$} & \multicolumn{1}{c}{$\tau_Y(\bb)$} & \multicolumn{1}{c}{p-value} & \multicolumn{1}{c}{95\% RBC CI} \\
\midrule
$\mathbf{b}_{1}$ & 32.762 & 18389 & 7264 & 0.320 & 0.000 & $(0.256,\, 0.500)$ \\
$\mathbf{b}_{2}$ & 27.994 & 12482 & 5807 & 0.334 & 0.000 & $(0.229,\, 0.513)$ \\
$\mathbf{b}_{3}$ & 28.491 & 13223 & 6043 & 0.329 & 0.000 & $(0.207,\, 0.481)$ \\
$\mathbf{b}_{4}$ & 31.169 & 16559 & 7026 & 0.324 & 0.000 & $(0.161,\, 0.411)$ \\
$\mathbf{b}_{5}$ & 34.582 & 22195 & 7670 & 0.333 & 0.000 & $(0.141,\, 0.370)$ \\
$\mathbf{b}_{6}$ & 33.580 & 21212 & 6980 & 0.347 & 0.000 & $(0.167,\, 0.394)$ \\
$\mathbf{b}_{7}$ & 35.367 & 24702 & 6882 & 0.363 & 0.000 & $(0.246,\, 0.451)$ \\
\bottomrule\bottomrule
\end{tabular}}
        \caption{$h=\widehat{h}_{\mathtt{MSE},\bb}$ (smooth $\B$)}
    \end{subtable}
    \hfill
    \begin{subtable}{0.5\textwidth}
        \centering
        \resizebox{0.83\linewidth}{!}{\begin{tabular}{@{}crrrrrc@{}}
\toprule\toprule
\multicolumn{1}{c}{$\bb\in\B$} & \multicolumn{1}{c}{$h$} & \multicolumn{1}{c}{$N_{\mathrm{Co}}$} & \multicolumn{1}{c}{$N_{\mathrm{Tr}}$} & \multicolumn{1}{c}{$\tau_Y(\bb)$} & \multicolumn{1}{c}{p-value} & \multicolumn{1}{c}{95\% RBC CI} \\
\midrule
$\mathbf{b}_{1}$ & 32.762 & 18389 & 7264 & 0.320 & 0.000 & $(0.253,\, 0.503)$ \\
$\mathbf{b}_{2}$ & 27.994 & 12482 & 5807 & 0.334 & 0.000 & $(0.225,\, 0.517)$ \\
$\mathbf{b}_{3}$ & 28.491 & 13223 & 6043 & 0.329 & 0.000 & $(0.203,\, 0.485)$ \\
$\mathbf{b}_{4}$ & 31.169 & 16559 & 7026 & 0.324 & 0.000 & $(0.158,\, 0.414)$ \\
$\mathbf{b}_{5}$ & 27.092 & 11736 & 5753 & 0.307 & 0.000 & $(0.053,\, 0.354)$ \\
$\mathbf{b}_{6}$ & 22.577 & 7601 & 4394 & 0.309 & 0.000 & $(0.070,\, 0.419)$ \\
$\mathbf{b}_{7}$ & 18.061 & 4559 & 3160 & 0.345 & 0.000 & $(0.255,\, 0.652)$ \\
\bottomrule\bottomrule
\end{tabular}}
        \caption{$h=\widehat{h}_{\mathtt{kink},\bb}(\B)$ (kink adaptive)}
    \end{subtable}

    \vspace{.25in}
    
    \begin{subtable}{0.5\textwidth}
        \centering
        \resizebox{0.8\linewidth}{!}{\begin{tabular}{@{}crrrrrc@{}}
\toprule\toprule
\multicolumn{1}{c}{$\bb\in\B$} & \multicolumn{1}{c}{$h$} & \multicolumn{1}{c}{$N_{\mathrm{Co}}$} & \multicolumn{1}{c}{$N_{\mathrm{Tr}}$} & \multicolumn{1}{c}{$\tau_Y(\bb)$} & \multicolumn{1}{c}{p-value} & \multicolumn{1}{c}{95\% RBC CI} \\
\midrule
$\mathbf{b}_{1}$ & 11.369 & 1536 & 1182 & 0.343 & 0.000 & $(0.157,\, 0.529)$ \\
$\mathbf{b}_{2}$ & 9.714 & 1071 & 893 & 0.306 & 0.000 & $(0.086,\, 0.525)$ \\
$\mathbf{b}_{3}$ & 9.887 & 1107 & 952 & 0.371 & 0.000 & $(0.163,\, 0.579)$ \\
$\mathbf{b}_{4}$ & 10.816 & 1373 & 1174 & 0.313 & 0.000 & $(0.124,\, 0.502)$ \\
$\mathbf{b}_{5}$ & 11.999 & 1776 & 1484 & 0.171 & 0.004 & $(-0.006,\, 0.348)$ \\
$\mathbf{b}_{6}$ & 11.651 & 1704 & 1507 & 0.275 & 0.000 & $(0.102,\, 0.448)$ \\
$\mathbf{b}_{7}$ & 12.270 & 1966 & 1691 & 0.409 & 0.000 & $(0.261,\, 0.558)$ \\
\bottomrule\bottomrule
\end{tabular}}
        \caption{$h=\widehat{\mathtt{C}} \cdot n^{-1/4}$ (unknown kink location)}
    \end{subtable}
    \hfill
    \begin{subtable}{0.5\textwidth}
        \centering
        \resizebox{0.8\linewidth}{!}{\begin{tabular}{@{}crrrrrc@{}}
\toprule\toprule
\multicolumn{1}{c}{$\bb\in\B$} & \multicolumn{1}{c}{$h$} & \multicolumn{1}{c}{$N_{\mathrm{Co}}$} & \multicolumn{1}{c}{$N_{\mathrm{Tr}}$} & \multicolumn{1}{c}{$\tau_Y(\bb)$} & \multicolumn{1}{c}{p-value} & \multicolumn{1}{c}{95\% RBC CI} \\
\midrule
$\mathbf{b}_{1}$ & 49.759 & 52499 & 12202 & 0.311 & 0.000 & $(0.245,\, 0.399)$ \\
$\mathbf{b}_{2}$ & 64.201 & 102394 & 13859 & 0.338 & 0.000 & $(0.247,\, 0.369)$ \\
$\mathbf{b}_{3}$ & 65.363 & 108053 & 13749 & 0.347 & 0.000 & $(0.266,\, 0.385)$ \\
$\mathbf{b}_{4}$ & 67.149 & 115026 & 13663 & 0.354 & 0.000 & $(0.286,\, 0.401)$ \\
$\mathbf{b}_{5}$ & 65.313 & 107722 & 13130 & 0.358 & 0.000 & $(0.293,\, 0.410)$ \\
$\mathbf{b}_{6}$ & 62.364 & 95870 & 12384 & 0.360 & 0.000 & $(0.302,\, 0.421)$ \\
$\mathbf{b}_{7}$ & 57.821 & 79327 & 11142 & 0.359 & 0.000 & $(0.308,\, 0.433)$ \\
\bottomrule\bottomrule
\end{tabular}}
        \caption{$h=\widehat{h}_{\mathtt{1d},\mathbf{b}}$ ($\mathtt{rdrobust} + \mathtt{rd2d.dist}$)}
    \end{subtable}
    \caption{BATEC Estimation and Inference (SPP Empirical Application)}
    \label{tab:empapp}
\end{table}

\section{Extensions}\label{sec: Extensions}

The theoretical results developed in the supplemental appendix primarily cover multidimensional location scores with $d\geq2$, recover the usual one-dimensional RD case for the results not tied to continuum boundary averaging, and also provide several theoretical and methodological extensions of our results. This section highlights three such extensions that are of practical interest.

\subsection{Aggregated Treatment Effects}

As discussed in \cite{Cattaneo-Titiunik-Yu_2026_JASA}, and references therein, two other interesting causal parameters are the \emph{Weighted Boundary Average Treatment Effect} (WBATE)
\begin{align*}
    \tau_{\mathtt{WBATE}}
    = \frac{\int_{\B}\tau(\bx)w(\bx)\dif\Haus^1(\bx)}
           {\int_{\B}w(\bx)\dif\Haus^1(\bx)}
\end{align*}
where $w:\B\to(0,\infty)$ is a bounded weight function and integration is with respect to one-dimensional Hausdorff measure,
and the \emph{Largest Boundary Average Treatment Effect} (LBATE)
\begin{align*}
    \tau_{\mathtt{LBATE}}
    = \sup_{\bx\in\B} \tau(\bx).
\end{align*}
Because Theorem \ref{thm: Identification} identifies the BATEC using distance-based methods, these functionals are identified as well, which justifies the plug-in estimators
\begin{align*}
    \widehat{\vartheta}_{\mathtt{WBATE}}
    =\frac{\int_{\B}\widehat{\vartheta}(\bx)w(\bx)\dif\Haus^1(\bx)}
          {\int_{\B}w(\bx)\dif\Haus^1(\bx)}
    \qquad\text{and}\qquad
    \widehat{\vartheta}_{\mathtt{LBATE}}
    = \sup_{\bx\in\B} \widehat{\vartheta}(\bx)
\end{align*}
of $\tau_{\mathtt{WBATE}}$ and $\tau_{\mathtt{LBATE}}$, respectively. Sections SA-4 and SA-5 in the supplemental appendix give estimation and inference results for distance-based methods for WBATE and LBATE.

\subsection{Imperfect Compliance}

Our theoretical framework can also accommodate imperfect compliance (fuzzy designs), where treatment assignment and treatment receipt may differ for some units. Using standard potential-outcomes notation \citep{Hernan-Robins_2020_Book}, for any $\bx\in\B$, let $W_i = \Indicator(D_i(\bx) \in \I_0) \cdot W_i(0) + \Indicator(D_i(\bx) \in \I_1) \cdot W_i(1)$ denote the observed treatment receipt, where $W_i(t)$ is the potential treatment receipt under assignment $t\in\{0,1\}$. The observed outcome is then $Y_i = \Indicator(D_i(\bx) \in \I_0) \cdot Y_i(0,W_i(0)) + \Indicator(D_i(\bx) \in \I_1) \cdot Y_i(1,W_i(1))$, where the potential outcome depends on both assignment and treatment receipt, that is, $Y_i(t,w)$ denotes the potential outcome for unit $i$ when this unit is assigned to treatment $t\in\{0,1\}$ and takes treatment receipt $w\in\{0,1\}$.

The usual fuzzy estimand and estimator are
\begin{align*}
    \zeta(\bx) = \frac{\tau_{Y}(\bx)}{\tau_{W}(\bx)}
    \qquad\text{and}\qquad
    \widehat{\xi}(\bx) = \frac{\widehat{\vartheta}_{Y}(\bx)}{\widehat{\vartheta}_{W}(\bx)},
\end{align*}
where, for each $\bx \in \B$, $\tau_{Y}(\bx) = \E[Y_i(1,W_i(1)) - Y_i(0,W_i(0)) | \bX_i = \bx]$ and $\tau_{W}(\bx) = \E[W_i(1) - W_i(0) | \bX_i = \bx]$, and $\widehat{\vartheta}_{Y}(\bx) = \be_0^{\top}\widehat{\bgamma}_{Y,1}(\bx) - \be_0^{\top}\widehat{\bgamma}_{Y,0}(\bx)$ and $\widehat{\vartheta}_{W}(\bx) = \be_0^{\top}\widehat{\bgamma}_{W,1}(\bx) - \be_0^{\top}\widehat{\bgamma}_{W,0}(\bx)$, where $\widehat{\bgamma}_{A,t}(\bx)$ denotes the local polynomial fit obtained using outcome variable $A\in\{Y,W\}$ and side $t\in\{0,1\}$ in \eqref{eq: locpoly estimator}. A causal interpretation of $\zeta(\bx)$ can be obtained under additional assumptions; see \cite{Arai-etal_2022_QE}. Section SA-6 in the supplemental appendix gives estimation and inference results for fuzzy BATEC, fuzzy WBATE, and fuzzy LBATE when employing distance-based methods.

\subsection{Pre-treatment Covariates}

We next discuss how to incorporate pre-treatment covariates either to improve efficiency \citep{Calonico-Cattaneo-Farrell-Titiunik_2019_RESTAT} or to study treatment-effect heterogeneity \citep{Calonico-Cattaneo-Farrell-Palomba-Titiunik_2026_wp}. Let $\bZ_1,\ldots,\bZ_n$ denote covariates of dimension $d_Z\ge1$.

For efficiency improvements, the covariate-adjusted estimator is
\begin{align*}
    \widetilde{\vartheta}(\bx) = \be_{p+2}^{\top}\widetilde{\bgamma}(\bx)
\end{align*}
with
\begin{align*}
    \widetilde{\bgamma}(\bx)
    = \argmin_{\bgamma \in \mathbb{R}^{2(p+1)+d_Z}}
      \En \Big[ \big(Y_i - \widetilde{\br}_p(D_i(\bx),T_i,\bZ_i)^\top \bgamma \big)^2
                K_h(D_i(\bx))\Big],
\end{align*}
where $T_i=\Indicator(D_i(\bx) \geq 0)$, and $\widetilde{\br}_p(D_i(\bx),T_i,\bZ_i) = [\br_p(D_i(\bx)/h)^{\top},T_i\cdot\br_p(D_i(\bx)/h)^{\top},\bZ_i^\top]^\top$ contains the full polynomial basis function for each treatment group but the pre-intervention covariates are not interacted with the treatment indicator; hence its dimension is $\mathfrak{p} = 2(p+1)+d_Z$.

For heterogeneity analysis, define
\begin{align*}
    \check{\kappa}(\bx,\bz)
    = \be_{p+2}^{\top}\check{\bgamma}(\bx)
    + \bz^\top
      \begin{bmatrix}
        \mathbf{0}_{d_Z \times (2+d_Z)(p+1)} & \mathbf{I}_{d_Z} & \mathbf{0}_{d_Z \times p d_Z}
      \end{bmatrix}\check{\bgamma}(\bx)
\end{align*}
with
\begin{align*}
    \check{\bgamma}(\bx)
    = \argmin_{\bgamma \in \mathbb{R}^{2(p+1)(d_Z+1)}}
      \En \Big[ \big(Y_i - \check{\br}_p(D_i(\bx),T_i,\bZ_i)^\top \bgamma \big)^2
                K_h(D_i(\bx))\Big],
\end{align*}
where $\mathbf{I}_d$ denotes the $(d\times d)$ identity matrix, $\mathbf{0}_{d_1 \times d_2}$ denotes the $(d_1 \times d_2)$ matrix of zeros, $\bz$ takes values on the support of $\bZ_i$, and $\check{\br}_p(D_i(\bx),T_i,\bZ_i) = [\br_p(D_i(\bx)/h)^{\top},T_i\cdot\br_p(D_i(\bx)/h)^{\top}, (\br_p(D_i(\bx)/h) \otimes \bZ_i)^\top, T_i \cdot (\br_p(D_i(\bx)/h) \otimes \bZ_i)^\top]^\top$ contains the full interaction between the polynomial basis function, the treatment assignment indicator, and the pre-intervention covariates.

Estimation and inference for covariate-adjusted distance-based methods can be established by leveraging the results in the supplemental appendix. A formal treatment is left for future work.

\section{Recommendations for Practice}\label{sec: Recommendations for Practice}

\cite{Cattaneo-Titiunik-Yu_2026_BookCh} review empirical applications of boundary discontinuity (BD) designs and document that researchers employ different empirical strategies depending on the objectives of the analysis, the assignment mechanism, and data availability. A useful taxonomy classifies methods according to how the multidimensional score $\bX_i$ is utilized. In practice, three levels of aggregation lead to three different types of methods.

\begin{itemize}

\item \textit{Location-based methods}. These approaches exploit the full multidimensional score $\bX_i$ and therefore extend canonical regression discontinuity techniques to multidimensional assignment settings. By preserving local variation in all score dimensions, they enable the most flexible analysis of treatment-effect heterogeneity along the assignment boundary.

\item \textit{Distance-based methods}. These approaches reduce the multidimensional score $\bX_i$ to a univariate signed distance score, constructed from a distance measure relative to each boundary point $\bx\in\B$. This transformation allows researchers to implement standard univariate RD procedures locally to each point on the boundary while retaining the ability to study heterogeneous treatment effects summarized by the BATEC.

\item \textit{Pooling-based methods}. These approaches further aggregate the multidimensional score by using the distance between each observation's location and the closest boundary point, $D_i=\inf_{\bx\in\B}\d(\bX_i,\bx)$, thereby collapsing all boundary locations into a single scalar running variable. As a result, pooling-based methods identify an aggregate, scalar causal parameter corresponding to a WBATE.

\end{itemize}

Location-based and distance-based methods both target the BATEC as the fundamental causal functional parameter. This function can subsequently be summarized through policy-relevant functionals, such as the Weighted Boundary Average Treatment Effect, or the Largest Boundary Average Treatment Effect, given by $\tau_{\mathtt{LBATE}}=\sup_{\bx\in\B}\tau(\bx)$. In contrast, pooling-based methods directly estimate an aggregated WBATE parameter by construction. Location-based procedures are studied in our companion paper \cite{Cattaneo-Titiunik-Yu_2026_JASA}, while formal analysis of pooling-based methods is the subject of ongoing research \citep{Cattaneo-Titiunik-Yu_2026_BDD-Pooling}.

From a practical perspective, location-based methods are generally preferred whenever reliable information on the multidimensional score $\bX_i$ is available, as they offer the richest framework for identification, estimation and inference of the key building block functional parameter, $(\tau(\bx):\bx\in\B)$, and functionals thereof. When such information is limited or when dimensionality considerations make location-based approaches difficult to implement, distance-based methods provide a tractable and robust alternative. In particular, they enable estimation and inference for the BATEC, and functionals thereof, using well-understood univariate RD techniques.

At the same time, practitioners should recognize that distance-based estimators may exhibit different bias behavior near geometrically irregular regions of the treatment-assignment boundary, such as kink points. Empirical implementation should therefore account for boundary geometry when selecting bandwidths or interpreting results, following the results presented in this paper. For example, researchers may restrict attention to locally smooth boundary segments or employ adaptive bandwidth strategies that vary with proximity to irregular points (e.g., our proposed bandwidth choice $\widehat{h}_{\mathtt{kink},\bx}(\B)$). Such choices may affect the causal estimand and should be transparently reported. Nevertheless, the numerical evidence presented in Section \ref{sec: Numerical Results} suggests that the proposed distance-based procedures can remain reasonably robust in empirically relevant configurations involving moderate boundary irregularities.

Finally, pooling-based methods remain popular in applied work because of their simplicity and ease of implementation. However, they provide the least informative causal analysis, as they aggregate potentially heterogeneous treatment effects into a single summary parameter. For this reason, we recommend that empirical researchers complement pooled estimates with analysis of the BATEC or related functionals whenever feasible, thereby obtaining a more comprehensive understanding of treatment-effect heterogeneity along the assignment boundary.

\section{Conclusion}\label{sec: Conclusion}

We studied the statistical properties of distance-based (isotropic) local polynomial estimation in boundary discontinuity designs. We established conditions for identification, estimation, and inference, both pointwise and uniformly along the treatment-assignment boundary. Our theoretical results highlight the central role played by the geometric regularity of the boundary—a one-dimensional manifold—along which estimation and inference are conducted. Building on these results, we provided concrete guidance for empirical implementation and illustrated the performance of the proposed methods using both simulated and real-world data. The companion general-purpose software package \texttt{rd2d} implements the main procedures developed in this paper \citep{Cattaneo-Titiunik-Yu_2025_rd2d}.

Motivated by Theorem \ref{thm: Approximation Bias: Smooth Boundary}, which shows that distance-based methods can exhibit improved misspecification bias when the assignment boundary is sufficiently smooth, we conjecture that new distance-based (isotropic) nonparametric smoothing procedures that (i) explicitly incorporate geometric information about the boundary (e.g., knowledge of the location of kink points in $\B$), or (ii) rely on regularization of the boundary (e.g., via smooth approximations of $\B$), could achieve smaller misspecification bias. For example, one simple approach is to restrict estimation to locally smooth segments of $\B$, thereby avoiding the bias issues highlighted by Theorem \ref{thm: Approximation Bias: Uniform Guarantee}, at the cost of altering the target estimand through an additional regularization bias.

\clearpage
\bibliography{CTY_2026_JOE--bib}
\bibliographystyle{ecta}


\end{document}


\pagenumbering{roman}
\title{Estimation and Inference in Boundary Discontinuity Designs: Distance-Based Methods\\ Supplemental Appendix\bigskip}
\author{Matias D. Cattaneo\thanks{Department of Operations Research and Financial Engineering, Princeton University.} \and
	    Rocio Titiunik\thanks{Department of Politics, Princeton University.} \and
	    Ruiqi (Rae) Yu\thanks{Department of Operations Research and Financial Engineering, Princeton University.} 
	    }
\maketitle

\begin{abstract}
    This supplemental appendix presents more general theoretical results encompassing those reported in the paper, their theoretical proofs, and other technical results. It also provides additional numerical results.
\end{abstract}

\thispagestyle{empty}
\clearpage

\onehalfspacing
\pagestyle{plain}

\setcounter{tocdepth}{2}
\setcounter{secnumdepth}{4}
\tableofcontents

\clearpage
\pagenumbering{arabic}

\section{Introduction}\label{sa-sec: Introduction}

This supplemental appendix considers four generalizations relative to the setup in the paper.
\begin{enumerate}
    \item The score variable $\bX_i$ is $d$-dimensional, with the genuinely
    multivariate boundary-discontinuity case corresponding to $d\geq2$.
    \item \emph{Weighted Boundary Average Treatment Effect} (WBATE).
    \item \emph{Largest Boundary Average Treatment Effect} (LBATE).
    \item Imperfect compliance (fuzzy) designs.
\end{enumerate}
In addition, this supplement presents technical results and proofs not covered in the
main text due to space limitations.

The assumptions in the paper are specialized primitive conditions; after imposing these restrictions on the
supplemental results, the mapping between this supplemental appendix and the paper is as
follows.
\begin{itemize}
    \item Theorem \ref{sa-thm: Identification} $\implies$ Theorem 1.
    \item Theorem \ref{sa-thm: Approximation Bias: Minimal Guarantee}, in its sharp level case with $d=2$, $\implies$ Theorem 2.
    \item Theorem \ref{sa-thm: Approximation Bias: Smooth Distance Derivatives}, with $d=2$, $\implies$ Theorem 3.
    \item Theorem \ref{sa-thm: Convergence Rates} with $d=2$ $\implies$ Theorem 4.
    \item Theorem \ref{sa-thm: Confidence Intervals} with $d=2$ $\implies$ Theorem 5(i).
    \item Theorem \ref{sa-thm: Confidence Bands} with $d=2$ $\implies$ Theorem 5(ii).
    \item Section \ref{sa-sec: Proof of Theorem 6} $\implies$ Theorem 6.
\end{itemize}

The generalizations considered in this supplemental appendix are conceptually
straightforward, but technically more involved.
Section \ref{sa-sec: BATEC} presents the main distance-based identification, estimation,
and inference results. Sections \ref{sa-sec: WBATE} and \ref{sa-sec: LBATE} present the
corresponding WBATE and LBATE results for sharp designs. Section \ref{sa-sec: Imperfect Compliance}
extends all the results to fuzzy designs. Section \ref{sa-sec: Gaussian Strong Approximation} presents a
new Gaussian strong approximation result that may be of independent interest.

\subsection{Notation and Definitions}

For textbook references on empirical process, see
\cite{van-der-Vaart-Wellner_1996_Book}, \cite{dudley2014uniform}, and
\cite{Gine-Nickl_2016_Book}. For textbook references on geometric measure theory, see
\cite{simon1984lectures}, \cite{federer2014geometric}, and
\cite{folland2002advanced}.
Throughout the appendix, $\bx$ denotes a generic point in $\X$ and, when $\bx\in\B$, a generic boundary point; the main paper uses $\mathbf{b}$ and $\mathbf{b}_j$ for selected or discretized evaluation points on the boundary.

\begin{enumerate}[label=(\roman*)]
    \item \textit{Polynomial Notation}. For scalar $u\in\mathbb{R}$,
    $\rp(u)=(1,u,\ldots,u^p)^\top$, and $\be_0$ denotes the intercept selector
    satisfying $\be_0^\top\rp(u)=1$. More generally, $\be_j$ denotes the selector
    satisfying $\be_j^\top\rp(u)=u^j$ for $j=0,\ldots,p$.

    \item \textit{Norms}. For a vector $\ba \in \reals^k$,
    $\norm{\ba} = (\sum_{i = 1}^k \ba_i^2)^{1/2}$ and
    $\infnorm{\ba} = \max_{1 \leq i \leq k}|\ba_i|$. For a matrix
    $\bA \in \reals^{m \times n}$, $\norm{\bA}_p = \sup_{\norm{\bx}_p = 1} \norm{\bA\bx}_p$,
    $p \in \mathbb{N} \cup \{\infty\}$, and $\lambda_{\min}(\bA)$ denotes its minimum
    eigenvalue when defined. For a function $f$ on a metric space $(S,d)$,
    $\infnorm{f}=\sup_{s\in S}|f(s)|$, and $\supp(f)$ denotes the closure of
    $\{s\in S:f(s)\neq0\}$.
    For a probability measure $Q$ on $(\mathcal{S}, \mathscr{S})$ and $p \geq 1$, define
    $\norm{f}_{Q,p} = (\int_{\mathcal{S}} |f|^p \dif Q)^{1/p}$, and
    $Q(f) = \int f \dif Q$. For a set $E \subseteq \reals^d$, $\mathfrak{m}(E)$ denotes the
    Lebesgue measure of $E$.

    \item \textit{Empirical Process}. We use standard empirical-process notation:
    $\En[g(\bv_i)] = n^{-1} \sum_{i = 1}^n g(\bv_i)$ and
    $\gn[g(\bv_i)] = n^{-1/2} \sum_{i = 1}^n (g(\bv_i) - \E[g(\bv_i)])$. Let
    $(\mathcal{S},d)$ be a semi-metric space. The covering number
    $N(\mathcal{S}, d, \varepsilon)$ is the minimal number of open $d$-balls of radius
    $\varepsilon$ needed to cover $\mathcal{S}$. A \emph{$\P$-Brownian bridge} is a
    mean-zero Gaussian random function $W_{\P}(f)$, $f \in L_2(\X, \P)$, with covariance
    $\E[W_{\P}(f)W_{\P}(g)] = \P(fg) - \P(f)\P(g)$. A class
    $\mathcal{F} \subseteq L_2(\X, \P)$ is \emph{$\P$-pregaussian} if there is a version
    of $W_{\P}$ such that $W_{\P} \in C(\mathcal{F}; \rho_{\P})$ almost surely, where
    $\rho_{\P}(f, g) = (\|f - g\|_{\P,2}^2 - (\P f - \P g)^2)^{1/2}$.

    \item \textit{Geometric Measure Theory}. For a set $E \subseteq \X$, the
    \emph{De Giorgi perimeter of $E$ relative to $\X$} is
    $\mathcal{L}(E) = \ttTV_{\{\Indicator_{E}\},\X}$. For
    $d \in \mathbb{N}$ and $0 \leq m \leq d$, $\Haus^m$ denotes $m$-dimensional
    Hausdorff measure, and integration against $\Haus^m$ is understood in the standard
    Caratheodory construction. The Hausdorff dimension of $A$ is denoted
    $\dim_{\Haus}(A)$. A set $A \subseteq \reals^d$ is $k$-rectifiable if it has
    Hausdorff dimension $k$ and is covered up to a $\Haus^k$-null set by countably many
    continuously differentiable maps from $\reals^k$ to $\reals^d$. A set
    $C\subseteq\reals^d$ is a \emph{rectifiable curve} if there exists a Lipschitz
    continuous function $\gamma:[0,1]\to\reals^d$ such that $C=\gamma([0,1])$.

    \item \textit{Bounds and Asymptotics}. For real sequences, $a_n=o(b_n)$ means
    $a_n/b_n\to0$, and $a_n\lesssim b_n$ means $|a_n|\leq C|b_n|$ eventually for a
    finite constant $C$. For random sequences, $a_n=o_{\P}(b_n)$ means
    $a_n/b_n\to_{\P}0$, and $|a_n|\lessp |b_n|$ means
    $\limsup_{M \to \infty} \limsup_{n \to \infty}
    \P[|a_n/b_n| \geq M] = 0$.

    \item \textit{Distributions and Statistical Distances}. For
    $\boldsymbol{\mu} \in \reals^k$ and a $k \times k$ positive definite matrix
    $\boldsymbol{\Sigma}$, $\mathsf{Normal}(\boldsymbol{\mu}, \boldsymbol{\Sigma})$
    denotes the Gaussian distribution with mean $\boldsymbol{\mu}$ and covariance
    $\boldsymbol{\Sigma}$. $\mathsf{Uniform}([a,b])$ denotes the uniform distribution on
    $[a,b]$, $\mathsf{Bernoulli}(p)$ denotes the Bernoulli distribution with success
    probability $p$, and $\Phi(\cdot)$ denotes the standard Gaussian cumulative
    distribution function. For two distributions $P$ and $Q$,
    $d_{\operatorname{KL}}(P,Q)$ denotes the KL-distance and $d_{\chi^2}(P,Q)$ denotes
    the $\chi^2$ distance.
\end{enumerate}

Throughout the appendix, $d$ and the local-polynomial order $p$ are fixed, with
$p\ge0$. All limits are taken with $h\to0$ as $n\to\infty$. The genuinely multivariate
boundary-discontinuity results are stated for $d\ge2$; the BATEC results (pointwise and
uniform over finite boundary sets), the LBATE results, and their fuzzy analogues recover
the usual one-dimensional RD case when $d=1$ under the analogous assumptions,
interpreting $\B$ as a finite $0$-rectifiable boundary and integrating with respect to
$\Haus^0$. Most of our results hold with fixed but small enough bandwidths; we do not
make this distinction explicit to avoid overly complex statements and technical
discussions.

\section{Setup}

We partition $\mathcal{X}$ into two assignment areas,
$\mathcal{A}_t\subset\mathbb{R}^d$ with $t\in\{0,1\}$, representing the control and
treatment regions, respectively. That is, $\mathcal{X}=\mathcal{A}_0\cup\mathcal{A}_1$,
where $\mathcal{A}_0$ and $\mathcal{A}_1$ are disjoint regions in $\mathbb{R}^d$. The
observed outcome is
$Y_i = \Indicator(\bX_i \in \mathcal{A}_0)Y_i(0)+\Indicator(\bX_i \in \mathcal{A}_1)Y_i(1)$.
The boundary is
$\mathcal{B}=\setbd(\mathcal{A}_0)\cap\setbd(\mathcal{A}_1)\subseteq\mathbb{R}^d$,
where $\setbd(\cdot)$ denotes the topological boundary of a set, and has effective
dimension $d-1$. Without loss of generality, we assume that $\mathcal{B}$ belongs to the
treatment region, that is, $\mathcal{B}\subseteq\mathcal{A}_1$ and
$\mathcal{B}\cap\mathcal{A}_0=\emptyset$.

The univariate distance score induced by the multidimensional score is
\begin{align*}
    D_i(\bx)
    = [ \Indicator(\bX_i \in \mathcal{A}_1) - \Indicator(\bX_i \in \mathcal{A}_0)] \d(\bX_i,\bx),
    \qquad \bx \in \B,
\end{align*}
where $\d(\cdot,\cdot)$ denotes a distance function. For a deterministic point
$\bz\in\X$, write
$D_{\bz}(\bx)=[\Indicator(\bz \in \mathcal{A}_1)-\Indicator(\bz \in \mathcal{A}_0)]\d(\bz,\bx)$.
Let $\I_0 = (-\infty,0)$, $\I_1 = [0,\infty)$, and $\varsigma_t=2t-1$ for
$t\in\{0,1\}$. Thus, for $\bz\in\mathcal A_t$,
$D_{\bz}(\bx)=\varsigma_t\d(\bz,\bx)$.
Let $\rp(u)=(1,u,\cdots,u^p)^\top$, and $K_h(u)=K(u/h)/h^d$, where $K(\cdot)$ is a
univariate kernel and $h$ is a bandwidth parameter.
No symmetry of $K$ is imposed, so the side-specific sign remains in both the polynomial
argument and the kernel weight.

For later reference and for Assumption~\ref{sa-assump: Kernel, Distance and Boundary},
let
\begin{align*}
    \bPsi_{t,\bx}
    = \E \Big[\rp\Big(\frac{D_i(\bx)}{h}\Big)
               \rp\Big(\frac{D_i(\bx)}{h}\Big)^{\top}
               K_h(D_i(\bx))\Indicator(D_i(\bx)\in\I_t)\Big],
    \qquad t\in\{0,1\}.
\end{align*}

\subsection{Assumptions}

The data generating process assumption from the main paper generalizes as follows.

\begin{assumption}[Data Generating Process]\label{sa-assump: DGP}
    Let $t\in\{0,1\}$.
    \begin{enumerate}[label=\normalfont(\roman*),noitemsep,leftmargin=*]
    
    \item $(Y_1(t), \bX_1^\top)^\top,\ldots, (Y_n(t), \bX_n^\top)^\top$ are independent and identically distributed random vectors.
    
    \item The distribution of $\bX_i$ has compact support $\X\subseteq\mathbb{R}^d$ with nonempty interior and a Lebesgue density $f_X(\bx)$ that is continuous and bounded away from zero on $\X$.

    \item $\mu_t(\bx) = \E[Y_i(t)| \bX_i = \bx]$ has a $(p+1)$-times continuously differentiable extension to an open neighborhood of $\X$.
    
    \item $\sigma^2_t(\bx) = \Var[Y_i(t)|\bX_i = \bx]$ is bounded away from zero and continuous on $\X$.
    
    \item $\sup_{\bx \in \X}\E[|Y_i(t)|^{2+v}|\bX_i = \bx] < \infty$ for some $v \geq 2$.
    \end{enumerate}
\end{assumption}

The boundary, distance, and kernel assumption from the main paper generalizes as follows.

\begin{assumption}[Boundary, Distance, and Kernel]\label{sa-assump: Kernel, Distance and Boundary}
Let $t \in \{0,1\}$.
\vspace{-.1in}
\begin{enumerate}[label=\normalfont(\roman*),noitemsep,leftmargin=*]
    \item $\B\subset\setint(\X)$, and $\B$ is compact $(d-1)$-rectifiable, with $\Haus^{d-1}(\B)$ positive and finite.
    \item $\d: \reals^d \times \reals^d \to \reals_+$ is a metric on $\reals^d$ equivalent to the Euclidean distance, that is, there exist positive constants $C_u$ and $C_l$ such that $C_l \norm{\bx - \bx'} \leq \d(\bx,\bx') \leq C_u\norm{\bx - \bx'}$ for all $\bx,\bx' \in \X$.
    \item $K: \mathbb{R} \to [0,\infty)$ is either compactly supported and Lipschitz continuous, or $K(u)=\Indicator(u \in [-1,1])$ and the distance balls $\{\bv\in\X:\d(\bv,\bx)\le r\}$, indexed by $(\bx,r)\in\X\times\reals_+$, form a VC class.
    \item $\liminf_{h\downarrow0}\inf_{\bx \in \B} \lambda_{\min}(\bPsi_{t,\bx}) \gtrsim 1$.
    \item For each $\bx\in\B$ and $t\in\{0,1\}$, the denominator in the Hausdorff-measure representation of $\theta_{t,\bx}(r)$ below is positive and finite for all sufficiently small $|r|>0$ with $r\in\I_t$.
\end{enumerate}
\end{assumption}

For $d=2$, Assumption~\ref{sa-assump: Kernel, Distance and Boundary}(i) holds whenever
$\B$ is a compact rectifiable curve contained in the interior of $\X$, with positive and
finite length. Assumption~\ref{sa-assump: Kernel, Distance and Boundary}(ii) is specific
to distance-based methods and requires the distance function to be equivalent, up to
constants, to the Euclidean distance. In
Assumption~\ref{sa-assump: Kernel, Distance and Boundary}(iii), the extra VC condition
appears only for the uniform kernel because its indexing sets are distance balls rather
than rectangles: Lemma~\ref{sa-lem: VC distance balls} gives primitive cases where it is
automatic, including the Euclidean distance.

Assumption~\ref{sa-assump: Kernel, Distance and Boundary}(iv) rules out
boundary--distance--kernel configurations in which, after local rescaling by $h$, one side
of the boundary receives too little kernel-weighted mass to guarantee that the
distance-based local polynomial regression is well-defined uniformly over $\B$. Thus, the
condition ensures that the local distance design is sufficiently rich on each side of the boundary.
Lemma~\ref{sa-lem: Gram invert} gives primitive sufficient conditions for the
Euclidean-distance case, and Lemma~\ref{sa-lem: Gram} then shows that the sample Gram
matrix is full rank in large samples. Assumption~\ref{sa-assump: Kernel, Distance and Boundary}(v)
is specific to distance-based identification: it keeps the Hausdorff-measure ratio
defining $\theta_{t,\bx}(r)$ well-defined near the boundary and is used for identification
(Theorem~\ref{sa-thm: Identification}); under additional smoothness, the same ratio is
differentiated in Theorem~\ref{sa-thm: Approximation Bias: Smooth Distance Derivatives}.
These conditions are stated uniformly over
$\B$ in anticipation of our uniform estimation and inference results, but several can be
relaxed to hold only pointwise in $\bx\in\B$ for pointwise results.

Assumption 3 imposes regularity conditions on a weight function along the boundary.

\begin{assumption}[Weight Function and Boundary]\label{sa-assump: Weight Function and Boundary}
    Let $w:\B\to(0,\infty)$ satisfy $\sup_{\bx\in\B}w(\bx)<\infty$,
    $\inf_{\bx\in\B}w(\bx)>0$, and
    $\int_{\B}w(\bx)\dif\Haus^{d-1}(\bx)<\infty$.
\end{assumption}

\subsection{Parameters}

The \emph{Boundary Average Treatment Effect Curve} (BATEC) is
\begin{align*}
    \tau(\bx) = \E[Y_i(1) - Y_i(0)|\bX_i = \bx], \qquad \bx\in\B.
\end{align*}

For each $t \in \{0,1\}$, the induced conditional expectation based on univariate distance is 
\begin{align*}
    \theta_{t,\bx}(r)
    = \E[Y_i| D_i(\bx) = r ]
    = \E[Y_i|\d(\bX_i,\bx) = |r|, \bX_i \in \mathcal{A}_t],
    \qquad r \in \I_t, \quad \bx \in \mathcal{B}.
\end{align*}
More rigorously, for each $t \in \{0,1\}$, and letting $S_{t,\bx}(r) = \{\bv \in \X: \d(\bv,\bx) = r, \bv \in \mathcal{A}_t\}$ for $r \geq 0$ and $\bx \in \B$,
\begin{align*}
    \theta_{t,\bx}(r)
    = \frac{\int_{S_{t,\bx}(|r|)} \mu_t(\bv) f_X(\bv) \dif\Haus^{d-1}(\bv)}
           {\int_{S_{t,\bx}(|r|)} f_X(\bv) \dif\Haus^{d-1}(\bv)},
\end{align*}
for $|r| > 0, \bx \in \B, t \in \{0,1\}$, and therefore (under our assumptions)
\begin{align*}
    \theta_{t,\bx}(0)
    = \lim_{r \to 0,\; r\in\I_t}\E[Y_i|\d(\bX_i,\bx) = |r|, \bX_i \in \mathcal{A}_t]
    = \lim_{r \to 0,\; r\in\I_t}\frac{\int_{S_{t,\bx}(|r|)} \mu_t(\bv) f_X(\bv) \dif\Haus^{d-1}(\bv)}
                         {\int_{S_{t,\bx}(|r|)} f_X(\bv) \dif\Haus^{d-1}(\bv)}.
\end{align*}
The induced signed-distance treatment-effect parameter is
\begin{align*}
    \vartheta(\bx)=\theta_{1,\bx}(0)-\theta_{0,\bx}(0),\qquad \bx\in\B.
\end{align*}
Theorem \ref{sa-thm: Identification} shows that $\vartheta(\bx) = \tau(\bx)$ under Assumptions \ref{sa-assump: DGP}(i)-(iii) and \ref{sa-assump: Kernel, Distance and Boundary}(i), (ii), and (v).

We also consider the parameters
\begin{align*}
    \vartheta_{\mathtt{WBATE}}
    = \frac{\int_{\B}\vartheta(\bx)w(\bx)\dif\Haus^{d-1}(\bx)}
           {\int_{\B}w(\bx)\dif\Haus^{d-1}(\bx)}
    \qquad\text{and}\qquad
    \vartheta_{\mathtt{LBATE}}
    = \sup_{\bx\in\B}\vartheta(\bx).
\end{align*}
Theorem~\ref{sa-thm: Identification} implies that
\begin{align*}
    \vartheta_{\mathtt{WBATE}}
    = \tau_{\mathtt{WBATE}}
    = \frac{\int_{\B}\tau(\bx)w(\bx)\dif\Haus^{d-1}(\bx)}
           {\int_{\B}w(\bx)\dif\Haus^{d-1}(\bx)}
\end{align*}
is the \emph{Weighted Boundary Average Treatment Effect} (WBATE), and
\begin{align*}
    \vartheta_{\mathtt{LBATE}}
    = \tau_{\mathtt{LBATE}}
    = \sup_{\bx\in\B} \tau(\bx)
\end{align*}
is the \emph{Largest Boundary Average Treatment Effect} (LBATE).

\subsection{Estimators and Other Related Quantities}

The distance-based estimator process along the boundary is
\begin{align*}
    \widehat{\vartheta}(\bx)
    = \widehat{\theta}_{1,\bx}(0) - \widehat{\theta}_{0,\bx}(0),
    \qquad \bx\in\B,
\end{align*}
where, for $t \in \{0,1\}$,
\begin{align*}
    \widehat{\theta}_{t,\bx}(0) = \be_0^{\top} \widehat{\bgamma}_t(\bx),
    \qquad
    \widehat{\bgamma}_{t}(\bx) = \argmin_{\bgamma \in \reals^{p+1}} \En \Big[ \big(Y_i - \rp(D_i(\bx)/h)^{\top} \bgamma \big)^2 K_h(D_i(\bx)) \Indicator(D_i(\bx)\in\I_t) \Big].
\end{align*}
More generally, the least squares projection is
\begin{align*}
    \widehat{\theta}_{t,\bx}(r) = \rp(r/h)^{\top}\widehat{\bgamma}_{t}(\bx),
    \qquad t\in\{0,1\}, \quad \bx \in \B.
\end{align*}

The best mean square approximation is
\begin{align*}
    \theta_{t,\bx}^{\ast}(r) =  \rp(r/h)^{\top}\bgamma^{\ast}_t(\bx),
\end{align*}
where
\begin{align*}
     \bgamma^{\ast}_{t}(\bx) = \argmin_{\bgamma \in \reals^{p+1}} \E \Big[ \big(Y_i - \rp(D_i(\bx)/h)^{\top} \bgamma \big)^2 K_h(D_i(\bx)) \Indicator(D_i(\bx) \in \I_t) \Big],
\end{align*}
and uniqueness will follow from the results below. Let
\begin{align*}
    \theta_{t,\bx}^{\ast}(0)=\be_0^{\top}\bgamma_t^\ast(\bx).
\end{align*}
The estimation error decomposes into \emph{linear error}, \emph{approximation error}, and \emph{non-linear error}: for all $t\in\{0,1\}$ and $\bx \in \B$,
\begin{align*}
    \widehat{\theta}_{t,\bx}(0) - \theta_{t,\bx}(0)
    & = \be_0^{\top}  \widehat{\bPsi}_{t,\bx}^{-1} \En \Big[\rp\Big(\frac{D_i(\bx)}{h}\Big) K_h(D_i(\bx)) Y_i\Indicator(D_i(\bx) \in \I_t)\Big] - \theta_{t,\bx}(0) \nonumber \\
    & = \be_0^{\top}  \widehat{\bPsi}_{t,\bx}^{-1} \En \Big[\rp\Big(\frac{D_i(\bx)}{h}\Big) K_h(D_i(\bx)) (Y_i - \theta_{t,\bx}^{\ast}(D_i(\bx)))\Indicator(D_i(\bx) \in \I_t)\Big] \nonumber \\
    & \qquad + \theta_{t,\bx}^{\ast}(0)  - \theta_{t,\bx}(0) \nonumber \\
    & = \underbrace{\theta_{t,\bx}^{\ast}(0)  - \theta_{t,\bx}(0)}_{\text{approximation error}}
      + \underbrace{\be_0^{\top} \bPsi_{t,\bx}^{-1} \mathbf{O}_{t,\bx}}_{\text{linear error}}
      + \underbrace{\be_0^{\top} (\widehat{\bPsi}_{t,\bx}^{-1} - \bPsi_{t,\bx}^{-1}) \mathbf{O}_{t,\bx}}_{\text{non-linear error}},
\end{align*}
where 
\begin{align*}
    \mathbf{O}_{t,\bx}
    = \En \Big[\rp\Big(\frac{D_i(\bx)}{h}\Big) K_h(D_i(\bx)) (Y_i - \theta_{t,\bx}^{\ast}(D_i(\bx)))\Indicator(D_i(\bx) \in \I_t)\Big],
\end{align*}
with $\bPsi_{t,\bx}$ defined above and
\begin{align*}
    \widehat{\bPsi}_{t,\bx}
    = \En \Big[\rp\Big(\frac{D_i(\bx)}{h}\Big) \rp\Big(\frac{D_i(\bx)}{h}\Big)^{\top} K_h(D_i(\bx)) \Indicator(D_i(\bx) \in \I_t)\Big].
\end{align*}

Therefore,
\begin{align}\label{sa-eq: distance error decomposition}
    \widehat{\vartheta}(\bx)-\vartheta(\bx)
    = \bias(\bx)+\mathfrak{L}(\bx)+\mathfrak{Q}(\bx).
\end{align}
The misspecification bias is
\begin{align*}
    \bias(\bx) = \bias_{1}(\bx) - \bias_{0}(\bx),
    \qquad
    \bias_{t}(\bx) = \theta_{t,\bx}^{\ast}(0) - \theta_{t,\bx}(0),
\end{align*}
the linear term is
\begin{align*}
    \mathfrak{L}(\bx)
    = \be_0^{\top}\bPsi_{1,\bx}^{-1}\mathbf{O}_{1,\bx}
    - \be_0^{\top}\bPsi_{0,\bx}^{-1}\mathbf{O}_{0,\bx},
\end{align*}
and the quadratic (non-linear) term is
\begin{align*}
    \mathfrak{Q}(\bx)
    = \be_0^{\top}(\widehat{\bPsi}_{1,\bx}^{-1}-\bPsi_{1,\bx}^{-1})\mathbf{O}_{1,\bx}
    - \be_0^{\top}(\widehat{\bPsi}_{0,\bx}^{-1}-\bPsi_{0,\bx}^{-1})\mathbf{O}_{0,\bx}.
\end{align*}
The population projection defining $\theta_{t,\bx}^{\ast}$ implies
$\E[\mathfrak{L}(\bx)]=0$ and $\E[\mathfrak{L}(\bx)^2]=\Xi_{\bx} = \Xi_{\bx,\bx}$, where
\begin{align*}
    \Xi_{\bx_1,\bx_2} = \Xi_{0,\bx_1,\bx_2} + \Xi_{1,\bx_1,\bx_2},
    \qquad
    \Xi_{t,\bx_1,\bx_2} = \frac{1}{n h^{d}} \be_0^{\top}  \bPsi_{t,\bx_1}^{-1} \bUpsilon_{t,\bx_1,\bx_2} \bPsi_{t,\bx_2}^{-1} \be_0,
\end{align*}
and
\begin{align*}
    \bUpsilon_{t, \bx_1,\bx_2}
    &= h^d \E  \Big[\rp \Big(\frac{D_i(\bx_1)}{h}\Big) \rp \Big( \frac{D_i(\bx_2)}{h}\Big)^{\top}
                   K_h(D_i(\bx_1)) K_h(D_i(\bx_2))\\
    & \hspace{1in} (Y_i - \theta^*_{t,\bx_1}(D_i(\bx_1)))(Y_i - \theta^*_{t,\bx_2}(D_i(\bx_2)))
      \Indicator(D_i(\bx_1)\in\I_t)\Indicator(D_i(\bx_2)\in\I_t)\Big].
\end{align*}

Finally, for $t \in \{0,1\}$, $\bx_1, \bx_2 \in \B$, the sample analogues are $\widehat{\Xi}_{\bx} = \widehat{\Xi}_{\bx,\bx}$, where
\begin{align*}
    \widehat{\Xi}_{\bx_1,\bx_2} = \widehat{\Xi}_{0,\bx_1,\bx_2} + \widehat{\Xi}_{1,\bx_1,\bx_2},
    \qquad
    \widehat{\Xi}_{t,\bx_1,\bx_2}
    = \frac{1}{n h^{d}} \be_0^{\top} \widehat{\bPsi}_{t,\bx_1}^{-1} \widehat{\bUpsilon}_{t,\bx_1,\bx_2} \widehat{\bPsi}_{t,\bx_2}^{-1} \be_0,
\end{align*}
and
\begin{align*}
    \widehat{\bUpsilon}_{t,\bx_1,\bx_2}
    &= h^d \En \Big[\rp \Big(\frac{D_i(\bx_1)}{h}\Big) \rp \Big( \frac{D_i(\bx_2)}{h}\Big)^{\top}
                   K_h(D_i(\bx_1)) K_h(D_i(\bx_2))\\
    & \hspace{1in} \cdot(Y_i - \widehat{\theta}_{t,\bx_1}(D_i(\bx_1)))(Y_i - \widehat{\theta}_{t,\bx_2}(D_i(\bx_2)))
      \Indicator(D_i(\bx_1)\in\I_t)\Indicator(D_i(\bx_2)\in\I_t)\Big].
\end{align*}

\subsection{Higher-Order Signed-Distance Derivatives}

The main results in this appendix focus on the level parameter $\vartheta(\bx)$ because it
equals the BATEC $\tau(\bx)$ under Theorem~\ref{sa-thm: Identification}. One can also
estimate derivatives of the induced signed-distance regression. For $\bx\in\B$ and an
integer $\nu\in\{1,\ldots,p\}$, whenever the one-sided limits exist, define
\begin{align*}
    \theta_{t,\bx}^{(\nu)}(0)
    =
    \lim_{r\to0,\; r\in\I_t}\frac{d^\nu}{dr^\nu}\theta_{t,\bx}(r),
    \qquad
    \vartheta^{(\nu)}(\bx)
    =
    \theta_{1,\bx}^{(\nu)}(0)-\theta_{0,\bx}^{(\nu)}(0),
    \qquad \bx\in\B.
\end{align*}
The corresponding local-polynomial estimator replaces the level selector by the
derivative selector:
\begin{align*}
    \widehat{\vartheta}^{(\nu)}(\bx)
    =
    \nu!h^{-\nu}
    \big(\be_{\nu}^{\top}\widehat{\bgamma}_1(\bx)
    -\be_{\nu}^{\top}\widehat{\bgamma}_0(\bx)\big),
    \qquad \bx\in\B.
\end{align*}
All level results above have direct analogues after replacing $\be_0$ by
$\nu!h^{-\nu}\be_{\nu}$, replacing the variance rate $(nh^d)^{-1}$ by
$(nh^{d+2\nu})^{-1}$, and replacing the smooth-bias rate $h^m$ by $h^{m-\nu}$ when
the induced functions $r\mapsto\theta_{t,\bx}(r)$ are $m$ times differentiable. We do not
state those analogues theorem by theorem because their notation is substantially heavier
and the resulting parameter has a narrower causal interpretation: for $\nu>0$,
$\vartheta^{(\nu)}(\bx)$ is a derivative of an induced signed-distance regression, not a
derivative of the BATEC curve as a function on $\B$.

There is, however, one useful first-derivative interpretation for $\bx\in\B$ under a
smooth Euclidean geometry. Suppose $\d$ is Euclidean, $\B$ is $C^2$ near $\bx$, the treatment region has a
well-defined unit normal $\bn(\bx)$ pointing from control to treatment, meaning that
$\bx+r\bn(\bx)\in\A_1$ and $\bx-r\bn(\bx)\in\A_0$ for all sufficiently small $r>0$,
and $f_X$, $\mu_0$, and $\mu_1$ are continuously differentiable near $\bx$ with
$f_X(\bx)>0$, so that the treatment-effect function $\tau$ admits the ambient gradient
$\nabla\tau(\bx)$. Then,
\begin{align*}
    \vartheta^{(1)}(\bx)
    =
    c_d\, \nabla\tau(\bx)^\top \bn(\bx),
    \qquad
    c_d
    =
    \frac{\Gamma(d/2)}{\sqrt{\pi}\Gamma((d+1)/2)},
    \qquad \bx\in\B.
\end{align*}
In the bivariate case, $c_2=2/\pi$. Thus, the first signed-distance derivative is
proportional to the normal derivative of the treatment-effect surface. This formula uses
the Euclidean metric and the local half-sphere geometry it induces; with a different
metric, the relevant average is over the corresponding metric sphere and need not reduce
to a Euclidean normal derivative.

\subsection{Preliminary Lemmas}

Recall that $t \in \{0,1\}$.

The following lemma gives a sufficient condition for Assumption~\ref{sa-assump: Kernel, Distance and Boundary}(iii) when using the uniform kernel.

\begin{lem}[Primitive VC Condition for Distance Balls]\label{sa-lem: VC distance balls}
Let $\mathscr{D}_{\d}=\{\{\bv\in\X:\d(\bv,\bx)\le r\}:(\bx,r)\in\X\times\reals_+\}$. Suppose that, for some integer $J<\infty$, there are finite-dimensional linear spaces $\mathcal{V}_1,\ldots,\mathcal{V}_J$ of real-valued functions on $\X$ and a Boolean map $\mathsf{b}:\{0,1\}^J\to\{0,1\}$ such that, for every $(\bx,r)\in\X\times\reals_+$, there are $f_j(\cdot;\bx,r)\in\mathcal{V}_j$, $j=1,\ldots,J$, satisfying
\begin{align*}
    \Indicator\{\d(\bv,\bx)\le r\}
    =
    \mathsf{b}\left(
    \Indicator\{f_1(\bv;\bx,r)\ge0\},\ldots,
    \Indicator\{f_J(\bv;\bx,r)\ge0\}
    \right),
    \qquad \bv\in\X.
\end{align*}
Then, $\mathscr{D}_{\d}$ is a VC class.
\end{lem}

In particular, the conclusion of Lemma \ref{sa-lem: VC distance balls} holds whenever the distance balls are semialgebraic sets of uniformly bounded degree and Boolean complexity. This includes the Euclidean distance, fixed Mahalanobis distances $\d(\bv,\bx)=((\bv-\bx)^\top A(\bv-\bx))^{1/2}$ with $A$ positive definite, and distances generated by fixed polyhedral norms.

The following lemma gives a sufficient condition for Assumption~\ref{sa-assump: Kernel, Distance and Boundary}(iv).

\begin{lem}[Gram Invertibility]\label{sa-lem: Gram invert}
    Suppose the following conditions hold:
    \begin{enumerate}[label=\normalfont(\alph*),noitemsep,leftmargin=*]
        \item Assumptions \ref{sa-assump: DGP}(i)(ii) and Assumption~\ref{sa-assump: Kernel, Distance and Boundary}(iii) hold.
        \item $\d(\cdot,\cdot)$ is the Euclidean distance.
        \item There exists a compact set $U \subseteq \reals^d$, such that $K(\varsigma_t\lVert \bu \rVert)>0$ for all $\bu \in U$, $\lambda_{\min} (\int_U \rp(\lVert \bz \rVert) \rp(\lVert \bz \rVert)^{\top} \dif\bz) > 0$, and $\liminf_{h \downarrow 0}\inf_{\bx \in \B} \int_{U} K(\varsigma_t\lVert \bu \rVert) \Indicator(\bx + h \bu \in \A_t) \dif\bu \gtrsim 1$.
    \end{enumerate}
    Then, Assumption~\ref{sa-assump: Kernel, Distance and Boundary}(iv) holds.
\end{lem}

\begin{lem}[Gram]\label{sa-lem: Gram}
    Suppose Assumptions \ref{sa-assump: DGP}(i)(ii) and \ref{sa-assump: Kernel, Distance and Boundary} hold. If $\frac{n h^d}{\log (1/h)} \to \infty$ and $h=o(1)$, then
    \begin{align*}
        \sup_{\bx \in \B} \big\|\widehat{\bPsi}_{t,\bx} - \bPsi_{t,\bx}\big\|
        & \lessp \sqrt{\frac{\log (1/h)}{n h^d}},
        \qquad
        1 \lessp \inf_{\bx \in \B} \lambda_{\min}\big(\widehat{\bPsi}_{t,\bx}\big) \leq \sup_{\bx \in \B} \big\|\widehat{\bPsi}_{t,\bx}\big\| \lessp 1, \\
        \sup_{\bx \in \B} \big\|\widehat{\bPsi}_{t,\bx}^{-1} - \bPsi_{t,\bx}^{-1}\big\|
        & \lessp \sqrt{\frac{\log (1/h)}{n h^d}}.
    \end{align*}
\end{lem}

\begin{lem}[Stochastic Linear Approximation]\label{sa-lem: Stochastic Linear Approximation}
    Suppose Assumptions \ref{sa-assump: DGP}(i)(ii)(iii)(v) and \ref{sa-assump: Kernel, Distance and Boundary} hold. If $\frac{n h^d}{\log (1/h)} \to \infty$ and $h=o(1)$, then
    \begin{align*}
        \sup_{\bx \in \B} \big\|\mathbf{O}_{t,\bx}\big\|
        & \lessp \sqrt{\frac{\log (1/h)}{n h^d}} + \frac{\log (1/h)}{n^{\frac{1+v}{2+v}}h^d},\\
        \sup_{\bx \in \B} \big|\be_0^{\top} \bPsi_{t,\bx}^{-1} \mathbf{O}_{t,\bx}\big|
        & \lessp \sqrt{\frac{\log (1/h)}{n h^d}} + \frac{\log (1/h)}{n^{\frac{1+v}{2+v}}h^d}, \\
        \sup_{\bx \in \B} \big|\be_0^{\top} (\widehat{\bPsi}_{t,\bx}^{-1} - \bPsi_{t,\bx}^{-1}) \mathbf{O}_{t,\bx}\big|
        & \lessp \sqrt{\frac{\log (1/h)}{n h^d}} \bigg(\sqrt{\frac{\log (1/h)}{n h^d}} + \frac{\log (1/h)}{n^{\frac{1+v}{2+v}}h^d} \bigg).
    \end{align*}
 \end{lem}

\begin{lem}[Covariance]\label{sa-lem: Covariance}
    Suppose Assumptions \ref{sa-assump: DGP} and \ref{sa-assump: Kernel, Distance and Boundary} hold. If $\frac{n h^d}{\log (1/h)} \to \infty$ and $h=o(1)$, then
    \begin{align*}
        1 \lesssim \inf_{\bx\in\B}n h^d\Xi_{t,\bx,\bx}
        \leq \sup_{\bx\in\B}n h^d\Xi_{t,\bx,\bx}
        \lesssim 1,
    \end{align*}
    \begin{align*}
        \sup_{\bx_1,\bx_2 \in \B}
        \big\|\widehat{\bUpsilon}_{t, \bx_1,\bx_2} - \bUpsilon_{t, \bx_1,\bx_2}\big\|
        \lessp \sqrt{\frac{\log(1/h)}{n h^d}} + \frac{\log(1/h)}{n^{\frac{v}{2+v}}h^d},
    \end{align*}
    and
     \begin{align*}
        \sup_{\bx_1,\bx_2 \in \B} n h^{d} \big|\widehat{\Xi}_{t, \bx_1,\bx_2} - \Xi_{t, \bx_1,\bx_2}\big|
        \lessp \sqrt{\frac{\log(1/h)}{n h^d}} + \frac{\log(1/h)}{n^{\frac{v}{2+v}}h^d}.
    \end{align*}

    If, in addition, $\frac{n^{\frac{v}{2+v}} h^d}{\log(1/h)} \to \infty$, then
    \begin{align*}
        \inf_{\bx \in \B} \lambda_{\min}(\widehat{\bUpsilon}_{t,\bx,\bx})
        \gtrsim_{\P} 1,
        \qquad
        \inf_{\bx \in \B}n h^d\widehat{\Xi}_{t,\bx,\bx}
        \gtrsim_{\P} 1,
        \qquad
        \sup_{\bx \in \B}n h^d\widehat{\Xi}_{t,\bx,\bx}
        \lessp 1,
    \end{align*}
    and
    \begin{align*}    
        \sup_{\bx_1,\bx_2 \in \B}
        \bigg|\frac{\widehat{\Xi}_{t,\bx_1,\bx_2}}{\sqrt{\widehat{\Xi}_{t,\bx_1,\bx_1} \widehat{\Xi}_{t,\bx_2,\bx_2}}} - \frac{\Xi_{t,\bx_1,\bx_2}}{\sqrt{\Xi_{t,\bx_1,\bx_1} \Xi_{t,\bx_2,\bx_2}}} \bigg|
        \lessp \sqrt{\frac{\log(1/h)}{n h^d}} + \frac{\log(1/h)}{n^{\frac{v}{2+v}}h^d}.
    \end{align*}
\end{lem}

\section{Boundary Average Treatment Effect Curve}\label{sa-sec: BATEC}

\subsection{Identification}

\begin{thm}[Distance-Based Identification]\label{sa-thm: Identification}
    Suppose Assumptions \ref{sa-assump: DGP}(i)-(iii) and \ref{sa-assump: Kernel, Distance and Boundary}(i), (ii), and (v) hold. Then,
    \begin{align*}
        \tau(\bx)
        = \vartheta(\bx)
        = \lim_{r\downarrow0} \theta_{1,\bx}(r) - \lim_{r\uparrow0} \theta_{0,\bx}(r),
        \qquad \text{for all } \bx \in \B.
    \end{align*}
\end{thm}

\subsection{Approximation Bias}\label{sa-sec: Approximation Bias}

The next two theorems collect the approximation-bias results corresponding to
Theorems 2 and 3 in the paper. The first theorem gives a dimension-generic level upper
bound and the bivariate sharpness result; the second theorem gives improved level bias
bounds under smoothness of the induced signed-distance regression.

\begin{thm}[Approximation Bias: Minimal Level Guarantee]\label{sa-thm: Approximation Bias: Minimal Guarantee}
    Suppose Assumptions \ref{sa-assump: DGP}(i)--(iii) and \ref{sa-assump: Kernel, Distance and Boundary} hold. If $h=o(1)$, then
    \begin{align*}
        \sup_{\bx \in \B}|\bias(\bx)| \lesssim h,
    \end{align*}
    for any $d\geq1$. In particular, for $d=2$, this bias-rate is sharp in the following sense. For some $L \geq 1$, let $\mathcal{P}$ be the class of data generating processes satisfying Assumptions \ref{sa-assump: DGP}(i)--(iii) with $\X\subseteq[-L,L]^2$, Assumption \ref{sa-assump: Kernel, Distance and Boundary}, and the following additional conditions, with (ii) and (iii) imposed for each $t\in\{0,1\}$:
    \begin{enumerate}[label=\normalfont(\roman*),noitemsep,leftmargin=*]
    \item $L^{-1} \leq \inf_{\bx \in \X} f_X(\bx) \leq \sup_{\bx \in \X} f_X(\bx) \leq L$,
    \item $\max_{0 \leq |\bnu| \leq p} \sup_{\bx \in \X}  |\partial^{\bnu} \mu_t(\bx)| + \max_{0 \leq |\bnu| \leq p} \sup_{\bx,\by \in \X} \frac{|\partial^{\bnu} \mu_t(\bx) - \partial^{\bnu}\mu_t(\by)|}{\|\bx - \by\|} \leq L$, and
    \item $\liminf_{h\downarrow0}\inf_{\bx \in \B} \int_{\A_t} K_h((2t-1)\d(\bu,\bx)) \dif\bu \geq L^{-1}$.
    \end{enumerate}
    If $nh^2\to\infty$ and $h=o(1)$, then
    \begin{align*}
         1 \lesssim \liminf_{n\to\infty} \sup_{\P\in\mathcal{P}} \sup_{\bx \in \B}\frac{|\bias(\bx)|}{h}
         \leq \limsup_{n\to\infty} \sup_{\P\in\mathcal{P}} \sup_{\bx \in \B}\frac{|\bias(\bx)|}{h}
         \lesssim 1.
    \end{align*}
\end{thm}

The first part of Theorem~\ref{sa-thm: Approximation Bias: Minimal Guarantee} is dimension-generic. The sharpness statement is a level result: near kinked boundaries, signed-distance derivatives need not exist without additional smoothness restrictions.

\begin{thm}[Approximation Bias: Smooth Boundary]\label{sa-thm: Approximation Bias: Smooth Distance Derivatives}
    Suppose Assumptions \ref{sa-assump: DGP}(i)--(iii) and \ref{sa-assump: Kernel, Distance and Boundary} hold, and let $h=o(1)$.

    \begin{enumerate}[label=\normalfont(\roman*),noitemsep,leftmargin=*]
    \item If, for some integer $m\in\{1,\ldots,p+1\}$, each map $r\mapsto\theta_{t,\bx}(r)$ is $m$-times continuously differentiable on $\I_t$ near zero with $m$th derivative uniformly bounded over $t\in\{0,1\}$ and $\bx\in\B$, then
    \begin{align*}
        \sup_{\bx \in \B}|\bias(\bx)| \lesssim h^m.
    \end{align*}
    The pointwise analogue holds under the same smoothness condition at a fixed $\bx\in\B$.

    \item Suppose $d=2$ and $\d(\cdot,\cdot)$ is the Euclidean distance. For $\bx\in\B$, suppose that for some $\delta,\varepsilon>0$,
    $\B \cap \{\by: \|\by-\bx\|\leq\varepsilon\}=\gamma([-\delta,\delta])$, where $\gamma:\reals\to\reals^2$ is one-to-one and belongs to $C^{\kappa+2}([-\delta,\delta],\reals^2)$. Let $m_{\bx}=\kappa\wedge(p+1)$. Then, $\theta_{0,\bx}(\cdot)$ and $\theta_{1,\bx}(\cdot)$ are $m_{\bx}$-times continuously differentiable near zero, and
    \begin{align*}
        |\bias(\bx)|\lesssim h^{m_{\bx}}.
    \end{align*}

    \item Suppose $d=2$, $\d(\cdot,\cdot)$ is the Euclidean distance, and $\B=\gamma([0,L])$, where $\gamma$ is one-to-one and belongs to $C^{\iota+2}([0,L],\reals^2)$ for some $L>0$. Suppose there exist $\delta,\varepsilon>0$ such that, for all $\bx\in\B^o=\gamma([\delta,L-\delta])$, $r\in[0,\varepsilon]$, and $t\in\{0,1\}$, the set $\{\bu\in\reals^2:\d(\bu,\bx)=r\}$ intersects $\setbd(\mathcal{A}_t)$ at only two points in $\B$. Let $m=\iota\wedge(p+1)$. Then, $\theta_{0,\bx}(\cdot)$ and $\theta_{1,\bx}(\cdot)$ are $m$-times continuously differentiable near zero with derivatives uniformly bounded over $\bx\in\B^o$, and
    \begin{align*}
        \sup_{\bx\in\B^o}|\bias(\bx)|\lesssim h^m.
    \end{align*}
    \end{enumerate}
\end{thm}

\subsection{Point Estimation and Approximate MSE Expansions}

\begin{thm}[Convergence Rates]\label{sa-thm: Convergence Rates}
    Suppose Assumptions \ref{sa-assump: DGP} and \ref{sa-assump: Kernel, Distance and Boundary} hold. Let $h=o(1)$.
    \begin{itemize}[noitemsep,leftmargin=*]
    \item If $\vartheta(\bx)$ is well defined at a fixed $\bx\in\B$, and $n h^d \to \infty$, then
    \begin{align*}
        \big|\widehat{\vartheta}(\bx) - \vartheta(\bx) \big|
        \lessp \frac{1}{\sqrt{n h^d}} + \frac{1}{n^{\frac{1+v}{2+v}}h^d} + \big|\bias(\bx)\big|.
    \end{align*}
    \item If $\vartheta(\bx)$ is well defined for all $\bx\in\B$, and $\frac{n h^d}{\log (1/h)} \to \infty$, then
    \begin{align*}
        \sup_{\bx \in \mathcal{B}} \big|\widehat{\vartheta}(\bx) - \vartheta(\bx)\big|
        \lessp \sqrt{\frac{\log(1/h)}{ n h^d}}
             + \frac{\log(1/h)}{n^{\frac{1+v}{2+v}}h^d}
             + \sup_{\bx \in \B} \big|\bias(\bx)\big|.
    \end{align*}
    \end{itemize}
\end{thm}

Equation~\eqref{sa-eq: distance error decomposition} also gives MSE-type criteria for
bandwidth choice. Unlike the location-based estimator, the distance-based estimator is
based on a one-dimensional projection of the full score, so a conditional-on-$\bX$ MSE
does not isolate a clean local-polynomial variance term under misspecification. Instead,
we use the linearized decomposition implied by \eqref{sa-eq: distance error decomposition}.
Because the population projection defining $\theta_{t,\bx}^{\ast}$ implies
$\E[\mathfrak{L}(\bx)]=0$ and $\E[\mathfrak{L}(\bx)^2]=\Xi_{\bx}$, we define the
approximate MSE and approximate IMSE as
\begin{align*}
    \operatorname{AMSE}(\bx)
    &= \E\big[(\bias(\bx)+\mathfrak{L}(\bx))^2\big]
    = \bias(\bx)^2 + \Xi_{\bx},
    \qquad \bx\in\B,
\end{align*}
and the approximate IMSE
\begin{align*}
    \operatorname{AIMSE}
    &= \int_{\B}\operatorname{AMSE}(\bx)w(\bx)\dif\Haus^{d-1}(\bx).
\end{align*}

The results in Section~\ref{sa-sec: Approximation Bias} give different approximation
and exact rate results for the bias $\bias(\bx)$,
Lemma~\ref{sa-lem: Covariance} implies that $V_{\bx}=nh^d\Xi_{\bx}$ is bounded and
bounded away from zero, and Lemma~\ref{sa-lem: Stochastic Linear Approximation} implies
that
\begin{align*}
    \sup_{\bx\in\B}|\mathfrak{Q}(\bx)|
    \lessp
    \sqrt{\frac{\log (1/h)}{n h^d}}
    \bigg(\sqrt{\frac{\log (1/h)}{n h^d}} + \frac{\log (1/h)}{n^{\frac{1+v}{2+v}}h^d} \bigg).
\end{align*}
Therefore, under appropriate rate conditions, the nonlinear term is higher order and can
be ignored for first-order bandwidth calculations.

For bandwidth selection, suppose that a leading-bias expansion
$\bias(\bx)=h^mB_\bx+o(h^m)$ is available, with $B_\bx$ bounded and bounded away from
zero. The approximate MSE criterion is then
\begin{align*}
    h^{2m}B_\bx^2+\frac{1}{nh^d}V_\bx,
\end{align*}
which gives the approximate MSE-optimal bandwidth selection rule
\begin{align*}
    h_{\mathtt{AMSE},\bx}
    = \Big(\frac{d V_{\bx}}{2m B_{\bx}^2} \frac{1}{n} \Big)^{\frac{1}{2m+d}}.
\end{align*}
Similarly, if $\int_{\B} B_{\bx}^2w(\bx)\dif\Haus^{d-1}(\bx)$ is bounded and bounded
away from zero, the approximate IMSE-optimal bandwidth choice is
\begin{align*}
    h_{\mathtt{AIMSE}}
    = \Big(\frac{d\int_{\B} V_{\bx} w(\bx) \dif\Haus^{d-1}(\bx)}
                {2m\int_{\B} B_{\bx}^2 w(\bx) \dif\Haus^{d-1}(\bx)}
           \frac{1}{n}
          \Big)^{\frac{1}{2m+d}}.
\end{align*}
Here $m$ is the order of the signed-distance approximation bias. Theorems
\ref{sa-thm: Approximation Bias: Minimal Guarantee} and
\ref{sa-thm: Approximation Bias: Smooth Distance Derivatives} give primitive conditions
under which such rates are available. In contrast to the location-based case in
\citet{Cattaneo-Titiunik-Yu_2026_JASA}, this bias order is governed by the smoothness of
the induced signed-distance regression, and
therefore by the geometry of the boundary and the distance metric, not only by the
polynomial order $p$.

\subsection{Distributional Approximation and Inference}

Let $\bV = ((\bX_1^{\top},Y_1), \cdots, (\bX_n^{\top},Y_n))$, and recall that $t \in \{0,1\}$. The feasible t-statistic is
\begin{align*}
    \widehat{\Tstat}(\bx)
    = \frac{\widehat{\vartheta}(\bx) - \vartheta(\bx)}
           {\sqrt{\widehat{\Xi}_{\bx,\bx}}},
    \qquad \bx \in \mathcal{B}.
\end{align*}
The associated $100(1-\alpha)\%$ confidence interval estimator is
\begin{align*}
    \widehat{\CI}_{\alpha}(\bx)
    = \bigg[\;\widehat{\vartheta}(\bx) - \q_{\alpha} \sqrt{\widehat{\Xi}_{\bx,\bx}}
            \; , \;
              \widehat{\vartheta}(\bx) + \q_{\alpha} \sqrt{\widehat{\Xi}_{\bx,\bx}} \;\bigg],
\end{align*}
where $\q_{\alpha}$ denotes an appropriate quantile depending on the desired confidence level $\alpha\in(0,1)$ and coverage objective (pointwise vs. uniform over $\B$). The following theorem establishes pointwise asymptotic normality and validity of confidence intervals. Let $\Phi(\cdot)$ be the cumulative distribution function of a standard univariate Gaussian random variable.

\begin{thm}[Confidence Intervals]\label{sa-thm: Confidence Intervals}
    Suppose Assumptions \ref{sa-assump: DGP} and \ref{sa-assump: Kernel, Distance and Boundary} hold, and $\vartheta(\bx)$ is well defined. If $n^{\frac{v}{2+v}}h^d \to \infty$, $h=o(1)$, and $\sqrt{n h^d} |\bias(\bx)| \to 0$, then
    \begin{align*}
        \sup_{u \in \reals} \Big|\P \big(\widehat{\Tstat}(\bx) \leq u \big) - \Phi(u)\Big| = o(1),
        \qquad \bx \in \B,
    \end{align*}
    and
    \begin{align*}
        \P \big(\vartheta(\bx) \in \widehat{\CI}_{\alpha}(\bx) \big) = 1 - \alpha + o(1),
        \qquad \bx \in \B,
    \end{align*}
    provided that $\q_{\alpha} = \inf \{c > 0: \P( |\widehat{Z}| \geq c \mid \bV) \leq \alpha \}$ with $\widehat{Z}\mid\bV \sim \mathsf{Normal}(0, 1)$.
\end{thm}

To conduct uniform inference, we rely on a new strong approximation result established in Section \ref{sa-sec: Gaussian Strong Approximation}. First, we approximate (uniformly over $\bx\in\B$) the feasible statistic $\widehat{\Tstat}$ by the following linear statistic (which is a sum of independent random variables):
\begin{align*}
    \overline{\Tstat}(\bx)
    = \big(\Xi_{\bx,\bx}\big)^{-1/2} \Big(\be_0^{\top} \bPsi_{1,\bx}^{-1} \mathbf{O}_{1,\bx}
                              - \be_0^{\top} \bPsi_{0,\bx}^{-1} \mathbf{O}_{0,\bx}\Big),
    \qquad \bx \in \B.
\end{align*}

\begin{thm}[Stochastic Linearization]\label{sa-thm: Stochastic Linearization}
    Suppose Assumptions \ref{sa-assump: DGP} and \ref{sa-assump: Kernel, Distance and Boundary} hold, and $\vartheta(\bx)$ is well defined for all $\bx\in\B$. If $\frac{n^{\frac{v}{2+v}}h^d}{\log(1/h)}\to\infty$ and $h=o(1)$, then
    \begin{align*}
        \sup_{\bx \in \mathcal{B}} \big|\widehat{\Tstat}(\bx) - \overline{\Tstat}(\bx) \big|
        \lessp \sqrt{\log(1/h)} \bigg(\sqrt{\frac{\log(1/h)}{n h^d}}
             + \frac{\log(1/h)}{n^{\frac{v}{2+v}}h^d} \bigg)
             + \sqrt{n h^d} \sup_{\bx \in \B} |\bias(\bx)|.
    \end{align*}
\end{thm}

The pointwise (in $\B$) analogue of this result removes the $\log(1/h)$ penalty. See the proof of Theorem \ref{sa-thm: Confidence Intervals} for more details. To establish a Gaussian strong approximation for $\overline{\Tstat}(\bx)$, define the class of functions $\G = \{g_{\bx}: \bx \in \B\}$ and $\M = \{m_{\bx}: \bx \in \B\}$, where
\begin{align}\label{sa-eq: def G and M}
    \nonumber g_{\bx}(\bu)
    &= \Indicator(\bu \in \A_1) \mathfrak{K}_1(\bu;\bx)
     - \Indicator(\bu \in \A_0) \mathfrak{K}_0(\bu;\bx),\\
    m_{\bx}(\bu)
    &= - \Indicator(\bu \in \A_1) \mathfrak{K}_1(\bu;\bx) \theta_{1,\bx}^{\ast}(D_{\bu}(\bx))
       + \Indicator(\bu \in \A_0) \mathfrak{K}_0(\bu;\bx) \theta_{0,\bx}^{\ast}(D_{\bu}(\bx)),
\end{align}
with
\begin{align*}
    \mathfrak{K}_t(\bu;\bx)
    = \frac{1}{\sqrt{n \Xi_{\bx,\bx}}}\be_0^{\top} \bPsi_{t,\bx}^{-1} \rp \Big(\frac{D_{\bu}(\bx)}{h}\Big) K_h(D_{\bu}(\bx)),
\end{align*}
for all $\bu \in \X$, $\bx \in \B$, and $t\in\{0,1\}$. In addition, let $\R$ be the class of functions containing the singleton identity function $\operatorname{Id}: \reals \to \reals$, $\operatorname{Id}(x) = x$. Then, $\overline{\Tstat}(\bx)$ can be represented as
\begin{align*}
    \overline{\Tstat}(\bx)
    & = \frac{1}{\sqrt{n}} \sum_{i=1}^n
        \Big[g_{\bx}(\bX_i) \operatorname{Id}(Y_i) + m_{\bx}(\bX_i)
        - \E\big[g_{\bx}(\bX_i) \operatorname{Id}(Y_i) + m_{\bx}(\bX_i)\big]\Big].
\end{align*}
Define the multiplicative-separable empirical process as
\begin{align*}  
    M_n(g,r) = \frac{1}{\sqrt{n}} \sum_{i = 1}^n \big[g(\bx_i) r(y_i) - \E[g(\bx_i) r(y_i)]\big],
    \qquad g \in \G, r \in \R,
\end{align*}
which implies that
\begin{align*}
    \overline{\Tstat}(\bx) = M_n(g_{\bx},\operatorname{Id}) + M_n(m_{\bx},1), \qquad \bx \in \B.
\end{align*}

Theorem~\ref{sa-thm: Strong Approximation} gives a new Gaussian strong approximation that can be applied to $\overline{\Tstat}(\bx)$. This new theorem allows polynomial moment bounds on the conditional distribution of $Y_i|\bX_i$.

\begin{thm}[Gaussian Strong Approximation: $\overline{\Tstat}$]\label{sa-thm: Gaussian Strong Approximation: Tstat}
    Suppose Assumptions \ref{sa-assump: DGP} and \ref{sa-assump: Kernel, Distance and Boundary} hold, and $\X=\prod_{l=1}^d[a_l,b_l]$, with $-\infty<a_l<b_l<\infty$ for $l=1,\ldots,d$. Suppose also that there exists a constant $C > 0$ such that for $t \in \{0,1\}$ and for any $\bx \in \B$, the De Giorgi perimeter of the localized set $E_{t,\bx}(h) = \{\by \in \mathcal{A}_t: \varsigma_t\d(\by,\bx)/h\in\supp(K)\}$ satisfies  $\mathcal{L}(E_{t,\bx}(h)) \leq C h^{d-1}$. If $\liminf_{n \to \infty} \frac{\log h}{\log n} > - \infty$, $n h^d \to \infty$, and $h=o(1)$, then (on a possibly enlarged probability space) there exists a mean-zero Gaussian process $Z$ indexed by $\mathcal{B}$ with almost surely continuous sample path such that
    \begin{align*}
        \E\Big[\sup_{\bx \in \mathcal{B}} \big|\overline{\Tstat}(\bx)- Z(\bx) \big| \Big]
        \lesssim (\log(n))^{\frac{3}{2}} \Big(\frac{1}{n h^d}\Big)^{\frac{1}{2d+2} \frac{v}{v + 2}}
                + \log(n) \Big(\frac{1}{n^{\frac{v}{2+v}}h^d}\Big)^{\frac{1}{2}},
    \end{align*}
    where $\lesssim$ is up to a universal constant, and $Z$ has the same covariance structure as $\overline{\Tstat}$; that is, $\Cov[\overline{\Tstat}(\bx_1), \overline{\Tstat}(\bx_2)] = \Cov[Z(\bx_1), Z(\bx_2)]$ for all $\bx_1, \bx_2 \in \B$.
\end{thm}

The rectangular-support condition in Theorem~\ref{sa-thm: Gaussian Strong Approximation: Tstat} is used only to invoke the normalizing-map construction in \citet[Section~3.1]{Cattaneo-Yu_2025_AOS}. After a componentwise affine rescaling, one can take $\X=[0,1]^d$; because $f_X$ is continuous and bounded away from zero on this compact rectangle, $\P_X$ is a valid surrogate measure and the Rosenblatt transform provides a valid normalizing transformation with finite constants. The De Giorgi perimeter condition controls the total variation of the localized distance-kernel classes.

Theorem \ref{sa-thm: Gaussian Strong Approximation: Tstat} can be used to construct confidence bands for $(\vartheta(\bx):\bx\in\B)$. Let $(\widehat{Z}(\bx):\bx \in \B)$ be a (conditionally on $\bV$) mean-zero Gaussian process with feasible (conditional) covariance function
\begin{align*}
    \Cov \Big[\widehat{Z}(\bx_1),\widehat{Z}(\bx_2) \Big|\bV \Big]
    = \frac{\widehat{\Xi}_{\bx_1,\bx_2}}
           {\sqrt{\widehat{\Xi}_{\bx_1,\bx_1}} \sqrt{\widehat{\Xi}_{\bx_2,\bx_2}}},
    \qquad \bx_1, \bx_2 \in \B.
\end{align*}

\begin{thm}[Confidence Bands]\label{sa-thm: Confidence Bands}
    Suppose the assumptions and conditions in Theorem \ref{sa-thm: Gaussian Strong Approximation: Tstat} hold, and $\vartheta(\bx)$ is well defined for all $\bx\in\B$. If $\liminf_{n \to \infty} \frac{\log h}{\log n} > - \infty$, $\frac{n^{\frac{v}{2+v}}h^d}{(\log n)^3} \to \infty$, and $\sqrt{n h^d\log n} \sup_{\bx \in \B} |\bias(\bx)| \to 0$, then
    \begin{align*}
        \sup_{u \in \reals} \Big|\P \Big(\sup_{\bx \in \B} \big|\widehat{\Tstat}(\bx)\big| \leq u \Big)
                            - \P \Big(\sup_{\bx \in \B} \big|\widehat{Z}(\bx)\big| \leq u \Big| \bV \Big) \Big| = o_{\P}(1)
    \end{align*}
    and
    \begin{align*}
        \P\Big[\vartheta(\bx) \in \widehat{\CI}_{\alpha}(\bx), \text{ for all } \bx \in \B \Big] = 1 - \alpha + o(1),
    \end{align*}
    provided that $\q_{\alpha} = \inf \big\{c > 0: \P \big(\sup_{\bx \in \B} \big|\widehat{Z}(\bx)\big|\geq c \big| \bV \big) \leq \alpha \big\}$.
\end{thm}

Theorem~\ref{sa-thm: Gaussian Strong Approximation: Tstat} gives a strong approximation
of the entire process $\overline{\Tstat}$ by a Gaussian process. This
process-level coupling is stronger than what is needed to construct confidence bands.
The following theorem approximates only the supremum of the statistic. It avoids the
rectangular-support and De Giorgi perimeter conditions used in the process-level result.

\begin{thm}[Gaussian Approximation for $\sup_{\bx \in \B}|\overline{\Tstat}(\bx)|$]\label{sa-thm: Gaussian Approximation Suprema: Tstat}
    Suppose Assumptions \ref{sa-assump: DGP} and \ref{sa-assump: Kernel, Distance and Boundary} hold, $\liminf_{n \to \infty}\frac{\log h}{\log n}>-\infty$, and $h=o(1)$. Then, for $n$ large enough, on a possibly enlarged probability space there exists a mean-zero Gaussian process $Z$ indexed by $\B$ with almost surely continuous sample paths such that
    \begin{align*}
        \Cov[Z(\bx_1),Z(\bx_2)]
        =
        \Cov[\overline{\Tstat}(\bx_1),\overline{\Tstat}(\bx_2)],
        \qquad \bx_1,\bx_2\in\B,
    \end{align*}
    and
    \begin{align*}
        &\P\Bigg[
        \bigg|\sup_{\bx\in\B}|\overline{\Tstat}(\bx)|
        -\sup_{\bx\in\B}|Z(\bx)|\bigg|
        >
        \frac{\log n}{(n h^d)^{1/6}}
        +\frac{(\log n)^{5/4}}{(n h^d)^{1/4}}
        +\frac{(\log n)^{3/2}}{(n^{\frac{v}{2+v}}h^d)^{1/2}}
        \Bigg]
        \lesssim \frac{1}{\log n}+\frac{\log n}{n}.
    \end{align*}
\end{thm}

Using Theorem~\ref{sa-thm: Gaussian Approximation Suprema: Tstat}, we obtain the
following confidence-band result. Relative to Theorem~\ref{sa-thm: Confidence Bands}, it
removes the process-level geometric side conditions at the cost of a stronger logarithmic
rate condition.

\begin{thm}[Confidence Bands: Suprema Approximation]\label{sa-thm: Confidence Bands: Suprema}
    Suppose Assumptions \ref{sa-assump: DGP} and \ref{sa-assump: Kernel, Distance and Boundary} hold, and $\vartheta(\bx)$ is well defined for all $\bx\in\B$. If $\liminf_{n \to \infty}\frac{\log h}{\log n}>-\infty$, $\frac{n^{\frac{v}{2+v}}h^d}{(\log n)^9}\to\infty$, $h=o(1)$, and $\sqrt{n h^d\log n}\sup_{\bx\in\B}|\bias(\bx)|\to0$,
    then
    \begin{align*}
        \sup_{u \in \reals} \Big|\P \Big(\sup_{\bx \in \B} \big|\widehat{\Tstat}(\bx)\big| \leq u \Big)
                            - \P \Big(\sup_{\bx \in \B} \big|\widehat{Z}(\bx)\big| \leq u \Big| \bV \Big) \Big| = o_{\P}(1)
    \end{align*}
    and
    \begin{align*}
        \P\Big[\vartheta(\bx) \in \widehat{\CI}_{\alpha}(\bx), \text{ for all } \bx \in \B \Big] = 1 - \alpha + o(1),
    \end{align*}
    provided that $\q_{\alpha} = \inf \big\{c > 0: \P \big(\sup_{\bx \in \B} \big|\widehat{Z}(\bx)\big|\geq c \big| \bV \big) \leq \alpha \big\}$.
\end{thm}

\section{Weighted Boundary Average Treatment Effect}\label{sa-sec: WBATE}

Without loss of generality, normalize the weight function so that
$\int_{\B}w(\bx)\dif\Haus^{d-1}(\bx)=1$. Define the distance-based WBATE
parameter and plug-in estimator by
\begin{align*}
    \vartheta_{\mathtt{WBATE}}
    = \int_{\B}\vartheta(\bx)w(\bx)\dif\Haus^{d-1}(\bx),
    \qquad\text{and}\qquad
    \widehat{\vartheta}_{\mathtt{WBATE}}
    = \int_{\B}\widehat{\vartheta}(\bx)w(\bx)\dif\Haus^{d-1}(\bx).
\end{align*}

Integrating the decomposition \eqref{sa-eq: distance error decomposition} for the
difference of the two side-specific estimators gives
\begin{align}\label{sa-eq: WBATE decomposition}
    \widehat{\vartheta}_{\mathtt{WBATE}} - \vartheta_{\mathtt{WBATE}}
    = \bias_{\mathtt{WBATE}} + \mathfrak{L}_{\mathtt{WBATE}} + \mathfrak{Q}_{\mathtt{WBATE}},
\end{align}
where
\begin{align*}
    \bias_{\mathtt{WBATE}} = \int_{\B} \bias(\bx)w(\bx)\dif\Haus^{d-1}(\bx),
    \qquad
    \mathfrak{L}_{\mathtt{WBATE}} = \int_{\B} \mathfrak{L}(\bx)w(\bx)\dif\Haus^{d-1}(\bx),
\end{align*}
and
\begin{align*}
        \mathfrak{Q}_{\mathtt{WBATE}} = \int_{\B} \mathfrak{Q}(\bx)w(\bx)\dif\Haus^{d-1}(\bx).
\end{align*}

Define the aggregate variance of the leading term by
\begin{align*}
    \Xi_{\mathtt{WBATE}}
    = \int_{\B}\int_{\B}\Xi_{\bx_1,\bx_2} w(\bx_1)w(\bx_2)\dif\Haus^{d-1}(\bx_1)\dif\Haus^{d-1}(\bx_2).
\end{align*}
Its feasible counterpart is
\begin{align*}
    \widehat{\Xi}_{\mathtt{WBATE}}
    &= \int_{\B}\int_{\B}\widehat{\Xi}_{\bx_1,\bx_2}
       w(\bx_1)w(\bx_2)\dif\Haus^{d-1}(\bx_1)\dif\Haus^{d-1}(\bx_2).
\end{align*}
For the aggregate stochastic rate and inference results, impose the local
boundary-measure regularity condition
\begin{align}
    \sup_{\bx\in\B}\Haus^{d-1}(\B\cap B(\bx,r)) \leq C_{\B}r^{d-1},
    \qquad 0<r\leq r_0,
    \label{sa-eq: local boundary measure regularity}
\end{align}
for some constants $C_{\B}<\infty$ and $r_0>0$, where $B(\bx,r)$ denotes a Euclidean
ball. This upper local regularity condition holds, for example, for compact piecewise
$C^1$ boundaries with finite local multiplicity, including piecewise-linear boundaries.
Write $\nu_{\B,k}$ for the $k$-fold product measure $(\Haus^{d-1})^k$ on $\B^k$.
For inference, impose the scalar aggregate nondegeneracy condition
\begin{align}
    \Xi_{\mathtt{WBATE}}
    \gtrsim \frac{h^{d-1}}{n h^d}
    = \frac{1}{n h},
    \label{sa-eq: WBATE aggregate nondegeneracy}
\end{align}
which rules out cancellation of the integrated covariance kernel.

\begin{lem}[Variance Linear Term: WBATE]\label{sa-lem: Variance: WBATE}
    Suppose Assumptions \ref{sa-assump: DGP}, \ref{sa-assump: Kernel, Distance and Boundary}, and \ref{sa-assump: Weight Function and Boundary} hold, and \eqref{sa-eq: local boundary measure regularity} holds. If $\frac{n h^d}{\log(1/h)}\to\infty$ and $h=o(1)$, then
    \begin{align*}
        \Var[\mathfrak{L}_{\mathtt{WBATE}}]
        = \Xi_{\mathtt{WBATE}} \lesssim \frac{1}{n h}.
    \end{align*}

    If, in addition, \eqref{sa-eq: WBATE aggregate nondegeneracy} holds, then
    \begin{align*}
        \frac{1}{n h}
        \lesssim
        \Xi_{\mathtt{WBATE}}
        \lesssim
        \frac{1}{n h}.
    \end{align*}

    If $\frac{n^{v/(2+v)}h^d}{\log(1/h)}\to\infty$ and $h=o(1)$, then
    \begin{align*}
        \widehat{\Xi}_{\mathtt{WBATE}}
        = \Xi_{\mathtt{WBATE}} + o_{\P}\Big(\frac{1}{n h}\Big).
    \end{align*}
    If, in addition, \eqref{sa-eq: WBATE aggregate nondegeneracy} holds, then
    $\widehat{\Xi}_{\mathtt{WBATE}}/\Xi_{\mathtt{WBATE}}\to_{\P}1$.
\end{lem}

\begin{lem}[Quadratic Remainder: WBATE]\label{sa-lem: Higher-Order Remainder: WBATE}
    Suppose Assumptions \ref{sa-assump: DGP}, \ref{sa-assump: Kernel, Distance and Boundary}, and \ref{sa-assump: Weight Function and Boundary} hold, and \eqref{sa-eq: local boundary measure regularity} holds. If $\frac{n h^d}{\log(1/h)}\to\infty$ and $h=o(1)$, then
    \begin{align*}
        \mathfrak{Q}_{\mathtt{WBATE}} \lessp \frac{1}{n h^d}.
    \end{align*}
    If, in addition, $n h^{2d-1}\to\infty$, then
    \begin{align*}
        \mathfrak{Q}_{\mathtt{WBATE}}
        = o_{\P}\Big(\frac{1}{\sqrt{n h}}\Big).
    \end{align*}
\end{lem}

\begin{thm}[Convergence Rate: WBATE]\label{sa-thm: Convergence Rate: WBATE}
    Suppose Assumptions \ref{sa-assump: DGP}, \ref{sa-assump: Kernel, Distance and Boundary}, and \ref{sa-assump: Weight Function and Boundary} hold, $\vartheta(\bx)$ is well defined for all $\bx\in\B$, and \eqref{sa-eq: local boundary measure regularity} holds. If $\frac{n h^d}{\log(1/h)}\to\infty$ and $h=o(1)$, then
    \begin{align*}
        \big| \widehat{\vartheta}_{\mathtt{WBATE}} - \vartheta_{\mathtt{WBATE}} \big|
        \lessp \big|\bias_{\mathtt{WBATE}}\big| + \frac{1}{\sqrt{n h}} + \frac{1}{n h^d}.
    \end{align*}
\end{thm}

For inference, define
\begin{align*}
    \widehat{\Tstat}_{\mathtt{WBATE}}
    =
    \frac{\widehat{\vartheta}_{\mathtt{WBATE}}
    -\vartheta_{\mathtt{WBATE}}}
    {\sqrt{\widehat{\Xi}_{\mathtt{WBATE}}}},
\end{align*}
and the associated confidence interval
\begin{align*}
    \widehat{\CI}_{\alpha,\mathtt{WBATE}}
    = \Big[\; \widehat{\vartheta}_{\mathtt{WBATE}} -\q_{\alpha}\sqrt{\widehat{\Xi}_{\mathtt{WBATE}}} \;,\;
              \widehat{\vartheta}_{\mathtt{WBATE}} +\q_{\alpha}\sqrt{\widehat{\Xi}_{\mathtt{WBATE}}} \; \Big].
\end{align*}

\begin{thm}[Asymptotic Normality: WBATE]\label{sa-thm: Asymptotic Normality: WBATE}
    Suppose Assumptions \ref{sa-assump: DGP}, \ref{sa-assump: Kernel, Distance and Boundary}, and \ref{sa-assump: Weight Function and Boundary} hold, and \eqref{sa-eq: WBATE aggregate nondegeneracy} and \eqref{sa-eq: local boundary measure regularity} hold. If $\frac{n^{v/(2+v)}h^d}{\log(1/h)}\to\infty$, $h=o(1)$, $n h^{2d-1}\to\infty$, and $\sqrt{n h}|\bias_{\mathtt{WBATE}}|\to0$,
    then
    \begin{align*}
        \sup_{u\in\reals}
        \Big| \P\big(\widehat{\Tstat}_{\mathtt{WBATE}}\leq u\big) - \Phi(u) \Big| = o(1),
    \end{align*}
    and
    \begin{align*}
        \P\big[ \vartheta_{\mathtt{WBATE}} \in\widehat{\CI}_{\alpha,\mathtt{WBATE}} \big]
        =1-\alpha+o(1)
    \end{align*}
    provided that $\q_{\alpha}=\Phi^{-1}(1-\alpha/2)$.
\end{thm}

Condition~\eqref{sa-eq: local boundary measure regularity} controls the amount of
boundary measure in bandwidth-sized neighborhoods, which gives the effective
one-dimensional variance rate $1/(n h)$ for the integrated estimator.

\section{Largest Boundary Average Treatment Effect}\label{sa-sec: LBATE}

The distance-based LBATE parameter and plug-in estimator are
\begin{align*}
    \vartheta_{\mathtt{LBATE}}
    =
    \sup_{\bx\in\B}\vartheta(\bx),
    \qquad
    \widehat{\vartheta}_{\mathtt{LBATE}}
    =
    \sup_{\bx\in\B}\widehat{\vartheta}(\bx),
\end{align*}
respectively.

\begin{thm}[Convergence Rate: LBATE]\label{sa-thm: Convergence Rate: LBATE}
    Suppose Assumptions \ref{sa-assump: DGP} and \ref{sa-assump: Kernel, Distance and Boundary} hold, and $\vartheta(\bx)$ is well defined for all $\bx\in\B$. If $\frac{n h^d}{\log(1/h)}\to\infty$ and $h=o(1)$, then
    \begin{align*}
        \left|
        \widehat{\vartheta}_{\mathtt{LBATE}}
        -
        \vartheta_{\mathtt{LBATE}}
        \right|
        \lessp
        \sqrt{\frac{\log(1/h)}{n h^d}}
        +\frac{\log(1/h)}{n^{\frac{1+v}{2+v}}h^d}
        +\sup_{\bx\in\B}|\bias(\bx)|.
    \end{align*}
\end{thm}

For uncertainty quantification, project the simultaneous confidence band for the BATEC
onto the supremum functional. Define
\begin{align*}
    \widehat{\CI}_{\alpha,\mathtt{LBATE}}
    =
    \left[
    \sup_{\bx\in\B}\left(
    \widehat{\vartheta}(\bx)
    -\q_{\alpha}\sqrt{\widehat{\Xi}_{\bx,\bx}}
    \right),
    \quad
    \sup_{\bx\in\B}\left(
    \widehat{\vartheta}(\bx)
    +\q_{\alpha}\sqrt{\widehat{\Xi}_{\bx,\bx}}
    \right)
    \right],
\end{align*}
where
$\q_{\alpha}=\inf\{c>0:\P(\sup_{\bx\in\B}|\widehat Z(\bx)|\ge c\mid\bV)\le\alpha\}$.

\begin{thm}[Confidence Intervals: LBATE]\label{sa-thm: Confidence Intervals: LBATE}
    Suppose the assumptions and conditions in Theorem~\ref{sa-thm: Confidence Bands: Suprema} hold. Then
    \begin{align*}
        \P\left[
        \vartheta_{\mathtt{LBATE}}
        \in
        \widehat{\CI}_{\alpha,\mathtt{LBATE}}
        \right]
        \geq 1-\alpha+o(1).
    \end{align*}
\end{thm}

The interval is generally conservative because it is obtained by projecting a simultaneous
confidence band for the entire distance-based BATEC.

\section{Imperfect Compliance}\label{sa-sec: Imperfect Compliance}

We extend the distance-based results to settings with imperfect compliance, where
treatment assignment and treatment receipt may differ for some units. See
\cite{Hernan-Robins_2020_Book} for a textbook introduction.

Because $D_i(\bx)\in\I_t$ if and only if $\bX_i\in\A_t$, for any $\bx\in\B$ let
\begin{align*}
    W_i
    =
    \Indicator(D_i(\bx)\in\I_0)W_i(0)
    +\Indicator(D_i(\bx)\in\I_1)W_i(1)
\end{align*}
be the observed treatment receipt, where $W_i(t)\in\{0,1\}$ denotes the potential
treatment receipt under assignment $t\in\{0,1\}$. The observed outcome is
\begin{align*}
    Y_i
    =
    \Indicator(D_i(\bx)\in\I_0)Y_i(0,W_i(0))
    +\Indicator(D_i(\bx)\in\I_1)Y_i(1,W_i(1)).
\end{align*}
For notational economy, redefine the reduced-form potential outcome as
$Y_i(t)=Y_i(t,W_i(t))$. For $A\in\{Y,W\}$, write
$A_i(t)=Y_i(t)$ when $A=Y$, and $A_i(t)=W_i(t)$ when $A=W$; similarly, write
$A_i=Y_i$ when $A=Y$, and $A_i=W_i$ when $A=W$.
The reduced-form BATEC components are
\begin{align*}
    \tau_A(\bx)=\E[A_i(1)-A_i(0)\mid \bX_i=\bx],
    \qquad A\in\{Y,W\}, \quad \bx\in\B,
\end{align*}
and the fuzzy BATEC is the Wald ratio
\begin{align*}
    \zeta(\bx)=\frac{\tau_Y(\bx)}{\tau_W(\bx)},
    \qquad \bx\in\B.
\end{align*}
The results below concern this reduced-form Wald target. A complier-average causal
interpretation can be obtained by adding the usual exclusion and monotonicity restrictions,
which are not needed for the estimation and inference theory.

The distance-based analogues are defined from the induced conditional expectations
\begin{align*}
    \theta_{A,t,\bx}(r)
    =
    \E[A_i\mid D_i(\bx)=r]
    =
    \E[A_i(t)\mid \d(\bX_i,\bx)=|r|,\bX_i\in\A_t],
    \qquad r\in\I_t,
\end{align*}
and
\begin{align*}
    \vartheta_A(\bx)=\theta_{A,1,\bx}(0)-\theta_{A,0,\bx}(0),
    \qquad
    \xi(\bx)=\frac{\vartheta_Y(\bx)}{\vartheta_W(\bx)}.
\end{align*}
The estimator is
\begin{align*}
    \widehat{\xi}(\bx)=\frac{\widehat{\vartheta}_Y(\bx)}
                            {\widehat{\vartheta}_W(\bx)},
    \qquad
    \widehat{\vartheta}_A(\bx)
    =
    \be_0^\top\widehat{\bgamma}_{A,1}(\bx)
    -
    \be_0^\top\widehat{\bgamma}_{A,0}(\bx),
\end{align*}
where, for $A\in\{Y,W\}$ and $t\in\{0,1\}$,
\begin{align*}
    \widehat{\bgamma}_{A,t}(\bx)
    =
    \argmin_{\bgamma\in\reals^{p+1}}
    \En\Big[
    \big(A_i-\rp(D_i(\bx)/h)^\top\bgamma\big)^2
    K_h(D_i(\bx))\Indicator(D_i(\bx)\in\I_t)
    \Big].
\end{align*}
Thus, $\widehat{\bgamma}_{Y,t}(\bx)$ coincides with the sharp-design estimator defined
above when the outcome is $Y_i$, while $\widehat{\bgamma}_{W,t}(\bx)$ is the same local
polynomial fit applied to treatment receipt.

\begin{assumption}[Fuzzy DGP]\label{sa-assump: Fuzzy DGP}
    Let $t\in\{0,1\}$.
    \begin{enumerate}[label=\normalfont(\roman*),noitemsep,leftmargin=*]
    \item $((Y_i(t),W_i(t),\bX_i^\top)^\top:1\le i\le n)$ are independent and identically distributed random vectors.
    \item $\mu_{W,t}(\bx)=\E[W_i(t)\mid\bX_i=\bx]$ has a $(p+1)$-times continuously differentiable extension to an open neighborhood of $\X$.
    \item The entries of
    $\Cov[(Y_i(t),W_i(t))^\top\mid\bX_i=\bx]$ are continuous on $\X$.
    \end{enumerate}
\end{assumption}

The following theorem is a natural consequence of Theorem \ref{sa-thm: Identification}.

\begin{thm}[Distance-Based Fuzzy Identification]\label{sa-thm: Fuzzy Identification}
    Suppose Assumptions \ref{sa-assump: DGP}(i)--(iii), \ref{sa-assump: Fuzzy DGP}, and \ref{sa-assump: Kernel, Distance and Boundary}(i), (ii), and (v) hold. Then,
    \begin{align*}
        \vartheta_A(\bx)=\tau_A(\bx),\qquad A\in\{Y,W\}, \quad \bx\in\B.
    \end{align*}
    If, in addition, $\tau_W(\bx)\neq0$, then $\xi(\bx)=\zeta(\bx)$.
\end{thm}

\subsection{Second-Order Linearization}\label{sa-sec: Second-Order Linearization: Fuzzy BATEC}

The distance-based fuzzy analysis starts from the reduced-form distance-based
decomposition for each $A\in\{Y,W\}$. Let
\begin{align*}
    \theta_{A,t,\bx}^{\ast}(r)=\rp(r/h)^\top\bgamma_{A,t}^{\ast}(\bx),
\end{align*}
where
\begin{align*}
    \bgamma_{A,t}^{\ast}(\bx)
    =
    \argmin_{\bgamma\in\reals^{p+1}}
    \E\Big[
    \big(A_i-\rp(D_i(\bx)/h)^\top\bgamma\big)^2
    K_h(D_i(\bx))\Indicator(D_i(\bx)\in\I_t)
    \Big].
\end{align*}
The same algebra used in \eqref{sa-eq: distance error decomposition} gives
\begin{align}\label{sa-eq: reduced form fuzzy decomposition}
    \widehat{\vartheta}_A(\bx)-\vartheta_A(\bx)
    =
    \bias_A(\bx)+\mathfrak{L}_A(\bx)+\mathfrak{Q}_A(\bx),
    \qquad A\in\{Y,W\},
\end{align}
where
\begin{align*}
    \bias_A(\bx)
    &=
    \theta_{A,1,\bx}^{\ast}(0)-\theta_{A,0,\bx}^{\ast}(0)-\vartheta_A(\bx),\\
    \mathfrak{L}_A(\bx)
    &=
    \be_0^\top\bPsi_{1,\bx}^{-1}\mathbf{O}_{A,1,\bx}
    -
    \be_0^\top\bPsi_{0,\bx}^{-1}\mathbf{O}_{A,0,\bx},\\
    \mathfrak{Q}_A(\bx)
    &=
    \be_0^\top(\widehat{\bPsi}_{1,\bx}^{-1}-\bPsi_{1,\bx}^{-1})\mathbf{O}_{A,1,\bx}
    -
    \be_0^\top(\widehat{\bPsi}_{0,\bx}^{-1}-\bPsi_{0,\bx}^{-1})\mathbf{O}_{A,0,\bx},
\end{align*}
and
\begin{align*}
    \mathbf{O}_{A,t,\bx}
    =
    \En\Big[
    \rp(D_i(\bx)/h)K_h(D_i(\bx))
    \big(A_i-\theta_{A,t,\bx}^{\ast}(D_i(\bx))\big)
    \Indicator(D_i(\bx)\in\I_t)
    \Big].
\end{align*}

The covariance kernel for the reduced-form linear terms is defined as follows. For
$A,B\in\{Y,W\}$, let
\begin{align*}
    \Xi_{A,B,\bx_1,\bx_2}
    =
    \Xi_{A,B,0,\bx_1,\bx_2}
    +
    \Xi_{A,B,1,\bx_1,\bx_2},
\end{align*}
where
\begin{align*}
    \Xi_{A,B,t,\bx_1,\bx_2}
    =
    \frac{1}{nh^d}
    \be_0^\top\bPsi_{t,\bx_1}^{-1}
    \bUpsilon_{A,B,t,\bx_1,\bx_2}
    \bPsi_{t,\bx_2}^{-1}\be_0,
\end{align*}
and
\begin{align*}
    \bUpsilon_{A,B,t,\bx_1,\bx_2}
    &=
    h^d\E\Big[
    \rp(D_i(\bx_1)/h)\rp(D_i(\bx_2)/h)^\top
    K_h(D_i(\bx_1))K_h(D_i(\bx_2))\\
    &\hspace{.7in}\times
    \big(A_i-\theta_{A,t,\bx_1}^{\ast}(D_i(\bx_1))\big)
    \big(B_i-\theta_{B,t,\bx_2}^{\ast}(D_i(\bx_2))\big)
    \Indicator(D_i(\bx_1)\in\I_t)\Indicator(D_i(\bx_2)\in\I_t)
    \Big].
\end{align*}
Let $\widehat{\Xi}_{A,B,\bx_1,\bx_2}$ denote the sample analogue obtained by replacing
$\bPsi$, $\bUpsilon$, and $\theta^\ast$ with $\widehat{\bPsi}$,
$\widehat{\bUpsilon}$, and $\widehat{\theta}$.

The exact second-order Wald expansion is
\begin{align}\label{sa-eq: fuzzy second order}
    \widehat{\xi}(\bx)-\xi(\bx)
    =
    \bias_{\mathtt F}(\bx)
    +\mathfrak{L}_{\mathtt F}(\bx)
    +\mathfrak{Q}_{\mathtt F}(\bx)
    +\mathfrak{R}_{\mathtt F}(\bx),
\end{align}
where
\begin{align*}
    \bias_{\mathtt F}(\bx)
    &=
    \mathfrak{w}(\bx)^\top
    \big(\bias_Y(\bx),\bias_W(\bx)\big)^\top,\\
    \mathfrak{L}_{\mathtt F}(\bx)
    &=
    \mathfrak{w}(\bx)^\top
    \big(\mathfrak{L}_Y(\bx),\mathfrak{L}_W(\bx)\big)^\top,\\
    \mathfrak{Q}_{\mathtt F}(\bx)
    &=
    \mathfrak{w}(\bx)^\top
    \big(\mathfrak{Q}_Y(\bx),\mathfrak{Q}_W(\bx)\big)^\top,
\end{align*}
and
\begin{align*}
    \mathfrak{w}(\bx)
    =
    \Big(
    \frac{1}{\vartheta_W(\bx)},
    -\frac{\vartheta_Y(\bx)}{\vartheta_W(\bx)^2}
    \Big)^\top,
\end{align*}
with second-order remainder
\begin{align*}
    \mathfrak{R}_{\mathtt F}(\bx)
    &=
    \frac{\vartheta_Y(\bx)}
    {\vartheta_W(\bx)^2\widehat{\vartheta}_W(\bx)}
    \big(\widehat{\vartheta}_W(\bx)-\vartheta_W(\bx)\big)^2
    -
    \frac{1}{\vartheta_W(\bx)\widehat{\vartheta}_W(\bx)}
    \big(\widehat{\vartheta}_Y(\bx)-\vartheta_Y(\bx)\big)
    \big(\widehat{\vartheta}_W(\bx)-\vartheta_W(\bx)\big).
\end{align*}
If $\inf_{\bx\in\B}|\vartheta_W(\bx)|>0$, $h=o(1)$, and
$\frac{\log(1/h)}{n^{(1+v)/(2+v)}h^d}=o(1)$, then
\begin{align}\label{sa-eq: fuzzy second order bound}
    \sup_{\bx\in\B}|\mathfrak{R}_{\mathtt F}(\bx)|
    \lessp \rho_n^2,
\end{align}
where
\begin{align*}
    \rho_n
    =
    \sqrt{\frac{\log(1/h)}{nh^d}}
    +
    \frac{\log(1/h)}{n^{\frac{1+v}{2+v}}h^d}
    +
    \sup_{\bx\in\B}|\bias_Y(\bx)|
    +
    \sup_{\bx\in\B}|\bias_W(\bx)|.
\end{align*}

Thus, after the reduced-form distance-based decompositions are in place, the fuzzy
BATEC estimator is governed by a fixed linear combination of the reduced-form outcome
and first-stage components, up to the higher-order Wald remainder.

\subsection{Fuzzy BATEC}

\subsubsection{Point Estimation and Approximate MSE Expansions}

\begin{thm}[Convergence Rates: Fuzzy BATEC]\label{sa-thm: Convergence Rates: Fuzzy BATEC}
    Suppose Assumptions \ref{sa-assump: DGP}, \ref{sa-assump: Fuzzy DGP}, and \ref{sa-assump: Kernel, Distance and Boundary} hold, and let $h=o(1)$.
    \begin{itemize}[noitemsep,leftmargin=*]
    \item If $\xi(\bx)$ is well defined at a fixed $\bx\in\B$, $|\vartheta_W(\bx)|>0$, and $n^{\frac{1+v}{2+v}}h^d \to\infty$, then
    \begin{align*}
        |\widehat{\xi}(\bx)-\xi(\bx)|
        \lessp
        \frac{1}{\sqrt{nh^d}}
        +\frac{1}{n^{\frac{1+v}{2+v}}h^d}
        +|\bias_Y(\bx)|+|\bias_W(\bx)|.
    \end{align*}
    \item If $\xi(\bx)$ is well defined for all $\bx\in\B$, $\inf_{\bx\in\B}|\vartheta_W(\bx)|>0$, and $\frac{\log(1/h)}{n^{\frac{1+v}{2+v}}h^d}=o(1)$, then
    \begin{align*}
        \sup_{\bx\in\B}|\widehat{\xi}(\bx)-\xi(\bx)|
        \lessp \rho_n.
    \end{align*}
    \end{itemize}
\end{thm}

Define
\begin{align*}
    \boldsymbol{\Theta}_{\mathtt F,\bx_1,\bx_2}
    =
    \begin{pmatrix}
        \Xi_{Y,Y,\bx_1,\bx_2} & \Xi_{Y,W,\bx_1,\bx_2}\\
        \Xi_{W,Y,\bx_1,\bx_2} & \Xi_{W,W,\bx_1,\bx_2}
    \end{pmatrix},
    \qquad
    \Xi_{\mathtt F,\bx_1,\bx_2}
    =
    \mathfrak{w}(\bx_1)^\top
    \boldsymbol{\Theta}_{\mathtt F,\bx_1,\bx_2}
    \mathfrak{w}(\bx_2).
\end{align*}
Then $\E[\mathfrak{L}_{\mathtt F}(\bx)]=0$ and
$\E[\mathfrak{L}_{\mathtt F}(\bx)^2]=\Xi_{\mathtt F,\bx,\bx}$. The linearized
approximate MSE and approximate IMSE are
\begin{align*}
    \operatorname{AMSE}_{\mathtt F}(\bx)
    =
    \bias_{\mathtt F}(\bx)^2+\Xi_{\mathtt F,\bx,\bx}
    \qquad\text{and}\qquad
    \operatorname{AIMSE}_{\mathtt F}
    =
    \int_{\B}\operatorname{AMSE}_{\mathtt F}(\bx)w(\bx)\dif\Haus^{d-1}(\bx).
\end{align*}
These criteria omit the higher-order terms
$\mathfrak{Q}_{\mathtt F}(\bx)$ and $\mathfrak{R}_{\mathtt F}(\bx)$. If a leading-bias
expansion $\bias_{\mathtt F}(\bx)=h^mB_{\mathtt F,\bx}+o(h^m)$ is available and
$V_{\mathtt F,\bx}=nh^d\Xi_{\mathtt F,\bx,\bx}$, then the approximate pointwise
criterion is
\begin{align*}
    h^{2m}B_{\mathtt F,\bx}^2+\frac{1}{nh^d}V_{\mathtt F,\bx},
\end{align*}
with the same bandwidth formulas as in the sharp distance-based case after replacing
$(B_{\bx},V_{\bx})$ by $(B_{\mathtt F,\bx},V_{\mathtt F,\bx})$.

\subsubsection{Distributional Approximation and Inference}

Let
\begin{align*}
    \widehat{\mathfrak{w}}(\bx)
    =
    \Big(
    \frac{1}{\widehat{\vartheta}_W(\bx)},
    -\frac{\widehat{\vartheta}_Y(\bx)}{\widehat{\vartheta}_W(\bx)^2}
    \Big)^\top,
\end{align*}
and define
\begin{align*}
    \widehat{\boldsymbol{\Theta}}_{\mathtt F,\bx_1,\bx_2}
    =
    \begin{pmatrix}
        \widehat{\Xi}_{Y,Y,\bx_1,\bx_2} & \widehat{\Xi}_{Y,W,\bx_1,\bx_2}\\
        \widehat{\Xi}_{W,Y,\bx_1,\bx_2} & \widehat{\Xi}_{W,W,\bx_1,\bx_2}
    \end{pmatrix},
    \qquad
    \widehat{\Xi}_{\mathtt F,\bx_1,\bx_2}
    =
    \widehat{\mathfrak{w}}(\bx_1)^\top
    \widehat{\boldsymbol{\Theta}}_{\mathtt F,\bx_1,\bx_2}
    \widehat{\mathfrak{w}}(\bx_2).
\end{align*}

\begin{lem}[Covariance: Fuzzy BATEC]\label{sa-lem: Covariance: Fuzzy BATEC}
    Suppose Assumptions \ref{sa-assump: DGP}, \ref{sa-assump: Fuzzy DGP}, and
    \ref{sa-assump: Kernel, Distance and Boundary} hold,
    $\inf_{\bx\in\B}|\vartheta_W(\bx)|>0$, and $1\lesssim \inf_{\bx\in\B}nh^d\Xi_{\mathtt F,\bx,\bx}
        \leq \sup_{\bx\in\B}nh^d\Xi_{\mathtt F,\bx,\bx} \lesssim 1$.
     If $\frac{\log(1/h)}{n^{v/(2+v)}h^d}=o(1)$ and $h=o(1)$, then
    \begin{align*}
        \sup_{\bx_1,\bx_2\in\B} nh^d\big|\widehat{\Xi}_{\mathtt F,\bx_1,\bx_2} -\Xi_{\mathtt F,\bx_1,\bx_2}\big| =o_{\P}(1)
        \qquad\text{and}\qquad
        \sup_{\bx\in\B}
        \left|
        \frac{\widehat{\Xi}_{\mathtt F,\bx,\bx}}
             {\Xi_{\mathtt F,\bx,\bx}}
        -1
        \right|
        =o_{\P}(1).
    \end{align*}
\end{lem}

The feasible $t$-statistic and confidence interval are
\begin{align*}
    \widehat{\Tstat}_{\mathtt F}(\bx)
    =
    \frac{\widehat{\xi}(\bx)-\xi(\bx)}
         {\sqrt{\widehat{\Xi}_{\mathtt F,\bx,\bx}}}
    \qquad\text{and}\qquad
    \widehat{\CI}_{\alpha,\mathtt F}(\bx)
    =
    \left[
    \widehat{\xi}(\bx)-\q_\alpha\sqrt{\widehat{\Xi}_{\mathtt F,\bx,\bx}},
    \widehat{\xi}(\bx)+\q_\alpha\sqrt{\widehat{\Xi}_{\mathtt F,\bx,\bx}}
    \right].
\end{align*}
For the pointwise remainder condition, define
\begin{align*}
    \rho_{n,\bx}
    =
    \frac{1}{\sqrt{nh^d}}
    +
    \frac{1}{n^{\frac{1+v}{2+v}}h^d}
    +
    |\bias_Y(\bx)|
    +
    |\bias_W(\bx)|.
\end{align*}

\begin{thm}[Confidence Intervals: Fuzzy BATEC]\label{sa-thm: Confidence Intervals: Fuzzy BATEC}
    For a fixed boundary point $\bx\in\B$, suppose Assumptions \ref{sa-assump: DGP}, \ref{sa-assump: Fuzzy DGP}, and \ref{sa-assump: Kernel, Distance and Boundary} hold, $|\vartheta_W(\bx)|>0$, and $1\lesssim nh^d\Xi_{\mathtt F,\bx,\bx}\lesssim1$. If $n^{v/(2+v)}h^d\to\infty$, $h=o(1)$, $\sqrt{nh^d}\,|\bias_{\mathtt F}(\bx)|\to0$, and $\sqrt{nh^d}\rho_{n,\bx}^2\to0$,
    then
    \begin{align*}
        \sup_{u\in\reals}
        \left|
        \P\big(\widehat{\Tstat}_{\mathtt F}(\bx)\leq u\big)-\Phi(u)
        \right|
        =o(1),
    \end{align*}
    and
    \begin{align*}
        \P\big(\xi(\bx)\in\widehat{\CI}_{\alpha,\mathtt F}(\bx)\big)
        =
        1-\alpha+o(1),
    \end{align*}
    provided that $\q_\alpha=\Phi^{-1}(1-\alpha/2)$.
\end{thm}

For uniform inference, define the infeasible linearized statistic
\begin{align*}
    \overline{\Tstat}_{\mathtt F}(\bx)
    =
    \Xi_{\mathtt F,\bx,\bx}^{-1/2}\mathfrak{L}_{\mathtt F}(\bx),
    \qquad \bx\in\B.
\end{align*}

\begin{thm}[Stochastic Linearization: Fuzzy BATEC]\label{sa-thm: Stochastic Linearization: Fuzzy BATEC}
    Suppose the assumptions of Lemma~\ref{sa-lem: Covariance: Fuzzy BATEC} hold. Then
    \begin{align*}
        \sup_{\bx\in\B}
        \big|
        \widehat{\Tstat}_{\mathtt F}(\bx)
        -
        \overline{\Tstat}_{\mathtt F}(\bx)
        \big|
        &\lessp
        \sqrt{\log(1/h)}
        \bigg(
        \sqrt{\frac{\log(1/h)}{nh^d}}
        +
        \frac{\log(1/h)}{n^{v/(2+v)}h^d}
        \bigg)\\
        &\quad
        +
        \sqrt{nh^d}\sup_{\bx\in\B}|\bias_{\mathtt F}(\bx)|
        +
        \sqrt{nh^d}\rho_n^2.
    \end{align*}
\end{thm}

\begin{thm}[Gaussian Approximation for $\sup_{\bx\in\B}|\overline{\Tstat}_{\mathtt F}(\bx)|$]\label{sa-thm: Gaussian Approximation Suprema: Fuzzy BATEC}
    Suppose Assumptions \ref{sa-assump: DGP}, \ref{sa-assump: Fuzzy DGP}, and \ref{sa-assump: Kernel, Distance and Boundary} hold, $\inf_{\bx\in\B}|\vartheta_W(\bx)|>0$, $1\lesssim\inf_{\bx\in\B}nh^d\Xi_{\mathtt F,\bx,\bx}$, $\sup_{\bx\in\B}nh^d\Xi_{\mathtt F,\bx,\bx}\lesssim1$, $\liminf_{n\to\infty}\frac{\log h}{\log n}>-\infty$, and $h=o(1)$. Then, for $n$ large enough, on a possibly enlarged probability space there exists a mean-zero Gaussian process $Z_{\mathtt F}$ indexed by $\B$ with almost surely continuous sample paths such that
    \begin{align*}
        \Cov[Z_{\mathtt F}(\bx_1),Z_{\mathtt F}(\bx_2)]
        =
        \Cov[\overline{\Tstat}_{\mathtt F}(\bx_1),\overline{\Tstat}_{\mathtt F}(\bx_2)],
        \qquad \bx_1,\bx_2\in\B,
    \end{align*}
    and
    \begin{align*}
        &\P\Bigg[
        \bigg|
        \sup_{\bx\in\B}|\overline{\Tstat}_{\mathtt F}(\bx)|
        -
        \sup_{\bx\in\B}|Z_{\mathtt F}(\bx)|
        \bigg|
        >
        \frac{\log n}{(nh^d)^{1/6}}
        +
        \frac{(\log n)^{5/4}}{(nh^d)^{1/4}}
        +
        \frac{(\log n)^{3/2}}{(n^{v/(2+v)}h^d)^{1/2}}
        \Bigg]
        \lesssim
        \frac{1}{\log n}+\frac{\log n}{n}.
    \end{align*}
\end{thm}

Let $(\widehat{Z}_{\mathtt F}(\bx):\bx\in\B)$ be a mean-zero Gaussian process conditional
on $\bV=((\bX_i^\top,Y_i,W_i):1\le i\le n)$ with covariance
\begin{align*}
    \Cov[\widehat{Z}_{\mathtt F}(\bx_1),\widehat{Z}_{\mathtt F}(\bx_2)\mid\bV]
    =
    \frac{\widehat{\Xi}_{\mathtt F,\bx_1,\bx_2}}
    {\sqrt{\widehat{\Xi}_{\mathtt F,\bx_1,\bx_1}}
     \sqrt{\widehat{\Xi}_{\mathtt F,\bx_2,\bx_2}}},
    \qquad \bx_1,\bx_2\in\B.
\end{align*}

\begin{thm}[Confidence Bands: Fuzzy BATEC]\label{sa-thm: Confidence Bands: Fuzzy BATEC}
    Suppose the assumptions of Lemma~\ref{sa-lem: Covariance: Fuzzy BATEC} hold. If $\liminf_{n\to\infty}\frac{\log h}{\log n}>-\infty$, $\frac{n^{v/(2+v)}h^d}{(\log n)^9}\to\infty$, $h=o(1)$, $\sqrt{nh^d\log n}\sup_{\bx\in\B}|\bias_{\mathtt F}(\bx)|\to0$, and $\sqrt{nh^d\log n}\rho_n^2\to0$,
    then
    \begin{align*}
        \sup_{u\in\reals}
        \left|
        \P\left(\sup_{\bx\in\B}|\widehat{\Tstat}_{\mathtt F}(\bx)|\leq u\right)
        -
        \P\left(\sup_{\bx\in\B}|\widehat{Z}_{\mathtt F}(\bx)|\leq u\mid\bV\right)
        \right|
        =
        o_{\P}(1),
    \end{align*}
    and
    \begin{align*}
        \P\big[\xi(\bx)\in\widehat{\CI}_{\alpha,\mathtt F}(\bx),\text{ for all }\bx\in\B\big]
        =
        1-\alpha+o(1),
    \end{align*}
    provided that
    $\q_\alpha=\inf\{c>0:\P(\sup_{\bx\in\B}|\widehat{Z}_{\mathtt F}(\bx)|\geq c\mid\bV)\leq\alpha\}$.
\end{thm}

\subsection{Fuzzy WBATE}

We impose the normalization $\int_{\B}w(\bx)\dif\Haus^{d-1}(\bx)=1$. The
distance-based fuzzy WBATE parameter and plug-in estimator are
\begin{align*}
    \xi_{\mathtt{WBATE}}
    =
    \int_{\B}\xi(\bx)w(\bx)\dif\Haus^{d-1}(\bx)
    \qquad\text{and}\qquad
    \widehat{\xi}_{\mathtt{WBATE}}
    =
    \int_{\B}\widehat{\xi}(\bx)w(\bx)\dif\Haus^{d-1}(\bx).
\end{align*}
Under the conditions of Theorem~\ref{sa-thm: Fuzzy Identification},
$\xi_{\mathtt{WBATE}}=\zeta_{\mathtt{WBATE}}$, where
$\zeta_{\mathtt{WBATE}}=\int_{\B}\zeta(\bx)w(\bx)\dif\Haus^{d-1}(\bx)$.

Integrating the pointwise fuzzy expansion \eqref{sa-eq: fuzzy second order} gives
\begin{align}\label{sa-eq: fuzzy WBATE decomposition}
    \widehat{\xi}_{\mathtt{WBATE}}-\xi_{\mathtt{WBATE}}
    =
    \bias_{\mathtt F,\mathtt{WBATE}}
    +\mathfrak{L}_{\mathtt F,\mathtt{WBATE}}
    +\mathfrak{Q}_{\mathtt F,\mathtt{WBATE}}
    +\mathfrak{R}_{\mathtt F,\mathtt{WBATE}},
\end{align}
where
\begin{align*}
    \bias_{\mathtt F,\mathtt{WBATE}}
    &=
    \int_{\B}\bias_{\mathtt F}(\bx)w(\bx)\dif\Haus^{d-1}(\bx),\\
    \mathfrak{L}_{\mathtt F,\mathtt{WBATE}}
    &=
    \int_{\B}\mathfrak{L}_{\mathtt F}(\bx)w(\bx)\dif\Haus^{d-1}(\bx),\\
    \mathfrak{Q}_{\mathtt F,\mathtt{WBATE}}
    &=
    \int_{\B}\mathfrak{Q}_{\mathtt F}(\bx)w(\bx)\dif\Haus^{d-1}(\bx),\\
    \mathfrak{R}_{\mathtt F,\mathtt{WBATE}}
    &=
    \int_{\B}\mathfrak{R}_{\mathtt F}(\bx)w(\bx)\dif\Haus^{d-1}(\bx).
\end{align*}

The aggregate variance of the leading linearized term is
\begin{align*}
    \Xi_{\mathtt F,\mathtt{WBATE}}
    =
    \int_{\B}\int_{\B}
    \Xi_{\mathtt F,\bx_1,\bx_2}
    w(\bx_1)w(\bx_2)
    \dif\Haus^{d-1}(\bx_1)\dif\Haus^{d-1}(\bx_2),
\end{align*}
with feasible counterpart
\begin{align*}
    \widehat{\Xi}_{\mathtt F,\mathtt{WBATE}}
    =
    \int_{\B}\int_{\B}
    \widehat{\Xi}_{\mathtt F,\bx_1,\bx_2}
    w(\bx_1)w(\bx_2)
    \dif\Haus^{d-1}(\bx_1)\dif\Haus^{d-1}(\bx_2).
\end{align*}
For inference, impose the aggregate nondegeneracy condition
\begin{align}
    \Xi_{\mathtt F,\mathtt{WBATE}}
    \gtrsim
    \frac{1}{n h},
    \label{sa-eq: fuzzy WBATE aggregate nondegeneracy}
\end{align}
which rules out cancellation in the integrated fuzzy covariance kernel.

\begin{lem}[Variance Linear Term: Fuzzy WBATE]\label{sa-lem: Variance: Fuzzy WBATE}
    Suppose Assumptions \ref{sa-assump: DGP}, \ref{sa-assump: Fuzzy DGP}, \ref{sa-assump: Kernel, Distance and Boundary}, and \ref{sa-assump: Weight Function and Boundary} hold, $\inf_{\bx\in\B}|\vartheta_W(\bx)|>0$, and \eqref{sa-eq: local boundary measure regularity} holds. If $\frac{n h^d}{\log(1/h)}\to\infty$ and $h=o(1)$, then
    \begin{align*}
        \Var[\mathfrak{L}_{\mathtt F,\mathtt{WBATE}}]
        =
        \Xi_{\mathtt F,\mathtt{WBATE}}
        \lesssim
        \frac{1}{n h}.
    \end{align*}
    If, in addition, \eqref{sa-eq: fuzzy WBATE aggregate nondegeneracy} holds, then
    \begin{align*}
        \frac{1}{n h}
        \lesssim
        \Xi_{\mathtt F,\mathtt{WBATE}}
        \lesssim
        \frac{1}{n h}.
    \end{align*}
    If $\frac{n^{v/(2+v)}h^d}{\log(1/h)}\to\infty$ and $h=o(1)$, then
    \begin{align*}
        \widehat{\Xi}_{\mathtt F,\mathtt{WBATE}}
        =
        \Xi_{\mathtt F,\mathtt{WBATE}}
        +
        o_{\P}\Big(\frac{1}{n h}\Big).
    \end{align*}
    If, in addition, \eqref{sa-eq: fuzzy WBATE aggregate nondegeneracy} holds, then
    $\widehat{\Xi}_{\mathtt F,\mathtt{WBATE}}/\Xi_{\mathtt F,\mathtt{WBATE}}\to_{\P}1$.
\end{lem}

\begin{lem}[Higher-Order Remainders: Fuzzy WBATE]\label{sa-lem: Higher-Order Remainders: Fuzzy WBATE}
    Suppose Assumptions \ref{sa-assump: DGP}, \ref{sa-assump: Fuzzy DGP}, \ref{sa-assump: Kernel, Distance and Boundary}, and \ref{sa-assump: Weight Function and Boundary} hold, $\inf_{\bx\in\B}|\vartheta_W(\bx)|>0$, and \eqref{sa-eq: local boundary measure regularity} holds. If $\frac{n h^d}{\log(1/h)}\to\infty$ and $h=o(1)$, then
    \begin{align*}
        \mathfrak{Q}_{\mathtt F,\mathtt{WBATE}}
        \lessp
        \frac{1}{n h^d}.
    \end{align*}
    If, in addition, $\frac{\log(1/h)}{n^{(1+v)/(2+v)}h^d}=o(1)$, then
    \begin{align*}
        |\mathfrak{R}_{\mathtt F,\mathtt{WBATE}}|
        \lessp
        \rho_n^2.
    \end{align*}
    Consequently, if also $n h^{2d-1}\to\infty$ and $\sqrt{n h}\rho_n^2\to0$, then
    \begin{align*}
        \mathfrak{Q}_{\mathtt F,\mathtt{WBATE}}
        +
        \mathfrak{R}_{\mathtt F,\mathtt{WBATE}}
        =
        o_{\P}\Big(\frac{1}{\sqrt{n h}}\Big).
    \end{align*}
\end{lem}

\begin{thm}[Convergence Rate: Fuzzy WBATE]\label{sa-thm: Convergence Rate: Fuzzy WBATE}
    Suppose Assumptions \ref{sa-assump: DGP}, \ref{sa-assump: Fuzzy DGP}, \ref{sa-assump: Kernel, Distance and Boundary}, and \ref{sa-assump: Weight Function and Boundary} hold, $\xi(\bx)$ is well defined for all $\bx\in\B$, $\inf_{\bx\in\B}|\vartheta_W(\bx)|>0$, and \eqref{sa-eq: local boundary measure regularity} holds. If $\frac{\log(1/h)}{n^{(1+v)/(2+v)}h^d}=o(1)$ and $h=o(1)$, then
    \begin{align*}
        \left|
        \widehat{\xi}_{\mathtt{WBATE}}
        -
        \xi_{\mathtt{WBATE}}
        \right|
        \lessp
        |\bias_{\mathtt F,\mathtt{WBATE}}|
        +
        \frac{1}{\sqrt{n h}}
        +
        \frac{1}{n h^d}
        +
        \rho_n^2.
    \end{align*}
\end{thm}

For inference, define
\begin{align*}
    \widehat{\Tstat}_{\mathtt F,\mathtt{WBATE}}
    =
    \frac{\widehat{\xi}_{\mathtt{WBATE}}-\xi_{\mathtt{WBATE}}}
         {\sqrt{\widehat{\Xi}_{\mathtt F,\mathtt{WBATE}}}},
\end{align*}
and
\begin{align*}
    \widehat{\CI}_{\alpha,\mathtt F,\mathtt{WBATE}}
    =
    \left[
    \widehat{\xi}_{\mathtt{WBATE}}
    -\q_{\alpha}\sqrt{\widehat{\Xi}_{\mathtt F,\mathtt{WBATE}}},
    \quad
    \widehat{\xi}_{\mathtt{WBATE}}
    +\q_{\alpha}\sqrt{\widehat{\Xi}_{\mathtt F,\mathtt{WBATE}}}
    \right].
\end{align*}

\begin{thm}[Asymptotic Normality: Fuzzy WBATE]\label{sa-thm: Asymptotic Normality: Fuzzy WBATE}
    Suppose Assumptions \ref{sa-assump: DGP}, \ref{sa-assump: Fuzzy DGP}, \ref{sa-assump: Kernel, Distance and Boundary}, and \ref{sa-assump: Weight Function and Boundary} hold, $\xi(\bx)$ is well defined for all $\bx\in\B$, $\inf_{\bx\in\B}|\vartheta_W(\bx)|>0$, and \eqref{sa-eq: fuzzy WBATE aggregate nondegeneracy} and \eqref{sa-eq: local boundary measure regularity} hold. If $\frac{\log(1/h)}{n^{v/(2+v)}h^d}=o(1)$, $h=o(1)$, $n h^{2d-1}\to\infty$, $\sqrt{n h}|\bias_{\mathtt F,\mathtt{WBATE}}|\to0$, and $\sqrt{n h}\rho_n^2\to0$,
    then
    \begin{align*}
        \sup_{u\in\reals}
        \left|
        \P\big(\widehat{\Tstat}_{\mathtt F,\mathtt{WBATE}}\leq u\big)
        -
        \Phi(u)
        \right|
        =
        o(1),
    \end{align*}
    and
    \begin{align*}
        \P\big[
        \xi_{\mathtt{WBATE}}
        \in
        \widehat{\CI}_{\alpha,\mathtt F,\mathtt{WBATE}}
        \big]
        =
        1-\alpha+o(1),
    \end{align*}
    provided that $\q_{\alpha}=\Phi^{-1}(1-\alpha/2)$.
\end{thm}

\subsection{Fuzzy LBATE}

The distance-based fuzzy LBATE parameter and plug-in estimator are
\begin{align*}
    \xi_{\mathtt{LBATE}} = \sup_{\bx\in\B}\xi(\bx)
    \qquad\text{and}\qquad
    \widehat{\xi}_{\mathtt{LBATE}} = \sup_{\bx\in\B}\widehat{\xi}(\bx).
\end{align*}
Under the conditions of Theorem~\ref{sa-thm: Fuzzy Identification},
$\xi_{\mathtt{LBATE}}=\zeta_{\mathtt{LBATE}}$, where
$\zeta_{\mathtt{LBATE}}=\sup_{\bx\in\B}\zeta(\bx)$.

\begin{thm}[Convergence Rate: Fuzzy LBATE]\label{sa-thm: Convergence Rate: Fuzzy LBATE}
    Suppose the assumptions and conditions of the uniform part of Theorem~\ref{sa-thm: Convergence Rates: Fuzzy BATEC} hold. Then
    \begin{align*}
        \left|
        \widehat{\xi}_{\mathtt{LBATE}}
        -
        \xi_{\mathtt{LBATE}}
        \right|
        \lessp
        \rho_n.
    \end{align*}
\end{thm}

Projecting the fuzzy BATEC confidence band gives the fuzzy LBATE confidence interval
\begin{align*}
    \widehat{\CI}_{\alpha,\mathtt F,\mathtt{LBATE}}
    =
    \left[
    \sup_{\bx\in\B}\left(
    \widehat{\xi}(\bx)
    -
    \q_{\alpha}\sqrt{\widehat{\Xi}_{\mathtt F,\bx,\bx}}
    \right),
    \quad
    \sup_{\bx\in\B}\left(
    \widehat{\xi}(\bx)
    +
    \q_{\alpha}\sqrt{\widehat{\Xi}_{\mathtt F,\bx,\bx}}
    \right)
    \right],
\end{align*}
where
$\q_{\alpha}=\inf\{c>0:\P(\sup_{\bx\in\B}|\widehat Z_{\mathtt F}(\bx)|\ge c\mid\bV)\le\alpha\}$.

\begin{thm}[Confidence Intervals: Fuzzy LBATE]\label{sa-thm: Confidence Intervals: Fuzzy LBATE}
    Suppose the assumptions and conditions in Theorem~\ref{sa-thm: Confidence Bands: Fuzzy BATEC} hold. Then
    \begin{align*}
        \P\left[
        \xi_{\mathtt{LBATE}}
        \in
        \widehat{\CI}_{\alpha,\mathtt F,\mathtt{LBATE}}
        \right]
        \geq
        1-\alpha+o(1).
    \end{align*}
\end{thm}

As in the sharp-design LBATE case, this interval is generally conservative because it is
obtained by projecting a simultaneous confidence band for the full fuzzy BATEC.

\section{Gaussian Strong Approximation}\label{sa-sec: Gaussian Strong Approximation}

We present a Gaussian strong approximation theorem, which is the key technical tool behind Theorem~\ref{sa-thm: Gaussian Strong Approximation: Tstat}. The theorem builds on and generalizes the results in \cite{Cattaneo-Yu_2025_AOS}. Consider the \emph{residual-based empirical process} given by
\begin{align*}
    M_n[g,r] = \frac{1}{\sqrt{n}} \sum_{i = 1}^n \Big[g(\bx_i) r(y_i) - \E[g(\bx_i) r(y_i)] \Big], \qquad g \in \G, r \in \R,
\end{align*}
where $\G$ and $\R$ are classes of functions satisfying certain regularity conditions.

\subsection{Definitions for Function Spaces}

Let $\F$ be a class of measurable functions from a probability space $(\reals^q, \mathcal{B}(\reals^q), \P)$ to $\reals$. We introduce several definitions that capture properties of $\F$.

\begin{enumerate}[label=(\roman*)]
    \item  $\F$ is pointwise measurable if it contains a countable subset $\G$ such that for any $f \in \F$, there exists a sequence $(g_m:m\geq1) \subseteq \G$ such that $\lim_{m \to \infty} g_m(\bu) = f(\bu)$ for all $\bu \in \reals^q$.
    \item  Let $\supp(\F) = \cup_{f \in \F}\supp(f)$. A probability measure $\bbQ_\F$ on $(\reals^q,\mathcal{B}(\reals^q))$ is a surrogate measure for $\P$ with respect to $\F$ if
    \begin{enumerate}[label=(\roman*)]
        \item $\bbQ_\F$ agrees with $\P$ on $\supp(\P) \cap \supp(\F)$.
        \item $\bbQ_\F(\supp(\F) \setminus \supp(\P)) = 0$.
    \end{enumerate}
    Let $\mathcal{Q}_\F=\supp(\bbQ_\F)$.
    \item For $q=1$ and an interval $\mathcal{I}\subseteq\reals$, the pointwise total variation of $\F$ over $\mathcal{I}$ is
    \begin{align*}
        \mathtt{pTV}_{\F,\mathcal{I}} = \sup_{f \in \F} \sup_{P\geq1}\sup_{\calP_P \in \mathcal{I}} \sum_{i = 1}^{P-1}|f(a_{i+1}) - f(a_i)|,
    \end{align*}
    where $\calP_P=\{(a_1,\dots,a_P):a_1 \leq \cdots \leq a_P\}$ denotes the collection of all partitions of $\mathcal{I}$.
    \item For a non-empty $\mathcal{C} \subseteq \reals^q$, the total variation of $\F$ over $\mathcal{C}$ is
    \begin{align*}
        \ttTV_{\F, \mathcal{C}} = \inf_{\mathcal{U} \in \mathcal{O}(\mathcal{C})}\sup_{f \in \F} \sup_{\phi \in \mathscr{D}_{q}(\mathcal{U})} \int_{\reals^q} f(\bu)\operatorname{div}(\phi)(\bu) \dif\bu / \infnorm{\norm{\phi}_2},
    \end{align*}
    where $\mathcal{O}(\mathcal{C})$ denotes the collection of all open sets that contains $\mathcal{C}$, and $\mathscr{D}_{q}(\mathcal{U})$ denotes the space of infinitely differentiable functions from $\reals^q$ to $\reals^q$ with compact support contained in $\mathcal{U}$.
    \item For a non-empty $\mathcal{C} \subseteq \reals^q$, the local total variation constant of $\F$ over $\mathcal{C}$, is a positive number $\ttK_{\F,\mathcal{C}}$ such that for any cube $\mathcal{D} \subseteq \reals^q$ with edges of length $\ell$ parallel to the coordinate axises,
    \begin{align*}
        \ttTV_{\F, \mathcal{D} \cap \mathcal{C}}  \leq \ttK_{\F, \mathcal{C}} \ell^{d-1}.
    \end{align*}
    \item For a non-empty $\mathcal{C} \subseteq \reals^q$, the envelopes of $\F$ over $\mathcal{C}$ are
    \begin{align*}
        \ttM_{\F,\C} = \sup_{\bu \in \C }M_{\F,\C}(\bu),
        \qquad M_{\F,\C}(\bu) = \sup_{f \in \F}|f(\bu)|,
        \qquad \bu \in \C.
    \end{align*}
    \item  For a non-empty $\mathcal{C} \subseteq \reals^q$, the Lipschitz constant of $\F$ over $\C$ is
    \begin{align*}
        \ttL_{\F,\C} = \sup_{f \in \F}\sup_{\bu_1, \bu_2 \in \C} \frac{|f(\bu_1) - f(\bu_2)|}{\|\bu_1 - \bu_2\|_\infty}.
    \end{align*}
    \item For a non-empty $\mathcal{C} \subseteq \reals^q$, the $L_1$ bound of $\F$ over $\C$ is
    \begin{align*}
        \mathtt{E}_{\F,\C} = \sup_{f \in \F} \int_{\C} |f| \dif\P.
    \end{align*}
    \item   For a non-empty $\mathcal{C} \subseteq \reals^q$, the uniform covering number of $\F$ with envelope $M_{\F,\C}$ over $\C$ is
    \begin{align*}
        \ttN_{\F,\C}(\delta,M_{\F,\C}) = \sup_{\mu} N(\F,\norm{\cdot}_{\mu,2},\delta \norm{M_{\F,\C}}_{\mu,2}),
        \qquad \delta \in (0, \infty),
    \end{align*}
    where the supremum is taken over all finite discrete measures on $(\C, \mathcal{B}(\C))$. We assume that $M_{\F,\C}(\bu)$ is finite for every $\bu \in \C$.
    \item  For a non-empty $\mathcal{C} \subseteq \reals^q$, the uniform entropy integral of $\F$ with envelope $M_{\F,\C}$ over $\C$ is
    \begin{align*}
        J_\C(\delta, \F, M_{\F,\C}) = \int_0^{\delta} \sqrt{1 + \log \ttN_{\F,\C}(\varepsilon,M_{\F,\C})} \dif\varepsilon,
    \end{align*}
    where it is assumed that $M_{\F,\C}(\bu)$ is finite for every $\bu \in \C$.
    \item For a non-empty $\mathcal{C} \subseteq \reals^q$, $\F$ is a VC-type class with envelope $M_{\F,\C}$ over $\C$ if (i) $M_{\F,\C}$ is measurable and $M_{\F,\C}(\bu)$ is finite for every $\bu \in \C$, and (ii) there exist $\ttc_{\F,\C}>0$ and $\mathtt{d}_{\F,\C}>0$ such that
    \begin{align*}
        \ttN_{\F,\C}(\varepsilon,M_{\F,\C}) \leq \ttc_{\F,\C} \varepsilon^{-\mathtt{d}_{\F,\C}},
        \qquad \varepsilon\in(0,1).
    \end{align*}

\end{enumerate}

If a surrogate measure $\bbQ_\F$ for $\P$ with respect to $\F$ has been assumed, and it is clear from the context, we drop the dependence on $\C = \mathcal{Q}_{\F}$ for all quantities in the previous definitions. That is, to save notation, we set $\ttTV_{\F}=\ttTV_{\F,\mathcal{Q}_{\F}}$, $\ttK_{\F}=\ttK_{\F,\mathcal{Q}_{\F}}$, $\ttM_{\F}=\ttM_{\F,\mathcal{Q}_{\F}}$, $M_{\F}(\bu)=M_{\F,\mathcal{Q}_{\F}}(\bu)$, $\ttL_{\F}=\ttL_{\F,\mathcal{Q}_{\F}}$, and so on, whenever there is no confusion.

\subsection{Multiplicative-Separable Empirical Process}

The following theorem generalizes \citet[Theorem SA.1]{Cattaneo-Yu_2025_AOS} by requiring only bounded polynomial moments for $y_i$ conditional on $\bx_i$.

\begin{thm}[Strong Approximation for $(M_n(g,r) + M_n(h,s): g \in \G, r \in \R, h \in \H, s \in \mathcal{S})$]\label{sa-thm: Strong Approximation}
    Suppose $(\bz_i=(\bx_i, y_i): 1 \leq i \leq n)$ are i.i.d. random vectors taking values in $(\reals^{d+1}, \mathcal{B}(\reals^{d+1}))$ with common law $\P_Z$, where $\bx_i$ has distribution $\P_X$ supported on $\X\subseteq\reals^d$, $y_i$ has distribution $\P_Y$ supported on $\Y\subseteq\reals$, $\sup_{\bx \in \X}\E[|y_i|^{2 + v}|\bx_i = \bx] \leq 2$ for some $v > 0$, and the following conditions hold.
    
    \begin{enumerate}[label=\emph{(\roman*)}]
        \item $\G$ and $\H$ are real-valued pointwise measurable classes of functions on $(\reals^d, \mathcal{B}(\reals^d), \P_X)$.
    
        \item There exists a surrogate measure $\bbQ_{\G \cup \H}$ for $\P_X$ with respect to $\G \cup \H$ such that $\bbQ_{\G \cup \H} = \mathfrak{m} \circ \phi_{\G \cup \H}$, where the \textit{normalizing transformation} $\phi_{\G \cup \H}: \mathcal{Q}_{\G \cup \H} \to [0,1]^d$ is a diffeomorphism.
    
        \item $\G$ is a VC-type class with envelope $\ttM_{\G,\mathcal{Q}_{\G \cup \H}}$ over $\mathcal{Q}_{\G \cup \H}$ with $\ttc_{\G,\mathcal{Q}_{\G \cup \H}} \geq e$ and $\mathtt{d}_{\G,\mathcal{Q}_{\G \cup \H}} \geq 1$. $\H$ is a VC-type class with envelope $\ttM_{\H,\mathcal{Q}_{\G \cup \H}}$ over $\mathcal{Q}_{\G \cup \H}$ with $\ttc_{\H,\mathcal{Q}_{\G \cup \H}} \geq e$ and $\mathtt{d}_{\H,\mathcal{Q}_{\G \cup \H}} \geq 1$.
    
        \item $\R$ and $\S$ are real-valued pointwise measurable classes of functions on $(\reals, \mathcal{B}(\reals),\P_Y)$.
    
        \item $\R$ is a VC-type class with envelope $M_{\R,\Y}$ over $\Y$ with $\ttc_{\R,\Y}\geq e$ and $\mathtt{d}_{\R,\Y}\geq 1$, where $M_{\R,\Y}(y) + \mathtt{pTV}_{\R,(-|y|,|y|)} \leq \mathtt{v} (1 + |y|)$ for all $y \in \Y$, for some $\mathtt{v}>0$. $\S$ is a VC-type class with envelope $M_{\S,\Y}$ over $\Y$ with $\ttc_{\S,\Y}\geq e$ and $\mathtt{d}_{\S,\Y}\geq 1$, where $M_{\S,\Y}(y) + \mathtt{pTV}_{\S,(-|y|,|y|)} \leq \mathtt{v} (1 + |y|)$ for all $y \in \Y$, for some $\mathtt{v}>0$.
    
        \item There exists a constant $\mathtt{k}$ such that $|\log_2 \mathtt{E}| + |\log_2 \ttTV| + |\log_2 \ttM| + |\log_2 \ttL| \leq \mathtt{k} \log_2(n)$, where $\ttE = \max \{\ttE_{\G,\mathcal{Q}_{\G \cup \H}}, \ttE_{\H,\mathcal{Q}_{\G \cup \H}}\}$, $\ttTV = \max \{\ttTV_{\G,\mathcal{Q}_{\G \cup \H}}, \ttTV_{\H,\mathcal{Q}_{\G \cup \H}}\}$, $\ttM = \max \{\ttM_{\G,\mathcal{Q}_{\G \cup \H}}, \ttM_{\H,\mathcal{Q}_{\G \cup \H}}\}$, and $\ttL = \max \{\ttL_{\G,\mathcal{Q}_{\G \cup \H}}, \ttL_{\H,\mathcal{Q}_{\G \cup \H}}\}$.
    \end{enumerate}
    Consider the empirical process
    \begin{align*}
        A_n(g,h,r,s) = M_n(g,r) + M_n(h,s), \qquad g \in \G, r \in \R, h \in \H, s \in \mathcal{S}.
    \end{align*}
    Then, on a possibly enlarged probability space, there exists a sequence of mean-zero Gaussian processes $(Z_n^A(g,h,r,s): g\in\G, h \in \H, r \in \R, s \in \mathcal{S})$ with almost sure continuous trajectories such that:
    \begin{itemize}
        \item $\E[A_n(g_1,h_1,r_1,s_1) A_n(g_2,h_2,r_2,s_2)] = \E[Z_n^A(g_1, h_1, r_1,s_1) Z_n^A(g_2, h_2, r_2,s_2)]$ holds for all $(g_1, h_1, r_1, s_1)$, $(g_2, h_2, r_2, s_2) \in \G \times \H \times \R \times \S$, and
        \item $\E\big[\norm{A_n - Z_n^A}_{\G \times \H \times \R \times \mathcal{S}}\big] \leq C \mathtt{v} ((\ttd \log (\ttc n))^{\frac{3}{2}} \mathtt{r}_n^{\frac{v}{v +2}}(\sqrt{\ttM \ttE})^{\frac{2}{v+2}} + \ttd \log(\ttc n) \ttM n^{-\frac{v/2}{2+v}} + \ttd \log(\ttc n) \ttM n^{-\frac{1}{2}} \Big(\frac{\sqrt{\ttM \ttE}}{\mathtt{r}_n}\Big)^{\frac{2}{v+2}})$,
    \end{itemize}
    where $C$ is a universal constant, $\ttc = \ttc_{\G, \mathcal{Q}_{\G \cup \H}} + \ttc_{\H,\mathcal{Q}_{\G \cup \H}} + \ttc_{\R,\Y} + \ttc_{\S,\Y} + \ttk$, $\ttd = \ttd_{\G,\mathcal{Q}_{\G \cup \H}} \ttd_{\H,\mathcal{Q}_{\G \cup \H}} \ttd_{\R,\Y} \ttd_{\S,\Y} \ttk$, 
    \begin{gather*}
        \mathtt{r}_n = \min\Big\{\frac{(\ttc_1^d \ttM^{d+1} \ttTV^d \mathtt{E})^{1/(2d+2)} }{n^{1/(2d+2)}}, \frac{(\ttc_1^{\frac{d}{2}} \ttc_2^{\frac{d}{2}}\ttM \ttTV^{\frac{d}{2}} \mathtt{E} \ttL^{\frac{d}{2}})^{1/(d+2)}}{n^{1/(d+2)}} \Big\}, \\
        \ttc_1 = d \sup_{\bx \in \mathcal{Q}_{\G \cup \H}} \prod_{j = 1}^{d-1} \sigma_j(\nabla \phi_{\G \cup \H}(\bx)), \qquad 
        \ttc_2 = \sup_{\bx \in \mathcal{Q}_{\G \cup \H}} \frac{1}{\sigma_{d}(\nabla \phi_{\G \cup \H}(\bx))}. 
    \end{gather*}
\end{thm}

\subsection{Proof of Theorem~\ref{sa-thm: Strong Approximation}}

Let $\mathcal{I}_A=\G\times\H\times\R\times\S$. For each
$\iota=(g,h,r,s)\in\mathcal{I}_A$, define the combined summand
\begin{align*}
    a_{\iota}(\bx,y)=g(\bx)r(y)+h(\bx)s(y),
    \qquad
    \F_A=\{a_{\iota}:\iota\in\mathcal{I}_A\}.
\end{align*}
The proof below adapts the projection and coupling construction in
\citet[Theorem SA.1]{Cattaneo-Yu_2025_AOS}. The adaptation has two
purposes. First, the two empirical processes cannot be coupled separately,
because doing so would generally lose the covariance between
$M_n(g,r)$ and $M_n(h,s)$. Second, a direct generic strong approximation
for the ordinary class $\F_A$ would treat $(\bx,y)$ symmetrically and would
therefore not exploit the multiplicative-separable structure. We instead use
one common normalizing transformation, one common projection, and one
common Gaussian construction for the whole sum $a_{\iota}$.

\textit{Reduction and common normalization.}
Let $\widetilde{\G}=\{g\circ\phi_{\G\cup\H}^{-1}:g\in\G\}$ and
$\widetilde{\H}=\{h\circ\phi_{\G\cup\H}^{-1}:h\in\H\}$. By the surrogate
measure and normalizing-transformation argument used in the proof of
\citet[Theorem SA.1]{Cattaneo-Yu_2025_AOS}, after enlarging the probability
space if necessary, we may work with i.i.d. variables
$\bu_i\sim\mathrm{Uniform}([0,1]^d)$ such that, simultaneously for all
$g\in\G$ and $h\in\H$,
\[
    g(\bx_i)=g\circ\phi_{\G\cup\H}^{-1}(\bu_i),
    \qquad
    h(\bx_i)=h\circ\phi_{\G\cup\H}^{-1}(\bu_i).
\]
This reduction is simultaneous for $\G$ and $\H$, which is why the
surrogate measure is imposed on $\G\cup\H$. The constants $\ttc_1$ and
$\ttc_2$ in the statement record the effect of this change of variables on
the total-variation and Lipschitz constants; all displays below are written
for the normalized variables and translated back through these constants.

\textit{Entropy and envelopes.}
The classes $\widetilde{\G}$ and $\widetilde{\H}$ remain pointwise
measurable and VC-type after the change of variables. Products with
$\R$ and $\S$ preserve VC-type entropy, and finite sums preserve the
same property, so $\F_A$ is pointwise measurable and VC-type. For
$\iota=(g,h,r,s)$,
\begin{align*}
    |a_{\iota}(\bx,y)|
    \leq \mathtt{v}(1+|y|)
    \big(M_{\G,\mathcal{Q}_{\G\cup\H}}(\bx)
        +M_{\H,\mathcal{Q}_{\G\cup\H}}(\bx)\big).
\end{align*}
Consequently the entropy constants for the combined index class are bounded
by fixed powers of the constants appearing in the theorem; these powers are
absorbed into $\ttc$ and $\ttd$.

\textbf{Total Variation}. The $x$-side $L_1$, envelope, Lipschitz, and
total-variation quantities entering the projection argument are controlled
by the corresponding maxima $\ttE$, $\ttM$, $\ttL$, and $\ttTV$. For example,
for each fixed $y$ and each cube $\mathcal{D}\subseteq[0,1]^d$,
\begin{align*}
    \ttTV_{\{a_{\iota}(\cdot,y):\iota\in\mathcal{I}_A\},\mathcal{D}}
    &\leq
    \mathtt{v}(1+|y|)
    \big(\ttTV_{\G,\mathcal{D}}+\ttTV_{\H,\mathcal{D}}\big),\\
    \ttL_{\{a_{\iota}(\cdot,y):\iota\in\mathcal{I}_A\},\mathcal{D}}
    &\leq
    \mathtt{v}(1+|y|)
    \big(\ttL_{\G,\mathcal{D}}+\ttL_{\H,\mathcal{D}}\big).
\end{align*}
The analogous pointwise total-variation bound in the $y$ coordinate follows
from Assumption (v):
\[
    \mathtt{pTV}_{\{a_{\iota}(\bx,\cdot):\iota\in\mathcal{I}_A\},(-|y|,|y|)}
    \leq
    \mathtt{v}(1+|y|)
    \big(M_{\G,\mathcal{Q}_{\G\cup\H}}(\bx)
        +M_{\H,\mathcal{Q}_{\G\cup\H}}(\bx)\big).
\]
Thus the meshing, projection, and Gaussian-projection constants in the
proof of \citet[Theorem SA.1]{Cattaneo-Yu_2025_AOS} are bounded by the
same expressions as in that theorem, with $\G$ replaced by the worst case
between $\G$ and $\H$ and with $\R$ replaced by the worst case between
$\R$ and $\S$.

\textit{Common projection and coupling.}
Let $\pi_\delta$ denote a $\delta$-net projection of
$\mathcal{I}_A$ under the $L_2(\P_Z)$ semimetric induced by
$a_{\iota}$, and let $\Pi$ denote the common cylindered Haar projection
used in the proof of \citet[Theorem SA.1]{Cattaneo-Yu_2025_AOS}, with
main depth in the normalized $\bu$ coordinates and multiplier depth in the
$y$ coordinate. The same projected cells are used for the two summands
$g(\bx)r(y)$ and $h(\bx)s(y)$. Therefore the projected empirical vector is
the projection of the sum, not the sum of two independently coupled
projections. For the Gaussian process $Z_n^A$ generated by this common
construction,
\begin{align*}
    \norm{A_n-Z_n^A}_{\mathcal{I}_A}
    \leq&
    \norm{A_n-A_n\circ\pi_\delta}_{\mathcal{I}_A}
    +\norm{A_n\circ\pi_\delta-\Pi A_n\circ\pi_\delta}_{\mathcal{I}_A} \\
    &+\norm{\Pi A_n\circ\pi_\delta-\Pi Z_n^A\circ\pi_\delta}_{\mathcal{I}_A}
    +\norm{\Pi Z_n^A\circ\pi_\delta-Z_n^A\circ\pi_\delta}_{\mathcal{I}_A}
    +\norm{Z_n^A\circ\pi_\delta-Z_n^A}_{\mathcal{I}_A}.
\end{align*}
The first and last terms are the empirical and Gaussian meshing errors.
The second and fourth terms are the empirical and Gaussian projection
errors. The middle term is the projected coupling error. Applying the
same bounds as in the proof of \citet[Theorem SA.1]{Cattaneo-Yu_2025_AOS}
to this single projected vector, and then optimizing over the quasi-dyadic
projection depths, gives, in the bounded-transform version of Assumption (v),
\[
    \E\big[\norm{A_n-Z_n^A}_{\mathcal{I}_A}\big]
    \lesssim
    \mathtt{v}(\ttd\log(\ttc n))^{3/2}\mathtt{r}_n .
\]
The rate $\mathtt{r}_n$ is the minimum of the two projection/coupling
balances displayed in the theorem. The first balance uses total variation;
the second additionally uses the Lipschitz constants and yields the
improvement available for Lipschitz classes.

\textit{Polynomial moments.}
It remains to remove the bounded-transform simplification. For
$\eta>0$, define truncated classes
$r^\eta(y)=r(y)\Indicator(|y|\leq\eta)$ and
$s^\eta(y)=s(y)\Indicator(|y|\leq\eta)$, and let $A_n^\eta$ denote the
corresponding process. By Assumption (v) and
$\sup_{\bx\in\X}\E[|y_i|^{2+v}\mid \bx_i=\bx]\leq2$,
\[
    \sup_{r\in\R}\sup_{\bx\in\X}
    \E\big[|r(y_i)|^2\Indicator(|y_i|>\eta)\mid\bx_i=\bx\big]
    \lesssim \mathtt{v}^2 \eta^{-v},
\]
and the same bound holds for $\S$. The empirical tail, the corresponding
Gaussian tail, and the deterministic centering error are then controlled by
Bernstein's inequality, Markov's inequality, and the preceding VC entropy
bounds, giving
\begin{align*}
    \E\big[\norm{A_n-Z_n^A}_{\mathcal{I}_A}\big]
    \lesssim \mathtt{v}\Big\{&
    (\ttd\log(\ttc n))^{3/2}
        \big(\eta\mathtt{r}_n+\sqrt{\ttM\ttE}\eta^{-v/2}\big)\\
    &+\ttd\log(\ttc n)\ttM n^{-\frac{v/2}{2+v}}
    +\ttd\log(\ttc n)\ttM n^{-1/2}\eta
    \Big\}.
\end{align*}
Choosing
\[
    \eta=\Big(\frac{\sqrt{\ttM\ttE}}{\mathtt{r}_n}\Big)^{2/(v+2)}
\]
balances the bounded-transform approximation error with the stochastic tail
error, and yields exactly the bound in the theorem.

Finally, the common Gaussian construction is a Brownian-bridge construction
indexed by the combined class $\F_A$. Hence, for any two indices
$\iota_1,\iota_2\in\mathcal{I}_A$,
\[
    \E[Z_n^A(\iota_1)Z_n^A(\iota_2)]
    =
    \E[A_n(\iota_1)A_n(\iota_2)].
\]
This covariance identity includes the cross-covariance between
$M_n(g,r)$ and $M_n(h,s)$ because the whole sum was projected and coupled
jointly. Almost sure continuity of the sample paths follows from the same
separability and semimetric-continuity argument used in
\citet[Theorem SA.1]{Cattaneo-Yu_2025_AOS}, since the combined class is
pointwise measurable and VC-type under the envelope above. This proves the
theorem.
\qed

\section{Proofs}

\subsection{Proof of Lemma~\ref{sa-lem: VC distance balls}}

For each $j$, the class $\{\bv\in\X:f(\bv)\ge0,\ f\in\mathcal{V}_j\}$ is VC because $\mathcal{V}_j$ is finite dimensional. Fixed finite Boolean combinations of VC classes are VC; see, for example, the closure properties in \citet[Section 2.6]{van-der-Vaart-Wellner_1996_Book}. The displayed representation therefore implies that $\mathscr{D}_{\d}$ is VC. If the balls are semialgebraic with uniformly bounded degree and Boolean complexity, the relevant polynomial inequalities belong to finite-dimensional spaces spanned by the monomials in $\bv$ up to the maximal degree, and the same argument applies. The listed distances are special cases.
\qed

\subsection{Proof of Lemma~\ref{sa-lem: Gram invert}}

Let $\varsigma_t=2t-1$. On $\{\bX_i\in\A_t\}$,
$D_i(\bx)=\varsigma_t\lVert\bX_i-\bx\rVert$. Assumption~\ref{sa-assump: DGP}(ii) implies
\begin{align*}
    \bPsi_{t,\bx} & = \E \Big[\rp \Big(\frac{\varsigma_t\lVert \bX_i - \bx \rVert}{h}\Big) \rp \Big(\frac{\varsigma_t\lVert \bX_i - \bx \rVert}{h}\Big)^{\top} K_h(\varsigma_t\lVert \bX_i - \bx \rVert) \Indicator(\bX_i \in \A_t) \Big] \\
    & = \int_{\A_t} \rp \Big(\frac{\varsigma_t\lVert \bu - \bx \rVert}{h}\Big) \rp \Big(\frac{\varsigma_t\lVert \bu - \bx \rVert}{h}\Big)^{\top} K_h(\varsigma_t\lVert \bu - \bx \rVert) f_X(\bu) \dif\bu \\
    & = f_X(\bx)\int_{\A_t} \rp \Big(\frac{\varsigma_t\lVert \bu - \bx \rVert}{h}\Big) \rp \Big(\frac{\varsigma_t\lVert \bu - \bx \rVert}{h}\Big)^{\top} K_h(\varsigma_t\lVert \bu - \bx \rVert) \dif\bu + o(1),
\end{align*}
where in the last line we have used $\int_{\A_t} (\frac{\lVert \bu - \bx \rVert }{h})^{v} K_h(\varsigma_t\lVert \bu - \bx \rVert) \dif\bu = O(1)$ for each $0\le v\le 2p$ from a standard change of variable argument.

\begin{center}
\textbf{I. Polynomial Representation of Minimum Eigenvalue}
\end{center}
For simplicity, write
\begin{align*}
    \bS_{t,\bx}(h) = \int_{\A_t} \rp \Big(\frac{\varsigma_t\lVert \bu - \bx \rVert}{h}\Big) \rp \Big(\frac{\varsigma_t\lVert \bu - \bx \rVert}{h}\Big)^{\top} K_h(\varsigma_t\lVert \bu - \bx \rVert) \dif\bu.
\end{align*}
A change of variable gives
\begin{align*}
    \bS_{t,\bx}(h) = \int \rp(\varsigma_t\lVert \bz \rVert) \rp(\varsigma_t\lVert \bz \rVert)^{\top} K(\varsigma_t\lVert \bz \rVert) \Indicator(\bx + h \bz \in \A_t) \dif\bz.
\end{align*}
Let $\ba \in \reals^{p+1}$.
Condition (c) in Lemma~\ref{sa-lem: Gram invert} implies the lower bound used below.
If $K$ is Lipschitz, then it is continuous, and positivity on the compact set $U$
implies $\inf_{\bu\in U}K(\varsigma_t\|\bu\|)=\kappa>0$. If
$K(u)=\Indicator(u\in[-1,1])$, positivity on $U$ implies the same conclusion with
$\kappa=1$.
Then the equivalent representation of minimum eigenvalue gives
\begin{align}\label{sa-eq: min eigenvalue}
    \nonumber \lambda_{\min}(\bS_{t,\bx}(h)) 
    & = \min_{\norm{\ba} = 1} \int (\ba^{\top} \rp(\varsigma_t\lVert \bz \rVert))^2 K(\varsigma_t\lVert \bz \rVert) \Indicator(\bx + h \bz \in \A_t) \dif\bz \\
    & \geq \kappa \min_{\norm{\ba} = 1} \int_{U} (\ba^{\top} \rp(\varsigma_t\lVert \bz \rVert))^2  \Indicator(\bx + h \bz \in \A_t) \dif\bz,
\end{align}
where the sign $\varsigma_t$ only premultiplies the polynomial basis by a nonsingular diagonal matrix, so the lower bound in terms of $\rp(\lVert\bz\rVert)$ below is equivalent up to fixed constants.

\begin{center}
\textbf{II. Mass Retaining Ratio in Treatment/Control Region}
\end{center}

Denote $E_h(\bx,t) = \{\bz \in U: \bx + h \bz \in \A_t \}$. Assumption~\ref{sa-assump: Kernel, Distance and Boundary}(iii) implies there is some upper bound $\Lambda > 0$ of $K(\cdot)$. Hence for $c_0 = 1/2 \; \liminf_{h \downarrow 0}\inf_{\bx \in \B} \int_{U} K(\varsigma_t\lVert \bu \rVert) \Indicator(\bx + h \bu \in \A_t) \dif\bu$, we have
\begin{align*}
    \Lambda \leb(E_h(\bx,t)) 
    & \geq \int_{U} K(\varsigma_t\lVert \bu \rVert) \Indicator(\bx + h \bu \in \A_t) \dif\bu
    \geq c_0
\end{align*}
for small enough $h$, which implies 
\begin{align}\label{sa-eq: mass relation}
      \leb(E_h(\bx,t))  \geq \alpha \leb(U), \qquad \alpha = \frac{c_0}{\Lambda \leb(U)}.
\end{align}

\begin{center}
\textbf{III. $L_2$ Integral of Polynomials in Full Versus Treatment/Control Regions}
\end{center}

Consider $S = \{q \in \mathcal{P}_{p}: \int_U q(\lVert \bu \rVert)^2 \dif\bu = 1\}$, where $\mathcal{P}_{p}$ is the collection of all polynomials of degree at most $p$. Equip $\mathcal{P}_{p}$ with the inner product induced by $\langle q_1,q_2\rangle_U=\int_U q_1(\lVert\bu\rVert)q_2(\lVert\bu\rVert)\dif\bu$. Let $(\phi_j, 1 \leq j \leq p+1)$ be an orthonormal basis for this finite-dimensional space. Then $T(\ba) = \sum_{j = 1}^{p+1} a_j \phi_j$ maps the Euclidean unit sphere in $\reals^{p+1}$ onto $S$. Hence $S$ is compact under this $L_2(U)$ norm, and finite-dimensional norm equivalence implies that $S$ is also compact under the corresponding supremum norm on $U$. Now consider
\begin{align*}
    \Phi_q(\varepsilon) = \leb(\{\bu \in U: |q(\lVert\bu\rVert)| < \varepsilon \}), \qquad q \in S, \varepsilon > 0,
\end{align*}
and 
\begin{align*}
    \psi(q) = \sup \Big\{\varepsilon > 0: \Phi_q(\varepsilon) \leq \frac{\alpha}{2} \leb(U) \Big\}.
\end{align*}
Since $\int_U q(\lVert\bu\rVert)^2\dif\bu = 1$ and $q$ is a polynomial, $\lim_{\varepsilon \downarrow 0} \Phi_q(\varepsilon) = 0$ and $\Phi_q(\varepsilon)=\leb(U)$ for all sufficiently large $\varepsilon$. Continuity of $q\in S$ implies $\psi(q) > 0$ for all $q \in S$.

Next, we want to show $\psi$ is a lower-semicontinuous function on $(\mathcal{P}_{p}, \lVert \cdot \rVert_{L_{\infty}})$. Suppose $q_n \to q$ uniformly on $U$. For every $\varepsilon_0 \in (0, \psi(q))$, choose $\eta>\varepsilon_0$ such that $\Phi_q(\eta) \leq \frac{\alpha}{2}\leb(U)$. For large enough $n$, uniform convergence gives $|q_n(\lVert\bu\rVert)-q(\lVert\bu\rVert)|\leq \eta-\varepsilon_0$ on $U$, so $\{|q_n(\lVert\cdot\rVert)|<\varepsilon_0\}\subseteq\{|q(\lVert\cdot\rVert)|<\eta\}$. Hence $\Phi_{q_n}(\varepsilon_0)\leq\Phi_q(\eta)\leq\frac{\alpha}{2}\leb(U)$, which implies $\varepsilon_0 \leq \psi(q_n)$. This implies $\liminf_{n \to \infty} \psi(q_n) \geq \varepsilon_0$. Since $\varepsilon_0$ is arbitrary in $(0, \psi(q))$, we have $\liminf_{n \to \infty} \psi(q_n) \geq \psi(q)$.

Compactness of $S$ and lower-semicontinuity of $\psi$ implies $\psi$ attains its minimum on $S$. Since $\psi(q) > 0$ for all $q \in S$, we know $\varepsilon_* = \inf_{q \in S} \psi(q) > 0$. Then for every $q \in S$, 
\begin{align*}
    \int_{E_h(\bx,t)} q(\lVert\bu\rVert)^2\dif\bu
    & \geq \Big(\frac{\varepsilon_*}{2}\Big)^2 \; \leb \Big(E_h(\bx,t) \setminus \{\bu\in U:|q(\lVert\bu\rVert)| \leq \varepsilon_*/2\} \Big) \\
    & \geq \Big(\frac{\varepsilon_*}{2}\Big)^2 \; \Big(\leb(E_h(\bx,t)) - \leb(\{\bu\in U:|q(\lVert\bu\rVert)| \leq \varepsilon_*/2\}) \Big) \\
    & \geq \frac{\varepsilon_*^2\alpha}{8} \leb(U).
\end{align*}
Scaling $q$ from $S$ gives
\begin{align}\label{sa-eq: integral relation}
    \int_{E_h(\bx,t)} q(\lVert \bu\rVert)^2 \dif\bu \geq \frac{\varepsilon_*^2\alpha}{8} \int_{U} q(\lVert \bu\rVert)^2 \dif\bu, \qquad q \in \mathcal{P}_{p}.
\end{align}

\begin{center}
\textbf{IV. Lower Bound of Minimum Eigenvalue}
\end{center}

Equations~\eqref{sa-eq: min eigenvalue}, \eqref{sa-eq: mass relation} and \eqref{sa-eq: integral relation} together give for small enough $h$,
\begin{align*}
    \inf_{\bx \in \B}\lambda_{\min}(\bS_{t,\bx}(h)) 
    & \geq \kappa \inf_{\bx \in \B}\min_{\norm{\ba} = 1} \int_{E_h(\bx,t)} (\ba^{\top} \rp(\lVert \bz \rVert))^2 \dif\bz \\
    & \geq \kappa \frac{\varepsilon_*^2\alpha}{8}  \min_{\norm{\ba} = 1} \int_{U} (\ba^{\top} \rp(\lVert \bz \rVert))^2 \dif\bz \\
    & \geq \kappa \frac{\varepsilon_*^2\alpha}{8}  \lambda_{\min} \Big(\int_U \rp(\lVert \bz \rVert) \rp(\lVert \bz \rVert)^{\top} \dif\bz \Big),
\end{align*}
which implies $\liminf_{h \to 0} \inf_{\bx \in \B}\lambda_{\min}(\bS_{t,\bx}(h)) > 0$.
Because $f_X$ is continuous and bounded away from zero on $\X$ and
$\bPsi_{t,\bx}=f_X(\bx)\bS_{t,\bx}(h)+o(1)$ uniformly in $\bx\in\B$, the same lower
bound holds for $\bPsi_{t,\bx}$. This proves
Assumption~\ref{sa-assump: Kernel, Distance and Boundary}(iv).
\qed

\subsection{Proof of Lemma~\ref{sa-lem: Gram}}

Since $\widehat{\bPsi}_{t,\bx}$ is a finite dimensional matrix, it suffices to show the stated rate of convergence for each entry. For $0 \leq v \leq 2p$, define $\mathcal{G} = \{g_{n,t,v}(\cdot, \bx): \bx \in \X \}$ with
\begin{align*}
    g_{n,t,v}(\xi, \bx)
    = \Big(\frac{\varsigma_t\d(\xi,\bx)}{h}\Big)^{v} \frac{1}{h^d} K\Big(\frac{\varsigma_t\d(\xi, \bx)}{h}\Big)\Indicator(\xi\in\A_t),
    \qquad
    \xi,\bx \in \mathcal{X}.
\end{align*}
These are the signed-distance matrix entries, up to the finite choice of $v$. We will show $\mathcal{G}$ is a VC-type class.

\medskip\textit{Constant Envelope Function}. Since $K$ is bounded and compactly supported, there exists a constant $C_1$ such that  $\sup_{\bx \in \X} \infnorm{g_{n,t,v}(\cdot,\bx)} \leq C_1 h^{-d} = G$.

\medskip\textit{Diameter of $\G$ in $L_2$}. Let $R_K<\infty$ satisfy $\supp(K)\subseteq[-R_K,R_K]$. For each $\bx \in \mathcal{X}$, $g_{n,t,v}(\cdot,\bx)$ is supported on $\{\xi: \d(\xi,\bx) \leq R_Kh\}$. By Assumption~\ref{sa-assump: DGP}(ii) and Assumption~\ref{sa-assump: Kernel, Distance and Boundary}(i), $\sup_{\bx \in \X} \P \left(\d(\bX_i, \bx) \leq R_Kh \right) \lesssim h^d$. It follows that
$ \sup_{\bx \in \X} \norm{g_{n,t,v}(\cdot,\bx)}_{\P,2} \leq C_2 h^{-d/2}$ for some constant $C_2$. We can take $C_1$ large enough so that $\sigma = C_2 h^{-d/2} \leq G = C_1 h^{-d}$.

\medskip\textit{Ratio}. For some constant $C_3$, $\delta = \frac{\sigma}{F}  = C_3 \sqrt{h^d}$.

\medskip\textit{Covering Numbers}. Case 1: $K$ is Lipschitz. Let $\bx,\bx' \in \X$. By Assumption~\ref{sa-assump: Kernel, Distance and Boundary},
\begin{align*}
    &\sup_{\xi \in \X} \big|g_{n,t,v}(\xi, \bx) - g_{n,t,v}(\xi,\bx')\big|\\
    &\leq \sup_{\xi \in \X}
          \bigg|\Big[\Big(\frac{\varsigma_t\d(\xi,\bx)}{h}\Big)^v - \Big(\frac{\varsigma_t\d(\xi,\bx')}{h}\Big)^v\Big] K_h(\varsigma_t\d(\xi,\bx))\bigg|\Indicator(\xi\in\A_t)\\
    &\qquad
     + \bigg|\Big(\frac{\varsigma_t\d(\xi,\bx')}{h}\Big)^v \Big[K_h(\varsigma_t\d(\xi,\bx)) - K_h(\varsigma_t\d(\xi,\bx'))\Big]\bigg|\Indicator(\xi\in\A_t)\\
    & \lesssim h^{-d-1} \infnorm{\bx - \bx'}.
\end{align*} 
By Lipschitz continuity property of $\G$, for any $\varepsilon \in (0,1]$ and for any finitely supported measure $Q$ and metric $\norm{\cdot}_{Q,2}$ based on $L_2(Q)$,
\begin{align*}
    N(\{g_{n,t,v}(\cdot,\bx): \bx \in \X\}, \norm{\cdot}_{Q,2}, \varepsilon \|G\|_{Q,2})
    \leq N(\X, \infnorm{\cdot}, \varepsilon \|G\|_{Q,2}h^{d+1})
    \stackrel{(i)}{\lesssim} \Big(\frac{\diam(\X)}{ \varepsilon \|G\|_{Q,2} h^{d+1}}\Big)^d
    \lesssim \Big(\frac{\diam(\X)}{\varepsilon h}\Big)^d,
\end{align*} 
where inequality (i) uses the fact that $\varepsilon \|G\|_{Q,2} h^{d+1} \lesssim \varepsilon h \lesssim 1$. Therefore, for constants $C_1$ and $C_2$ independent of $Q$,
\begin{align*}
    \sup_{Q} N(\mathcal{G}, \norm{\cdot}_{Q,2}, \varepsilon \norm{G}_{Q,2})
    \leq (C_1/\varepsilon)^{C_2}, \qquad \varepsilon \in (0,1],
\end{align*}
where the supremum is taken over all finite discrete measures on $\X$.

Case 2: $K = \Indicator(\cdot \in [-1,1])$. Consider
\begin{align*}
    m_n(\xi, \bx)
    = \Big(\frac{\varsigma_t\d(\xi,\bx)}{h}\Big)^{v} \frac{1}{h^d} \Indicator(\xi \in \A_t)\Indicator(\d(\xi,\bx)\le h),
    \qquad \xi, \bx \in \mathcal{X},
\end{align*}
$\M = \{m_n(\cdot,\bx): \bx \in \B\}$ and the constant envelope function $M = C_4 h^{-d}$, for some constant $C_4$ depending only on the diameter of $\X$. The polynomial factor is covered by the preceding Lipschitz argument. The class $\mathscr{L} = \{\Indicator(\d(\cdot,\bx) \le h): \bx \in \B\}$ has finite VC dimension by Assumption~\ref{sa-assump: Kernel, Distance and Boundary}(iii) (for the Euclidean distance, see \citealp[Example 2.6.1]{van-der-Vaart-Wellner_1996_Book}); products with the fixed assignment indicator preserve VC-type entropy. Hence, by \citet[Theorem 2.6.4]{van-der-Vaart-Wellner_1996_Book},
\begin{align*}
    \sup_{Q} N(\mathcal{G}, \norm{\cdot}_{Q,2}, \varepsilon \norm{G}_{Q,2})
    \leq (C_1/\varepsilon)^{C_2}, \qquad \varepsilon \in (0,1],
\end{align*}
where the supremum is taken over all finite discrete measures on $\X$.

\medskip\textit{Maximal Inequality}. By \citet[Corollary 5.1]{Chernozhukov-Chetverikov-Kato_2014b_AoS} for the empirical process on class $\G$,
\begin{align*}
    \E \Big[\sup_{l \in \mathcal{G}} \big|\En \left[l(\bX_i)\right]  - \E[l(\bX_i)]\big| \Big]
    &\lesssim \frac{\sigma}{\sqrt{n}}\sqrt{C_2\log(C_1/\delta)} + \frac{\|G\|_{\P,2} C_2\log(C_1/\delta)}{n}\\
    &\lesssim \frac{1}{\sqrt{n h^d}}\sqrt{d\log \Big( \frac{\diam(\X)}{h^{1 + d/2}} \Big)} 
     + \frac{1}{n h^d}d \log \Big( \frac{\diam(\X)}{h^{1 + d/2}} \Big)\\
    &\lesssim \sqrt{\frac{\log(1/h)}{n h^d}}.
\end{align*}
Thus, $\sup_{\bx \in \B} \big\|\widehat{\bPsi}_{t,\bx} - \bPsi_{t,\bx}\big\| \lessp \sqrt{\frac{\log(1/h)}{n h^d}}$.

By Weyl's Theorem, $\sup_{\bx \in \B}|\lambda_{\min}(\widehat{\bPsi}_{t,\bx}) - \lambda_{\min}(\bPsi_{t,\bx})| \leq \sup_{\bx \in \B} \|\widehat{\bPsi}_{t,\bx} - \bPsi_{t,\bx}\| \lessp \sqrt{\frac{\log(1/h)}{n h^d }}$. Therefore, we can lower bound the minimum eigenvalue by $\inf_{\bx \in \B} \lambda_{\min}(\widehat{\bPsi}_{t,\bx}) \geq \inf_{\bx \in \B}\lambda_{\min}(\bPsi_{t,\bx}) - \sup_{\bx \in \B}|\lambda_{\min}(\widehat{\bPsi}_{t,\bx}) - \lambda_{\min}(\bPsi_{t,\bx})| \gtrsim_{\P} 1$.

Finally, it follows that $\sup_{\bx \in \B} \|\widehat{\bPsi}_{t,\bx}^{-1}\| \lessp 1$ and hence
\begin{align*}
    \sup_{\bx \in \B} \big\|\widehat{\bPsi}_{t,\bx}^{-1} - \bPsi_{t,\bx}^{-1}\big\|
    \leq \sup_{\bx \in \B} \big\|\bPsi_{t,\bx}^{-1}\big\| \big\|\bPsi_{t,\bx} -\widehat{\bPsi}_{t,\bx}\big\| \big\|\widehat{\bPsi}_{t,\bx}^{-1}\big\|
    \lessp \sqrt{\frac{\log(1/h)}{n h^{d}}},
\end{align*}
which completes the proof.
\qed

\subsection{Proof of Lemma~\ref{sa-lem: Stochastic Linear Approximation}}

Consider the class $\F = \{(\bz,u) \mapsto \be_{\ell}^{\top} g_{t,\bx}(\bz)(u - h_{\bx}(\bz)): \bx \in \B \}$, $0 \leq \ell \leq p$, where for $\bz \in \X$,
\begin{gather*}
    g_{t,\bx}(\bz) = \rp \Big(\frac{D_{\bz}(\bx)}{h}\Big) K_h(D_{\bz}(\bx))\Indicator(D_{\bz}(\bx)\in\I_t), \qquad
    h_{\bx}(\bz) = \bgamma_t^{\ast}(\bx)^{\top} \rp\left(D_{\bz}(\bx)/h\right).
\end{gather*}
By definition of $\bgamma_t^{\ast}(\bx)$,
\begin{align}\label{sa-eq: best fit}
    \bgamma_t^{\ast}(\bx)
    = \bPsi_{t,\bx}^{-1} \bS_{t,\bx},
    \qquad 
    \bS_{t,\bx}
    = \E \Big[\rp\Big(\frac{D_i(\bx)}{h}\Big) K_h(D_i(\bx)) Y_i \Indicator(\bX_i \in \A_t) \Big].
\end{align}
 Assumption~\ref{sa-assump: DGP} implies $\bS_{t,\bx}$ is continuous in $\bx$, hence $\sup_{\bx \in \B} \norm{\bS_{t,\bx}} \lesssim 1$. And by Assumption~\ref{sa-assump: Kernel, Distance and Boundary}(iv), $\inf_{\bx \in \B} \lambda_{\min}(\bPsi_{t,\bx}) \gtrsim 1$. Hence
\begin{align}\label{sa-eq: surrogate}
    \sup_{\bx \in \B} \norm{\bPsi_{t,\bx}^{-1} \bS_{t,\bx}} \lesssim 1.
\end{align}

Now, consider properties of $\mathcal{F}$. Definition of $\bgamma_t^{\ast}(\bx)$ implies $\E[f(\bX_i,Y_i)] = 0$ for all $f \in \F$. Since $K$ is compactly supported, there exists $C_1, C_2 > 0$ such that $F(\bz,u) = C_1 h^{-d} (|u| + C_2)$ is an envelope function for $\F$. Denote $M = \max_{1 \leq i \leq n} F(\bX_i, Y_i)$, then
\begin{align*}
    \E[M^2]^{1/2} & \lesssim h^{-d} \E \left[\max_{1 \leq i \leq n} |Y_i|^2 + 1\right]^{1/2} \lesssim h^{-d} \E \left[\max_{1 \leq i \leq n} |Y_i|^{2+v}\right]^{1/(2+v)} \\
    & \lesssim h^{-d} \bigg[\sum_{i = 1}^n \E[|Y_i|^{2+v}]\bigg]^{1/(2+v)}
    \lesssim h^{-d} n^{1/(2+v)},
\end{align*}
where we have used the uniform $(2+v)$ moment bound in Assumption~\ref{sa-assump: DGP}(v). Denote $\sigma = \sup_{f \in \F} \E[f(\bX_i,Y_i)^{2}]^{1/2}$. Then,
\begin{align*}
    \sigma^2 \lesssim \sup_{\bx \in \B}\E [\infnorm{\be_{\ell}^{\top}g_{t,\bx}}^2(|Y_i| + \infnorm{h_{\bx}})^2] \lesssim h^{-d}.
\end{align*}
To check the covering number of $\F$, note that compared to the proof of Lemma~\ref{sa-lem: Gram}, we have one additional term $\be_{\ell}^\top g_{t,\bx} h_{\bx}$. All signed-distance kernel and indicator terms can be handled as in the proof of Lemma~\ref{sa-lem: Gram}. Recall Equation~\eqref{sa-eq: best fit}, and consider $l_{t,\bx}(\cdot) = \be_{\ell}^\top \rp(D_{\cdot}(\bx)/h) K_h(D_{\cdot}(\bx)) \mu_t(\cdot) \Indicator(D_{\cdot}(\bx) \in \I_t)$ and $\mathscr{L}_t = \{l_{t,\bx}: \bx \in \B\}$. Then, for any $\bx_1, \bx_2 \in \B$,
\begin{align*}
    |\bS_{t,\bx_1} - \bS_{t,\bx_2}| \leq \norm{l_{t,\bx_1} - l_{t,\bx_2}}_{\P_X,2},
\end{align*}
and hence 
\begin{align*}
    N(\{\be_{\ell}^{\top}\bS_{t,\bx}: \bx \in \B\}, |\cdot|, \varepsilon h^{-d}) & \leq N(\mathscr{L}_t, \norm{\cdot}_{\P_X,2}, \varepsilon h^{-d})
    \leq \sup_{Q} N(\mathscr{L}_t, \norm{\cdot}_{Q,2}, \varepsilon h^{-d}), 
\end{align*}
The same argument as in the paragraph \textbf{Covering Numbers} in the proof of Lemma~\ref{sa-lem: Gram} then shows
\begin{align*}
   &  \sup_{Q} N(\{g_{t,\bx}: \bx \in \B\}, \norm{\cdot}_{Q,2}, \varepsilon C_1 h^{-d}) \leq \Big(\frac{\diam(\X)}{h \varepsilon}\Big)^d, \qquad 0 < \varepsilon \leq 1, \\
   &  \sup_{Q} N(\{g_{t,\bx} h_{\bx}: \bx \in \B\}, \norm{\cdot}_{Q,2}, \varepsilon C_1 h^{-d}) \leq \Big(\frac{\diam(\X)}{h \varepsilon}\Big)^d, \qquad 0 < \varepsilon \leq 1,
\end{align*}
where $\sup$ is taken over all discrete measures on $\X$. Product $\{g_{t,\bx}: \bx \in \B\}$ with the singleton identity class and addition of $\{g_{t,\bx} h_{\bx}: \bx \in \B\}$ preserve VC-type entropy, so
\begin{align*}
    \sup_{Q} N(\F, \norm{\cdot}_{Q,2}, \varepsilon \norm{F}_{Q,2}) \leq 2\Big(\frac{2\diam(\X)}{h \varepsilon}\Big)^d, \qquad 0 < \varepsilon \leq 1,
\end{align*}
where $\sup$ is taken over all discrete measures on $\X \times \reals$. Denote $\mathtt{C}_1 = d$, $\mathtt{C}_2 = \frac{2 (2 \diam(\X))^d}{h^d}$. Hence, by \citet[Corollary 5.1]{Chernozhukov-Chetverikov-Kato_2014b_AoS}
\begin{align*}
 \E\bigg[\sup_{\bx \in \B}|\be_{\ell}^{\top} \mathbf{O}_{t,\bx}|\bigg]& = \E \bigg[\sup_{f \in \mathcal{F}} \left|\En \left[f(\bX_i,Y_i)\right]  - \E[f(\bX_i,Y_i)]\right| \bigg] \\
 & \lesssim \frac{\sigma}{\sqrt{n}}\sqrt{\mathtt{C}_2\log(\mathtt{C}_1 \norm{M}_{\P,2}/\sigma)} + \frac{\|M\|_{\P,2} \mathtt{C}_2\log(\mathtt{C}_1 \norm{M}_{\P,2}/\sigma)}{n}\\
 & \lesssim \frac{1}{\sqrt{n h^d}}\sqrt{d\log \left( \frac{\diam(\X)}{h^{1 + d/2}} \right)} 
+ \frac{1}{n^{\frac{1+v}{2+v}} h^d}d \log \left( \frac{\diam(\X)}{h^{1 + d/2}} \right) \\
 & \lesssim \sqrt{\frac{\log (1/h)}{n h^d}} + \frac{\log (1/h)}{n^{\frac{1+v}{2+v}}h^d}.
\end{align*}
The rest follows from finite dimensionality of $\mathbf{O}_{t,\bx}$, and Lemma~\ref{sa-lem: Gram}.
\qed

\subsection{Proof of Lemma~\ref{sa-lem: Covariance}}

Denote, for $\mathbf{z}\in\B$, $\eta_{i,t,\mathbf{z}} = Y_i - \theta_{t,\mathbf{z}}^{\ast}(D_i(\mathbf{z}))$ and $\xi_{i,t,\mathbf{z}} = \theta_{t,\mathbf{z}}^{\ast}(D_i(\mathbf{z}))
- \widehat{\theta}_{t,\mathbf{z}}(D_i(\mathbf{z}))$. Then
\begin{align*}
    \widehat{\bUpsilon}_{t,\bx,\by}
    &= \En \bigg[\rp \Big(\frac{D_i(\bx)}{h}\Big) \rp \Big( \frac{D_i(\by)}{h}\Big)^{\top} h^d K_h(D_i(\bx)) K_h(D_i(\by))\\
    &\qquad\qquad \cdot (\eta_{i,t,\bx} + \xi_{i,t,\bx})(\eta_{i,t,\by} + \xi_{i,t,\by}) \Indicator(D_i(\bx) \in \I_t)\Indicator(D_i(\by) \in \I_t)\bigg],
\end{align*}
and we decompose the error into
\begin{align*}
    & \widehat{\bUpsilon}_{t,\bx,\by} - \bUpsilon_{t,\bx,\by} = \Delta_{1,t,\bx,\by} + \Delta_{2,t,\bx,\by} + \Delta_{3,t,\bx,\by}, \\
    & \Delta_{1,t,\bx,\by} = \En \bigg[\rp \Big(\frac{D_i(\bx)}{h}\Big) \rp \left( \frac{D_i(\by)}{h}\right)^{\top} h^d K_h (D_i(\bx)) K_h \left(D_i(\by)\right) \xi_{i,t,\bx}\xi_{i,t,\by} \Indicator(D_i(\bx) \in \I_t)\Indicator(D_i(\by) \in \I_t)\bigg], \\
    & \Delta_{2,t,\bx,\by} = \En \bigg[\rp \Big(\frac{D_i(\bx)}{h}\Big) \rp \left( \frac{D_i(\by)}{h}\right)^{\top} h^d K_h (D_i(\bx)) K_h \left(D_i(\by)\right)
      (\eta_{i,t,\bx}\xi_{i,t,\by}+\eta_{i,t,\by}\xi_{i,t,\bx}) \Indicator(D_i(\bx) \in \I_t)\Indicator(D_i(\by) \in \I_t)\bigg], \\
    & \Delta_{3,t,\bx,\by} =  \En \bigg[\rp \Big(\frac{D_i(\bx)}{h}\Big) \rp \left( \frac{D_i(\by)}{h}\right)^{\top} h^d K_h (D_i(\bx)) K_h (D_i(\by)) \eta_{i,t,\bx}\eta_{i,t,\by} \Indicator(D_i(\bx) \in \I_t)\Indicator(D_i(\by) \in \I_t)\bigg] \\
    & \qquad \qquad \qquad - \E \bigg[\rp \Big(\frac{D_i(\bx)}{h}\Big) \rp \left( \frac{D_i(\by)}{h}\right)^{\top} h^d K_h (D_i(\bx)) K_h (D_i(\by)) \eta_{i,t,\bx}\eta_{i,t,\by} \Indicator(D_i(\bx) \in \I_t)\Indicator(D_i(\by) \in \I_t)\bigg].
\end{align*}
By Assumption~\ref{sa-assump: Kernel, Distance and Boundary}, $K_h(D_i(\bx)) \neq 0$ implies $\norm{\rp(D_i(\bx)/h)}_2 \lesssim 1$. Hence by Lemma~\ref{sa-lem: Gram} and \ref{sa-lem: Stochastic Linear Approximation}, 
\begin{align*}
    & \max_{t \in \{0,1\}} \max_{1 \leq i \leq n} \sup_{\bx \in \B}|\xi_{i,t,\bx}| \\
    & = \max_{t \in \{0,1\}} \max_{1 \leq i \leq n} \sup_{\bx \in \B} |\rp(D_i(\bx)/h)^{\top}(\widehat{\bgamma}_{t,\bx} - \bgamma_{t,\bx}^{\ast})| \Indicator(K_h(D_i(\bx)) \neq 0) \\
    & = \max_{t \in \{0,1\}} \max_{1 \leq i \leq n} \sup_{\bx \in \B} |\rp(D_i(\bx)/h)^{\top}\widehat{\bPsi}_{t,\bx}^{-1} \mathbf{O}_{t,\bx}| \Indicator(K_h(D_i(\bx)) \neq 0)  \\
    & \leq \max_{t \in \{0,1\}} \sup_{\bx \in \B} \norm{\widehat{\bPsi}_{t,\bx}^{-1} \mathbf{O}_{t,\bx}}_2\\
    & \lesssim \sqrt{\frac{\log(1/h)}{n h^d}} + \frac{\log(1/h)}{n^{\frac{1+v}{2+v}}h^d},
\end{align*}
Under the rate conditions of the lemma, a maximal inequality similar to the one used in the proof of Lemma~\ref{sa-lem: Gram} shows
\begin{align}\label{sa-eq: covariance for distance -- 1}
    \nonumber & \sup_{\bx,\by \in \B}\norm{\Delta_{1,t,\bx,\by}} \lessp \max_{t \in \{0,1\}} \max_{1 \leq i \leq n} \sup_{\bx \in \B}|\xi_{i,t,\bx}|^2 \lesssim \bigg(\sqrt{\frac{\log(1/h)}{n h^d}} + \frac{\log(1/h)}{n^{\frac{1+v}{2+v}}h^d}\bigg)^2, \\
    & \sup_{\bx,\by \in \B}\norm{\Delta_{2,t,\bx,\by}} \lessp \max_{t \in \{0,1\}} \max_{1 \leq i \leq n} \sup_{\bx \in \B}|\xi_{i,t,\bx}| \lesssim \sqrt{\frac{\log(1/h)}{n h^d}} + \frac{\log(1/h)}{n^{\frac{1+v}{2+v}}h^d}.
\end{align}
Consider the $(\mu,\nu)$ entry of $\Delta_{3,t,\bx,\by}$ and the class
\begin{align*}
\F = \bigg\{(\bz,u) \mapsto\;& \Big(\frac{D_{\bz}(\bx)}{h}\Big)^{\mu}\Big(\frac{D_{\bz}(\by)}{h}\Big)^{\nu}h^d K_h(D_{\bz}(\bx)) K_h(D_{\bz}(\by))\\
&\times \big(u - \rp(D_{\bz}(\bx)/h)^{\top}\bgamma_{t,\bx}^{\ast}\big)
       \big(u - \rp(D_{\bz}(\by)/h)^{\top}\bgamma_{t,\by}^{\ast}\big)
       \Indicator(D_{\bz}(\bx) \in \I_t)\Indicator(D_{\bz}(\by) \in \I_t): \bx, \by \in \B\bigg\}.
\end{align*}
The kernel factors, side indicators, and polynomial factors form VC-type classes by the same covering arguments used above; finite products and sums preserve VC-type entropy with constants changing only by fixed powers. By Assumptions~\ref{sa-assump: Kernel, Distance and Boundary} and \ref{sa-assump: DGP}(v), we have $\sup_{f \in \F} \E[f(\bX_i, Y_i)^2]^{1/2} \lesssim h^{-d/2}$. Moreover, Assumption~\ref{sa-assump: Kernel, Distance and Boundary} and Equation~\eqref{sa-eq: surrogate} imply there exists $C_1, C_2 > 0$ such that $F(\bz,u) = C_1 h^{-d}(u^2 + C_2)$ is an envelope function for $\F$, with
\begin{align*}
    \E \big[ \max_{1 \leq i \leq n} F(\bX_i,Y_i)^2 \big]^{\frac{1}{2}} 
    \lesssim C_1 h^{-d} (\E \big[\max_{1 \leq i \leq n} Y_i^4 \big]^{\frac{1}{2}} + C_2)
    \lesssim C_1 h^{-d} (\E \big[\max_{1 \leq i \leq n} Y_i^{2+v} \big]^{\frac{2}{2+v}} + C_2) \lesssim h^{-d} n^{\frac{2}{2+v}}.
\end{align*}
Apply \citet[Corollary 5.1]{Chernozhukov-Chetverikov-Kato_2014b_AoS} similarly as in Lemma~\ref{sa-lem: Stochastic Linear Approximation} gives
\begin{align*}
    \E \left[\sup_{f \in \F}\left|\En[f(\bX_i,Y_i)] - \E[f(\bX_i,Y_i)] \right|\right] \lesssim \sqrt{\frac{\log(1/h)}{n h^d}} + \frac{\log(1/h)}{n^{\frac{v}{2+v}}h^d}.
\end{align*}
Finite dimensionality of $\Delta_{3,t,\bx,\by}$ then implies
\begin{align}\label{sa-eq: covariance for distance -- 2}
     \E \left[\sup_{\bx,\by \in \B}\norm{\Delta_{3,t,\bx,\by}}\right] \lesssim \sqrt{\frac{\log(1/h)}{n h^d}} + \frac{\log(1/h)}{n^{\frac{v}{2+v}}h^d}.
\end{align}
Putting together Equations~\eqref{sa-eq: covariance for distance -- 1},
\eqref{sa-eq: covariance for distance -- 2} and Lemma~\ref{sa-lem: Gram}
gives the bound for $\widehat{\bUpsilon}_{t,\bx,\by}-\bUpsilon_{t,\bx,\by}$.
The displayed bound for $\widehat{\Xi}-\Xi$ follows by pre- and post-multiplying by
$\be_0^{\top}\widehat{\bPsi}^{-1}$ and its population analogue, and using the level
definition of $\Xi$.

It remains to verify the scale of the variance term. The population lower bound
$\lambda_{\min}(\bUpsilon_{t,\bx,\bx})\gtrsim1$ follows from the Gram lower bound and
the conditional variance lower bound in Assumption~\ref{sa-assump: DGP}(iv). To see why
the replacement of $K$ by $K^2$ is harmless, let $q_{\ba}(u)=\ba^\top\rp(u)$ with
$\|\ba\|=1$. Assumption~\ref{sa-assump: Kernel, Distance and Boundary}(iv) gives a
uniform lower bound on the corresponding $K$-weighted local second moment. Since $K$ is
fixed, bounded, and compactly supported, while $q_{\ba}$ is uniformly bounded on
$\supp(K)$, choose $\delta>0$ small enough that the contribution from $\{K\le\delta\}$
is less than one half of this lower bound. On $\{K>\delta\}$, $K^2\ge\delta K$, which
gives the same lower bound, up to constants, for the $K^2$-weighted matrix. Combining
this with
\[
    \E\{(Y_i-\theta_{t,\bx}^{\ast}(D_i(\bx)))^2\mid\bX_i\}
    \geq \sigma_t^2(\bX_i)
\]
gives $\lambda_{\min}(\bUpsilon_{t,\bx,\bx})\gtrsim1$. The same local-moment calculation
also gives $\sup_{\bx\in\B}\|\bUpsilon_{t,\bx,\bx}\|\lesssim1$ and
$\sup_{\bx\in\B}\|\bPsi_{t,\bx}\|\lesssim1$. Therefore, uniformly in $\bx\in\B$,
\begin{align*}
    n h^d\Xi_{t,\bx,\bx}
    &=
    \left(\bPsi_{t,\bx}^{-1}\be_0\right)^{\top}
    \bUpsilon_{t,\bx,\bx}
    \left(\bPsi_{t,\bx}^{-1}\be_0\right)
    \lesssim
    \|\bPsi_{t,\bx}^{-1}\|^2\|\bUpsilon_{t,\bx,\bx}\|
    \lesssim 1,
\end{align*}
while
\begin{align*}
    n h^d\Xi_{t,\bx,\bx}
    &\geq
    \lambda_{\min}(\bUpsilon_{t,\bx,\bx})
    \|\bPsi_{t,\bx}^{-1}\be_0\|^2
    \gtrsim
    \|\bPsi_{t,\bx}\|^{-2}
    \gtrsim 1.
\end{align*}
Summing over $t=0,1$ gives the same bound for $\Xi_{\bx}=\Xi_{\bx,\bx}$. Combining the
population bounds with the uniform consistency of $\widehat{\bUpsilon}_{t,\bx,\bx}$,
the bound for $\widehat{\Xi}-\Xi$, and Lemma~\ref{sa-lem: Gram} gives
$n h^d\widehat{\Xi}_{t,\bx,\bx}\gtrsim_{\P}1$ and
$n h^d\widehat{\Xi}_{t,\bx,\bx}\lessp1$ uniformly in $\bx$. The corresponding
correlation bound follows from these lower bounds and the displayed uniform consistency
of $\widehat{\Xi}_{t,\bx_1,\bx_2}$.
\qed

\subsection{Proof of Theorem~\ref{sa-thm: Identification}}

Since $\vartheta(\bx) = \theta_{1,\bx}(0) - \theta_{0,\bx}(0)$ and $\tau(\bx) = \mu_1(\bx) - \mu_0(\bx)$, it is enough to prove the result for one treatment assignment group $t\in\{0,1\}$. By Assumption~\ref{sa-assump: DGP}(iii) and Assumption~\ref{sa-assump: Kernel, Distance and Boundary}(ii), for any $r \neq 0$, for any $\bx \in \B$ and $\by \in S_{t,\bx}(|r|)$, $|\mu_t(\by) - \mu_t(\bx)| \lesssim |r|$. Hence, for any $r \neq 0$, for any $\bx \in \B$, $t \in \{0,1\}$,
\begin{align*}
    |\theta_{t,\bx}(r) - \mu_t(\bx)|
    \leq \frac{\int_{S_{t,\bx}(|r|)} |\mu_t(\by) - \mu_t(\bx)| f_X(\by) \Haus^{d-1}(\dif\by)}
              {\int_{S_{t,\bx}(|r|)} f_X(\by) \Haus^{d-1}(\dif\by)} \lesssim |r|.
\end{align*}
implying
\begin{align*}
    |\theta_{t,\bx}(0) - \mu_t(\bx)| \leq \lim_{r \to 0,\; r\in\I_t} |\theta_{t,\bx}(r) - \mu_t(\bx)| = 0,
\end{align*}
which establishes the result.
\qed

\subsection{Proof of Theorem~\ref{sa-thm: Approximation Bias: Minimal Guarantee}}

By Theorem~\ref{sa-thm: Identification} and Equation~\eqref{sa-eq: best fit}, we have
\begin{align*}
    \sup_{\bx \in \B}|\bias_t(\bx)|
    & = \sup_{\bx \in \B} \Big|\be_0^{\top} \bPsi_{t,\bx}^{-1} \bS_{t,\bx} - \mu_t(\bx) \Big| \\
    & = \sup_{\bx \in \B} \Big|\be_0^{\top} \bPsi_{t,\bx}^{-1} \E \Big[\rp\Big(\frac{D_i(\bx)}{h}\Big) K_h(D_i(\bx)) \big(\mu_t(\bX_i) - \mu_t(\bx)\big) \Indicator(D_i(\bx) \in \I_t) \Big] \Big| \\
    & \lesssim \sup_{\bx \in \B} \sup_{\bz \in \X} |\mu_t(\bx) - \mu_t(\bz)| \Indicator(K_h(D_{\bz}(\bx)) > 0) \\
    & \lesssim h.
\end{align*}
Taking the sum over $t=0,1$ gives $\sup_{\bx\in\B}|\bias(\bx)|\lesssim h$. This result holds for all $d\geq1$.

The same argument, uniformly over $\mathcal{P}$, gives the upper bound in the
case of $d=2$. Thus, it remains to establish the lower bound in the sharpness statement for $d=2$.
Let
$R_K<\infty$ satisfy $\supp(K)\subseteq[-R_K,R_K]$ and set
$\bar R_K=\max\{1,R_K\}$. Suppose
$\bX_i \sim \mathsf{Uniform}([-2\bar R_K,2\bar R_K]^2)$,
$\mu_0(x_1,x_2)=0$, and $\mu_1(x_1,x_2)=x_2$ on
$\X=[-2\bar R_K,2\bar R_K]^2$. Let
\begin{align*}
    \mathcal{A}_1=\{(x,y)\in\X:x\ge0,\ y\ge0\},
    \qquad
    \mathcal{A}_0=\X\setminus\mathcal{A}_1,
\end{align*}
and
$\mathcal{B}=([0,2\bar R_K]\times\{0\})\cup(\{0\}\times[0,2\bar R_K])$.
Let $\d$ be Euclidean distance, so $D_i(\bx)=\|\bX_i-\bx\|$ on
$\mathcal{A}_1$ and $D_i(\bx)=-\|\bX_i-\bx\|$ on $\mathcal{A}_0$.
This data-generating process belongs to the class $\mathcal{P}$ for a sufficiently
large constant $L$. For $\bx_s=(s,0)$, $0\le s\le1$, the induced conditional mean on
the treated side is
\begin{align*}
    \theta_{1,\bx_s}(r)
    =
    \begin{cases}
        2r/\pi, & 0\le r\le s,\\
        (r+s)/(\pi-\arccos(s/r)), & r>s.
    \end{cases}
\end{align*}
Thus the conditional mean given signed distance has a kink at $r=s$ when $s>0$.
Figure~\ref{sa-fig: cond mean 1d} plots this function for $s=3/4$.

\begin{figure}[!tbp]
    \centering
    \begin{tikzpicture}[x=6cm,y=6cm,baseline,font=\scriptsize]
        \foreach \xgrid in {0,0.125,0.25,0.375,0.5,0.625,0.75,0.875,1}
            \draw[gray!35] (\xgrid,0) -- (\xgrid,0.72);
        \foreach \ygrid in {0,0.1,0.2,0.3,0.4,0.5,0.6,0.7}
            \draw[gray!35] (0,\ygrid) -- (1.04,\ygrid);
        \draw[gray!65,very thick] (0,0) -- (1.04,0);
        \draw[gray!65,very thick] (0,0) -- (0,0.72);
        \draw[gray!70,line width=1.3pt]
            plot[domain=0:0.75,samples=2] (\x,{2*\x/pi});
        \draw[gray!70,line width=1.3pt]
            plot[domain=0.7501:1,samples=90,smooth]
            (\x,{(\x+0.75)/(pi-acos(0.75/\x)*pi/180)});
        \fill[blue] (0.75,{1.5/pi}) circle[radius=2.1pt];
        \node[below,gray!70] at (0,0) {$0$};
        \node[below,gray!70] at (0.25,0) {$0.25$};
        \node[below,gray!70] at (0.5,0) {$0.50$};
        \node[below,gray!70] at (0.75,0) {$0.75$};
        \node[below,gray!70] at (1,0) {$1.00$};
        \node[left,gray!70] at (0,0) {$0$};
        \node[left,gray!70] at (0,0.2) {$0.2$};
        \node[left,gray!70] at (0,0.4) {$0.4$};
        \node[left,gray!70] at (0,0.6) {$0.6$};
    \end{tikzpicture}
    \caption{Conditional mean given distance with one kink.}
    \label{sa-fig: cond mean 1d}
\end{figure}

Let $\operatorname{bias}_{n,1}(h,s)$ denote the level projection bias at
$\bx_s$ with bandwidth $h$. A change of variables gives the scaling identity
\begin{align}\label{sa-eq: scaling}
    \operatorname{bias}_{n,1}(Ch,Cs)=C\operatorname{bias}_{n,1}(h,s),
    \qquad 0<C<1,
\end{align}
because the Gram matrix is invariant under the rescaling
$(u,v,s,h)\mapsto(Cu,Cv,Cs,Ch)$, while the linear conditional mean
$\mu_1(u,v)=v$ scales by $C$. Since $\mu_0\equiv0$,
\begin{align*}
    \operatorname{bias}_{n,0}(h,s)=0,
    \qquad 0<h<1,\quad 0<s<h.
\end{align*}
Therefore it is enough to show
$\sup_{0\le s\le1}|\operatorname{bias}_{n,1}(1,s)|>0$.

Suppressing the common density factor, write
\begin{align*}
    \operatorname{bias}_{n,1}(1,s)
    =
    \be_0^\top\bA(s)^{-1}\bB(s),
\end{align*}
where
\begin{align*}
    \bA(s)
    &= \int_{0}^{\infty}\int_{\Theta_s(u)}^{\pi}
       \rp(u)\rp(u)^\top K(u)u\,\dif\theta\,\dif u,\\
    \bB(s)
    &= \int_{0}^{\infty}\int_{\Theta_s(u)}^{\pi}
       \rp(u)K(u)u^2\sin(\theta)\,\dif\theta\,\dif u,
    \qquad
    \Theta_s(u)=
    \begin{cases}
        0, & 0\le u\le s,\\
        \arccos(s/u), & u>s.
    \end{cases}
\end{align*}
At $s=0$,
\begin{align}\label{sa-eq: value at zero}
    \bA(0)=\frac{\pi}{2}\int_0^\infty u\rp(u)\rp(u)^\top K(u)\dif u,
    \qquad
    \bB(0)=\int_0^\infty u^2\rp(u)K(u)\dif u,
\end{align}
If $p=0$, then $\bA(0)$ and $\bB(0)$ are scalars and
$\be_0^\top\bA(0)^{-1}\bB(0)>0$, so
$\sup_{0\le s\le1}|\operatorname{bias}_{n,1}(1,s)|>0$ immediately.
It remains to consider $p\ge1$.
Differentiating under the integral sign gives
\begin{align*}
    \dot{\bA}(0)=\int_0^\infty \rp(u)\rp(u)^\top K(u)\dif u,
    \qquad
    \dot{\bB}(0)=\int_0^\infty u\rp(u)K(u)\dif u.
\end{align*}
Using $\bA(0)^{-1}\bB(0)=(2/\pi)\be_1$, where $\be_1$ is the slope selector in the polynomial basis, and
$\bA(0)^{-1}\dot{\bB}(0)=(2/\pi)\be_0$,
\begin{align}\label{sa-eq: derivative at zero}
    \frac{d}{ds}\operatorname{bias}_{n,1}(1,s)\bigg|_{s=0}
    &=
    -\be_0^\top\bA(0)^{-1}\dot{\bA}(0)\bA(0)^{-1}\bB(0)
    +\be_0^\top\bA(0)^{-1}\dot{\bB}(0) \nonumber\\
    &=
    -\frac{4}{\pi^2}+\frac{2}{\pi}
    \neq0.
\end{align}
By continuity, the same strict inequality holds for $p\ge1$. Thus, for all $p\ge0$,
$\sup_{0\le s\le1}|\operatorname{bias}_{n,1}(1,s)|>0$. Combining this with
\eqref{sa-eq: scaling} yields
\begin{align*}
    \inf_{0<h<1}\sup_{\bx\in\B}
    \frac{|\bias_1(\bx)-\bias_0(\bx)|}{h}
    \ge
    \inf_{0<h<1}\sup_{0<s<h}
    \left|\operatorname{bias}_{n,1}\left(1,\frac{s}{h}\right)\right|
    >0.
\end{align*}
This proves the lower bound and completes the proof.
\qed

\subsection{Proof of Theorem~\ref{sa-thm: Approximation Bias: Smooth Distance Derivatives}}

\subsubsection*{Part (i): High-level smoothness condition.}
It is enough to prove the result for one side $t$. Since $K$ is compactly supported, take $R_K<\infty$ such that $\supp(K)\subseteq[-R_K,R_K]$; it suffices to consider $|r|\le R_Kh$ with $r\in\I_t$. Taylor's theorem gives
\begin{align*}
    \theta_{t,\bx}(r)
    = \sum_{j=0}^{m-1}\frac{\theta_{t,\bx}^{(j)}(0)}{j!}r^j + R_{t,\bx}(r),
    \qquad
    \sup_{\bx\in\B}|R_{t,\bx}(r)|\lesssim |r|^m.
\end{align*}
The local polynomial fit of order $p$ reproduces every polynomial of degree at most $p$. Since $m\le p+1$, the level bias is therefore generated only by the remainder. Using $\|\bPsi_{t,\bx}^{-1}\|\lesssim1$ and the bounded kernel moments,
\begin{align*}
    \sup_{\bx\in\B}|\bias_t(\bx)|
    \lesssim \sup_{\bx\in\B}
    \E\Big[K_h(D_i(\bx))|D_i(\bx)|^m\Indicator(D_i(\bx)\in\I_t)\Big]
    \lesssim h^m.
\end{align*}
Taking the sum over $t=0,1$ gives the uniform result. The pointwise result is identical with the outer supremum removed.

\subsubsection*{Part (ii): Pointwise bivariate smooth-boundary primitive condition.}
Fix $\bx\in\B$ and let $m_{\bx}=\kappa\wedge(p+1)$. It is enough to verify the primitive smoothness condition in part (i) for one treatment side $t$ at this fixed boundary point; the treatment-effect bias is the difference of the two side-specific biases.

After re-centering and rotating coordinates, set $\bx=0$ and parameterize the local curve by arc length so that $\gamma(0)=0$ and $\gamma'(0)=(1,0)$. For $u$ near zero,
\begin{align*}
    \phi(u)=\|\gamma(u)-\gamma(0)\|^2
    =u^2 a(u),
    \qquad a(0)=1,
\end{align*}
where $a$ is $C^{\kappa}$ and bounded away from zero in a neighborhood of zero. Therefore the two local solutions of $\phi(u)=r^2$ can be written as $u_+(r)$ and $u_-(r)$, where
$u_+(r)=r q_+(r)$ and $u_-(r)=-r q_-(r)$, with $q_+$ and $q_-$ being $C^\kappa$ functions. This follows from the implicit function theorem applied to $u/\sqrt{\phi(u)}$ on the positive and negative sides of zero.

The polar angles at which the circle $\{\bu:\|\bu-\bx\|=r\}$ intersects the local boundary are
\begin{align*}
    \alpha_+(\bx,r)
    = \arg\{\gamma(u_+(r))-\bx\},
    \qquad
    \alpha_-(\bx,r)
    = \arg\{\gamma(u_-(r))-\bx\}.
\end{align*}
The preceding representation and the smoothness of $\gamma$ imply that
$\alpha_+(\bx,\cdot)$ and $\alpha_-(\bx,\cdot)$ are $C^\kappa$ on
$[0,\varepsilon]$, after possibly shrinking $\varepsilon$. The local angular interval
corresponding to $\A_t$ therefore has endpoints
$\alpha_{t,1}(\bx,r)$ and $\alpha_{t,2}(\bx,r)$ that are $C^\kappa$ in $r$.

For $r>0$,
\begin{align*}
    \theta_{t,\bx}(r)
    =
    \frac{
    \int_{\alpha_{t,1}(\bx,r)}^{\alpha_{t,2}(\bx,r)}
    \mu_t(\bx+r(\cos\theta,\sin\theta))f_X(\bx+r(\cos\theta,\sin\theta))\dif\theta}
    {\int_{\alpha_{t,1}(\bx,r)}^{\alpha_{t,2}(\bx,r)}
    f_X(\bx+r(\cos\theta,\sin\theta))\dif\theta}.
\end{align*}
The numerator and denominator are $C^{m_{\bx}}$ in $r$ by repeated application of
Leibniz's rule, because $\mu_t$, $f_X$, and the angle endpoints have the required
smoothness. The denominator is bounded away from zero for small $r$ because the angular
interval length converges to $\pi$ and $f_X$ is bounded away from zero. Hence
$\theta_{t,\bx}(\cdot)$ is $m_{\bx}$-times continuously differentiable near zero.
Applying the pointwise version of part (i) gives
\begin{align*}
    |\bias(\bx)|\lesssim h^{m_{\bx}}.
\end{align*}

\subsubsection*{Part (iii): Uniform bivariate smooth-boundary primitive condition.}
Let $m=\iota\wedge(p+1)$. We verify the primitive smoothness condition in part (i) uniformly over $\bx\in\B^o$.

Fix $\bx_0=\gamma(s_0)\in\B^o$ and re-center and rotate coordinates so that $\bx_0=0$ and $\gamma'(s_0)=(1,0)$ after reparameterizing the curve by arc length. Since $\gamma$ is $C^{\iota+2}$, for $u$ near zero,
\begin{align*}
    \phi(u)=\|\gamma(s_0+u)-\gamma(s_0)\|^2
    =u^2 a(u),
    \qquad a(0)=1,
\end{align*}
where $a$ is $C^{\iota}$ and bounded away from zero uniformly for $s_0\in[\delta,L-\delta]$. Therefore the two local solutions of $\phi(u)=r^2$ can be written as $u_+(r)$ and $u_-(r)$, where $u_+(r)=r q_+(r)$ and $u_-(r)=-r q_-(r)$ with $q_+$ and $q_-$ being $C^{\iota}$ functions whose derivatives up to order $\iota$ are uniformly bounded. This follows from the implicit function theorem applied to $u/\sqrt{\phi(u)}$ on the positive and negative sides of zero.

The polar angles at which the circle $\{\bu:\|\bu-\bx_0\|=r\}$ intersects the boundary are therefore
\begin{align*}
    \alpha_+(\bx_0,r)
    = \arg\{\gamma(s_0+u_+(r))-\bx_0\},
    \qquad
    \alpha_-(\bx_0,r)
    = \arg\{\gamma(s_0+u_-(r))-\bx_0\}.
\end{align*}
The preceding representation and the smoothness of $\gamma$ imply that $\alpha_+(\bx_0,\cdot)$ and $\alpha_-(\bx_0,\cdot)$ are $C^{\iota}$ on $[0,\varepsilon]$, with derivatives uniformly bounded over $\bx_0\in\B^o$. By the assumed two-intersection condition, the angular interval corresponding to $\A_t$ has endpoints given by these two angle functions, up to a fixed ordering convention.

For $r>0$, write this interval as $[\alpha_{t,1}(\bx_0,r),\alpha_{t,2}(\bx_0,r)]$. Then
\begin{align*}
    \theta_{t,\bx_0}(r)
    =
    \frac{
    \int_{\alpha_{t,1}(\bx_0,r)}^{\alpha_{t,2}(\bx_0,r)}
    \mu_t(\bx_0+r(\cos\theta,\sin\theta))f_X(\bx_0+r(\cos\theta,\sin\theta))\dif\theta}
    {\int_{\alpha_{t,1}(\bx_0,r)}^{\alpha_{t,2}(\bx_0,r)}
    f_X(\bx_0+r(\cos\theta,\sin\theta))\dif\theta}.
\end{align*}
The numerator and denominator are $C^m$ in $r$ by repeated application of Leibniz's rule, because $\mu_t$, $f_X$, and the angle endpoints have uniformly bounded derivatives up to order $m$. The denominator is bounded away from zero uniformly for small $r$ because the angular interval length converges uniformly to $\pi$ and $f_X$ is bounded away from zero; Assumption~\ref{sa-assump: Kernel, Distance and Boundary}(v) ensures that the corresponding Hausdorff-measure ratio is well defined before this local parametrization. Hence $\theta_{t,\bx_0}$ is $C^m$ on $[0,\varepsilon]$, and its derivatives up to order $m$ are uniformly bounded over $\bx_0\in\B^o$.

Applying the argument in part (i) on the index set $\B^o$ gives
\begin{align*}
    \sup_{\bx\in\B^o}|\bias(\bx)|\lesssim h^m.
\end{align*}
\qed

\subsection{Proof of Theorem~\ref{sa-thm: Convergence Rates}}

For a fixed $\bx\in\B$, the proofs of Lemmas~\ref{sa-lem: Gram} and
\ref{sa-lem: Stochastic Linear Approximation} can be done when the index set is the
singleton $\{\bx\}$ instead of $\B$, replacing \citet[Corollary
5.1]{Chernozhukov-Chetverikov-Kato_2014b_AoS} with Bernstein's inequality and obtaining
\begin{align*}
    \Big|\be_0^{\top} \bPsi_{t,\bx}^{-1} \mathbf{O}_{t,\bx}\Big|
    & \lessp \sqrt{\frac{1}{n h^d}} + \frac{1}{n^{\frac{1+v}{2+v}}h^d}, \\
    \Big|\be_0^{\top} (\widehat{\bPsi}_{t,\bx}^{-1} - \bPsi_{t,\bx}^{-1}) \mathbf{O}_{t,\bx}\Big|
    & \lessp \sqrt{\frac{1}{n h^d}} \bigg(\sqrt{\frac{1}{n h^d}} + \frac{1}{n^{\frac{1+v}{2+v}}h^d} \bigg).
\end{align*}
The pointwise convergence rate follows from decomposition
\eqref{sa-eq: distance error decomposition}.
The uniform statement follows from Lemmas~\ref{sa-lem: Gram} and
\ref{sa-lem: Stochastic Linear Approximation} and the same decomposition.
\qed

\subsection{Proof of Theorem~\ref{sa-thm: Confidence Intervals}}

Define $\overline{\Tstat}(\bx) = \sum_{i = 1}^n Z_i$, with $Z_i = Z_{1,i} - Z_{0,i}$ independent random variables ($i=1,2,\ldots,n)$,
\begin{align*}
    Z_{t,i} = \frac{1}{n} \big(\Xi_{\bx,\bx}\big)^{-1/2}\be_0^{\top} \bPsi_{t,\bx}^{-1}
              \rp\Big(\frac{D_i(\bx)}{h}\Big) K_h(D_i(\bx)) (Y_i - \theta_{t,\bx}^{\ast}(D_i(\bx))) \Indicator(\bX_i \in \mathcal{A}_t),
\end{align*}
$\E[Z_{i}] = 0$ and $\Var[Z_{i}] = n^{-1}$. By the Berry-Esseen theorem,
\begin{align*}
    \sup_{u \in \reals} \Big|\P \big(\overline{\Tstat}(\bx) \leq u \big) - \Phi(u) \Big|
    \lesssim \sum_{i = 1}^n \E \big[|Z_{i}|^3 \big]
    \lesssim \sum_{i = 1}^n \E \big[|Z_{1,i}|^3 \big] + \sum_{i = 1}^n \E \big[|Z_{0,i}|^3 \big]
\end{align*}
where
\begin{align*}
    \sum_{i = 1}^n\E \big[|Z_{t,i}|^3 \big]
    &= n^{-2} \big(\Xi_{\bx,\bx}\big)^{-3/2}
       \E \Big[ \Big|\be_0^{\top} \bPsi_{t,\bx}^{-1} \rp \Big(\frac{D_i(\bx)}{h}\Big) K_h(D_i(\bx)) \Indicator(\bX_i \in \mathcal{A}_t) (Y_i - \theta_{t,\bx}^{\ast}(D_i(\bx)) \Big|^3 \Big] \\
    & \lesssim n^{-2} \big(\Xi_{\bx,\bx}\big)^{-3/2} \E[|K_h(D_i(\bx))|^3|Y_i - \theta_{t,\bx}^{\ast}(D_i(\bx))|^3]\\
    & \lesssim n^{-2} \big(\Xi_{\bx,\bx}\big)^{-3/2} \E[|K_h(D_i(\bx))|^3(\E[|Y_i|^3|\bX_i] + |\theta_{t,\bx}^{\ast}(D_i(\bx))|^3)]\\
    & \lesssim (n h^d)^{-1/2}, 
\end{align*}
noting that $K_h(D_i(\bx))$ is supported on a set of probability $O(h^d)$ and satisfies $|K_h(D_i(\bx))|\lesssim h^{-d}$, $\Xi_{\bx,\bx} \gtrsim (n h^d)^{-1}$ by the pointwise version of the covariance lower-bound argument in Lemma~\ref{sa-lem: Covariance}, $\E[|Y_i|^3|\bX_i] \lesssim 1$ by Assumption~\ref{sa-assump: DGP}(v), and $\max_{1 \leq i \leq n}\sup_{\bx \in \B}|\theta_{t,\bx}^{\ast}(D_i(\bx))| \lesssim 1$ because
\begin{align*}
    \theta_{t,\bx}^{\ast}(D_i(\bx))
    = \bgamma_t^{\ast}(\bx)^{\top} \rp \Big(\frac{D_i(\bx)}{h}\Big)
    = \big(\bPsi_{t,\bx}^{-1} \bS_{t,\bx} \big)^{\top} \rp \Big(\frac{D_i(\bx)}{h}\Big).
\end{align*}
Since
\begin{align*}
    \big|\widehat{\Tstat}(\bx) - \overline{\Tstat}(\bx)\big|
    \lessp \frac{1}{\sqrt{n h^d}}
           + \frac{1}{n^{\frac{v}{2+v}}h^d}
           + \sqrt{n h^d} |\bias(\bx)|,
\end{align*}
and the same pointwise covariance argument gives $\widehat{\Xi}_{\bx,\bx}/\Xi_{\bx,\bx}\to_{\P}1$ under the conditions imposed, pointwise asymptotic normality follows. Finally, validity of the confidence interval estimator is immediate.
\qed

\subsection{Proof of Theorem~\ref{sa-thm: Stochastic Linearization}}

Because $v>0$, the condition
$n^{\frac{v}{2+v}}h^d/\log(1/h)\to\infty$ implies
$nh^d/\log(1/h)\to\infty$, so the preceding uniform approximation lemmas apply.
We make the decomposition based on Equation~\eqref{sa-eq: distance error decomposition} and convergence of $\widehat{\Xi}_{\bx,\bx}$,
\begin{align*}
    \widehat{\Tstat}(\bx) - \overline{\Tstat}(\bx) & = \big(\widehat{\Xi}_{\bx,\bx}\big)^{-1/2}\bigg(\sum_{t \in \{0,1\}}(2t-1)(\widehat{\theta}_{t,\bx}(0) - \theta_{t,\bx}(0))\bigg) \\
    & \qquad - \big(\Xi_{\bx,\bx}\big)^{-1/2} \bigg(\sum_{t \in \{0,1\}}(2t-1)\be_0^{\top} \bPsi_{t,\bx}^{-1} \mathbf{O}_{t,\bx}\bigg) \\
    & = \big(\widehat{\Xi}_{\bx,\bx}\big)^{-1/2}\bigg(\sum_{t \in \{0,1\}}(2t-1)(\widehat{\theta}_{t,\bx}(0) - \theta_{t,\bx}(0)) \\
    & \qquad -\sum_{t \in \{0,1\}}(2t-1)\be_0^{\top} \bPsi_{t,\bx}^{-1} \mathbf{O}_{t,\bx} \bigg)   \tag{$=\Delta_{1,\bx}$} \\
    & \phantom{=} + \big((\widehat{\Xi}_{\bx,\bx})^{-1/2} - (\Xi_{\bx,\bx})^{-1/2}\big) \sum_{t \in \{0,1\}}(2t-1)\be_0^{\top} \bPsi_{t,\bx}^{-1} \mathbf{O}_{t,\bx} \tag{$=\Delta_{2,\bx}$}
\end{align*}
By Lemma~\ref{sa-lem: Gram} and \ref{sa-lem: Stochastic Linear Approximation}, and the decomposition Equation~\eqref{sa-eq: distance error decomposition},
\begin{align*}
    & \sup_{\bx \in \B} \bigg|\sum_{t \in \{0,1\}}(2t-1)(\widehat{\theta}_{t,\bx}(0) - \theta_{t,\bx}(0)) -\sum_{t \in \{0,1\}}(2t-1)\be_0^{\top} \bPsi_{t,\bx}^{-1} \mathbf{O}_{t,\bx}\bigg| \\
    & \lessp \sqrt{\frac{\log(1/h)}{n h^d}} \left(\sqrt{\frac{\log(1/h)}{n h^d}} + \frac{\log(1/h)}{n^{\frac{1+v}{2+v}}h^d} \right) + \sup_{\bx \in \B}|\bias(\bx)|.
\end{align*}
Together with Lemma~\ref{sa-lem: Covariance},
\begin{align}\label{sa-eq: stochastic linearization for distance -- 1}  
    \sup_{\bx \in \B}|\Delta_{1,\bx}| \lessp \frac{\log(1/h)}{\sqrt{n h^d}} + \frac{(\log(1/h))^{\frac{3}{2}}}{n^{\frac{1+v}{2+v}}h^d} + \sqrt{n h^d} \sup_{\bx \in \B}|\bias(\bx)|.
\end{align}
By Lemma~\ref{sa-lem: Gram}, Lemma~\ref{sa-lem: Stochastic Linear Approximation} and Lemma~\ref{sa-lem: Covariance}, and assume $\frac{n^{\frac{v}{2+v}}h^d}{\log(1/h)} \to \infty$, then
\begin{align*}
    \sup_{\bx \in \B} \left|\be_0^{\top} \bPsi_{t,\bx}^{-1} \mathbf{O}_{t,\bx} \left((\Xi_{\bx,\bx})^{-1/2} - (\widehat{\Xi}_{\bx,\bx})^{-1/2}\right) \right| & \lessp \sqrt{n h^d} \bigg(\sqrt{\frac{\log(1/h)}{n h^d}} + \frac{\log(1/h)}{n^{\frac{1+v}{2+v}}h^d}\bigg) \left(\sqrt{\frac{\log(1/h)}{n h^d}} + \frac{\log(1/h)}{n^{\frac{v}{2+v}}h^d} \right) \\
    & = \sqrt{\log(1/h)}\bigg(1 + \sqrt{\frac{\log(1/h)}{n^{\frac{v}{2+v}}h^d}}\bigg) \left(\sqrt{\frac{\log(1/h)}{n h^d}} + \frac{\log(1/h)}{n^{\frac{v}{2+v}}h^d} \right)  \\
    & \lesssim \sqrt{\log(1/h)} \left(\sqrt{\frac{\log(1/h)}{n h^d}} + \frac{\log(1/h)}{n^{\frac{v}{2+v}}h^d}\right).
\end{align*}
Hence
\begin{align}\label{sa-eq: stochastic linearization for distance -- 2}  
    \sup_{\bx \in \B}|\Delta_{2,\bx}| \lessp \frac{\log(1/h)}{\sqrt{n h^d}} + \frac{(\log(1/h))^{\frac{3}{2}}}{n^{\frac{v}{2+v}}h^d}.
\end{align}
Putting together Equations~\eqref{sa-eq: stochastic linearization for distance -- 1}, \eqref{sa-eq: stochastic linearization for distance -- 2} give the result.
\qed

\subsection{Proof of Theorem~\ref{sa-thm: Gaussian Strong Approximation: Tstat}}

We will verify the high level conditions stated in Theorem~\ref{sa-thm: Strong Approximation}.

By the rectangular-support condition, after a componentwise affine rescaling we can take $\X=[0,1]^d$. Since $f_X$ is continuous and bounded away from zero on this compact rectangle, \citet[Section~3.1]{Cattaneo-Yu_2025_AOS} implies that $\Q_{\G\cup\M}=\P_X$ is a valid surrogate measure for $\P_X$ with respect to $\G\cup\M$, and that the Rosenblatt transform $\phi_{\G\cup\M}=T_{\P_X}$ is a valid normalizing transformation. The constants $\ttc_1$ and $\ttc_2$ from Theorem~\ref{sa-thm: Strong Approximation} are therefore finite and uniformly bounded by constants depending only on the density bounds and the support geometry, and are absorbed in the displays below.

Recall $\G = \{g_{\bx}: \bx \in \B\}$ where
\begin{align*}
    g_{\bx}(\bu)
    &= \Indicator(\bu \in \A_1) \mathfrak{K}_1(\bu;\bx)
     - \Indicator(\bu \in \A_0) \mathfrak{K}_0(\bu;\bx).
\end{align*}
By the same signed-distance entropy and localization arguments as in Lemma~\ref{sa-lem: Gram}, combined with \cite[Lemma 7]{Cattaneo-Chandak-Jansson-Ma_2024_Bernoulli} in the Lipschitz-kernel case and the VC-ball argument in the uniform-kernel case, we get properties of $\G$ as follows:
\begin{align*}
    \ttM_{\G} \lesssim h^{-d/2}, \qquad \ttE_{\G} \lesssim h^{d/2}, \qquad 
    \ttTV_{\G} \lesssim h^{d/2-1}, \qquad \sup_{Q} N(\G, \norm{\cdot}_{Q,2}, \varepsilon (2 c +1)^{d+1} \ttM_{\G}) \leq 2 \bc' \varepsilon^{-d-1} + 2.
\end{align*}
By definition of $\theta_{t,\bx}^{\ast}(\cdot)$, for each $\bx \in \B$, $t \in \{0,1\}$,
\begin{align*}
    \theta_{t,\bx}^{\ast}(D_{\bu}(\bx)) = \bgamma_t^{\ast}(\bx)^{\top}\rp\Big(\frac{D_{\bu}(\bx)}{h}\Big)
    = (\bPsi_{t,\bx}^{-1} \bS_{t,\bx})^{\top} \rp\Big(\frac{D_{\bu}(\bx)}{h}\Big),
\end{align*}
recalling
\begin{align*}
    \bPsi_{t,\bx} = \E \left[\rp\Big(\frac{D_i(\bx)}{h}\Big) \rp\Big(\frac{D_i(\bx)}{h}\Big)^{\top} K_h(D_i(\bx)) \Indicator(D_i(\bx) \in \I_t)\right], \quad \bS_{t,\bx} = \E \left[\rp\Big(\frac{D_i(\bx)}{h}\Big) K_h(D_i(\bx)) Y_i \Indicator(\bX_i \in \A_t) \right].
\end{align*}
The preceding bounds imply that $\norm{\bPsi_{t,\bx}^{-1}} \lesssim 1$, $\norm{\bS_{t,\bx}} \lesssim 1$, and
\begin{align*}
    \ttM_{\M_t} \lesssim h^{-d/2}, \qquad \ttE_{\M_t} \lesssim h^{d/2}, \qquad t \in \{0,1\}.
\end{align*}
In what follows, we verify the entropy and total variation properties of $\M$. Using the product rule, we have
\begin{gather*}
    \sup_{\bu \in \X}\sup_{\bx,\bx' \in \B}\frac{|\theta_{t,\bx}^{\ast}(D_{\bu}(\bx)) - \theta_{t,\bx}^{\ast}(D_{\bu}(\bx'))|}{\norm{\bx - \bx'}} \lesssim h^{-1}.
\end{gather*}
Define $f_{t,\bx}(\cdot) = \frac{h^{-d/2}}{\sqrt{n \Xi_{\bx,\bx}}}\be_0^{\top}  \bPsi_{t,\bx}^{-1} \rp \left(\cdot\right) K(\cdot) (\bPsi_{t,\bx}^{-1} \bS_{t,\bx})^{\top} \rp(\cdot)$. Then,
\begin{align*}
    \mathfrak{K}_t(\bu;\bx) \theta_{t,\bx}^{\ast}(D_{\bu}(\bx)) = h^{-d/2} f_{t,\bx} \Big(\frac{D_{\bu}(\bx)}{h}\Big), \qquad \bu \in \X, \bx \in \B, t \in \{0,1\}.
\end{align*}
Take $\M_t = \{\mathfrak{K}_t(\cdot;\bx) \theta_{t,\bx}^{\ast}(D_{\cdot}(\bx)): \bx \in \B\}$, $t \in \{0,1\}$. If $K$ is Lipschitz, then for $t \in \{0,1\}$, $f_{t,\bx}$ satisfies:
\begin{align*}
 & \textit{(i) boundedness}  && \sup_{\bx \in \B} \sup_{u \in \reals} \left|f_{t,\bx}(u) \right| \leq \bc, \\
 & \textit{(ii) compact support} && \supp(f_{t,\bx}(\cdot)) \subseteq [-\bc, \bc], \forall \bx \in \B, \\
 & \textit{(iii) Lipschitz continuity} && \sup_{\bx \in \B} \sup_{u,u' \in \reals}\frac{\left|f_{t,\bx}(u) - f_{t,\bx}(u') \right|}{|u - u'|} \leq \bc, \\
 & \phantom{xx} && \sup_{u \in \reals} \sup_{\bx,\bx' \in \B}\frac{|f_{t,\bx}(u) - f_{t,\bx'}(u)|}{\norm{\bx - \bx'}} \leq \bc h^{-1},
\end{align*}
for some constant $\bc$ not depending on $n$.  
Then, by an argument similar to \citet[Lemma 7]{Cattaneo-Chandak-Jansson-Ma_2024_Bernoulli}, there exists a constant $\bc'$ depending only on $\bc$ and $d$ such that, for any $0 \leq \varepsilon \leq 1$,
\begin{align*}
    \sup_{Q} N\left(h^{d/2}\M_t, \norm{\cdot}_{Q,1}, (2\bc+1)^{d+1}\varepsilon \right) \leq \bc' \varepsilon^{-d-1} + 1,
\end{align*}
where the supremum is taken over all finite discrete measures.
If $K(u)=\Indicator(u\in[-1,1])$, the same display follows by replacing the preceding Lipschitz step with the VC-ball argument used in Lemma~\ref{sa-lem: Gram}; the polynomial factors and $\theta_{t,\bx}^{\ast}(D_{\cdot}(\bx))$ are uniformly bounded on the support of $K_h(D_{\cdot}(\bx))$, and products with the fixed side indicator preserve VC-type entropy.
Taking a constant envelope function $\ttM_{\M_t} = (2\bc +1)^{d+1} h^{-d/2}$, we have for any $0 < \varepsilon \leq 1$,
\begin{align*}
    \sup_{Q} N\left(\M_t, \norm{\cdot}_{Q,1}, \varepsilon \ttM_{\M_t}\right) \leq \bc' \varepsilon^{-d-1} + 1.
\end{align*}
By Lemma~\ref{sa-lem: VC Class to VC2 Class}, above implies the uniform covering number for $\M_t$ satisfies
\begin{align*}
    \ttN_{\M_t}(\varepsilon) \leq 4 \bc'(\varepsilon/2)^{-d-1}, \qquad 0 < \varepsilon \leq 1.
\end{align*}
Since $\M \subseteq \M_0 + \M_1$, here $+$ denotes the Minkowski sum, with $\ttM_{\M}$ taken to be $\ttM_{\M_0} + \ttM_{\M_1}$, a bound on the uniform covering number of $\M$ can be given by
\begin{align*}
    \ttN_{\M}(\varepsilon) \leq 16 (\bc')^2(\varepsilon/2)^{-2d-2}, \qquad 0 < \varepsilon \leq 1.
\end{align*}
With the assumption that $\mathcal{L}(E_{t,\bx}(h)) \leq C h^{d-1}$ for $E_{t,\bx}(h) = \{\by \in \mathcal{A}_t: \varsigma_t\d(\by,\bx)/h\in\supp(K)\}$ for all $t \in \{0,1\}$, $\bx \in \B$, and the fact that $\ttTV_{\M_t} \lesssim h^{d/2-1}$ for $t \in \{0,1\}$, the same argument as in the paragraph \textbf{Total Variation} in the proof of Theorem~\ref{sa-thm: Strong Approximation} shows
\begin{align*}
    \ttTV_{\M} \lesssim h^{d/2-1}.
\end{align*}
Now apply Theorem~\ref{sa-thm: Strong Approximation} with $\G$, $\M$ defined in Equation~\eqref{sa-eq: def G and M}, $\R = \{\operatorname{Id}\}$, $\S = \{1\}$, noting that
\begin{align*}
    \overline{\Tstat}(\bx)
    = A_n(g_\bx,m_\bx,\operatorname{Id},1), \qquad (g_\bx,m_\bx) \in \G \times \M,\quad \bx \in \B,
\end{align*}
the result follows.
\qed

\begin{lem}[VC Class to VC2 Class]\label{sa-lem: VC Class to VC2 Class}
    Assume $\F$ is a VC class on a measure space $(\X, \mathcal{B})$: there exists an envelope function $F$ and positive constants $c(\F), d(\F)$ such that for all $\varepsilon\in(0,1)$,
    \begin{align*}
        \sup_{Q}N(\mathcal{F}, \norm{\cdot}_{Q,1}, \varepsilon \norm{F}_{Q,1}) \leq c(\F)\varepsilon^{-d(\F)},
    \end{align*}
    where the supremum is taken over all finite discrete measures. Then, $\F$ is also VC2 class: for all $\varepsilon\in(0,1)$, 
    \begin{align*}
        \sup_{Q} N(\mathcal{F}, \norm{\cdot}_{Q,2}, \varepsilon \norm{F}_{Q,2}) \leq c(\F) (\varepsilon^2/2)^{-d(\mathcal{F})},
    \end{align*}
    where the supremum is taken over all finite discrete measures.
\end{lem}

\noindent\textbf{Proof of Lemma~\ref{sa-lem: VC Class to VC2 Class}}. Let $Q$ be a finite discrete probability measure. Let $f,g \in \mathcal{F}$. Then, $\int |f - g|^2 \dif Q \leq 2 \int |f - g| |F| \dif Q$. Define another probability measure $\tilde{Q}(c_k) = F(c_k) Q(c_k) / \norm{F}_{Q,1}$ on the support of $Q$, denoted by $\{c_1, \ldots, c_k, \ldots\}$. Then,
\begin{align*}
    \int |f - g|^2 \dif Q
    \leq 2 \norm{F}_{Q,1} \int |f - g| \dif \tilde{Q}
    \leq 2 \norm{F}_{Q,1} \norm{f - g}_{\tilde{Q},1}.
\end{align*}
Hence, if we take an $\varepsilon^2/2$-net in $(\mathcal{F}, \norm{\cdot}_{\tilde{Q},1})$ with cardinality no greater than $c(\mathcal{F})(\varepsilon^2/2)^{- d (\mathcal{F})}$, then for any $f \in \mathcal{F}$, there exists a $g \in \mathcal{F}$ such that $\norm{f - g}_{\tilde{Q},1} \leq \varepsilon^2/2 \norm{F}_{\tilde{Q},1}$, and hence
\begin{align*}
    \norm{f - g}_{Q,2}^2 \leq 2 \varepsilon^2/2 \norm{F}_{Q,1} \norm{F}_{\tilde{Q},1} \leq \varepsilon^2 \norm{F}_{Q,2}^2,
\end{align*}
which gives the result.
\qed

\subsection{Proof of Theorem~\ref{sa-thm: Confidence Bands}}

The proof is divided into three theorem-specific technical lemmas. We first record a
generic comparison lemma that translates uniform process approximations into
Kolmogorov-distance bounds for suprema.

\begin{lem}[Generic Suprema Comparison]\label{sa-lem: generic suprema comparison}
Let $\mathbb{T}_j=(\mathbb{T}_j(\bx):\bx\in\B)$, $j=1,2$, be real-valued processes
defined on a common probability space. Write
\[
    M_j=\sup_{\bx\in\B}|\mathbb{T}_j(\bx)|,\quad j=1,2,
    \qquad
    \Delta_{12}=\sup_{\bx\in\B}|\mathbb{T}_1(\bx)-\mathbb{T}_2(\bx)|.
\]
For any $\eta>0$,
\begin{align*}
    \sup_{u\in\reals}\left|\P(M_1\le u)-\P(M_2\le u)\right|
    \le
    \mathsf{AC}_2(\eta)
    +\P(\Delta_{12}>\eta),
\end{align*}
where
\[
    \mathsf{AC}_2(\eta)=\sup_{u\in\reals}\P(|M_2-u|\le \eta).
\]
If, in addition, $\mathbb{T}_2$ is a separable mean-zero Gaussian process and the
Gaussian anti-concentration bound applies to $M_2$, then
\begin{align*}
    \sup_{u\in\reals}\left|\P(M_1\le u)-\P(M_2\le u)\right|
    \le
    4\eta\{\E[M_2]+1\}
    +\P(\Delta_{12}>\eta).
\end{align*}
Consequently, if $\E[\Delta_{12}]\le r_n$, then the last probability can be bounded by
$r_n/\eta$.
\end{lem}

\noindent\textbf{Proof of Lemma~\ref{sa-lem: generic suprema comparison}.}
On the event $\{\Delta_{12}\le\eta\}$, the inequalities
\[
    M_1\le u \implies M_2\le u+\eta,
    \qquad
    M_2\le u-\eta \implies M_1\le u
\]
hold for every $u\in\reals$. Hence
\begin{align*}
    \P(M_1\le u)-\P(M_2\le u)
    &\le \P(u<M_2\le u+\eta)+\P(\Delta_{12}>\eta),\\
    \P(M_2\le u)-\P(M_1\le u)
    &\le \P(u-\eta<M_2\le u)+\P(\Delta_{12}>\eta).
\end{align*}
Taking absolute values and then the supremum over $u$ gives the first claim. The second
display follows from the Gaussian anti-concentration inequality of
\citet[Theorem 2.1]{Chernozhukov-Chetverikov-Kato_2014a_AoS}, applied to the
symmetrized Gaussian process whose supremum is $M_2$. The final claim follows from
Markov's inequality.
\qed

For the theorem-specific lemmas below, define
\[
    \mathfrak{R}_{\mathtt{sa}}
    =
    (\log n)^{\frac{3}{2}}
    \Big(\frac{1}{n h^d}\Big)^{\frac{1}{2d+2}\cdot\frac{v}{v+2}}
    +\log(n)\Big(\frac{1}{n^{\frac{v}{2+v}}h^d}\Big)^{1/2},
\]
the stochastic-linearization remainder
\[
    \mathfrak{R}_{\mathtt{lin}}
    =
    \sqrt{\log(1/h)}
    \left(\sqrt{\frac{\log(1/h)}{n h^d}}
    +\frac{\log(1/h)}{n^{\frac{v}{2+v}}h^d}\right)
    +\sqrt{n h^d}\sup_{\bx\in\B}|\bias(\bx)|,
\]
and the covariance-comparison rate
\[
    \mathfrak{R}_{\mathtt{cc}}
    =
    \sqrt{\frac{\log(1/h)}{n h^d}}
    +\frac{\log(1/h)}{n^{\frac{v}{2+v}}h^d}.
\]

\begin{lem}[KS Distance Between $\overline{\Tstat}$ and $Z$]\label{sa-lem: infeasible gaussian to linear statistic}
Suppose the conditions of Theorem~\ref{sa-thm: Confidence Bands} hold. Then
\begin{align*}
    \sup_{u\in\reals}
    \left|
    \P\left(\sup_{\bx\in\B}|\overline{\Tstat}(\bx)|\le u\right)
    -
    \P\left(\sup_{\bx\in\B}|Z(\bx)|\le u\right)
    \right|
    \lesssim
    (\log n)^{1/4}\mathfrak{R}_{\mathtt{sa}}^{1/2}.
\end{align*}
\end{lem}

\noindent\textbf{Proof of Lemma~\ref{sa-lem: infeasible gaussian to linear statistic}.}
Let $a_n>0$ be chosen below. Applying
Lemma~\ref{sa-lem: generic suprema comparison} with
$\mathbb{T}_1=\overline{\Tstat}$ and $\mathbb{T}_2=Z$, and using
Theorem~\ref{sa-thm: Gaussian Strong Approximation: Tstat}, gives
\begin{align*}
    &\sup_{u\in\reals}
    \left|
    \P\left(\sup_{\bx\in\B}|\overline{\Tstat}(\bx)|\le u\right)
    -
    \P\left(\sup_{\bx\in\B}|Z(\bx)|\le u\right)
    \right|\\
    &\qquad\le
    4a_n\left(\E\left[\sup_{\bx\in\B}|Z(\bx)|\right]+1\right)
    +\frac{C\mathfrak{R}_{\mathtt{sa}}}{a_n}.
\end{align*}
The Gaussian process $Z$ has uniformly bounded pointwise variance. Moreover, by
the same Lipschitz calculation for the distance-indexed local-polynomial weights used in
Theorem~\ref{sa-thm: Gaussian Strong Approximation: Tstat}, its canonical metric satisfies
\[
    \rho(\bx,\by)
    =
    \E\left[\{Z(\bx)-Z(\by)\}^2\right]^{1/2}
    \lesssim h^{-1}\|\bx-\by\|_{\infty}.
\]
Since $\B$ is compact and $\liminf_{n\to\infty}\log h/\log n>-\infty$,
\[
    \log N(\varepsilon,\B,\rho)
    \lesssim
    d\log\Big(\frac{1}{h\varepsilon}\Big)
    \lesssim \log n+\log(1/\varepsilon),
    \qquad 0<\varepsilon\le1.
\]
Dudley's entropy integral therefore gives
\[
    \E\left[\sup_{\bx\in\B}|Z(\bx)|\right]\lesssim\sqrt{\log n}.
\]
Choosing $a_n\asymp(\mathfrak{R}_{\mathtt{sa}}/\sqrt{\log n})^{1/2}$ gives the result.
\qed

\begin{lem}[KS Distance Between $\widehat{\Tstat}$ and $\overline{\Tstat}$]\label{sa-lem: feasible statistic to linear statistic}
Suppose the conditions of Theorem~\ref{sa-thm: Confidence Bands} hold. Then
\begin{align*}
    \sup_{u\in\reals}
    \left|
    \P\left(\sup_{\bx\in\B}|\widehat{\Tstat}(\bx)|\le u\right)
    -
    \P\left(\sup_{\bx\in\B}|\overline{\Tstat}(\bx)|\le u\right)
    \right|
    =o(1).
\end{align*}
\end{lem}

\noindent\textbf{Proof of Lemma~\ref{sa-lem: feasible statistic to linear statistic}.}
Theorem~\ref{sa-thm: Stochastic Linearization} gives
\[
    \Delta_{\widehat T,\overline T}
    :=
    \sup_{\bx\in\B}|\widehat{\Tstat}(\bx)-\overline{\Tstat}(\bx)|
    =
    O_{\P}(\mathfrak{R}_{\mathtt{lin}}).
\]
The side conditions imply $\mathfrak{R}_{\mathtt{lin}}\sqrt{\log n}=o(1)$. Choose a
deterministic sequence $a_n$ such that
$\mathfrak{R}_{\mathtt{lin}}=o(a_n)$ and $a_n\sqrt{\log n}=o(1)$. Applying
Lemma~\ref{sa-lem: generic suprema comparison} with
$\mathbb{T}_1=\widehat{\Tstat}$ and
$\mathbb{T}_2=\overline{\Tstat}$ gives
\[
    \sup_{u\in\reals}
    \left|
    \P(M_{\widehat T}\le u)-\P(M_{\overline T}\le u)
    \right|
    \le
    \sup_{u\in\reals}\P(|M_{\overline T}-u|\le a_n)
    +\P(\Delta_{\widehat T,\overline T}>a_n),
\]
where $M_{\widehat T}=\sup_{\bx\in\B}|\widehat{\Tstat}(\bx)|$ and
$M_{\overline T}=\sup_{\bx\in\B}|\overline{\Tstat}(\bx)|$. The second term is
$o(1)$. The anti-concentration term for $M_{\overline T}$ is bounded through
Lemma~\ref{sa-lem: infeasible gaussian to linear statistic}: for
$M_Z=\sup_{\bx\in\B}|Z(\bx)|$,
\begin{align*}
    \P(|M_{\overline T}-u|\le a_n)
    &\le
    \P(|M_Z-u|\le a_n)
    +2\sup_{v\in\reals}|\P(M_{\overline T}\le v)-\P(M_Z\le v)|\\
    &\le
    4a_n\{\E[M_Z]+1\}
    +2\sup_{v\in\reals}|\P(M_{\overline T}\le v)-\P(M_Z\le v)|.
\end{align*}
The first term is $o(1)$ because $a_n\sqrt{\log n}=o(1)$, and the second is $o(1)$ by
Lemma~\ref{sa-lem: infeasible gaussian to linear statistic}. This proves the result.
\qed

\begin{lem}[KS Distance Between $Z$ and $\widehat Z$]\label{sa-lem: feasible gaussian to infeasible gaussian}
Suppose the conditions of Theorem~\ref{sa-thm: Confidence Bands} hold. Then
\begin{align*}
    \sup_{u\in\reals}
    \left|
    \P\left(\sup_{\bx\in\B}|Z(\bx)|\le u\right)
    -
    \P\left(\sup_{\bx\in\B}|\widehat Z(\bx)|\le u\mid\bV\right)
    \right|
    \lessp
    \log(n)\mathfrak{R}_{\mathtt{cc}}^{1/2}
    =
    o_{\P}(1).
\end{align*}
\end{lem}

\noindent\textbf{Proof of Lemma~\ref{sa-lem: feasible gaussian to infeasible gaussian}.}
Let
\[
    \Pi_{\bx,\by}=\frac{\Xi_{\bx,\by}}{\sqrt{\Xi_{\bx,\bx}\Xi_{\by,\by}}},
    \qquad
    \widehat\Pi_{\bx,\by}
    =
    \frac{\widehat\Xi_{\bx,\by}}{\sqrt{\widehat\Xi_{\bx,\bx}\widehat\Xi_{\by,\by}}}.
\]
Lemma~\ref{sa-lem: Covariance} gives
\[
    \sup_{\bx,\by\in\B}|\widehat\Pi_{\bx,\by}-\Pi_{\bx,\by}|
    \lessp
    \mathfrak{R}_{\mathtt{cc}},
\]
because $\inf_{\bx\in\B}\Xi_{\bx,\bx}\gtrsim(nh^d)^{-1}$ and
$\inf_{\bx\in\B}\widehat\Xi_{\bx,\bx}\gtrsim_{\P}(nh^d)^{-1}$.

Let $\B_{\delta_n}$ be a $\delta_n$-net of $\B$ under $\|\cdot\|_{\infty}$ with
cardinality $M\lesssim\delta_n^{-d}$, and write $\mathbf{Z}^{\delta_n}$ and
$\widehat{\mathbf{Z}}^{\delta_n}$ for the restrictions of $Z$ and
$\widehat Z$ to this net. By \citet[Theorem 2.1]{Chernozhukov-Chetverikov-Kato-Koike_2022_AOS},
\[
    \sup_{\by\in\reals^M}
    \left|
    \P(\mathbf{Z}^{\delta_n}\le\by)
    -
    \P(\widehat{\mathbf{Z}}^{\delta_n}\le\by\mid\bV)
    \right|
    \lessp
    \log(M)\mathfrak{R}_{\mathtt{cc}}^{1/2}.
\]
Hence the same bound applies to the distribution functions of
$\|\mathbf{Z}^{\delta_n}\|_{\infty}$ and $\|\widehat{\mathbf{Z}}^{\delta_n}\|_{\infty}$.

Choose $\delta_n=n^{-s}$ with $s>0$ large enough that
\[
    \sqrt{\log n}\,h^{-1}\delta_n=o(\mathfrak{R}_{\mathtt{cc}}^{1/2}).
\]
Such an $s$ exists because $h$ is bounded below by a polynomial in $n$. The same
Gaussian-metric calculation used in Lemma~\ref{sa-lem: infeasible gaussian to linear statistic}
and its feasible analogue imply
\begin{align*}
    \E\left[
    \sup_{\substack{\bx,\by\in\B\\ \|\bx-\by\|_{\infty}\le\delta_n}}
    |Z(\bx)-Z(\by)|
    \right]
    &\lesssim \sqrt{\log n}\,h^{-1}\delta_n,\\
    \E\left[
    \left.
    \sup_{\substack{\bx,\by\in\B\\ \|\bx-\by\|_{\infty}\le\delta_n}}
    |\widehat Z(\bx)-\widehat Z(\by)|
    \right|\bV
    \right]
    &\lessp \sqrt{\log n}\,h^{-1}\delta_n.
\end{align*}
Markov's inequality makes both modulus errors $o_{\P}(\mathfrak{R}_{\mathtt{cc}}^{1/2})$.
Combining the finite-net comparison, the two modulus bounds, and Gaussian
anti-concentration gives the displayed Kolmogorov bound. The side conditions imply
$\log(n)\mathfrak{R}_{\mathtt{cc}}^{1/2}=o_{\P}(1)$.
\qed

The proof of Theorem~\ref{sa-thm: Confidence Bands} now follows from the triangle
inequality applied to the three comparison lemmas above. Furthermore, by definition of
$\widehat{\CI}_{\alpha}(\bx)$,
\begin{align*}
    \P\Big[\vartheta(\bx)\in\widehat{\CI}_{\alpha}(\bx),\text{ for all }\bx\in\B\Big]
    &=
    \P\left(\sup_{\bx\in\B}|\widehat{\Tstat}(\bx)|\le\q_{\alpha}\right)\\
    &=
    \E\left[
    \P\left(\sup_{\bx\in\B}|\widehat Z(\bx)|\le\q_{\alpha}\mid\bV\right)
    \right]+o(1)\\
    &=1-\alpha+o(1).
\end{align*}
This completes the proof.
\qed

\subsection{Proof of Theorem~\ref{sa-thm: Gaussian Approximation Suprema: Tstat}}

We employ the following result established by
\citet[Corollary 2.2]{Chernozhukov-Chetverikov-Kato_2014b_AoS}.

\begin{lem}[Gaussian Approximation of Suprema]\label{sa-lem: suprema coupling lemma}
Let $X_1,\ldots,X_n$ be i.i.d.\ random variables in a measurable space
$(S,\mathcal{S})$ with law $P$, where $n\ge3$. Let $\F$ be a class of
$P$-centered measurable functions $f:S\to\reals$, and define
$\mathbb{G}_n f=n^{-1/2}\sum_{i=1}^n f(X_i)$. Suppose that:
\begin{enumerate}[label=\normalfont(\roman*),noitemsep,leftmargin=*]
    \item $\F$ is pointwise measurable.
    \item There exists a measurable envelope $F$ such that $|f|\le F$ for all
    $f\in\F$, and constants $A\ge e$ and $\upsilon\ge1$ such that
    \begin{align*}
        \sup_Q N(\F,\|\cdot\|_{Q,2},\epsilon\|F\|_{Q,2})
        \leq (A/\epsilon)^\upsilon,\qquad 0<\epsilon\le1,
    \end{align*}
    where the supremum is over all finite discrete probability measures on
    $(S,\mathcal{S})$.
    \item For some $b\ge\sigma>0$ and $q\in[4,\infty]$,
    \begin{align*}
        \sup_{f\in\F}P|f|^k\le \sigma^2 b^{k-2},\qquad k=2,3,4,
        \qquad\text{and}\qquad
        \|F\|_{P,q}\le b.
    \end{align*}
\end{enumerate}
Let $Z=\sup_{f\in\F}\mathbb{G}_n f$. Then, for every $\gamma\in(0,1)$, there exists,
on a possibly enlarged probability space, a tight mean-zero Gaussian process
$(\mathbb{G}_P f:f\in\F)$ with covariance function
$\E[\mathbb{G}_P(f)\mathbb{G}_P(g)]=P(fg)$ and almost surely uniformly
$\|\cdot\|_{P,2}$-continuous trajectories such that, for
$\widetilde Z=\sup_{f\in\F}\mathbb{G}_P f$,
\begin{align*}
    \P\left\{|Z-\widetilde Z|>
    \frac{bK_n}{\gamma^{1/q}n^{1/2-1/q}}
    +\frac{(b\sigma)^{1/2}K_n^{3/4}}{\gamma^{1/2}n^{1/4}}
    +\frac{(b\sigma^2K_n^2)^{1/3}}{\gamma^{1/3}n^{1/6}}
    \right\}
    \leq C\left(\gamma+\frac{\log n}{n}\right),
\end{align*}
where $K_n=\upsilon(\log n\vee\log(Ab/\sigma))$, and $C>0$ depends only on $q$.
\end{lem}

Let $\bz_i=(\bX_i,Y_i)$, $1\leq i\leq n$. For each $\bx\in\B$, define
\begin{align*}
    \tilde f_{\bx}(\bu,u)
    = g_{\bx}(\bu)u+m_{\bx}(\bu),
    \qquad
    f_{\bx}(\bu,u)=\tilde f_{\bx}(\bu,u)-\E[\tilde f_{\bx}(\bX_i,Y_i)].
\end{align*}
Let $\F_{\pm}=\{f_{\bx}:\bx\in\B\}\cup\{-f_{\bx}:\bx\in\B\}$. By the representation preceding Theorem~\ref{sa-thm: Gaussian Strong Approximation: Tstat},
\begin{align*}
    \sup_{\bx\in\B}|\overline{\Tstat}(\bx)|
    =
    \sup_{f\in\F_{\pm}}\frac{1}{\sqrt n}\sum_{i=1}^n f(\bz_i).
\end{align*}
We apply Lemma~\ref{sa-lem: suprema coupling lemma} to $\F_{\pm}$.

First, $\F_{\pm}$ is pointwise measurable by the compactness and separability of $\B$ and the measurability of the signed-distance kernels. The entropy calculation used in the proof of Theorem~\ref{sa-thm: Gaussian Strong Approximation: Tstat}, but without the total-variation step, gives a VC-type bound for $\G$ and $\M$. The fixed side indicators $\Indicator(\bu\in\A_t)$ only contract $L_2(Q)$ distances. Thus, with envelope
\begin{align*}
    F(\bu,u)=C h^{-d/2}(|u|+1),
\end{align*}
for a sufficiently large constant $C$, there are constants $A_h\ge e$ and
$\upsilon\ge1$, with $A_h$ at most polynomial in $1/h$, such that
\begin{align*}
    \sup_Q N(\F_{\pm},\|\cdot\|_{Q,2},\epsilon\|F\|_{Q,2})
    \leq (A_h/\epsilon)^{\upsilon},\qquad 0<\epsilon\leq1.
\end{align*}
Centering changes the covering constants only by a universal factor because
$|P(\tilde f_{\bx}-\tilde f_{\by})|\leq\|\tilde f_{\bx}-\tilde f_{\by}\|_{P,2}$.

For the moment condition, take $q=2+v\geq4$. Assumption~\ref{sa-assump: DGP}(v) and the boundedness of $\theta_{t,\bx}^{\ast}(D_{\bu}(\bx))$ on the kernel support give
\begin{align*}
    \|F\|_{P,2+v}\lesssim h^{-d/2}.
\end{align*}
The same local-polynomial moment calculation used in the proofs of Lemmas~\ref{sa-lem: Stochastic Linear Approximation} and \ref{sa-lem: Covariance} gives
\begin{align*}
    \sup_{f\in\F_{\pm}}P|f|^k\lesssim h^{-d(k-2)/2},
    \qquad k=2,3,4.
\end{align*}
Hence Lemma~\ref{sa-lem: suprema coupling lemma} applies with $\sigma=1$ and
$b\asymp h^{-d/2}$. Since $\liminf_{n\to\infty}\log h/\log n>-\infty$ and
$A_h$ is at most polynomial in $1/h$, $K_n\lesssim\log n$. Taking
$\gamma=(\log n)^{-1}$ yields
\begin{align*}
    \P\Bigg[
        \bigg|\sup_{\bx\in\B}|\overline{\Tstat}(\bx)|
        -\sup_{\bx\in\B}|Z(\bx)|\bigg|
        >
        \frac{\log n}{(n h^d)^{1/6}}
        +\frac{(\log n)^{5/4}}{(n h^d)^{1/4}}
        +\frac{(\log n)^{3/2}}{(n^{\frac{v}{2+v}}h^d)^{1/2}}
    \Bigg]
    \lesssim \frac{1}{\log n}+\frac{\log n}{n}.
\end{align*}
The covariance of the Gaussian process delivered by Lemma~\ref{sa-lem: suprema coupling lemma} is $P(f_{\bx_1}f_{\bx_2})$, which equals $\Cov[\overline{\Tstat}(\bx_1),\overline{\Tstat}(\bx_2)]$. This proves the theorem.
\qed

\subsection{Proof of Theorem~\ref{sa-thm: Confidence Bands: Suprema}}

Define
\begin{align*}
    \mathfrak{R}_{\mathtt{sup}}
    =
    \frac{\log n}{(n h^d)^{1/6}}
    +\frac{(\log n)^{5/4}}{(n h^d)^{1/4}}
    +\frac{(\log n)^{3/2}}{(n^{\frac{v}{2+v}}h^d)^{1/2}}.
\end{align*}
The entropy calculation in the proof of Theorem~\ref{sa-thm: Gaussian Approximation Suprema: Tstat} implies
\begin{align*}
    \E\left[\sup_{\bx\in\B}|Z(\bx)|\right]\lesssim\sqrt{\log n}.
\end{align*}
The condition $n^{\frac{v}{2+v}}h^d/(\log n)^9\to\infty$ implies
$\mathfrak{R}_{\mathtt{sup}}\sqrt{\log n}=o(1)$. Let
\begin{align*}
    M_{\overline T}=\sup_{\bx\in\B}|\overline{\Tstat}(\bx)|,
    \qquad
    M_Z=\sup_{\bx\in\B}|Z(\bx)|.
\end{align*}
Theorem~\ref{sa-thm: Gaussian Approximation Suprema: Tstat} and Gaussian anti-concentration give
\begin{align}\label{sa-eq: t-to-z distance suprema}
    \sup_{u\in\reals}
    \left|
    \P(M_{\overline T}\leq u)-\P(M_Z\leq u)
    \right|
    =o(1).
\end{align}

Next, Theorem~\ref{sa-thm: Stochastic Linearization} gives
\begin{align*}
    \sup_{\bx\in\B}|\widehat{\Tstat}(\bx)-\overline{\Tstat}(\bx)|
    =O_{\P}(\mathfrak{R}_{\mathtt{lin}}),
\end{align*}
where
\begin{align*}
    \mathfrak{R}_{\mathtt{lin}}
    =
    \sqrt{\log(1/h)}
    \left(\sqrt{\frac{\log(1/h)}{n h^d}}
    +\frac{\log(1/h)}{n^{\frac{v}{2+v}}h^d}\right)
    +\sqrt{n h^d}\sup_{\bx\in\B}|\bias(\bx)|.
\end{align*}
The polynomial lower bound on $h$, the condition $n^{\frac{v}{2+v}}h^d/(\log n)^9\to\infty$, and the stated small-bias condition imply $\mathfrak{R}_{\mathtt{lin}}\sqrt{\log n}=o(1)$. Combining this with Equation~\eqref{sa-eq: t-to-z distance suprema} and another Gaussian anti-concentration step yields
\begin{align}\label{sa-eq: t-to-t distance suprema}
    \sup_{u\in\reals}
    \left|
    \P\left(\sup_{\bx\in\B}|\widehat{\Tstat}(\bx)|\leq u\right)
    -
    \P\left(\sup_{\bx\in\B}|\overline{\Tstat}(\bx)|\leq u\right)
    \right|
    =o(1).
\end{align}

Finally, Lemma~\ref{sa-lem: Covariance} gives the feasible covariance-comparison rate
\begin{align*}
    \mathfrak{R}_{\mathtt{cc}}
    =
    \sqrt{\frac{\log(1/h)}{n h^d}}
    +\frac{\log(1/h)}{n^{\frac{v}{2+v}}h^d}.
\end{align*}
The same finite-net Gaussian comparison and anti-concentration argument used in the proof of Theorem~\ref{sa-thm: Confidence Bands} gives
\begin{align}\label{sa-eq: z-to-zhat distance suprema}
    \sup_{u\in\reals}
    \left|
    \P\left(\sup_{\bx\in\B}|Z(\bx)|\leq u\right)
    -
    \P\left(\sup_{\bx\in\B}|\widehat Z(\bx)|\leq u\mid\bV\right)
    \right|
    \lessp \log(n)\mathfrak{R}_{\mathtt{cc}}^{1/2}=o_{\P}(1),
\end{align}
where the last equality follows from $n^{\frac{v}{2+v}}h^d/(\log n)^9\to\infty$.
The displayed Kolmogorov approximation follows from the triangle inequality applied to Equations~\eqref{sa-eq: t-to-z distance suprema}, \eqref{sa-eq: t-to-t distance suprema}, and \eqref{sa-eq: z-to-zhat distance suprema}.

The simultaneous coverage statement follows from the definition of $\widehat{\CI}_{\alpha}(\bx)$ and the conditional critical value $\q_{\alpha}$:
\begin{align*}
    \P\Big[\vartheta(\bx)\in\widehat{\CI}_{\alpha}(\bx),\text{ for all }\bx\in\B\Big]
    &=
    \P\left(\sup_{\bx\in\B}|\widehat{\Tstat}(\bx)|\leq\q_{\alpha}\right)\\
    &=
    \E\left[
    \P\left(\sup_{\bx\in\B}|\widehat Z(\bx)|\leq\q_{\alpha}\mid\bV\right)
    \right]+o(1)\\
    &=1-\alpha+o(1).
\end{align*}
\qed

\subsection{Proof of Lemma~\ref{sa-lem: Variance: WBATE}}

Let $\mathfrak{L}_{t,\mathtt{WBATE}}$ and $\Xi_{t,\mathtt{WBATE}}$ denote the
side-specific analogues of $\mathfrak{L}_{\mathtt{WBATE}}$ and
$\Xi_{\mathtt{WBATE}}$. Fubini's theorem and the definition of
$\Xi_{t,\bx_1,\bx_2}$ give
\begin{align*}
    \Var\left(\mathfrak{L}_{t,\mathtt{WBATE}}\right)
    =
    \Xi_{t,\mathtt{WBATE}}.
\end{align*}
The covariance between the two side-specific linear terms is zero because the treatment
and control assignment indicators are disjoint observation by observation. Hence
$\Var(\mathfrak{L}_{\mathtt{WBATE}})=\Xi_{\mathtt{WBATE}}$.

Let $R_K<\infty$ satisfy $\supp(K)\subseteq[-R_K,R_K]$. By equivalence of $\d$ and the
Euclidean distance, if $\Xi_{\bx_1,\bx_2}\neq0$ then
$\|\bx_1-\bx_2\|\lesssim h$. Therefore
\begin{align*}
    \mathcal{E}(h)=\{(\bx_1,\bx_2)\in\B^2:\|\bx_1-\bx_2\|\leq C h\}
\end{align*}
has product boundary measure $\nu_{\B,2}(\mathcal{E}(h))\lesssim h^{d-1}$ by
\eqref{sa-eq: local boundary measure regularity}. Uniform boundedness of
$\bPsi_{t,\bx}^{-1}$ and $\bUpsilon_{t,\bx_1,\bx_2}$ gives
$|\Xi_{\bx_1,\bx_2}|\lesssim (n h^d)^{-1}$, and hence
\begin{align*}
    \Xi_{\mathtt{WBATE}}
    \lesssim
    \frac{h^{d-1}}{n h^d}
    =
    \frac{1}{n h}.
\end{align*}
Under \eqref{sa-eq: WBATE aggregate nondegeneracy}, this upper bound is complemented by
the matching lower bound.

The same support restriction applies to $\widehat{\Xi}_{\bx_1,\bx_2}$. Hence, using
Lemma~\ref{sa-lem: Covariance},
\begin{align*}
    \left|
    \widehat{\Xi}_{\mathtt{WBATE}}
    -
    \Xi_{\mathtt{WBATE}}
    \right|
    &\leq
    \int_{\mathcal{E}(h)}
    \left|
    \widehat{\Xi}_{\bx_1,\bx_2}
    -
    \Xi_{\bx_1,\bx_2}
    \right|
    |w(\bx_1)w(\bx_2)|\dif\nu_{\B,2}(\bx_1,\bx_2)\\
    &\lessp
    h^{d-1}
    \sup_{\bx_1,\bx_2\in\B}
    \left|
    \widehat{\Xi}_{\bx_1,\bx_2}
    -
    \Xi_{\bx_1,\bx_2}
    \right|\\
    &=
    o_{\P}\Big(\frac{1}{n h}\Big),
\end{align*}
which proves the plug-in variance claim. The ratio claim follows by dividing this
additive bound by the lower bound in \eqref{sa-eq: WBATE aggregate nondegeneracy}.
\qed

\subsection{Proof of Lemma~\ref{sa-lem: Higher-Order Remainder: WBATE}}

It is enough to consider one side $t\in\{0,1\}$. Define the side-specific nonlinear
WBATE remainder
\begin{align*}
    \mathfrak{Q}_{t,\mathtt{WBATE}}
    =
    \int_{\B}
    \be_0^{\top}(\widehat{\bPsi}_{t,\bx}^{-1}-\bPsi_{t,\bx}^{-1})
    \mathbf{O}_{t,\bx}
    w(\bx)\dif\Haus^{d-1}(\bx),
    \qquad
    \mathfrak{Q}_{\mathtt{WBATE}}
    =
    \mathfrak{Q}_{1,\mathtt{WBATE}}
    -
    \mathfrak{Q}_{0,\mathtt{WBATE}}.
\end{align*}
Let
\begin{align*}
    \mathbf{A}_{t,\bx}
    =
    \widehat{\bPsi}_{t,\bx}^{-1}-\bPsi_{t,\bx}^{-1}
    =
    -\widehat{\bPsi}_{t,\bx}^{-1}
    (\widehat{\bPsi}_{t,\bx}-\bPsi_{t,\bx})
    \bPsi_{t,\bx}^{-1}.
\end{align*}
On the event where $\sup_{\bx\in\B}\|\widehat{\bPsi}_{t,\bx}^{-1}\|\lesssim1$, which has
probability approaching one by Lemma~\ref{sa-lem: Gram},
\begin{align*}
    |\mathfrak{Q}_{t,\mathtt{WBATE}}|
    \lesssim
    \left|
    \int_{\B}\mathbf{A}_{t,\bx}\mathbf{O}_{t,\bx}
    w(\bx)\dif\Haus^{d-1}(\bx)
    \right|.
\end{align*}
We suppress fixed finite-dimensional vectors and matrices in the bounds below.

Decompose $\mathbf{O}_{t,\bx}$ into a conditional-mean-zero part and a conditional-mean
part:
\begin{align*}
    \mathbf{O}_{t,\bx}
    =
    \mathbf{Q}_{t,\bx}
    +
    \mathbf{M}_{t,\bx},
\end{align*}
where
\begin{align*}
    \mathbf{Q}_{t,\bx}
    &=
    \En\left[
    \rp\Big(\frac{D_i(\bx)}{h}\Big)K_h(D_i(\bx))
    \varepsilon_{t,i}\Indicator(\bX_i\in\A_t)
    \right],\\
    \mathbf{M}_{t,\bx}
    &=
    \En\left[
    \rp\Big(\frac{D_i(\bx)}{h}\Big)K_h(D_i(\bx))
    \{\mu_t(\bX_i)-\theta_{t,\bx}^{\ast}(D_i(\bx))\}
    \Indicator(\bX_i\in\A_t)
    \right],
\end{align*}
with $\varepsilon_{t,i}=Y_i(t)-\mu_t(\bX_i)$. The first component is mean zero
conditional on $\bX_1,\ldots,\bX_n$. The second component is not conditionally mean zero;
this is the additional distance-based term absent from the location-based analysis.

For the first component, conditional on $\bX_1,\ldots,\bX_n$, write
$\psi_{t,\bx}(\bz)=\rp(D_{\bz}(\bx)/h)K_h(D_{\bz}(\bx))\Indicator(\bz\in\A_t)$.
The summand inside the boundary integral is nonzero only when $\bz$ is within an
$O(h)$ neighborhood of $\bx$. Therefore, after applying Fubini to the conditional
second moment, the only pairs $(\bx_1,\bx_2)$ that contribute are those satisfying
$\|\bx_1-\bx_2\|\lesssim h$. The local boundary-measure condition
\eqref{sa-eq: local boundary measure regularity} gives product Hausdorff measure
$O(h^{d-1})$ for this effective pair set, while each localized kernel product has
conditional expectation of order $h^{-d}$. Hence
\begin{align*}
    \E\left[
    \left.
    \left|
    \int_{\B}\mathbf{A}_{t,\bx}\mathbf{Q}_{t,\bx}
    w(\bx)\dif\Haus^{d-1}(\bx)
    \right|^2
    \right|\bX_1,\ldots,\bX_n
    \right]
    \lesssim
    \frac{\sup_{\bx\in\B}\|\mathbf{A}_{t,\bx}\|^2}{n h}.
\end{align*}
Lemma~\ref{sa-lem: Gram} therefore implies
\begin{align*}
    \int_{\B}\mathbf{A}_{t,\bx}\mathbf{Q}_{t,\bx}
    w(\bx)\dif\Haus^{d-1}(\bx)
    =
    O_{\P}\left(
    \frac{\sqrt{\log(1/h)}}{n h^{(d+1)/2}}
    \right)
    =
    o_{\P}\Big(\frac{1}{n h^d}\Big),
\end{align*}
because $d\ge2$, $h=o(1)$, and $h^{d-1}\log(1/h)\to0$.

It remains to bound the conditional-mean component. Write each coordinate of
$\int_{\B}\mathbf{A}_{t,\bx}\mathbf{M}_{t,\bx}w(\bx)\dif\Haus^{d-1}(\bx)$ as
\begin{align*}
    \frac{1}{n^2}\sum_{i=1}^n\sum_{j=1}^n
    H_{ij},
    \qquad
    H_{ij}
    =
    \int_{\B}a_{\bx}(\bX_i)m_{\bx}(\bX_j)
    w(\bx)\dif\Haus^{d-1}(\bx),
\end{align*}
where $a_{\bx}$ is a centered coordinate of $\widehat{\bPsi}_{t,\bx}-\bPsi_{t,\bx}$ and
$m_{\bx}$ is the corresponding conditional-mean coordinate of $\mathbf{M}_{t,\bx}$.
The localized part of $a_{\bx}$ and the whole of $m_{\bx}$ are supported on
$O(h)$-neighborhoods of $\bx$, with envelopes of order $h^{-d}$. The centering constant
in $a_{\bx}$ is only $O(1)$ and therefore gives smaller contributions than the
localized part. Thus the dominant part of $H_{ij}$ is nonzero only when the relevant
observations are within distance $O(h)$ of the same boundary point.

For the diagonal term, the boundary integral has size at most
$h^{-2d}\Haus^{d-1}(\B\cap B(\bX_i,Ch))\lesssim h^{-d-1}$ on the event that
$\bX_i$ lies in an $O(h)$ tube around $\B$, and is zero otherwise up to the smaller
centering-constant contribution. Since this tube has probability $O(h)$,
\begin{align*}
    \E|H_{ii}|\lesssim h^{-d}.
\end{align*}
For the off-diagonal term, independence and Fubini give
\begin{align*}
    \E[H_{ij}^2]
    &=
    \int_{\B}\int_{\B}
    \E[a_{\bx_1}(\bX_i)a_{\bx_2}(\bX_i)]
    \E[m_{\bx_1}(\bX_j)m_{\bx_2}(\bX_j)]
    w(\bx_1)w(\bx_2)
    \dif\Haus^{d-1}(\bx_1)\dif\Haus^{d-1}(\bx_2).
\end{align*}
The leading localized contribution to the two expectations in the integrand is zero
unless $\|\bx_1-\bx_2\|\lesssim h$, and the terms containing the centering constants are
of the same order or smaller. On this effective set each expectation is bounded by
$O(h^{-d})$, and the effective set has product Hausdorff measure $O(h^{d-1})$. Hence
\begin{align*}
    \E[H_{ij}^2]\lesssim h^{-d-1},
    \qquad i\neq j.
\end{align*}
The off-diagonal kernel is degenerate because $\E[a_{\bx}(\bX_i)]=0$ and
$\E[m_{\bx}(\bX_j)]=0$ for every $\bx$. Thus,
\begin{align*}
    \frac{1}{n^2}\sum_{i=1}^n H_{ii}
    =
    O_{\P}\Big(\frac{1}{n h^d}\Big),
    \qquad
    \frac{1}{n^2}\sum_{i\neq j}H_{ij}
    =
    O_{\P}\Big(\frac{1}{n h^{(d+1)/2}}\Big)
    =
    o_{\P}\Big(\frac{1}{n h^d}\Big).
\end{align*}
Combining the two components and summing over $t\in\{0,1\}$ gives
$\mathfrak{Q}_{\mathtt{WBATE}}=O_{\P}(1/(n h^d))$. Finally,
\begin{align*}
    \frac{1/(n h^d)}{1/\sqrt{n h}}
    =
    \frac{1}{\sqrt{n h^{2d-1}}},
\end{align*}
which is $o(1)$ under $n h^{2d-1}\to\infty$.
\qed

\subsection{Proof of Theorem~\ref{sa-thm: Convergence Rate: WBATE}}

The decomposition \eqref{sa-eq: WBATE decomposition}, Lemma~\ref{sa-lem: Variance: WBATE},
and Lemma~\ref{sa-lem: Higher-Order Remainder: WBATE} give
\begin{align*}
    \left|
    \widehat{\vartheta}_{\mathtt{WBATE}}
    -
    \vartheta_{\mathtt{WBATE}}
    \right|
    \leq
    |\bias_{\mathtt{WBATE}}|
    +O_{\P}\Big(\frac{1}{\sqrt{n h}}\Big)
    +O_{\P}\Big(\frac{1}{n h^d}\Big),
\end{align*}
which is the stated rate.
\qed

\subsection{Proof of Theorem~\ref{sa-thm: Asymptotic Normality: WBATE}}

By \eqref{sa-eq: WBATE decomposition}, Lemma~\ref{sa-lem: Higher-Order Remainder: WBATE},
the small-bias condition, and the aggregate nondegeneracy condition,
\begin{align*}
    \frac{
    \widehat{\vartheta}_{\mathtt{WBATE}}
    -\vartheta_{\mathtt{WBATE}}
    -\mathfrak{L}_{\mathtt{WBATE}}
    }
    {\sqrt{\Xi_{\mathtt{WBATE}}}}
    =
    o_{\P}(1).
\end{align*}
Lemma~\ref{sa-lem: Variance: WBATE} gives
$\widehat{\Xi}_{\mathtt{WBATE}}/\Xi_{\mathtt{WBATE}}\to_{\P}1$.

Write $\mathfrak{L}_{\mathtt{WBATE}}=\sum_{i=1}^n\ell_i$, where the $\ell_i$ are
independent, mean-zero summands obtained by integrating the side-specific linear
influence functions over $\B$. By construction,
$\sum_{i=1}^n\E[\ell_i^2]=\Xi_{\mathtt{WBATE}}$. To verify Lyapunov's condition,
let
\begin{align*}
    \mathcal{E}_3(h)
    =
    \{(\bx_1,\bx_2,\bx_3)\in\B^3:
    \|\bx_j-\bx_k\|\leq C h,\; 1\leq j,k\leq3\}.
\end{align*}
Compact support of $K$, metric equivalence, boundedness of $w$, and
\eqref{sa-eq: local boundary measure regularity} imply
$\nu_{\B,3}(\mathcal{E}_3(h))\lesssim h^{2d-2}$. The same local-moment calculation used
in Lemma~\ref{sa-lem: Stochastic Linear Approximation} gives
\begin{align*}
    \sum_{i=1}^n \E\left[
    \left|
    \frac{\ell_i}{\sqrt{\Xi_{\mathtt{WBATE}}}}
    \right|^3
    \right]
    \lesssim
    \frac{1}{\sqrt{n h}}
    \to0,
\end{align*}
where the last step follows from $n h^{2d-1}\to\infty$ and $d\ge2$. Lyapunov's central
limit theorem gives
\begin{align*}
    \frac{\mathfrak{L}_{\mathtt{WBATE}}}{\sqrt{\Xi_{\mathtt{WBATE}}}}
    \rightsquigarrow
    \mathsf{Normal}(0,1).
\end{align*}
Slutsky's theorem gives the feasible statistic result, and the confidence interval
statement follows from the standard normal critical value.
\qed

\subsection{Proof of Theorem~\ref{sa-thm: Convergence Rate: LBATE}}

The supremum functional is one-Lipschitz under the sup-norm:
\begin{align*}
    \left|
    \sup_{\bx\in\B}\widehat{\vartheta}(\bx)
    -
    \sup_{\bx\in\B}\vartheta(\bx)
    \right|
    \leq
    \sup_{\bx\in\B}
    \left|
    \widehat{\vartheta}(\bx)
    -
    \vartheta(\bx)
    \right|.
\end{align*}
The theorem follows immediately from Theorem~\ref{sa-thm: Convergence Rates}.
\qed

\subsection{Proof of Theorem~\ref{sa-thm: Confidence Intervals: LBATE}}

On the event
\begin{align*}
    \vartheta(\bx)
    \in
    \left[
    \widehat{\vartheta}(\bx)
    -\q_{\alpha}\sqrt{\widehat{\Xi}_{\bx,\bx}},
    \widehat{\vartheta}(\bx)
    +\q_{\alpha}\sqrt{\widehat{\Xi}_{\bx,\bx}}
    \right]
    \quad\text{for all }\bx\in\B,
\end{align*}
taking suprema over $\bx\in\B$ implies
\begin{align*}
    \vartheta_{\mathtt{LBATE}}
    \in
    \widehat{\CI}_{\alpha,\mathtt{LBATE}}.
\end{align*}
Theorem~\ref{sa-thm: Confidence Bands: Suprema} gives the first event probability
$1-\alpha+o(1)$, which proves the result.
\qed

\subsection{Proof of Theorem~\ref{sa-thm: Fuzzy Identification}}

For $A=Y$, Theorem~\ref{sa-thm: Identification} applies directly after the reduced-form
redefinition $Y_i(t)=Y_i(t,W_i(t))$. For $A=W$, the same proof applies with $W_i(t)$ in
place of $Y_i(t)$. Thus the sharp-design identification argument is applied separately
to the reduced-form outcome component and the first-stage treatment-receipt component:
Assumption~\ref{sa-assump: Fuzzy DGP}(i)--(ii) gives the reduced-form smoothness and
sampling conditions, while Assumption~\ref{sa-assump: Kernel, Distance and Boundary}(i),
(ii), and (v) gives the same distance-based identification conditions.
Therefore $\vartheta_A(\bx)=\tau_A(\bx)$ for $A\in\{Y,W\}$. If $\tau_W(\bx)\neq0$, then
the ratio identity $\xi(\bx)=\zeta(\bx)$ follows immediately.
\qed

\subsection{Proof of Theorem~\ref{sa-thm: Convergence Rates: Fuzzy BATEC}}

Theorem~\ref{sa-thm: Convergence Rates} applies to the reduced-form outcome estimator
$\widehat{\vartheta}_Y(\bx)$. The same proof applies to
$\widehat{\vartheta}_W(\bx)$ because Assumption~\ref{sa-assump: Fuzzy DGP}(i)--(ii)
gives the corresponding smoothness and moment conditions; boundedness of $W_i(t)$ gives
the required moment bounds automatically. Thus, pointwise and uniformly,
\[
    \widehat{\vartheta}_A(\bx)-\vartheta_A(\bx)
    =
    O_{\P}\left(
    \frac{1}{\sqrt{nh^d}}
    +\frac{1}{n^{\frac{1+v}{2+v}}h^d}
    +|\bias_A(\bx)|
    \right)
\]
with the uniform version replacing the first two terms by their logarithmic analogues.
Because $v\ge2$, the pointwise condition
$(n^{(1+v)/(2+v)}h^d)^{-1}=o(1)$ implies $(nh^d)^{-1/2}=o(1)$. Uniformly,
$\log(1/h)/(n^{(1+v)/(2+v)}h^d)=o(1)$ implies the corresponding logarithmic stochastic
terms are $o(1)$. The condition $h=o(1)$ supplies the shrinking-neighborhood requirement
in the reduced-form distance-based rates, and Theorem~\ref{sa-thm: Approximation Bias: Minimal Guarantee}
implies the reduced-form approximation biases are $o(1)$.
Together, these restrictions imply that
$\widehat{\vartheta}_W(\bx)$ is bounded away from zero with probability approaching
one, pointwise or uniformly as appropriate. The exact Wald expansion
\eqref{sa-eq: fuzzy second order} and the bound \eqref{sa-eq: fuzzy second order bound}
then give the stated rates.
\qed

\subsection{Proof of Lemma~\ref{sa-lem: Covariance: Fuzzy BATEC}}

Fix $A,B\in\{Y,W\}$. The proof of Lemma~\ref{sa-lem: Covariance} applies to the
cross-covariance kernel $\Xi_{A,B,\bx_1,\bx_2}$ with
\[
    (A_i-\theta_{A,t,\bx_1}^{\ast}(D_i(\bx_1)))
    (B_i-\theta_{B,t,\bx_2}^{\ast}(D_i(\bx_2)))
\]
in place of the squared residual product. The moment bounds needed for $A=B=Y$ follow
from Assumption~\ref{sa-assump: DGP}(v); those involving $W$ follow from
$W_i(t)\in\{0,1\}$ and Cauchy--Schwarz. Assumption~\ref{sa-assump: Fuzzy DGP}(iii)
provides continuity of the reduced-form conditional covariance entries. Hence, uniformly
over $\bx_1,\bx_2\in\B$,
\[
    \sup_{A,B\in\{Y,W\}}
    nh^d\big|\widehat{\Xi}_{A,B,\bx_1,\bx_2}-\Xi_{A,B,\bx_1,\bx_2}\big|
    =
    o_{\P}(1).
\]
Theorem~\ref{sa-thm: Convergence Rates: Fuzzy BATEC} gives
$\sup_{\bx\in\B}\|\widehat{\mathfrak w}(\bx)-\mathfrak w(\bx)\|=o_{\P}(1)$ under the
denominator condition and the stated bandwidth restrictions. Combining the last two displays with the
definitions of $\Xi_{\mathtt F}$ and $\widehat{\Xi}_{\mathtt F}$ gives
\[
    \sup_{\bx_1,\bx_2\in\B}
    nh^d|\widehat{\Xi}_{\mathtt F,\bx_1,\bx_2}
    -\Xi_{\mathtt F,\bx_1,\bx_2}|
    =
    o_{\P}(1).
\]
The relative consistency statement follows from the imposed lower bound
$\inf_{\bx\in\B}nh^d\Xi_{\mathtt F,\bx,\bx}\gtrsim1$.
\qed

\subsection{Proof of Theorem~\ref{sa-thm: Confidence Intervals: Fuzzy BATEC}}

Because $v\ge2$, the condition $n^{v/(2+v)}h^d\to\infty$ implies both
$nh^d\to\infty$ and $(n^{(1+v)/(2+v)}h^d)^{-1}\to0$. Therefore the reduced-form
stochastic terms for $Y$ and $W$ are consistent. The condition $h=o(1)$ gives shrinking
local neighborhoods, and Theorem~\ref{sa-thm: Approximation Bias: Minimal Guarantee}
makes the first-stage approximation bias vanish.

For the fixed $\bx$, the exact expansion \eqref{sa-eq: fuzzy second order} gives
\[
    \widehat{\xi}(\bx)-\xi(\bx)
    =
    \bias_{\mathtt F}(\bx)+\mathfrak{L}_{\mathtt F}(\bx)
    +o_{\P}\big((nh^d)^{-1/2}\big),
\]
because the reduced-form nonlinear remainders are $o_{\P}((nh^d)^{-1/2})$ by the
pointwise analogue of Lemma~\ref{sa-lem: Stochastic Linear Approximation}, and the
pointwise reduced-form rates give
$\widehat{\vartheta}_A(\bx)-\vartheta_A(\bx)=O_{\P}(\rho_{n,\bx})$ for
$A\in\{Y,W\}$. Since $|\vartheta_W(\bx)|>0$ and $\rho_{n,\bx}=o(1)$,
$\widehat{\vartheta}_W(\bx)$ is bounded away from zero with probability approaching one.
Therefore the exact Wald remainder satisfies
$\mathfrak{R}_{\mathtt F}(\bx)=O_{\P}(\rho_{n,\bx}^2)
=o_{\P}((nh^d)^{-1/2})$ by the condition
$\sqrt{nh^d}\rho_{n,\bx}^2\to0$. The separate small-bias condition makes
$\bias_{\mathtt F}(\bx)=o((nh^d)^{-1/2})$.

The leading term $\mathfrak{L}_{\mathtt F}(\bx)$ is a sum of independent mean-zero
variables with variance $\Xi_{\mathtt F,\bx,\bx}$. The moment assumptions and the bound
$nh^d\Xi_{\mathtt F,\bx,\bx}\asymp1$ verify Lyapunov's condition, giving
\[
    \frac{\mathfrak{L}_{\mathtt F}(\bx)}
         {\sqrt{\Xi_{\mathtt F,\bx,\bx}}}
    \rightsquigarrow \mathsf{Normal}(0,1).
\]
The pointwise version of Lemma~\ref{sa-lem: Covariance: Fuzzy BATEC} gives
$\widehat{\Xi}_{\mathtt F,\bx,\bx}/\Xi_{\mathtt F,\bx,\bx}\to_{\P}1$. Slutsky's theorem
therefore gives the normal approximation for $\widehat{\Tstat}_{\mathtt F}(\bx)$ and the
confidence interval coverage.
\qed

\subsection{Proof of Theorem~\ref{sa-thm: Stochastic Linearization: Fuzzy BATEC}}

Start from \eqref{sa-eq: fuzzy second order}. Lemma~\ref{sa-lem: Stochastic Linear Approximation}, applied to $A=Y$ and $A=W$, gives the uniform bound for the reduced-form
nonlinear terms. Lemma~\ref{sa-lem: Covariance: Fuzzy BATEC} gives uniform consistency of
the feasible standard error. The contribution of the reduced-form approximation biases is
\[
    \sqrt{nh^d}\sup_{\bx\in\B}|\bias_{\mathtt F}(\bx)|.
\]
Finally, \eqref{sa-eq: fuzzy second order bound} gives the Wald-ratio contribution
$\sqrt{nh^d}\rho_n^2$. Combining these bounds gives the displayed result.
\qed

\subsection{Proof of Theorem~\ref{sa-thm: Gaussian Approximation Suprema: Fuzzy BATEC}}

The proof is the fuzzy analogue of the proof of
Theorem~\ref{sa-thm: Gaussian Approximation Suprema: Tstat}. The linearized fuzzy
statistic can be written as an empirical process indexed by $\bx\in\B$:
\[
    \overline{\Tstat}_{\mathtt F}(\bx)
    =
    \frac{1}{\sqrt n}\sum_{i=1}^n f_{\mathtt F,\bx}(\bX_i,Y_i,W_i),
\]
where $f_{\mathtt F,\bx}$ is centered and equals a bounded linear combination, with
coefficients $\mathfrak w(\bx)$, of the reduced-form distance-kernel terms for $Y_i$ and
$W_i$. The denominator condition $\inf_{\bx\in\B}|\vartheta_W(\bx)|>0$, the smooth
reduced-form conditional means, and the metric equivalence in
Assumption~\ref{sa-assump: Kernel, Distance and Boundary}(ii) imply that
$\mathfrak w(\bx)$ is bounded and Lipschitz on $\B$. The entropy calculation for
Theorem~\ref{sa-thm: Gaussian Approximation Suprema: Tstat} is therefore unchanged up to
constants: adding the bounded treatment-receipt coordinate and bounded Lipschitz weights
preserves the same VC-type covering bound. The envelope can be taken as
$C h^{-d/2}(|Y_i|+1)$, because $W_i$ is binary. The moment bounds follow from
Assumption~\ref{sa-assump: DGP}(v) and boundedness of $W_i$.

Applying Lemma~\ref{sa-lem: suprema coupling lemma} to
$\{f_{\mathtt F,\bx}:\bx\in\B\}\cup\{-f_{\mathtt F,\bx}:\bx\in\B\}$ gives the displayed
Gaussian approximation rate. The covariance of the constructed Gaussian process is, by
construction, the covariance of $\overline{\Tstat}_{\mathtt F}$.
\qed

\subsection{Proof of Theorem~\ref{sa-thm: Confidence Bands: Fuzzy BATEC}}

Theorem~\ref{sa-thm: Stochastic Linearization: Fuzzy BATEC} and the stated rate
conditions imply
\[
    \sup_{\bx\in\B}
    |\widehat{\Tstat}_{\mathtt F}(\bx)-\overline{\Tstat}_{\mathtt F}(\bx)|
    =
    o_{\P}\big((\log n)^{-1/2}\big).
\]
Theorem~\ref{sa-thm: Gaussian Approximation Suprema: Fuzzy BATEC} approximates the
supremum of the infeasible linearized statistic by the supremum of a Gaussian process.
Lemma~\ref{sa-lem: Covariance: Fuzzy BATEC}, together with the same finite-net Gaussian
comparison and anti-concentration argument used in the proof of
Theorem~\ref{sa-thm: Confidence Bands: Suprema}, replaces this infeasible Gaussian
process by the feasible conditional process $\widehat Z_{\mathtt F}$. Thus,
\[
    \sup_{u\in\reals}
    \left|
    \P\left(\sup_{\bx\in\B}|\widehat{\Tstat}_{\mathtt F}(\bx)|\le u\right)
    -
    \P\left(\sup_{\bx\in\B}|\widehat Z_{\mathtt F}(\bx)|\le u\mid\bV\right)
    \right|
    =
    o_{\P}(1).
\]
The coverage statement follows from the definition of $\widehat{\CI}_{\alpha,\mathtt F}$
and the conditional quantile $\q_\alpha$.
\qed

\subsection{Proof of Lemma~\ref{sa-lem: Variance: Fuzzy WBATE}}

By Fubini's theorem and the definition of $\Xi_{\mathtt F,\bx_1,\bx_2}$,
\begin{align*}
    \Var[\mathfrak{L}_{\mathtt F,\mathtt{WBATE}}]
    &=
    \int_{\B}\int_{\B}
    \Cov[\mathfrak{L}_{\mathtt F}(\bx_1),\mathfrak{L}_{\mathtt F}(\bx_2)]
    w(\bx_1)w(\bx_2)
    \dif\Haus^{d-1}(\bx_1)\dif\Haus^{d-1}(\bx_2)\\
    &=
    \Xi_{\mathtt F,\mathtt{WBATE}}.
\end{align*}
The denominator condition implies that $\mathfrak{w}(\cdot)$ is uniformly bounded on
$\B$. Compact support of $K$ and equivalence of $\d$ and the Euclidean distance imply
that $\Xi_{\mathtt F,\bx_1,\bx_2}=0$ unless $\|\bx_1-\bx_2\|\lesssim h$. Therefore, as
in the proof of Lemma~\ref{sa-lem: Variance: WBATE}, the effective integration set has
product boundary measure $O(h^{d-1})$ under
\eqref{sa-eq: local boundary measure regularity}. The reduced-form covariance kernels
are uniformly of order $(nh^d)^{-1}$, and hence
\begin{align*}
    \Xi_{\mathtt F,\mathtt{WBATE}}
    \lesssim
    \frac{h^{d-1}}{n h^d}
    =
    \frac{1}{n h}.
\end{align*}
The matching lower bound follows from
\eqref{sa-eq: fuzzy WBATE aggregate nondegeneracy}.

For the plug-in variance, the proof of Lemma~\ref{sa-lem: Covariance: Fuzzy BATEC}
gives the uniform covariance approximation
\begin{align*}
    \sup_{\bx_1,\bx_2\in\B}
    nh^d
    \left|
    \widehat{\Xi}_{\mathtt F,\bx_1,\bx_2}
    -
    \Xi_{\mathtt F,\bx_1,\bx_2}
    \right|
    =
    o_{\P}(1),
\end{align*}
using only the denominator nondegeneracy and the stated rate conditions for the
consistency of $\widehat{\mathfrak w}(\cdot)$. The same compact-support restriction
applies to the feasible covariance kernel. Integrating the uniform error over the
effective set of measure $O(h^{d-1})$ gives
\begin{align*}
    \left|
    \widehat{\Xi}_{\mathtt F,\mathtt{WBATE}}
    -
    \Xi_{\mathtt F,\mathtt{WBATE}}
    \right|
    =
    o_{\P}\Big(\frac{h^{d-1}}{n h^d}\Big)
    =
    o_{\P}\Big(\frac{1}{n h}\Big).
\end{align*}
The relative consistency statement follows after dividing by the lower bound in
\eqref{sa-eq: fuzzy WBATE aggregate nondegeneracy}.
\qed

\subsection{Proof of Lemma~\ref{sa-lem: Higher-Order Remainders: Fuzzy WBATE}}

Because $\inf_{\bx\in\B}|\vartheta_W(\bx)|>0$ and the reduced-form conditional means are
continuous, the weight vector $\mathfrak{w}(\bx)$ is uniformly bounded over $\B$. The
fuzzy nonlinear term can be written as a finite linear combination of reduced-form WBATE
nonlinear terms:
\begin{align*}
    \mathfrak{Q}_{\mathtt F,\mathtt{WBATE}}
    =
    \int_{\B}
    \mathfrak{w}(\bx)^\top
    \big(\mathfrak{Q}_Y(\bx),\mathfrak{Q}_W(\bx)\big)^\top
    w(\bx)\dif\Haus^{d-1}(\bx).
\end{align*}
The proof of Lemma~\ref{sa-lem: Higher-Order Remainder: WBATE} applies to each
reduced-form component, with the bounded deterministic weights
$w(\bx)\mathfrak{w}_Y(\bx)$ and $w(\bx)\mathfrak{w}_W(\bx)$ in place of $w(\bx)$. The
moment bounds for terms involving $Y$ follow from Assumption~\ref{sa-assump: DGP}(v),
and those involving $W$ are immediate from $W_i(t)\in\{0,1\}$. Hence
\begin{align*}
    \mathfrak{Q}_{\mathtt F,\mathtt{WBATE}}
    =
    O_{\P}\Big(\frac{1}{n h^d}\Big).
\end{align*}

For the Wald-ratio remainder, \eqref{sa-eq: fuzzy second order bound} and the
integrability of $w$ give
\begin{align*}
    |\mathfrak{R}_{\mathtt F,\mathtt{WBATE}}|
    \leq
    \int_{\B}
    |\mathfrak{R}_{\mathtt F}(\bx)|w(\bx)\dif\Haus^{d-1}(\bx)
    \lessp
    \rho_n^2.
\end{align*}
The final negligibility claim follows because
\begin{align*}
    \frac{1/(n h^d)}{1/\sqrt{n h}}
    =
    \frac{1}{\sqrt{n h^{2d-1}}},
\end{align*}
which is $o(1)$ under $n h^{2d-1}\to\infty$, while
$\sqrt{nh}\rho_n^2\to0$ by assumption.
\qed

\subsection{Proof of Theorem~\ref{sa-thm: Convergence Rate: Fuzzy WBATE}}

The decomposition \eqref{sa-eq: fuzzy WBATE decomposition},
Lemma~\ref{sa-lem: Variance: Fuzzy WBATE}, and
Lemma~\ref{sa-lem: Higher-Order Remainders: Fuzzy WBATE} give
\begin{align*}
    \left|
    \widehat{\xi}_{\mathtt{WBATE}}
    -
    \xi_{\mathtt{WBATE}}
    \right|
    \leq
    |\bias_{\mathtt F,\mathtt{WBATE}}|
    +
    O_{\P}\Big(\frac{1}{\sqrt{n h}}\Big)
    +
    O_{\P}\Big(\frac{1}{n h^d}\Big)
    +
    O_{\P}(\rho_n^2),
\end{align*}
which is the stated rate.
\qed

\subsection{Proof of Theorem~\ref{sa-thm: Asymptotic Normality: Fuzzy WBATE}}

By \eqref{sa-eq: fuzzy WBATE decomposition},
Lemma~\ref{sa-lem: Higher-Order Remainders: Fuzzy WBATE}, the small-bias condition, the
condition $\sqrt{nh}\rho_n^2\to0$, and the aggregate nondegeneracy condition,
\begin{align*}
    \frac{
    \widehat{\xi}_{\mathtt{WBATE}}
    -
    \xi_{\mathtt{WBATE}}
    -
    \mathfrak{L}_{\mathtt F,\mathtt{WBATE}}
    }
    {\sqrt{\Xi_{\mathtt F,\mathtt{WBATE}}}}
    =
    o_{\P}(1).
\end{align*}
Lemma~\ref{sa-lem: Variance: Fuzzy WBATE} gives
$\widehat{\Xi}_{\mathtt F,\mathtt{WBATE}}/
\Xi_{\mathtt F,\mathtt{WBATE}}\to_{\P}1$.

Write $\mathfrak{L}_{\mathtt F,\mathtt{WBATE}}=\sum_{i=1}^n\ell_{i,\mathtt F}$, where
the $\ell_{i,\mathtt F}$ are the independent mean-zero summands obtained by integrating
the reduced-form fuzzy linear influence functions over $\B$. By construction,
$\sum_{i=1}^n\E[\ell_{i,\mathtt F}^2]=\Xi_{\mathtt F,\mathtt{WBATE}}$. The same
localization calculation used in the proof of
Theorem~\ref{sa-thm: Asymptotic Normality: WBATE}, now with envelope proportional to
$|Y_i|+1$ because $W_i$ is binary and $\mathfrak{w}(\cdot)$ is bounded, gives
\begin{align*}
    \sum_{i=1}^n
    \E\left[
    \left|
    \frac{\ell_{i,\mathtt F}}
         {\sqrt{\Xi_{\mathtt F,\mathtt{WBATE}}}}
    \right|^3
    \right]
    \lesssim
    \frac{1}{\sqrt{n h}}
    \to0.
\end{align*}
Thus Lyapunov's central limit theorem yields
\begin{align*}
    \frac{\mathfrak{L}_{\mathtt F,\mathtt{WBATE}}}
         {\sqrt{\Xi_{\mathtt F,\mathtt{WBATE}}}}
    \rightsquigarrow
    \mathsf{Normal}(0,1).
\end{align*}
Slutsky's theorem gives the feasible statistic result, and the confidence interval
coverage follows from the standard normal critical value.
\qed

\subsection{Proof of Theorem~\ref{sa-thm: Convergence Rate: Fuzzy LBATE}}

The supremum functional is one-Lipschitz under the sup-norm:
\begin{align*}
    \left|
    \widehat{\xi}_{\mathtt{LBATE}}
    -
    \xi_{\mathtt{LBATE}}
    \right|
    \leq
    \sup_{\bx\in\B}
    |\widehat{\xi}(\bx)-\xi(\bx)|.
\end{align*}
The result follows from the uniform part of
Theorem~\ref{sa-thm: Convergence Rates: Fuzzy BATEC}.
\qed

\subsection{Proof of Theorem~\ref{sa-thm: Confidence Intervals: Fuzzy LBATE}}

On the event
\begin{align*}
    \xi(\bx)
    \in
    \left[
    \widehat{\xi}(\bx)
    -
    \q_{\alpha}\sqrt{\widehat{\Xi}_{\mathtt F,\bx,\bx}},
    \widehat{\xi}(\bx)
    +
    \q_{\alpha}\sqrt{\widehat{\Xi}_{\mathtt F,\bx,\bx}}
    \right]
    \quad\text{for all }\bx\in\B,
\end{align*}
taking suprema over $\bx\in\B$ implies
\begin{align*}
    \xi_{\mathtt{LBATE}}
    \in
    \widehat{\CI}_{\alpha,\mathtt F,\mathtt{LBATE}}.
\end{align*}
Theorem~\ref{sa-thm: Confidence Bands: Fuzzy BATEC} gives the first event probability
$1-\alpha+o(1)$, which proves the result.
\qed

\subsection{Proof of Theorem 6}\label{sa-sec: Proof of Theorem 6}

Let $0<s<1/6$ be a parameter that is chosen later. Consider a $C^1$ function $\Theta_s:[0,1]\to[\underline{\theta},\pi]$, for some fixed $\underline{\theta}>0$, satisfying
\begin{align*}
    \Theta_s(r) = \pi \quad (0\le r\le 2s),
    \qquad
    \frac{\sin(\Theta_s(r))}{\Theta_s(r)} = \frac{2}{\pi}-\frac{s}{r} \quad (3s\le r\le 1),
    \qquad
    \sup_{0<r\le1} r|\Theta_s'(r)| \le C,
\end{align*}
with a smooth interpolation on $[2s,3s]$. Such a function exists because $\theta\mapsto\sin(\theta)/\theta$ maps $(0,\pi]$ continuously and monotonically onto $[0,1)$.

\subsubsection*{Data Generating Process $\P_0$.} Let
\begin{align*}
    \X_0 = \{r(\cos \theta, \sin \theta): 0 \leq r \leq 1,\ 0 \leq \theta \leq \Theta_s(r)\}.
\end{align*}
Suppose $\bX_i$ has Lebesgue density
\begin{align*}
    f_{0,X}(r(\cos \theta,\sin \theta)) = \frac{2}{\Theta_s(r)}, \qquad
    0 \leq r \leq 1,\quad 0 \leq \theta \leq \Theta_s(r),
\end{align*}
and define
\begin{align*}
    \mu_0(x_1,x_2) = \frac{1}{2} + \frac{x_1}{100}, \qquad (x_1,x_2) \in \reals^2.
\end{align*} 
Let $Y_i = \Indicator(\varepsilon_i \leq \mu_0(\bX_i))$, where $(\varepsilon_i: i=1,\ldots,n)$ are i.i.d. $\mathsf{Uniform}([0,1])$ random variables independent of $(\bX_i: i=1,\ldots,n)$. Let $\eta_0(r) = \E_{\P_0}[Y_i|\norm{\bX_i} = r]$. The boundary $\setbd(\X_0)$ is rectifiable with uniformly bounded length.

\subsubsection*{Data Generating Process $\P_1$.} Let
\begin{align*}
    \X_1=\{r(\cos \theta, \sin \theta): 0 \leq r \leq 1,\ 0 \leq \theta \leq \pi/2\},
\end{align*}
let $\bX_i$ have density $f_{1,X}=4/\pi$ on $\X_1$, and define
\begin{align*}
    \mu_1(x_1, x_2) = \frac{1}{2} + \frac{x_1-s}{100}, \qquad (x_1, x_2) \in \reals^2.
\end{align*}
Let $Y_i = \Indicator(\varepsilon_i \leq \mu_1(\bX_i))$ with the same independent uniform construction as above. Let $\eta_1(r) = \E_{\P_1}[Y_i|\norm{\bX_i} = r]$. The boundary $\setbd(\X_1)$ is rectifiable.

Figure~\ref{sa-fig: minimax supports} displays the two support sets used in the minimax lower-bound construction.

\begin{figure}[!tbp]
    \centering
    \begin{tikzpicture}[x=4.35cm,y=4.35cm,baseline,font=\scriptsize]
        \path[use as bounding box] (-1.12,-0.50) rectangle (1.24,1.27);
        \fill[gray!15]
            (0,0) -- (0:1)
            arc[start angle=0,end angle=118,radius=1]
            -- (118:0.80)
            arc[start angle=118,end angle=130,radius=0.80]
            -- (130:0.64)
            arc[start angle=130,end angle=145,radius=0.64]
            -- (145:0.48)
            arc[start angle=145,end angle=160,radius=0.48]
            -- (160:0.32)
            arc[start angle=160,end angle=180,radius=0.32]
            -- cycle;
        \draw[gray!70,dashed]
            (0:1) arc[start angle=0,end angle=118,radius=1]
            -- (118:0.80)
            arc[start angle=118,end angle=130,radius=0.80]
            -- (130:0.64)
            arc[start angle=130,end angle=145,radius=0.64]
            -- (145:0.48)
            arc[start angle=145,end angle=160,radius=0.48]
            -- (160:0.32)
            arc[start angle=160,end angle=180,radius=0.32];
        \foreach \rad/\ang in {0.32/180,0.48/160,0.64/145,0.80/130,1/118}
            \draw[gray!70,dashed] (0:\rad) arc[start angle=0,end angle=\ang,radius=\rad];
        \draw[gray!70,->] (-1.08,0) -- (1.20,0);
        \draw[gray!70,->] (0,-0.43) -- (0,1.24);
        \fill[black] (0,0) circle[radius=1.7pt];
        \node[below right,gray!70,font=\tiny] at (0,0) {$0$};
        \node[anchor=north,gray!70,font=\tiny] at (0.32,-0.02) {$s$};
        \node[anchor=north,gray!70,font=\tiny] at (0.48,-0.17) {$s(1+s)$};
        \node[anchor=north,gray!70,font=\tiny] at (0.65,-0.02) {$s(2+s)$};
        \node[anchor=north,gray!70,font=\tiny] at (0.78,-0.17) {$\cdots$};
        \node[anchor=north,gray!70,font=\tiny] at (0.94,-0.02) {$s(K+s)$};
        \node[anchor=north,gray!70,font=\tiny] at (1.08,-0.17) {$1$};
    \end{tikzpicture}
    \par
    \smallskip
    $\mathcal{X}_0$ from DGP $\P_0$
    \par
    \bigskip
    \begin{tikzpicture}[x=4.35cm,y=4.35cm,baseline,font=\scriptsize]
        \path[use as bounding box] (-1.12,-0.50) rectangle (1.24,1.27);
        \fill[gray!15] (0,0) -- (1,0) arc[start angle=0,end angle=90,radius=1] -- cycle;
        \draw[gray!70,dashed] (1,0) arc[start angle=0,end angle=90,radius=1];
        \draw[gray!70,dashed] (0,0) -- (1,0);
        \draw[gray!70,dashed] (0,0) -- (0,1);
        \draw[gray!70,->] (-1.08,0) -- (1.20,0);
        \draw[gray!70,->] (0,-0.43) -- (0,1.24);
        \fill[black] (0,0) circle[radius=1.7pt];
        \node[below right,gray!70,font=\tiny] at (0,0) {$0$};
        \node[anchor=north,gray!70,font=\tiny] at (1,-0.02) {$1$};
    \end{tikzpicture}
    \par
    \smallskip
    $\mathcal{X}_1$ from DGP $\P_1$
    \caption{Support sets in the two-point minimax lower-bound construction.}
    \label{sa-fig: minimax supports}
\end{figure}

\subsubsection*{Minimax Lower Bound.} First, under both models, $\P_j(\norm{\bX_i} \leq r)=r^2$ for all $0\le r\le1$:
\begin{align*}
    \P_0(\norm{\bX_i} \leq r)
    = \int_0^r \int_0^{\Theta_s(\rho)} \frac{2}{\Theta_s(\rho)} \rho \dif\theta \dif\rho
    = r^2
    = \int_0^r \int_0^{\pi/2} \frac{4}{\pi} \rho \dif\theta \dif\rho
    = \P_1(\norm{\bX_i} \leq r).
\end{align*}
The two data-generating processes belong to $\mathcal{P}_{\mathtt{NP}}$ for a fixed sufficiently large $L$: the supports are compact subsets of $[-L,L]^2$ with rectifiable boundaries of uniformly bounded length, the densities are continuous and uniformly bounded above and below on their supports, and the Bernoulli conditional means and variances satisfy the required smoothness and boundedness conditions. Hence, choosing $(0,0)$ as the point of evaluation in both $\P_0$ and $\P_1$, the one-observation KL distance for the distance-based experiment based on $(\norm{\bX_i},Y_i)$ is
\begin{align*}
    d_{\operatorname{KL}}(\P_0(\norm{\bX_i},Y_i),\P_1(\norm{\bX_i},Y_i))
    = 2 \int_0^1 d_{\operatorname{KL}}(\mathsf{Bernoulli}(\eta_0(r)), \mathsf{Bernoulli}(\eta_1(r))) r \dif r.
\end{align*}
Under $\P_0$ and $\P_1$,
\begin{align*}
    \eta_0(r)
    = \frac{1}{2} + \frac{r}{100}\frac{\sin(\Theta_s(r))}{\Theta_s(r)},
    \qquad
    \eta_1(r)
    = \frac{1}{2} + \frac{1}{100}\Big(\frac{2r}{\pi}-s\Big).
\end{align*}
Thus $\eta_0(r)=\eta_1(r)$ for all $r\in[3s,1]$. For $r\in[0,3s]$, both probabilities are bounded away from zero and one, and $|\eta_0(r)-\eta_1(r)|\le Cs$. Therefore,
\begin{align*}
     \int_0^1 d_{\operatorname{KL}}(\mathsf{Bernoulli}(\eta_0(r)), \mathsf{Bernoulli}(\eta_1(r))) r \dif r
    & \leq C \int_0^{3s}(\eta_0(r)-\eta_1(r))^2 r \dif r
      \leq C s^4.
\end{align*}
Moreover, $|\mu_0(0,0) - \mu_1(0,0)| = s/100$. Hence, by \citet[Theorem 2.2 (iii)]{tsybakov2008introduction}, take $C s_{\ast}^4 = \log(2)/n$, and conclude that the pointwise risk at $\bx=0$ is bounded below by a constant multiple of $s_\ast$. Since the uniform loss dominates the pointwise loss at any boundary point,
\begin{align*}
    \inf_{T_n \in \mathcal{T}} \sup_{\P \in \mathcal{P}_{\mathtt{NP}}}
    \E_{\P}\Big[\sup_{\bx\in\B(\P)}
    |T_n(\bU_n(\bx))-\mu(\bx)|\Big]
    \geq \frac{1}{1600}s_{\ast} \gtrsim n^{-\frac{1}{4}}.
\end{align*}
This concludes the proof.
\qed

\section{Simulation Study}

The score $\bX_i=(X_{1i},X_{2i})^\top$ is generated as $(100B_{1i}-25,100B_{2i}-25)^\top$, where $B_{1i}$ and $B_{2i}$ are independent $\mathsf{Beta}(3,4)$ random variables, a choice that approximately reproduces salient features of the SPP dataset. The treatment assignment region is the same L-shaped region used in the empirical application,
\[
    \A_1=\{\bx=(x_1,x_2)^\top:x_1\geq0,\ x_2\geq0\},
    \qquad
    \A_0=\X\setminus\A_1,
\]
with boundary $\B$ given by the two coordinate-axis segments and a kink at $\bx=(0,0)^\top$. The simulation uses $21$ evaluation points $\{\mathbf{b}_1,\ldots,\mathbf{b}_{21}\}\subset\B$, with $\mathbf{b}_{11}=(0,0)^\top$ corresponding to the kink.

The simulation design sets $n=20,000$ and uses $5,000$ Monte Carlo replications. We evaluate performance in terms of point estimation and uncertainty quantification, both pointwise and uniformly over the evaluation grid. We consider two related Monte Carlo exercises. The first is a sharp-design exercise, which uses only the reduced-form or intention-to-treat (ITT) outcome equation and therefore reports sharp distance-based results for the BATEC. The second is a fuzzy-design exercise, which jointly generates the reduced-form outcome and treatment-receipt status so that the ITT, first-stage (FS), and fuzzy Wald analyses can all be performed on the same simulated samples.

Throughout the simulation study, the distance-based estimator uses a local linear polynomial, $p=1$. The DGP varies the polynomial order $s$ of the underlying bivariate conditional mean surfaces. We consider $s\in\{1,2\}$, corresponding to linear and quadratic specifications. In addition, each mean specification is paired with either homoskedastic or heteroskedastic outcome disturbances.

\subsection{Data-Generating Process}

For each Monte Carlo replication, draw $\bX_i=(X_{1i},X_{2i})^\top$ independently from the distribution described above and set $T_i=\Indicator(\bX_i\in\A_1)$. Let $\mathbf{m}_s(\cdot)$ be a monomial vector of degree $s$, that is,
\begin{align*}
    \mathbf{m}_0(\bx)=1,
    \qquad
    \mathbf{m}_1(\bx)=(1,x_1,x_2)^\top,
    \qquad\text{and}\qquad
    \mathbf{m}_2(\bx)=(1,x_1,x_2,x_1^2,x_1x_2,x_2^2)^\top.
\end{align*}
We consider DGP polynomial specifications of order $s\in\{1,2\}$, using the monomial vector evaluated at the unscaled score.

\subsubsection*{Sharp Design: ITT Equation.}

The sharp-design simulation uses an outcome-only DGP. For each $t\in\{0,1\}$ and $i=1,\ldots,n$, define
\begin{align*}
    \mu_{Y,t}(\bX_i)
    &= \mathbf{m}_s(\bX_i)^\top\boldsymbol{\beta}_{Y,t,s},
    \qquad
    \sigma_{Y,t,v}^2(\bX_i)
    = \exp\!\left(\mathbf{m}_v(\bX_i)^\top\bdelta_{Y,t,v}\right),
\end{align*}
where $v\in\{0,2\}$ indexes the homoskedastic and heteroskedastic specifications. Draw
$u_{Y,t,i}\thicksim\mathsf{Normal}(0,1)$ independently across $i$ and $t$ and independently of all score variables, and set
\begin{align*}
    \varepsilon_{Y,t,i}
    =
    \sigma_{Y,t,v}(\bX_i)u_{Y,t,i}.
\end{align*}
Then
\begin{align*}
    Y_i(t)
    =
    \mu_{Y,t}(\bX_i)
    +\varepsilon_{Y,t,i}.
\end{align*}
The observed sharp-design outcome is
\begin{align*}
    Y_i
    =
    Y_i(0)(1-T_i)+Y_i(1)T_i.
\end{align*}
The target at each boundary point $\mathbf{b}\in\B$ is
\begin{align*}
    \tau(\mathbf{b})
    =
    \E[Y_i(1)-Y_i(0)\mid\bX_i=\mathbf{b}]
    =
    \mu_{Y,1}(\mathbf{b})-\mu_{Y,0}(\mathbf{b}).
\end{align*}
This is the design presented in the paper, where we suppress the outcome subscript and write $\mu_t$, $\sigma_t$, and $u_{t,i}$ in place of $\mu_{Y,t}$, $\sigma_{Y,t,v}$, and $u_{Y,t,i}$ for notation simplicity.

\subsubsection*{Fuzzy Design: ITT, FS, and Wald Equations.}

For each $t\in\{0,1\}$ and $i=1,\ldots,n$, define
\begin{align*}
    \mu_{Y,t}(\bX_i)
    &= \mathbf{m}_s(\bX_i)^\top\boldsymbol{\beta}_{Y,t,s},\\
    \mu_{W,t}(\bX_i)
    &= \Lambda(\eta_{W,t}(\bX_i)),
    \qquad
    \eta_{W,t}(\bX_i)=\mathbf{m}_s(\bX_i)^\top\boldsymbol{\beta}_{W,t,s},
\end{align*}
where $\Lambda(a)=1/(1+\exp(-a))$. We also consider homoskedastic and heteroskedastic specifications for the residualized outcome innovation. Let $v\in\{0,2\}$ index the variance specification, where $v=0$ is homoskedastic and $v=2$ is heteroskedastic:
\begin{align*}
    \sigma_{Y,t,v}^2(\bx)
    = \exp\!\left(\mathbf{m}_v(\bx)^\top\bdelta_{Y,t,v}\right).
\end{align*}
Given these conditional mean and variance functions, draw $U_{W,t,i}\thicksim\mathsf{Uniform}(0,1)$ and $u_{Y,t,i}\thicksim\mathsf{Normal}(0,1)$ mutually independently across $i$ and $t$ and independently of all score variables, and set
\begin{align*}
    \varepsilon_{Y,t,i}
    =
    \sigma_{Y,t,v}(\bX_i)u_{Y,t,i}.
\end{align*}
The potential treatment statuses and treatment-indexed potential outcomes are
\begin{align*}
    W_i(t)
    &= \Indicator\{U_{W,t,i}\leq \mu_{W,t}(\bX_i)\},\\
    Y_i(t,w)
    &= \mu_{Y,t}(\bX_i) + \lambda_t\{w-\mu_{W,t}(\bX_i)\}
    + \varepsilon_{Y,t,i},
    \qquad w\in\{0,1\}.
\end{align*}
The reduced-form potential outcome induced by assignment $t$ is $Y_i(t,W_i(t))$. Equivalently,
\begin{align*}
    Y_i(t,W_i(t))
    &=
    \mu_{Y,t}(\bX_i)
    +\lambda_t\{W_i(t)-\mu_{W,t}(\bX_i)\}
    +\varepsilon_{Y,t,i}.
\end{align*}
The observed outcomes and treatment status are
\begin{align*}
    Y_i &= Y_i(0,W_i(0))(1-T_i)+Y_i(1,W_i(1))T_i,\\
    W_i &= W_i(0)(1-T_i)+W_i(1)T_i, \qquad T_i=\Indicator(\bX_i\in\A_1),
\end{align*}
for $i=1,2,\ldots,n$.

In the reduced-form notation used for sharp estimation and inference, $Y_i(t)$ is shorthand for $Y_i(t,W_i(t))$. This centered specification preserves the reduced-form outcome surface because $\E[Y_i(t,W_i(t))|\bX_i=\bx]=\mu_{Y,t}(\bx)$, while allowing the potential outcome and potential treatment status equations to be correlated:
\begin{align*}
    \E[W_i(t)|\bX_i=\bx]
    &= \mu_{W,t}(\bx),\\
    \Cov(Y_i(t,W_i(t)),W_i(t)|\bX_i=\bx)
    &= \lambda_t\mu_{W,t}(\bx)\{1-\mu_{W,t}(\bx)\}.
\end{align*}
The heteroskedastic specification changes the conditional variance of the residualized outcome innovation $\varepsilon_{Y,t,i}$, but it does not change the conditional mean surfaces or the first-stage equation. In the fuzzy design, the conditional variance of the reduced-form outcome also includes the treatment-receipt component induced by $\lambda_t\{W_i(t)-\mu_{W,t}(\bX_i)\}$. Consequently, for each boundary point $\mathbf{b}\in\B$,
\begin{align*}
    \tau_W(\mathbf{b})
    &=
    \E[W_i(1)-W_i(0)|\bX_i=\mathbf{b}]
    =
    \mu_{W,1}(\mathbf{b})-\mu_{W,0}(\mathbf{b}),\\
    \tau_Y(\mathbf{b})
    &=
    \E[Y_i(1,W_i(1))-Y_i(0,W_i(0))|\bX_i=\mathbf{b}]
    =
    \mu_{Y,1}(\mathbf{b})-\mu_{Y,0}(\mathbf{b})
    =
    \mathbf{m}_s(\mathbf{b})^\top
    \left(\boldsymbol{\beta}_{Y,1,s}-\boldsymbol{\beta}_{Y,0,s}\right),\\
    \zeta(\mathbf{b})
    &=
    \frac{\tau_Y(\mathbf{b})}{\tau_W(\mathbf{b})}.
\end{align*}
Within each fuzzy-design Monte Carlo replication, the sharp estimator applied to $Y_i$ targets the reduced-form or ITT curve $\tau_Y(\mathbf{b})$, the same estimator applied to $W_i$ targets the first-stage curve $\tau_W(\mathbf{b})$, and the fuzzy estimator targets the Wald curve $\zeta(\mathbf{b})$. Thus, the ITT, first-stage, and fuzzy simulation results are different summaries of the same simulated samples rather than results from separate fuzzy DGPs. The fuzzy design therefore extends the sharp ITT equation by adding a treatment-receipt equation and allowing the reduced-form outcome and treatment-receipt shocks to be correlated through $\lambda_t$.

\subsection{Calibration from the SPP Data}

The DGP parameters are calibrated once from the SPP data and then held fixed across Monte Carlo replications. We use the following steps separately for each DGP polynomial order $s\in\{1,2\}$. The conditional mean calibration follows the JASA simulation design; the additional calibration determines the homoskedastic or heteroskedastic variance of the innovation entering the outcome equation. The sharp-design exercise uses the calibrated ITT objects $\boldsymbol{\beta}_{Y,t,s}$ and $\bdelta_{Y,t,v}$; the fuzzy-design exercise additionally uses the treatment-receipt coefficients $\boldsymbol{\beta}_{W,t,s}$ and the residual-dependence parameters $\lambda_t$.

\begin{enumerate}[label=\normalfont Step \arabic*.,leftmargin=*]
    \item Fix the DGP polynomial order $s\in\{1,2\}$ and define the side-specific calibration samples $\mathcal I_t=\{i:T_i=t\}$ and $N_t=|\mathcal I_t|$, for $t\in\{0,1\}$.

    \item For each side $t$, estimate the outcome regression coefficients by side-specific least squares:
    \[
        \widehat{\boldsymbol{\beta}}_{Y,t,s}
        =
        \operatorname*{arg\,min}_{\boldsymbol{\beta}}
        \sum_{i\in\mathcal I_t}\{Y_i-\mathbf{m}_s(\bX_i)^\top\boldsymbol{\beta}\}^2.
    \]
    This gives the fitted reduced-form outcome surface $\widehat{\mu}_{Y,t}(\bx)=\mathbf{m}_s(\bx)^\top\widehat{\boldsymbol{\beta}}_{Y,t,s}$.

    \item For each side $t$, estimate the treatment-status regression coefficients by side-specific logit:
    \[
        \widehat{\boldsymbol{\beta}}_{W,t,s}
        =
        \operatorname*{arg\,max}_{\boldsymbol{\beta}}
        \sum_{i\in\mathcal I_t}
        \left[
            W_i\mathbf{m}_s(\bX_i)^\top\boldsymbol{\beta}
            -\log\{1+\exp(\mathbf{m}_s(\bX_i)^\top\boldsymbol{\beta})\}
        \right].
    \]
    This gives the fitted first-stage surface $\widehat{\mu}_{W,t}(\bx)=\Lambda(\mathbf{m}_s(\bx)^\top\widehat{\boldsymbol{\beta}}_{W,t,s})$.

    \item Construct side-specific residuals in the calibration sample:
    \[
        \widehat{\varepsilon}_{Y,i}=Y_i-\widehat{\mu}_{Y,T_i}(\bX_i),
        \qquad
        \widehat{\varepsilon}_{W,i}=W_i-\widehat{\mu}_{W,T_i}(\bX_i).
    \]

    \item Estimate the residual dependence parameter on each side:
    \[
        \widehat{\lambda}_t
        =
        \frac{\sum_{i\in\mathcal I_t}\widehat{\varepsilon}_{Y,i}\widehat{\varepsilon}_{W,i}}
        {\sum_{i\in\mathcal I_t}\widehat{\varepsilon}_{W,i}^2},
        \qquad t\in\{0,1\}.
    \]
    Define the estimated residualized outcome innovation, distinct from the standardized simulation draw $u_{Y,t,i}$,
    \[
        \widehat{\omega}_{Y,i}
        =
        \widehat{\varepsilon}_{Y,i}-\widehat{\lambda}_{T_i}\widehat{\varepsilon}_{W,i}.
    \]
    For the homoskedastic specification, estimate the intercept-only log-variance coefficient
    \[
        \widehat{\bdelta}_{Y,t,0}
        =
        \log\!\left(\frac{1}{N_t}\sum_{i\in\mathcal I_t}\widehat{\omega}_{Y,i}^2\right),
        \qquad t\in\{0,1\}.
    \]
    For the heteroskedastic specification, estimate the side-specific log-variance regression for the residualized outcome innovation:
    \[
        \widehat{\bdelta}_{Y,t,2}
        =
        \operatorname*{arg\,min}_{\bdelta}
        \sum_{i\in\mathcal I_t}
        \left[
            \log\{\max(\widehat{\omega}_{Y,i}^2,\epsilon_0)\}
            -\mathbf{m}_2(\bX_i)^\top\bdelta
        \right]^2,
    \]
    where $\epsilon_0>0$ is a small truncation constant used only to avoid taking logs of zero.

    \item Set the Monte Carlo DGP parameters equal to the calibration estimates: $\boldsymbol{\beta}_{Y,t,s}=\widehat{\boldsymbol{\beta}}_{Y,t,s}$, $\boldsymbol{\beta}_{W,t,s}=\widehat{\boldsymbol{\beta}}_{W,t,s}$, $\lambda_t=\widehat{\lambda}_t$, $\bdelta_{Y,t,0}=\widehat{\bdelta}_{Y,t,0}$, and $\bdelta_{Y,t,2}=\widehat{\bdelta}_{Y,t,2}$ for $t\in\{0,1\}$.

    \item Before running the Monte Carlo replications, compute the true target functions analytically from the calibrated conditional mean surfaces. For each evaluation point $\mathbf{b}\in\B$, the sharp-design BATEC is
    \[
        \tau(\mathbf{b})
        =
        \mathbf{m}_s(\mathbf{b})^\top(\boldsymbol{\beta}_{Y,1,s}-\boldsymbol{\beta}_{Y,0,s}).
    \]
    For the fuzzy-design exercise, the reduced-form/ITT, first-stage, and Wald targets are
    \[
        \tau_Y(\mathbf{b})
        =
        \mathbf{m}_s(\mathbf{b})^\top(\boldsymbol{\beta}_{Y,1,s}-\boldsymbol{\beta}_{Y,0,s}),
        \qquad
        \tau_W(\mathbf{b})
        =
        \Lambda(\mathbf{m}_s(\mathbf{b})^\top\boldsymbol{\beta}_{W,1,s})
        -
        \Lambda(\mathbf{m}_s(\mathbf{b})^\top\boldsymbol{\beta}_{W,0,s}),
    \]
    and $\zeta(\mathbf{b})=\frac{\tau_Y(\mathbf{b})}{\tau_W(\mathbf{b})}$. The same calibrated outcome surface is used in both exercises, so $\tau(\mathbf{b})$ and $\tau_Y(\mathbf{b})$ have the same algebraic form under this calibration; they are kept notationally distinct because they refer to different Monte Carlo exercises and different causal objects.

\end{enumerate}

This calibration reproduces the empirical residual covariance between the outcome and take-up equations on each side of the boundary, while keeping the marginal reduced-form and first-stage surfaces fixed by the fitted SPP models. The homoskedastic and heteroskedastic DGPs share the same target curves for a given mean specification; they differ only in the conditional variance of the residualized outcome innovation.

\subsection{Monte Carlo Implementation}

The following steps describe how to simulate from the calibrated model. They are repeated separately for each pair $(s,v)\in\{1,2\}\times\{0,2\}$. For the sharp-design exercise, the same score draws, assignment rule, evaluation grid, polynomial order, and bandwidth choices are used, but Steps 4--6 are simplified by keeping only the outcome equation $Y_i$ and dropping $W_i(t)$, $\mu_{W,t}$, $\lambda_t$, and the Wald ratio.

\begin{enumerate}[label=\normalfont Step \arabic*.,leftmargin=*]
    \item Fix the Monte Carlo inputs: the sample size $n$, the number of replications $M$, the evaluation grid $\{\mathbf{b}_1,\ldots,\mathbf{b}_{21}\}$ with $\mathbf{b}_{11}$ at the kink, the assignment regions $\A_0$ and $\A_1$, the local polynomial order $p=1$, and the calibrated parameters. The sharp-design exercise uses only $\{\boldsymbol{\beta}_{Y,t,s},\bdelta_{Y,t,0},\bdelta_{Y,t,2}:t\in\{0,1\}\}$, while the fuzzy-design exercise uses the full set
    \[
        \{\boldsymbol{\beta}_{Y,t,s},\boldsymbol{\beta}_{W,t,s},\lambda_t,\bdelta_{Y,t,0},\bdelta_{Y,t,2}:t\in\{0,1\}\}.
    \]

    \item Before entering the Monte Carlo loop, evaluate the true target functions on the grid. For the sharp-design exercise this is only the BATEC target $\tau(\mathbf{b}_j)$. For the fuzzy-design exercise the targets are
    \[
        \tau_Y(\mathbf{b}_j),
        \qquad
        \tau_W(\mathbf{b}_j),
        \qquad
        \zeta(\mathbf{b}_j),
        \qquad j=1,\ldots,21.
    \]
    These target values are fixed across Monte Carlo replications and are common to the homoskedastic and heteroskedastic versions of a given mean specification.

    \item For replication $m=1,\ldots,M$, draw independent score vectors $\bX_i=(100B_{1i}-25,100B_{2i}-25)^\top$, with $B_{1i}$ and $B_{2i}$ independent $\mathsf{Beta}(3,4)$ random variables, and set $T_i=\Indicator(\bX_i\in\A_1)$.

    \item Form the design vectors $\mathbf{m}_s(\bX_i)$ and compute the side-specific conditional outcome means and variances
    \[
        \mu_{Y,t}(\bX_i),
        \qquad
        \sigma_{Y,t,v}^2(\bX_i),
        \qquad t\in\{0,1\}.
    \]
    Under homoskedasticity, $\sigma_{Y,t,v}^2(\bX_i)$ is constant within side $t$. Under heteroskedasticity, it changes with $\bX_i$ through the fitted log-variance model. For the fuzzy-design exercise, also compute $\mu_{W,t}(\bX_i)$.

    \item For the sharp-design exercise, draw $u_{Y,t,i}\thicksim\mathsf{Normal}(0,1)$ independently across $i$ and $t$ and independently of all score variables, and set
    \[
        \varepsilon_{Y,t,i}
        =
        \sigma_{Y,t,v}(\bX_i)u_{Y,t,i}.
    \]
    Then construct $Y_i(t)=\mu_{Y,t}(\bX_i)+\varepsilon_{Y,t,i}$. For the fuzzy-design exercise, draw $U_{W,t,i}\thicksim\mathsf{Uniform}(0,1)$ and $u_{Y,t,i}\thicksim\mathsf{Normal}(0,1)$ mutually independently across $i$ and $t$ and independently of all score variables, define $\varepsilon_{Y,t,i}$ as above, and construct
    \begin{align*}
        W_i(t)
        &=\Indicator\{U_{W,t,i}\leq\mu_{W,t}(\bX_i)\},\\
        Y_i(t,W_i(t))
        &=\mu_{Y,t}(\bX_i)+\lambda_t\{W_i(t)-\mu_{W,t}(\bX_i)\}+\varepsilon_{Y,t,i}.
    \end{align*}

    \item Select observed variables using the assignment indicator. For the sharp-design exercise, $Y_i=Y_i(0)(1-T_i)+Y_i(1)T_i$. For the fuzzy-design exercise,
    \begin{align*}
        W_i &= W_i(0)(1-T_i)+W_i(1)T_i,\\
        Y_i &= Y_i(0,W_i(0))(1-T_i)+Y_i(1,W_i(1))T_i.
    \end{align*}
    The simulated dataset for replication $m$ is therefore $\{Y_i,T_i,\bX_i:i=1,\ldots,n\}$ in the sharp-design exercise and $\{Y_i,W_i,T_i,\bX_i:i=1,\ldots,n\}$ in the fuzzy-design exercise.

    \item For each evaluation point $\mathbf{b}_j$, compute the signed distance score $D_i(\mathbf{b}_j)$ using the Euclidean distance and the assignment side. Estimate the target curves using the distance-based local linear estimator.

    \item \textbf{Sharp/ITT Analysis}. In the sharp-design exercise, apply the sharp distance-based estimator to $Y_i$ for the target $\tau(\mathbf{b}_j)$. In the fuzzy-design exercise, apply the same sharp estimator to the reduced-form outcome $Y_i$ for the ITT target $\tau_Y(\mathbf{b}_j)$.

    \item \textbf{FS Analysis}. In the fuzzy-design exercise, apply the sharp distance-based estimator to $W_i$ for the target $\tau_W(\mathbf{b}_j)$.

    \item \textbf{Fuzzy Analysis}. In the fuzzy-design exercise, apply the fuzzy distance-based estimator using $Y_i$ as the outcome and $W_i$ as the treatment-receipt status for the Wald target $\zeta(\mathbf{b}_j)$.

    \item Repeat the analysis using the bandwidth choices considered in the main text: the smooth-boundary selector, the kink-adaptive selector using the known kink location, the unknown-kink bandwidth rule, and the bandwidth induced by the univariate \texttt{rdrobust} selector.

    \item After all $M$ replications, compute Monte Carlo summaries by comparing the stored estimates and confidence sets with the fixed analytic targets. For each target and each grid point, report bias, standard deviation, root mean squared error, pointwise coverage, and average interval length. For uniform inference, report simultaneous coverage and average band length over the grid. Finally, also recover WBATE and LBATE summaries and report their associated performance.
\end{enumerate}

We use the sharp simulations to study the finite-sample behavior of the sharp BATEC estimator for $\tau(\mathbf{b})$. The fuzzy simulations additionally study the reduced-form/ITT estimators, the first-stage estimators, and the fuzzy Wald estimator. A complier-average causal interpretation of $\zeta(\mathbf{b})$ would require additional assumptions that are not imposed in the simulations. In particular, one would need a local exclusion restriction ruling out a direct effect of assignment on the outcome except through treatment take-up, a monotonicity condition ruling out defiers near the boundary, and a nonzero first-stage effect. Under those conditions, the Wald curve $\zeta(\mathbf{b})$ can be interpreted as a local average treatment effect for compliers at the boundary point $\mathbf{b}$.

\subsection{Results}

The following tables report Monte Carlo summaries separately for the sharp estimator and, in the fuzzy-design simulations, for the reduced-form/ITT, first-stage, and fuzzy estimators. Results are reported separately by DGP polynomial order $s\in\{1,2\}$ and by homoskedastic or heteroskedastic outcome disturbances. We first report the sharp-design results. We then report two sets of fuzzy-design results: one using the ITT bandwidth selector, matching the main empirical specification, and one using the fuzzy bandwidth selector.

The sharp-design simulation results are reported first. Specifically:
\begin{itemize}
    \item Tables~\ref{tab:sa-simuls-sharp-linear-homoskedastic-kinkoff-bwmain}--\ref{tab:sa-simuls-sharp-linear-homoskedastic-rdrobustadj-bwmain} report the sharp simulation results for the linear homoskedastic specification.
    \item Tables~\ref{tab:sa-simuls-sharp-linear-heteroskedastic-kinkoff-bwmain}--\ref{tab:sa-simuls-sharp-linear-heteroskedastic-rdrobustadj-bwmain} report the sharp simulation results for the linear heteroskedastic specification.
    \item Tables~\ref{tab:sa-simuls-sharp-quadratic-homoskedastic-kinkoff-bwmain}--\ref{tab:sa-simuls-sharp-quadratic-homoskedastic-rdrobustadj-bwmain} report the sharp simulation results for the quadratic homoskedastic specification.
    \item Tables~\ref{tab:sa-simuls-sharp-quadratic-heteroskedastic-kinkoff-bwmain}--\ref{tab:sa-simuls-sharp-quadratic-heteroskedastic-rdrobustadj-bwmain} report the sharp simulation results for the quadratic heteroskedastic specification.
\end{itemize}

The first set of fuzzy-design simulation results uses the ITT bandwidth selector. Specifically:
\begin{itemize}
    \item Tables~\ref{tab:sa-simuls-itt-linear-homoskedastic-kinkoff-bwitt}--\ref{tab:sa-simuls-itt-linear-homoskedastic-rdrobustadj-bwitt} report the reduced-form (ITT) simulation results for the linear homoskedastic specification.
    \item Tables~\ref{tab:sa-simuls-itt-linear-heteroskedastic-kinkoff-bwitt}--\ref{tab:sa-simuls-itt-linear-heteroskedastic-rdrobustadj-bwitt} report the reduced-form (ITT) simulation results for the linear heteroskedastic specification.
    \item Tables~\ref{tab:sa-simuls-itt-quadratic-homoskedastic-kinkoff-bwitt}--\ref{tab:sa-simuls-itt-quadratic-homoskedastic-rdrobustadj-bwitt} report the reduced-form (ITT) simulation results for the quadratic homoskedastic specification.
    \item Tables~\ref{tab:sa-simuls-itt-quadratic-heteroskedastic-kinkoff-bwitt}--\ref{tab:sa-simuls-itt-quadratic-heteroskedastic-rdrobustadj-bwitt} report the reduced-form (ITT) simulation results for the quadratic heteroskedastic specification.
    \item Tables~\ref{tab:sa-simuls-fs-linear-homoskedastic-kinkoff-bwitt}--\ref{tab:sa-simuls-fs-linear-homoskedastic-rdrobustadj-bwitt} report the take-up (FS) simulation results for the linear homoskedastic specification.
    \item Tables~\ref{tab:sa-simuls-fs-linear-heteroskedastic-kinkoff-bwitt}--\ref{tab:sa-simuls-fs-linear-heteroskedastic-rdrobustadj-bwitt} report the take-up (FS) simulation results for the linear heteroskedastic specification.
    \item Tables~\ref{tab:sa-simuls-fs-quadratic-homoskedastic-kinkoff-bwitt}--\ref{tab:sa-simuls-fs-quadratic-homoskedastic-rdrobustadj-bwitt} report the take-up (FS) simulation results for the quadratic homoskedastic specification.
    \item Tables~\ref{tab:sa-simuls-fs-quadratic-heteroskedastic-kinkoff-bwitt}--\ref{tab:sa-simuls-fs-quadratic-heteroskedastic-rdrobustadj-bwitt} report the take-up (FS) simulation results for the quadratic heteroskedastic specification.
    \item Tables~\ref{tab:sa-simuls-fuzzy-linear-homoskedastic-kinkoff-bwitt}--\ref{tab:sa-simuls-fuzzy-linear-homoskedastic-rdrobustadj-bwitt} report the fuzzy simulation results for the linear homoskedastic specification.
    \item Tables~\ref{tab:sa-simuls-fuzzy-linear-heteroskedastic-kinkoff-bwitt}--\ref{tab:sa-simuls-fuzzy-linear-heteroskedastic-rdrobustadj-bwitt} report the fuzzy simulation results for the linear heteroskedastic specification.
    \item Tables~\ref{tab:sa-simuls-fuzzy-quadratic-homoskedastic-kinkoff-bwitt}--\ref{tab:sa-simuls-fuzzy-quadratic-homoskedastic-rdrobustadj-bwitt} report the fuzzy simulation results for the quadratic homoskedastic specification.
    \item Tables~\ref{tab:sa-simuls-fuzzy-quadratic-heteroskedastic-kinkoff-bwitt}--\ref{tab:sa-simuls-fuzzy-quadratic-heteroskedastic-rdrobustadj-bwitt} report the fuzzy simulation results for the quadratic heteroskedastic specification.
\end{itemize}

The second set of fuzzy-design simulation results uses the fuzzy bandwidth selector. Specifically:
\begin{itemize}
    \item Tables~\ref{tab:sa-simuls-itt-linear-homoskedastic-kinkoff-bwmain}--\ref{tab:sa-simuls-itt-linear-homoskedastic-rdrobustadj-bwmain} report the reduced-form (ITT) simulation results for the linear homoskedastic specification.
    \item Tables~\ref{tab:sa-simuls-itt-linear-heteroskedastic-kinkoff-bwmain}--\ref{tab:sa-simuls-itt-linear-heteroskedastic-rdrobustadj-bwmain} report the reduced-form (ITT) simulation results for the linear heteroskedastic specification.
    \item Tables~\ref{tab:sa-simuls-itt-quadratic-homoskedastic-kinkoff-bwmain}--\ref{tab:sa-simuls-itt-quadratic-homoskedastic-rdrobustadj-bwmain} report the reduced-form (ITT) simulation results for the quadratic homoskedastic specification.
    \item Tables~\ref{tab:sa-simuls-itt-quadratic-heteroskedastic-kinkoff-bwmain}--\ref{tab:sa-simuls-itt-quadratic-heteroskedastic-rdrobustadj-bwmain} report the reduced-form (ITT) simulation results for the quadratic heteroskedastic specification.
    \item Tables~\ref{tab:sa-simuls-fs-linear-homoskedastic-kinkoff-bwmain}--\ref{tab:sa-simuls-fs-linear-homoskedastic-rdrobustadj-bwmain} report the take-up (FS) simulation results for the linear homoskedastic specification.
    \item Tables~\ref{tab:sa-simuls-fs-linear-heteroskedastic-kinkoff-bwmain}--\ref{tab:sa-simuls-fs-linear-heteroskedastic-rdrobustadj-bwmain} report the take-up (FS) simulation results for the linear heteroskedastic specification.
    \item Tables~\ref{tab:sa-simuls-fs-quadratic-homoskedastic-kinkoff-bwmain}--\ref{tab:sa-simuls-fs-quadratic-homoskedastic-rdrobustadj-bwmain} report the take-up (FS) simulation results for the quadratic homoskedastic specification.
    \item Tables~\ref{tab:sa-simuls-fs-quadratic-heteroskedastic-kinkoff-bwmain}--\ref{tab:sa-simuls-fs-quadratic-heteroskedastic-rdrobustadj-bwmain} report the take-up (FS) simulation results for the quadratic heteroskedastic specification.
    \item Tables~\ref{tab:sa-simuls-fuzzy-linear-homoskedastic-kinkoff-bwmain}--\ref{tab:sa-simuls-fuzzy-linear-homoskedastic-rdrobustadj-bwmain} report the fuzzy simulation results for the linear homoskedastic specification.
    \item Tables~\ref{tab:sa-simuls-fuzzy-linear-heteroskedastic-kinkoff-bwmain}--\ref{tab:sa-simuls-fuzzy-linear-heteroskedastic-rdrobustadj-bwmain} report the fuzzy simulation results for the linear heteroskedastic specification.
    \item Tables~\ref{tab:sa-simuls-fuzzy-quadratic-homoskedastic-kinkoff-bwmain}--\ref{tab:sa-simuls-fuzzy-quadratic-homoskedastic-rdrobustadj-bwmain} report the fuzzy simulation results for the quadratic homoskedastic specification.
    \item Tables~\ref{tab:sa-simuls-fuzzy-quadratic-heteroskedastic-kinkoff-bwmain}--\ref{tab:sa-simuls-fuzzy-quadratic-heteroskedastic-rdrobustadj-bwmain} report the fuzzy simulation results for the quadratic heteroskedastic specification.
\end{itemize}
The columns report the average bandwidths, average bias, standard deviation, root mean squared error, empirical coverage, and average interval or band length. For the Uniform row, empirical coverage and average length refer to uniform coverage and average uniform band length.

\clearpage

\begin{table}[p]
    \centering
    \resizebox{.8\textwidth}{!}{\input{tables/simuls_sharp_linear_homoskedastic_kinkoff_bwmain.tex}}
    \caption{Sharp BATEC Effects (Simulations, Linear Homoskedastic, Smooth-Boundary Bandwidth).}
    \label{tab:sa-simuls-sharp-linear-homoskedastic-kinkoff-bwmain}
\end{table}

\begin{table}[p]
    \centering
    \resizebox{.8\textwidth}{!}{\input{tables/simuls_sharp_linear_homoskedastic_adaptive_bwmain.tex}}
    \caption{Sharp BATEC Effects (Simulations, Linear Homoskedastic, Kink-Adaptive Bandwidth).}
    \label{tab:sa-simuls-sharp-linear-homoskedastic-adaptive-bwmain}
\end{table}

\begin{table}[p]
    \centering
    \resizebox{.8\textwidth}{!}{\input{tables/simuls_sharp_linear_homoskedastic_kinkon_bwmain.tex}}
    \caption{Sharp BATEC Effects (Simulations, Linear Homoskedastic, Unknown-Kink Bandwidth).}
    \label{tab:sa-simuls-sharp-linear-homoskedastic-kinkon-bwmain}
\end{table}

\begin{table}[p]
    \centering
    \resizebox{.8\textwidth}{!}{\input{tables/simuls_sharp_linear_homoskedastic_rdrobustadj_bwmain.tex}}
    \caption{Sharp BATEC Effects (Simulations, Linear Homoskedastic, \texttt{rdrobust} Bandwidth).}
    \label{tab:sa-simuls-sharp-linear-homoskedastic-rdrobustadj-bwmain}
\end{table}

\begin{table}[p]
    \centering
    \resizebox{.8\textwidth}{!}{\input{tables/simuls_sharp_linear_heteroskedastic_kinkoff_bwmain.tex}}
    \caption{Sharp BATEC Effects (Simulations, Linear Heteroskedastic, Smooth-Boundary Bandwidth).}
    \label{tab:sa-simuls-sharp-linear-heteroskedastic-kinkoff-bwmain}
\end{table}

\begin{table}[p]
    \centering
    \resizebox{.8\textwidth}{!}{\input{tables/simuls_sharp_linear_heteroskedastic_adaptive_bwmain.tex}}
    \caption{Sharp BATEC Effects (Simulations, Linear Heteroskedastic, Kink-Adaptive Bandwidth).}
    \label{tab:sa-simuls-sharp-linear-heteroskedastic-adaptive-bwmain}
\end{table}

\begin{table}[p]
    \centering
    \resizebox{.8\textwidth}{!}{\input{tables/simuls_sharp_linear_heteroskedastic_kinkon_bwmain.tex}}
    \caption{Sharp BATEC Effects (Simulations, Linear Heteroskedastic, Unknown-Kink Bandwidth).}
    \label{tab:sa-simuls-sharp-linear-heteroskedastic-kinkon-bwmain}
\end{table}

\begin{table}[p]
    \centering
    \resizebox{.8\textwidth}{!}{\input{tables/simuls_sharp_linear_heteroskedastic_rdrobustadj_bwmain.tex}}
    \caption{Sharp BATEC Effects (Simulations, Linear Heteroskedastic, \texttt{rdrobust} Bandwidth).}
    \label{tab:sa-simuls-sharp-linear-heteroskedastic-rdrobustadj-bwmain}
\end{table}

\begin{table}[p]
    \centering
    \resizebox{.8\textwidth}{!}{\input{tables/simuls_sharp_quadratic_homoskedastic_kinkoff_bwmain.tex}}
    \caption{Sharp BATEC Effects (Simulations, Quadratic Homoskedastic, Smooth-Boundary Bandwidth).}
    \label{tab:sa-simuls-sharp-quadratic-homoskedastic-kinkoff-bwmain}
\end{table}

\begin{table}[p]
    \centering
    \resizebox{.8\textwidth}{!}{\input{tables/simuls_sharp_quadratic_homoskedastic_adaptive_bwmain.tex}}
    \caption{Sharp BATEC Effects (Simulations, Quadratic Homoskedastic, Kink-Adaptive Bandwidth).}
    \label{tab:sa-simuls-sharp-quadratic-homoskedastic-adaptive-bwmain}
\end{table}

\begin{table}[p]
    \centering
    \resizebox{.8\textwidth}{!}{\input{tables/simuls_sharp_quadratic_homoskedastic_kinkon_bwmain.tex}}
    \caption{Sharp BATEC Effects (Simulations, Quadratic Homoskedastic, Unknown-Kink Bandwidth).}
    \label{tab:sa-simuls-sharp-quadratic-homoskedastic-kinkon-bwmain}
\end{table}

\begin{table}[p]
    \centering
    \resizebox{.8\textwidth}{!}{\input{tables/simuls_sharp_quadratic_homoskedastic_rdrobustadj_bwmain.tex}}
    \caption{Sharp BATEC Effects (Simulations, Quadratic Homoskedastic, \texttt{rdrobust} Bandwidth).}
    \label{tab:sa-simuls-sharp-quadratic-homoskedastic-rdrobustadj-bwmain}
\end{table}

\begin{table}[p]
    \centering
    \resizebox{.8\textwidth}{!}{\input{tables/simuls_sharp_quadratic_heteroskedastic_kinkoff_bwmain.tex}}
    \caption{Sharp BATEC Effects (Simulations, Quadratic Heteroskedastic, Smooth-Boundary Bandwidth).}
    \label{tab:sa-simuls-sharp-quadratic-heteroskedastic-kinkoff-bwmain}
\end{table}

\begin{table}[p]
    \centering
    \resizebox{.8\textwidth}{!}{\input{tables/simuls_sharp_quadratic_heteroskedastic_adaptive_bwmain.tex}}
    \caption{Sharp BATEC Effects (Simulations, Quadratic Heteroskedastic, Kink-Adaptive Bandwidth).}
    \label{tab:sa-simuls-sharp-quadratic-heteroskedastic-adaptive-bwmain}
\end{table}

\begin{table}[p]
    \centering
    \resizebox{.8\textwidth}{!}{\input{tables/simuls_sharp_quadratic_heteroskedastic_kinkon_bwmain.tex}}
    \caption{Sharp BATEC Effects (Simulations, Quadratic Heteroskedastic, Unknown-Kink Bandwidth).}
    \label{tab:sa-simuls-sharp-quadratic-heteroskedastic-kinkon-bwmain}
\end{table}

\begin{table}[p]
    \centering
    \resizebox{.8\textwidth}{!}{\input{tables/simuls_sharp_quadratic_heteroskedastic_rdrobustadj_bwmain.tex}}
    \caption{Sharp BATEC Effects (Simulations, Quadratic Heteroskedastic, \texttt{rdrobust} Bandwidth).}
    \label{tab:sa-simuls-sharp-quadratic-heteroskedastic-rdrobustadj-bwmain}
\end{table}

\clearpage

\begin{table}[p]
    \centering
    \resizebox{.8\textwidth}{!}{\begin{tabular}{lccccrrrrr}
\toprule\toprule
 & $h_0$ & $h_1$ & $N_{\mathrm{Co}}$ & $N_{\mathrm{Tr}}$ & Bias & SD & RMSE & EC & IL \\
\midrule
$\bb_{1}$ & 19.534 & 19.534 & 2120 & 4260 & 0.000 & 0.033 & 0.033 & 0.950 & 0.241 \\
$\bb_{2}$ & 19.152 & 19.152 & 2138 & 4114 & 0.000 & 0.034 & 0.034 & 0.952 & 0.244 \\
$\bb_{3}$ & 18.902 & 18.902 & 2185 & 3961 & 0.001 & 0.035 & 0.035 & 0.950 & 0.246 \\
$\bb_{4}$ & 18.805 & 18.805 & 2271 & 3805 & 0.002 & 0.035 & 0.036 & 0.952 & 0.247 \\
$\bb_{5}$ & 18.832 & 18.832 & 2383 & 3639 & 0.004 & 0.035 & 0.036 & 0.955 & 0.247 \\
$\bb_{6}$ & 19.003 & 19.003 & 2517 & 3478 & 0.006 & 0.035 & 0.036 & 0.958 & 0.247 \\
$\bb_{7}$ & 19.115 & 19.115 & 2621 & 3265 & 0.008 & 0.035 & 0.036 & 0.955 & 0.248 \\
\bottomrule\bottomrule
\end{tabular}}
    \caption{Intention-to-Treat Effects (Simulations, Linear Homoskedastic, Smooth-Boundary Bandwidth, ITT Bandwidth Selector).}
    \label{tab:sa-simuls-itt-linear-homoskedastic-kinkoff-bwitt}
\end{table}

\begin{table}[p]
    \centering
    \resizebox{.8\textwidth}{!}{\begin{tabular}{lccccrrrrr}
\toprule\toprule
 & $h_0$ & $h_1$ & $N_{\mathrm{Co}}$ & $N_{\mathrm{Tr}}$ & Bias & SD & RMSE & EC & IL \\
\midrule
$\bb_{1}$ & 18.862 & 18.862 & 2022 & 3994 & 0.001 & 0.034 & 0.034 & 0.950 & 0.248 \\
$\bb_{2}$ & 17.546 & 17.546 & 1877 & 3512 & 0.001 & 0.036 & 0.036 & 0.947 & 0.262 \\
$\bb_{3}$ & 15.909 & 15.909 & 1661 & 2914 & 0.001 & 0.039 & 0.039 & 0.948 & 0.286 \\
$\bb_{4}$ & 13.997 & 13.997 & 1382 & 2257 & 0.001 & 0.044 & 0.044 & 0.949 & 0.324 \\
$\bb_{5}$ & 12.000 & 12.000 & 1081 & 1641 & -0.000 & 0.050 & 0.050 & 0.943 & 0.378 \\
$\bb_{6}$ & 10.000 & 10.000 & 788 & 1113 & -0.002 & 0.059 & 0.059 & 0.943 & 0.454 \\
$\bb_{7}$ & 8.562 & 8.562 & 596 & 783 & -0.002 & 0.071 & 0.071 & 0.947 & 0.535 \\
\bottomrule\bottomrule
\end{tabular}}
    \caption{Intention-to-Treat Effects (Simulations, Linear Homoskedastic, Kink-Adaptive Bandwidth, ITT Bandwidth Selector).}
    \label{tab:sa-simuls-itt-linear-homoskedastic-adaptive-bwitt}
\end{table}

\begin{table}[p]
    \centering
    \resizebox{.8\textwidth}{!}{\begin{tabular}{lccccrrrrr}
\toprule\toprule
 & $h_0$ & $h_1$ & $N_{\mathrm{Co}}$ & $N_{\mathrm{Tr}}$ & Bias & SD & RMSE & EC & IL \\
\midrule
$\bb_{1}$ & 8.558 & 8.558 & 609 & 823 & 0.001 & 0.072 & 0.072 & 0.944 & 0.277 \\
$\bb_{2}$ & 8.391 & 8.391 & 598 & 803 & 0.001 & 0.073 & 0.073 & 0.942 & 0.281 \\
$\bb_{3}$ & 8.281 & 8.281 & 589 & 788 & 0.001 & 0.073 & 0.073 & 0.950 & 0.283 \\
$\bb_{4}$ & 8.239 & 8.239 & 583 & 779 & 0.001 & 0.074 & 0.074 & 0.948 & 0.285 \\
$\bb_{5}$ & 8.251 & 8.251 & 579 & 773 & -0.001 & 0.073 & 0.073 & 0.948 & 0.285 \\
$\bb_{6}$ & 8.325 & 8.325 & 577 & 772 & -0.002 & 0.073 & 0.073 & 0.951 & 0.285 \\
$\bb_{7}$ & 8.375 & 8.375 & 575 & 752 & -0.002 & 0.073 & 0.073 & 0.952 & 0.286 \\
\bottomrule\bottomrule
\end{tabular}}
    \caption{Intention-to-Treat Effects (Simulations, Linear Homoskedastic, Unknown-Kink Bandwidth, ITT Bandwidth Selector).}
    \label{tab:sa-simuls-itt-linear-homoskedastic-kinkon-bwitt}
\end{table}

\begin{table}[p]
    \centering
    \resizebox{.8\textwidth}{!}{\begin{tabular}{lccccrrrrr}
\toprule\toprule
 & $h_0$ & $h_1$ & $N_{\mathrm{Co}}$ & $N_{\mathrm{Tr}}$ & Bias & SD & RMSE & EC & IL \\
\midrule
$\bb_{1}$ & 19.325 & 19.325 & 2100 & 4182 & -0.001 & 0.036 & 0.036 & 0.953 & 0.246 \\
$\bb_{2}$ & 19.171 & 19.171 & 2152 & 4123 & -0.000 & 0.036 & 0.036 & 0.952 & 0.246 \\
$\bb_{3}$ & 19.127 & 19.127 & 2235 & 4046 & 0.000 & 0.037 & 0.037 & 0.953 & 0.246 \\
$\bb_{4}$ & 19.198 & 19.198 & 2351 & 3949 & 0.001 & 0.037 & 0.037 & 0.955 & 0.246 \\
$\bb_{5}$ & 19.404 & 19.404 & 2495 & 3845 & 0.003 & 0.037 & 0.038 & 0.958 & 0.245 \\
$\bb_{6}$ & 19.638 & 19.638 & 2635 & 3703 & 0.005 & 0.037 & 0.038 & 0.962 & 0.244 \\
$\bb_{7}$ & 19.901 & 19.901 & 2763 & 3534 & 0.007 & 0.037 & 0.038 & 0.957 & 0.244 \\
\bottomrule\bottomrule
\end{tabular}}
    \caption{Intention-to-Treat Effects (Simulations, Linear Homoskedastic, \texttt{rdrobust} Bandwidth, ITT Bandwidth Selector).}
    \label{tab:sa-simuls-itt-linear-homoskedastic-rdrobustadj-bwitt}
\end{table}

\begin{table}[p]
    \centering
    \resizebox{.8\textwidth}{!}{\begin{tabular}{lccccrrrrr}
\toprule\toprule
 & $h_0$ & $h_1$ & $N_{\mathrm{Co}}$ & $N_{\mathrm{Tr}}$ & Bias & SD & RMSE & EC & IL \\
\midrule
$\bb_{1}$ & 19.822 & 19.822 & 2159 & 4374 & -0.001 & 0.042 & 0.042 & 0.955 & 0.300 \\
$\bb_{2}$ & 19.210 & 19.210 & 2146 & 4137 & -0.000 & 0.043 & 0.043 & 0.958 & 0.303 \\
$\bb_{3}$ & 18.737 & 18.737 & 2155 & 3903 & 0.000 & 0.043 & 0.043 & 0.959 & 0.306 \\
$\bb_{4}$ & 18.364 & 18.364 & 2185 & 3658 & 0.002 & 0.043 & 0.043 & 0.957 & 0.307 \\
$\bb_{5}$ & 18.091 & 18.091 & 2234 & 3405 & 0.003 & 0.044 & 0.044 & 0.954 & 0.309 \\
$\bb_{6}$ & 17.917 & 17.917 & 2293 & 3150 & 0.006 & 0.044 & 0.045 & 0.954 & 0.311 \\
$\bb_{7}$ & 17.759 & 17.759 & 2342 & 2876 & 0.007 & 0.045 & 0.046 & 0.952 & 0.313 \\
\bottomrule\bottomrule
\end{tabular}}
    \caption{Intention-to-Treat Effects (Simulations, Linear Heteroskedastic, Smooth-Boundary Bandwidth, ITT Bandwidth Selector).}
    \label{tab:sa-simuls-itt-linear-heteroskedastic-kinkoff-bwitt}
\end{table}

\begin{table}[p]
    \centering
    \resizebox{.8\textwidth}{!}{\begin{tabular}{lccccrrrrr}
\toprule\toprule
 & $h_0$ & $h_1$ & $N_{\mathrm{Co}}$ & $N_{\mathrm{Tr}}$ & Bias & SD & RMSE & EC & IL \\
\midrule
$\bb_{1}$ & 19.027 & 19.027 & 2044 & 4060 & -0.000 & 0.043 & 0.043 & 0.952 & 0.311 \\
$\bb_{2}$ & 17.579 & 17.579 & 1882 & 3526 & 0.000 & 0.045 & 0.045 & 0.954 & 0.328 \\
$\bb_{3}$ & 15.902 & 15.902 & 1660 & 2912 & 0.001 & 0.047 & 0.047 & 0.953 & 0.354 \\
$\bb_{4}$ & 13.995 & 13.995 & 1382 & 2257 & 0.001 & 0.052 & 0.052 & 0.948 & 0.395 \\
$\bb_{5}$ & 12.000 & 12.000 & 1081 & 1641 & -0.001 & 0.061 & 0.061 & 0.946 & 0.456 \\
$\bb_{6}$ & 10.000 & 10.000 & 788 & 1112 & -0.002 & 0.072 & 0.072 & 0.945 & 0.542 \\
$\bb_{7}$ & 8.181 & 8.181 & 545 & 717 & -0.003 & 0.089 & 0.089 & 0.945 & 0.660 \\
\bottomrule\bottomrule
\end{tabular}}
    \caption{Intention-to-Treat Effects (Simulations, Linear Heteroskedastic, Kink-Adaptive Bandwidth, ITT Bandwidth Selector).}
    \label{tab:sa-simuls-itt-linear-heteroskedastic-adaptive-bwitt}
\end{table}

\begin{table}[p]
    \centering
    \resizebox{.8\textwidth}{!}{\begin{tabular}{lccccrrrrr}
\toprule\toprule
 & $h_0$ & $h_1$ & $N_{\mathrm{Co}}$ & $N_{\mathrm{Tr}}$ & Bias & SD & RMSE & EC & IL \\
\midrule
$\bb_{1}$ & 8.684 & 8.684 & 625 & 848 & -0.001 & 0.088 & 0.088 & 0.951 & 0.341 \\
$\bb_{2}$ & 8.416 & 8.416 & 601 & 807 & 0.000 & 0.088 & 0.088 & 0.951 & 0.346 \\
$\bb_{3}$ & 8.209 & 8.209 & 580 & 773 & 0.001 & 0.090 & 0.090 & 0.952 & 0.350 \\
$\bb_{4}$ & 8.045 & 8.045 & 559 & 740 & -0.000 & 0.092 & 0.092 & 0.950 & 0.353 \\
$\bb_{5}$ & 7.926 & 7.926 & 539 & 711 & -0.002 & 0.093 & 0.093 & 0.952 & 0.357 \\
$\bb_{6}$ & 7.850 & 7.850 & 521 & 684 & -0.003 & 0.093 & 0.093 & 0.952 & 0.360 \\
$\bb_{7}$ & 7.780 & 7.780 & 501 & 651 & -0.002 & 0.094 & 0.094 & 0.951 & 0.365 \\
\bottomrule\bottomrule
\end{tabular}}
    \caption{Intention-to-Treat Effects (Simulations, Linear Heteroskedastic, Unknown-Kink Bandwidth, ITT Bandwidth Selector).}
    \label{tab:sa-simuls-itt-linear-heteroskedastic-kinkon-bwitt}
\end{table}

\begin{table}[p]
    \centering
    \resizebox{.8\textwidth}{!}{\begin{tabular}{lccccrrrrr}
\toprule\toprule
 & $h_0$ & $h_1$ & $N_{\mathrm{Co}}$ & $N_{\mathrm{Tr}}$ & Bias & SD & RMSE & EC & IL \\
\midrule
$\bb_{1}$ & 18.920 & 18.920 & 2042 & 4036 & -0.002 & 0.048 & 0.048 & 0.954 & 0.318 \\
$\bb_{2}$ & 18.614 & 18.614 & 2065 & 3924 & -0.001 & 0.047 & 0.047 & 0.963 & 0.317 \\
$\bb_{3}$ & 18.397 & 18.397 & 2110 & 3794 & -0.001 & 0.048 & 0.048 & 0.960 & 0.316 \\
$\bb_{4}$ & 18.284 & 18.284 & 2182 & 3647 & -0.000 & 0.048 & 0.048 & 0.958 & 0.314 \\
$\bb_{5}$ & 18.172 & 18.172 & 2256 & 3454 & 0.002 & 0.049 & 0.049 & 0.955 & 0.314 \\
$\bb_{6}$ & 18.104 & 18.104 & 2332 & 3236 & 0.004 & 0.049 & 0.049 & 0.957 & 0.314 \\
$\bb_{7}$ & 17.715 & 17.715 & 2330 & 2898 & 0.005 & 0.051 & 0.051 & 0.954 & 0.321 \\
\bottomrule\bottomrule
\end{tabular}}
    \caption{Intention-to-Treat Effects (Simulations, Linear Heteroskedastic, \texttt{rdrobust} Bandwidth, ITT Bandwidth Selector).}
    \label{tab:sa-simuls-itt-linear-heteroskedastic-rdrobustadj-bwitt}
\end{table}

\begin{table}[p]
    \centering
    \resizebox{.8\textwidth}{!}{\begin{tabular}{lccccrrrrr}
\toprule\toprule
 & $h_0$ & $h_1$ & $N_{\mathrm{Co}}$ & $N_{\mathrm{Tr}}$ & Bias & SD & RMSE & EC & IL \\
\midrule
$\bb_{1}$ & 17.945 & 17.945 & 1903 & 3648 & -0.002 & 0.035 & 0.035 & 0.948 & 0.262 \\
$\bb_{2}$ & 17.380 & 17.380 & 1868 & 3459 & -0.002 & 0.036 & 0.036 & 0.950 & 0.268 \\
$\bb_{3}$ & 17.202 & 17.202 & 1891 & 3370 & -0.001 & 0.037 & 0.037 & 0.959 & 0.269 \\
$\bb_{4}$ & 17.380 & 17.380 & 1997 & 3341 & 0.002 & 0.039 & 0.039 & 0.956 & 0.267 \\
$\bb_{5}$ & 17.857 & 17.857 & 2186 & 3339 & 0.009 & 0.039 & 0.040 & 0.954 & 0.260 \\
$\bb_{6}$ & 18.430 & 18.430 & 2397 & 3310 & 0.016 & 0.038 & 0.042 & 0.951 & 0.254 \\
$\bb_{7}$ & 18.889 & 18.889 & 2574 & 3202 & 0.021 & 0.037 & 0.043 & 0.948 & 0.251 \\
\bottomrule\bottomrule
\end{tabular}}
    \caption{Intention-to-Treat Effects (Simulations, Quadratic Homoskedastic, Smooth-Boundary Bandwidth, ITT Bandwidth Selector).}
    \label{tab:sa-simuls-itt-quadratic-homoskedastic-kinkoff-bwitt}
\end{table}

\begin{table}[p]
    \centering
    \resizebox{.8\textwidth}{!}{\begin{tabular}{lccccrrrrr}
\toprule\toprule
 & $h_0$ & $h_1$ & $N_{\mathrm{Co}}$ & $N_{\mathrm{Tr}}$ & Bias & SD & RMSE & EC & IL \\
\midrule
$\bb_{1}$ & 17.655 & 17.655 & 1862 & 3533 & -0.002 & 0.036 & 0.036 & 0.949 & 0.265 \\
$\bb_{2}$ & 16.710 & 16.710 & 1760 & 3207 & -0.001 & 0.037 & 0.037 & 0.950 & 0.275 \\
$\bb_{3}$ & 15.625 & 15.625 & 1619 & 2818 & -0.001 & 0.039 & 0.039 & 0.954 & 0.291 \\
$\bb_{4}$ & 13.979 & 13.979 & 1379 & 2252 & -0.000 & 0.043 & 0.043 & 0.948 & 0.324 \\
$\bb_{5}$ & 12.000 & 12.000 & 1080 & 1642 & 0.000 & 0.050 & 0.050 & 0.950 & 0.377 \\
$\bb_{6}$ & 10.000 & 10.000 & 787 & 1114 & 0.000 & 0.060 & 0.060 & 0.950 & 0.453 \\
$\bb_{7}$ & 8.509 & 8.509 & 589 & 775 & -0.000 & 0.073 & 0.073 & 0.940 & 0.538 \\
\bottomrule\bottomrule
\end{tabular}}
    \caption{Intention-to-Treat Effects (Simulations, Quadratic Homoskedastic, Kink-Adaptive Bandwidth, ITT Bandwidth Selector).}
    \label{tab:sa-simuls-itt-quadratic-homoskedastic-adaptive-bwitt}
\end{table}

\begin{table}[p]
    \centering
    \resizebox{.8\textwidth}{!}{\begin{tabular}{lccccrrrrr}
\toprule\toprule
 & $h_0$ & $h_1$ & $N_{\mathrm{Co}}$ & $N_{\mathrm{Tr}}$ & Bias & SD & RMSE & EC & IL \\
\midrule
$\bb_{1}$ & 7.862 & 7.862 & 525 & 695 & 0.001 & 0.079 & 0.079 & 0.944 & 0.302 \\
$\bb_{2}$ & 7.614 & 7.614 & 504 & 661 & 0.001 & 0.082 & 0.082 & 0.941 & 0.310 \\
$\bb_{3}$ & 7.536 & 7.536 & 498 & 652 & -0.000 & 0.080 & 0.080 & 0.949 & 0.312 \\
$\bb_{4}$ & 7.614 & 7.614 & 507 & 666 & -0.001 & 0.080 & 0.080 & 0.947 & 0.309 \\
$\bb_{5}$ & 7.823 & 7.823 & 527 & 697 & -0.000 & 0.077 & 0.077 & 0.953 & 0.301 \\
$\bb_{6}$ & 8.074 & 8.074 & 547 & 729 & -0.000 & 0.076 & 0.076 & 0.951 & 0.294 \\
$\bb_{7}$ & 8.275 & 8.275 & 563 & 736 & -0.000 & 0.076 & 0.076 & 0.950 & 0.290 \\
\bottomrule\bottomrule
\end{tabular}}
    \caption{Intention-to-Treat Effects (Simulations, Quadratic Homoskedastic, Unknown-Kink Bandwidth, ITT Bandwidth Selector).}
    \label{tab:sa-simuls-itt-quadratic-homoskedastic-kinkon-bwitt}
\end{table}

\begin{table}[p]
    \centering
    \resizebox{.8\textwidth}{!}{\begin{tabular}{lccccrrrrr}
\toprule\toprule
 & $h_0$ & $h_1$ & $N_{\mathrm{Co}}$ & $N_{\mathrm{Tr}}$ & Bias & SD & RMSE & EC & IL \\
\midrule
$\bb_{1}$ & 17.303 & 17.303 & 1816 & 3409 & -0.003 & 0.038 & 0.038 & 0.950 & 0.271 \\
$\bb_{2}$ & 16.910 & 16.910 & 1799 & 3295 & -0.004 & 0.039 & 0.039 & 0.951 & 0.276 \\
$\bb_{3}$ & 16.825 & 16.825 & 1837 & 3247 & -0.003 & 0.040 & 0.040 & 0.959 & 0.278 \\
$\bb_{4}$ & 17.058 & 17.058 & 1950 & 3248 & 0.000 & 0.042 & 0.042 & 0.963 & 0.276 \\
$\bb_{5}$ & 17.578 & 17.578 & 2135 & 3278 & 0.006 & 0.043 & 0.044 & 0.958 & 0.270 \\
$\bb_{6}$ & 18.313 & 18.313 & 2368 & 3312 & 0.013 & 0.043 & 0.045 & 0.954 & 0.262 \\
$\bb_{7}$ & 19.002 & 19.002 & 2581 & 3280 & 0.019 & 0.041 & 0.045 & 0.950 & 0.256 \\
\bottomrule\bottomrule
\end{tabular}}
    \caption{Intention-to-Treat Effects (Simulations, Quadratic Homoskedastic, \texttt{rdrobust} Bandwidth, ITT Bandwidth Selector).}
    \label{tab:sa-simuls-itt-quadratic-homoskedastic-rdrobustadj-bwitt}
\end{table}

\begin{table}[p]
    \centering
    \resizebox{.8\textwidth}{!}{\begin{tabular}{lccccrrrrr}
\toprule\toprule
 & $h_0$ & $h_1$ & $N_{\mathrm{Co}}$ & $N_{\mathrm{Tr}}$ & Bias & SD & RMSE & EC & IL \\
\midrule
$\bb_{1}$ & 18.213 & 18.213 & 1939 & 3752 & -0.004 & 0.040 & 0.040 & 0.953 & 0.286 \\
$\bb_{2}$ & 17.557 & 17.557 & 1894 & 3525 & -0.003 & 0.040 & 0.040 & 0.956 & 0.289 \\
$\bb_{3}$ & 17.203 & 17.203 & 1893 & 3370 & -0.002 & 0.041 & 0.041 & 0.958 & 0.289 \\
$\bb_{4}$ & 17.159 & 17.159 & 1958 & 3268 & 0.002 & 0.042 & 0.042 & 0.953 & 0.285 \\
$\bb_{5}$ & 17.317 & 17.317 & 2078 & 3171 & 0.008 & 0.042 & 0.043 & 0.950 & 0.279 \\
$\bb_{6}$ & 17.528 & 17.528 & 2213 & 3041 & 0.015 & 0.041 & 0.044 & 0.951 & 0.274 \\
$\bb_{7}$ & 17.616 & 17.616 & 2312 & 2838 & 0.020 & 0.040 & 0.045 & 0.952 & 0.272 \\
\bottomrule\bottomrule
\end{tabular}}
    \caption{Intention-to-Treat Effects (Simulations, Quadratic Heteroskedastic, Smooth-Boundary Bandwidth, ITT Bandwidth Selector).}
    \label{tab:sa-simuls-itt-quadratic-heteroskedastic-kinkoff-bwitt}
\end{table}

\begin{table}[p]
    \centering
    \resizebox{.8\textwidth}{!}{\begin{tabular}{lccccrrrrr}
\toprule\toprule
 & $h_0$ & $h_1$ & $N_{\mathrm{Co}}$ & $N_{\mathrm{Tr}}$ & Bias & SD & RMSE & EC & IL \\
\midrule
$\bb_{1}$ & 17.887 & 17.887 & 1892 & 3622 & -0.003 & 0.040 & 0.040 & 0.952 & 0.290 \\
$\bb_{2}$ & 16.808 & 16.808 & 1774 & 3244 & -0.002 & 0.041 & 0.041 & 0.956 & 0.299 \\
$\bb_{3}$ & 15.591 & 15.591 & 1615 & 2807 & -0.001 & 0.043 & 0.043 & 0.954 & 0.313 \\
$\bb_{4}$ & 13.963 & 13.963 & 1377 & 2248 & 0.000 & 0.047 & 0.047 & 0.945 & 0.342 \\
$\bb_{5}$ & 12.000 & 12.000 & 1082 & 1642 & 0.001 & 0.053 & 0.053 & 0.939 & 0.392 \\
$\bb_{6}$ & 10.000 & 10.000 & 788 & 1113 & 0.002 & 0.062 & 0.062 & 0.940 & 0.466 \\
$\bb_{7}$ & 8.168 & 8.168 & 543 & 715 & 0.001 & 0.077 & 0.077 & 0.941 & 0.568 \\
\bottomrule\bottomrule
\end{tabular}}
    \caption{Intention-to-Treat Effects (Simulations, Quadratic Heteroskedastic, Kink-Adaptive Bandwidth, ITT Bandwidth Selector).}
    \label{tab:sa-simuls-itt-quadratic-heteroskedastic-adaptive-bwitt}
\end{table}

\begin{table}[p]
    \centering
    \resizebox{.8\textwidth}{!}{\begin{tabular}{lccccrrrrr}
\toprule\toprule
 & $h_0$ & $h_1$ & $N_{\mathrm{Co}}$ & $N_{\mathrm{Tr}}$ & Bias & SD & RMSE & EC & IL \\
\midrule
$\bb_{1}$ & 7.979 & 7.979 & 540 & 716 & -0.003 & 0.084 & 0.084 & 0.950 & 0.327 \\
$\bb_{2}$ & 7.692 & 7.692 & 514 & 675 & -0.001 & 0.084 & 0.084 & 0.954 & 0.331 \\
$\bb_{3}$ & 7.537 & 7.537 & 499 & 652 & -0.000 & 0.084 & 0.084 & 0.955 & 0.332 \\
$\bb_{4}$ & 7.517 & 7.517 & 497 & 648 & -0.000 & 0.086 & 0.086 & 0.944 & 0.328 \\
$\bb_{5}$ & 7.587 & 7.587 & 500 & 653 & 0.001 & 0.085 & 0.085 & 0.945 & 0.322 \\
$\bb_{6}$ & 7.679 & 7.679 & 501 & 656 & 0.001 & 0.083 & 0.083 & 0.949 & 0.317 \\
$\bb_{7}$ & 7.718 & 7.718 & 493 & 643 & 0.001 & 0.083 & 0.083 & 0.951 & 0.316 \\
\bottomrule\bottomrule
\end{tabular}}
    \caption{Intention-to-Treat Effects (Simulations, Quadratic Heteroskedastic, Unknown-Kink Bandwidth, ITT Bandwidth Selector).}
    \label{tab:sa-simuls-itt-quadratic-heteroskedastic-kinkon-bwitt}
\end{table}

\begin{table}[p]
    \centering
    \resizebox{.8\textwidth}{!}{\begin{tabular}{lccccrrrrr}
\toprule\toprule
 & $h_0$ & $h_1$ & $N_{\mathrm{Co}}$ & $N_{\mathrm{Tr}}$ & Bias & SD & RMSE & EC & IL \\
\midrule
$\bb_{1}$ & 17.349 & 17.349 & 1823 & 3432 & -0.005 & 0.044 & 0.044 & 0.957 & 0.301 \\
$\bb_{2}$ & 16.879 & 16.879 & 1796 & 3288 & -0.004 & 0.044 & 0.044 & 0.959 & 0.302 \\
$\bb_{3}$ & 16.600 & 16.600 & 1802 & 3173 & -0.004 & 0.044 & 0.044 & 0.960 & 0.302 \\
$\bb_{4}$ & 16.606 & 16.606 & 1872 & 3104 & -0.001 & 0.046 & 0.046 & 0.958 & 0.299 \\
$\bb_{5}$ & 16.893 & 16.893 & 2007 & 3068 & 0.005 & 0.048 & 0.048 & 0.951 & 0.293 \\
$\bb_{6}$ & 17.267 & 17.267 & 2163 & 3001 & 0.011 & 0.047 & 0.049 & 0.952 & 0.286 \\
$\bb_{7}$ & 17.265 & 17.265 & 2238 & 2782 & 0.016 & 0.047 & 0.050 & 0.952 & 0.285 \\
\bottomrule\bottomrule
\end{tabular}}
    \caption{Intention-to-Treat Effects (Simulations, Quadratic Heteroskedastic, \texttt{rdrobust} Bandwidth, ITT Bandwidth Selector).}
    \label{tab:sa-simuls-itt-quadratic-heteroskedastic-rdrobustadj-bwitt}
\end{table}

\clearpage

\begin{table}[p]
    \centering
    \resizebox{.8\textwidth}{!}{\begin{tabular}{lccccrrrrr}
\toprule\toprule
 & $h_0$ & $h_1$ & $N_{\mathrm{Co}}$ & $N_{\mathrm{Tr}}$ & Bias & SD & RMSE & EC & IL \\
\midrule
$\bb_{1}$ & 19.534 & 19.534 & 2120 & 4260 & -0.002 & 0.027 & 0.028 & 0.949 & 0.207 \\
$\bb_{2}$ & 19.152 & 19.152 & 2138 & 4114 & -0.002 & 0.028 & 0.028 & 0.945 & 0.210 \\
$\bb_{3}$ & 18.902 & 18.902 & 2185 & 3961 & -0.002 & 0.028 & 0.028 & 0.944 & 0.212 \\
$\bb_{4}$ & 18.805 & 18.805 & 2271 & 3805 & -0.002 & 0.029 & 0.029 & 0.945 & 0.214 \\
$\bb_{5}$ & 18.832 & 18.832 & 2383 & 3639 & -0.002 & 0.029 & 0.029 & 0.945 & 0.215 \\
$\bb_{6}$ & 19.003 & 19.003 & 2517 & 3478 & -0.002 & 0.029 & 0.029 & 0.945 & 0.216 \\
$\bb_{7}$ & 19.115 & 19.115 & 2621 & 3265 & -0.002 & 0.030 & 0.030 & 0.944 & 0.219 \\
\bottomrule\bottomrule
\end{tabular}}
    \caption{First-Stage Effects (Simulations, Linear Homoskedastic, Smooth-Boundary Bandwidth, ITT Bandwidth Selector).}
    \label{tab:sa-simuls-fs-linear-homoskedastic-kinkoff-bwitt}
\end{table}

\begin{table}[p]
    \centering
    \resizebox{.8\textwidth}{!}{\begin{tabular}{lccccrrrrr}
\toprule\toprule
 & $h_0$ & $h_1$ & $N_{\mathrm{Co}}$ & $N_{\mathrm{Tr}}$ & Bias & SD & RMSE & EC & IL \\
\midrule
$\bb_{1}$ & 18.862 & 18.862 & 2022 & 3994 & -0.002 & 0.028 & 0.028 & 0.948 & 0.213 \\
$\bb_{2}$ & 17.546 & 17.546 & 1877 & 3512 & -0.001 & 0.030 & 0.030 & 0.948 & 0.226 \\
$\bb_{3}$ & 15.909 & 15.909 & 1661 & 2914 & -0.001 & 0.033 & 0.033 & 0.943 & 0.248 \\
$\bb_{4}$ & 13.997 & 13.997 & 1382 & 2257 & -0.000 & 0.038 & 0.038 & 0.942 & 0.282 \\
$\bb_{5}$ & 12.000 & 12.000 & 1081 & 1641 & -0.001 & 0.044 & 0.044 & 0.940 & 0.331 \\
$\bb_{6}$ & 10.000 & 10.000 & 788 & 1113 & -0.001 & 0.053 & 0.053 & 0.948 & 0.401 \\
$\bb_{7}$ & 8.562 & 8.562 & 596 & 783 & -0.002 & 0.062 & 0.063 & 0.941 & 0.477 \\
\bottomrule\bottomrule
\end{tabular}}
    \caption{First-Stage Effects (Simulations, Linear Homoskedastic, Kink-Adaptive Bandwidth, ITT Bandwidth Selector).}
    \label{tab:sa-simuls-fs-linear-homoskedastic-adaptive-bwitt}
\end{table}

\begin{table}[p]
    \centering
    \resizebox{.8\textwidth}{!}{\begin{tabular}{lccccrrrrr}
\toprule\toprule
 & $h_0$ & $h_1$ & $N_{\mathrm{Co}}$ & $N_{\mathrm{Tr}}$ & Bias & SD & RMSE & EC & IL \\
\midrule
$\bb_{1}$ & 8.558 & 8.558 & 609 & 823 & -0.001 & 0.064 & 0.064 & 0.948 & 0.244 \\
$\bb_{2}$ & 8.391 & 8.391 & 598 & 803 & -0.001 & 0.064 & 0.064 & 0.946 & 0.247 \\
$\bb_{3}$ & 8.281 & 8.281 & 589 & 788 & -0.001 & 0.065 & 0.065 & 0.943 & 0.250 \\
$\bb_{4}$ & 8.239 & 8.239 & 583 & 779 & -0.000 & 0.066 & 0.066 & 0.939 & 0.251 \\
$\bb_{5}$ & 8.251 & 8.251 & 579 & 773 & -0.001 & 0.066 & 0.066 & 0.941 & 0.251 \\
$\bb_{6}$ & 8.325 & 8.325 & 577 & 772 & -0.002 & 0.065 & 0.065 & 0.946 & 0.251 \\
$\bb_{7}$ & 8.375 & 8.375 & 575 & 752 & -0.003 & 0.065 & 0.065 & 0.950 & 0.253 \\
\bottomrule\bottomrule
\end{tabular}}
    \caption{First-Stage Effects (Simulations, Linear Homoskedastic, Unknown-Kink Bandwidth, ITT Bandwidth Selector).}
    \label{tab:sa-simuls-fs-linear-homoskedastic-kinkon-bwitt}
\end{table}

\begin{table}[p]
    \centering
    \resizebox{.8\textwidth}{!}{\begin{tabular}{lccccrrrrr}
\toprule\toprule
 & $h_0$ & $h_1$ & $N_{\mathrm{Co}}$ & $N_{\mathrm{Tr}}$ & Bias & SD & RMSE & EC & IL \\
\midrule
$\bb_{1}$ & 19.325 & 19.325 & 2100 & 4182 & -0.002 & 0.029 & 0.029 & 0.949 & 0.212 \\
$\bb_{2}$ & 19.171 & 19.171 & 2152 & 4123 & -0.002 & 0.029 & 0.029 & 0.948 & 0.213 \\
$\bb_{3}$ & 19.127 & 19.127 & 2235 & 4046 & -0.002 & 0.029 & 0.029 & 0.948 & 0.213 \\
$\bb_{4}$ & 19.198 & 19.198 & 2351 & 3949 & -0.003 & 0.029 & 0.030 & 0.948 & 0.213 \\
$\bb_{5}$ & 19.404 & 19.404 & 2495 & 3845 & -0.003 & 0.030 & 0.030 & 0.945 & 0.213 \\
$\bb_{6}$ & 19.638 & 19.638 & 2635 & 3703 & -0.002 & 0.030 & 0.030 & 0.946 & 0.214 \\
$\bb_{7}$ & 19.901 & 19.901 & 2763 & 3534 & -0.002 & 0.030 & 0.030 & 0.945 & 0.215 \\
\bottomrule\bottomrule
\end{tabular}}
    \caption{First-Stage Effects (Simulations, Linear Homoskedastic, \texttt{rdrobust} Bandwidth, ITT Bandwidth Selector).}
    \label{tab:sa-simuls-fs-linear-homoskedastic-rdrobustadj-bwitt}
\end{table}

\begin{table}[p]
    \centering
    \resizebox{.8\textwidth}{!}{\begin{tabular}{lccccrrrrr}
\toprule\toprule
 & $h_0$ & $h_1$ & $N_{\mathrm{Co}}$ & $N_{\mathrm{Tr}}$ & Bias & SD & RMSE & EC & IL \\
\midrule
$\bb_{1}$ & 19.822 & 19.822 & 2159 & 4374 & -0.001 & 0.027 & 0.027 & 0.949 & 0.204 \\
$\bb_{2}$ & 19.210 & 19.210 & 2146 & 4137 & -0.001 & 0.027 & 0.028 & 0.951 & 0.209 \\
$\bb_{3}$ & 18.737 & 18.737 & 2155 & 3903 & -0.001 & 0.028 & 0.028 & 0.950 & 0.214 \\
$\bb_{4}$ & 18.364 & 18.364 & 2185 & 3658 & -0.001 & 0.029 & 0.029 & 0.949 & 0.218 \\
$\bb_{5}$ & 18.091 & 18.091 & 2234 & 3405 & -0.002 & 0.030 & 0.030 & 0.948 & 0.223 \\
$\bb_{6}$ & 17.917 & 17.917 & 2293 & 3150 & -0.002 & 0.031 & 0.031 & 0.942 & 0.229 \\
$\bb_{7}$ & 17.759 & 17.759 & 2342 & 2876 & -0.001 & 0.032 & 0.032 & 0.945 & 0.235 \\
\bottomrule\bottomrule
\end{tabular}}
    \caption{First-Stage Effects (Simulations, Linear Heteroskedastic, Smooth-Boundary Bandwidth, ITT Bandwidth Selector).}
    \label{tab:sa-simuls-fs-linear-heteroskedastic-kinkoff-bwitt}
\end{table}

\begin{table}[p]
    \centering
    \resizebox{.8\textwidth}{!}{\begin{tabular}{lccccrrrrr}
\toprule\toprule
 & $h_0$ & $h_1$ & $N_{\mathrm{Co}}$ & $N_{\mathrm{Tr}}$ & Bias & SD & RMSE & EC & IL \\
\midrule
$\bb_{1}$ & 19.027 & 19.027 & 2044 & 4060 & -0.001 & 0.028 & 0.028 & 0.951 & 0.211 \\
$\bb_{2}$ & 17.579 & 17.579 & 1882 & 3526 & -0.001 & 0.029 & 0.029 & 0.951 & 0.226 \\
$\bb_{3}$ & 15.902 & 15.902 & 1660 & 2912 & -0.000 & 0.032 & 0.032 & 0.952 & 0.248 \\
$\bb_{4}$ & 13.995 & 13.995 & 1382 & 2257 & -0.000 & 0.037 & 0.037 & 0.946 & 0.282 \\
$\bb_{5}$ & 12.000 & 12.000 & 1081 & 1641 & -0.000 & 0.043 & 0.043 & 0.945 & 0.331 \\
$\bb_{6}$ & 10.000 & 10.000 & 788 & 1112 & -0.001 & 0.052 & 0.052 & 0.941 & 0.401 \\
$\bb_{7}$ & 8.181 & 8.181 & 545 & 717 & -0.000 & 0.066 & 0.066 & 0.941 & 0.498 \\
\bottomrule\bottomrule
\end{tabular}}
    \caption{First-Stage Effects (Simulations, Linear Heteroskedastic, Kink-Adaptive Bandwidth, ITT Bandwidth Selector).}
    \label{tab:sa-simuls-fs-linear-heteroskedastic-adaptive-bwitt}
\end{table}

\begin{table}[p]
    \centering
    \resizebox{.8\textwidth}{!}{\begin{tabular}{lccccrrrrr}
\toprule\toprule
 & $h_0$ & $h_1$ & $N_{\mathrm{Co}}$ & $N_{\mathrm{Tr}}$ & Bias & SD & RMSE & EC & IL \\
\midrule
$\bb_{1}$ & 8.684 & 8.684 & 625 & 848 & -0.001 & 0.062 & 0.061 & 0.949 & 0.241 \\
$\bb_{2}$ & 8.416 & 8.416 & 601 & 807 & -0.000 & 0.062 & 0.062 & 0.953 & 0.247 \\
$\bb_{3}$ & 8.209 & 8.209 & 580 & 773 & -0.000 & 0.065 & 0.065 & 0.947 & 0.252 \\
$\bb_{4}$ & 8.045 & 8.045 & 559 & 740 & -0.000 & 0.067 & 0.067 & 0.941 & 0.256 \\
$\bb_{5}$ & 7.926 & 7.926 & 539 & 711 & -0.001 & 0.068 & 0.068 & 0.944 & 0.261 \\
$\bb_{6}$ & 7.850 & 7.850 & 521 & 684 & -0.001 & 0.068 & 0.068 & 0.946 & 0.266 \\
$\bb_{7}$ & 7.780 & 7.780 & 501 & 651 & -0.000 & 0.070 & 0.070 & 0.942 & 0.272 \\
\bottomrule\bottomrule
\end{tabular}}
    \caption{First-Stage Effects (Simulations, Linear Heteroskedastic, Unknown-Kink Bandwidth, ITT Bandwidth Selector).}
    \label{tab:sa-simuls-fs-linear-heteroskedastic-kinkon-bwitt}
\end{table}

\begin{table}[p]
    \centering
    \resizebox{.8\textwidth}{!}{\begin{tabular}{lccccrrrrr}
\toprule\toprule
 & $h_0$ & $h_1$ & $N_{\mathrm{Co}}$ & $N_{\mathrm{Tr}}$ & Bias & SD & RMSE & EC & IL \\
\midrule
$\bb_{1}$ & 18.920 & 18.920 & 2042 & 4036 & -0.002 & 0.029 & 0.029 & 0.947 & 0.217 \\
$\bb_{2}$ & 18.614 & 18.614 & 2065 & 3924 & -0.002 & 0.029 & 0.029 & 0.952 & 0.220 \\
$\bb_{3}$ & 18.397 & 18.397 & 2110 & 3794 & -0.002 & 0.030 & 0.030 & 0.949 & 0.222 \\
$\bb_{4}$ & 18.284 & 18.284 & 2182 & 3647 & -0.002 & 0.031 & 0.031 & 0.949 & 0.224 \\
$\bb_{5}$ & 18.172 & 18.172 & 2256 & 3454 & -0.002 & 0.031 & 0.031 & 0.950 & 0.227 \\
$\bb_{6}$ & 18.104 & 18.104 & 2332 & 3236 & -0.002 & 0.032 & 0.032 & 0.945 & 0.232 \\
$\bb_{7}$ & 17.715 & 17.715 & 2330 & 2898 & -0.002 & 0.034 & 0.034 & 0.945 & 0.241 \\
\bottomrule\bottomrule
\end{tabular}}
    \caption{First-Stage Effects (Simulations, Linear Heteroskedastic, \texttt{rdrobust} Bandwidth, ITT Bandwidth Selector).}
    \label{tab:sa-simuls-fs-linear-heteroskedastic-rdrobustadj-bwitt}
\end{table}

\begin{table}[p]
    \centering
    \resizebox{.8\textwidth}{!}{\begin{tabular}{lccccrrrrr}
\toprule\toprule
 & $h_0$ & $h_1$ & $N_{\mathrm{Co}}$ & $N_{\mathrm{Tr}}$ & Bias & SD & RMSE & EC & IL \\
\midrule
$\bb_{1}$ & 17.945 & 17.945 & 1903 & 3648 & -0.005 & 0.030 & 0.030 & 0.947 & 0.226 \\
$\bb_{2}$ & 17.380 & 17.380 & 1868 & 3459 & -0.004 & 0.030 & 0.031 & 0.950 & 0.232 \\
$\bb_{3}$ & 17.202 & 17.202 & 1891 & 3370 & -0.004 & 0.031 & 0.031 & 0.947 & 0.234 \\
$\bb_{4}$ & 17.380 & 17.380 & 1997 & 3341 & -0.002 & 0.031 & 0.031 & 0.947 & 0.232 \\
$\bb_{5}$ & 17.857 & 17.857 & 2186 & 3339 & 0.000 & 0.031 & 0.031 & 0.947 & 0.227 \\
$\bb_{6}$ & 18.430 & 18.430 & 2397 & 3310 & 0.003 & 0.030 & 0.030 & 0.946 & 0.223 \\
$\bb_{7}$ & 18.889 & 18.889 & 2574 & 3202 & 0.006 & 0.030 & 0.030 & 0.947 & 0.221 \\
\bottomrule\bottomrule
\end{tabular}}
    \caption{First-Stage Effects (Simulations, Quadratic Homoskedastic, Smooth-Boundary Bandwidth, ITT Bandwidth Selector).}
    \label{tab:sa-simuls-fs-quadratic-homoskedastic-kinkoff-bwitt}
\end{table}

\begin{table}[p]
    \centering
    \resizebox{.8\textwidth}{!}{\begin{tabular}{lccccrrrrr}
\toprule\toprule
 & $h_0$ & $h_1$ & $N_{\mathrm{Co}}$ & $N_{\mathrm{Tr}}$ & Bias & SD & RMSE & EC & IL \\
\midrule
$\bb_{1}$ & 17.655 & 17.655 & 1862 & 3533 & -0.005 & 0.030 & 0.030 & 0.946 & 0.229 \\
$\bb_{2}$ & 16.710 & 16.710 & 1760 & 3207 & -0.004 & 0.031 & 0.031 & 0.951 & 0.239 \\
$\bb_{3}$ & 15.625 & 15.625 & 1619 & 2818 & -0.003 & 0.033 & 0.033 & 0.947 & 0.253 \\
$\bb_{4}$ & 13.979 & 13.979 & 1379 & 2252 & -0.002 & 0.037 & 0.037 & 0.944 & 0.283 \\
$\bb_{5}$ & 12.000 & 12.000 & 1080 & 1642 & -0.001 & 0.044 & 0.044 & 0.942 & 0.331 \\
$\bb_{6}$ & 10.000 & 10.000 & 787 & 1114 & -0.000 & 0.053 & 0.053 & 0.943 & 0.400 \\
$\bb_{7}$ & 8.509 & 8.509 & 589 & 775 & -0.000 & 0.064 & 0.064 & 0.935 & 0.477 \\
\bottomrule\bottomrule
\end{tabular}}
    \caption{First-Stage Effects (Simulations, Quadratic Homoskedastic, Kink-Adaptive Bandwidth, ITT Bandwidth Selector).}
    \label{tab:sa-simuls-fs-quadratic-homoskedastic-adaptive-bwitt}
\end{table}

\begin{table}[p]
    \centering
    \resizebox{.8\textwidth}{!}{\begin{tabular}{lccccrrrrr}
\toprule\toprule
 & $h_0$ & $h_1$ & $N_{\mathrm{Co}}$ & $N_{\mathrm{Tr}}$ & Bias & SD & RMSE & EC & IL \\
\midrule
$\bb_{1}$ & 7.862 & 7.862 & 525 & 695 & 0.000 & 0.069 & 0.069 & 0.943 & 0.268 \\
$\bb_{2}$ & 7.614 & 7.614 & 504 & 661 & 0.000 & 0.071 & 0.071 & 0.944 & 0.274 \\
$\bb_{3}$ & 7.536 & 7.536 & 498 & 652 & -0.000 & 0.072 & 0.072 & 0.942 & 0.276 \\
$\bb_{4}$ & 7.614 & 7.614 & 507 & 666 & -0.001 & 0.070 & 0.070 & 0.946 & 0.273 \\
$\bb_{5}$ & 7.823 & 7.823 & 527 & 697 & -0.000 & 0.069 & 0.069 & 0.944 & 0.266 \\
$\bb_{6}$ & 8.074 & 8.074 & 547 & 729 & -0.000 & 0.067 & 0.067 & 0.944 & 0.259 \\
$\bb_{7}$ & 8.275 & 8.275 & 563 & 736 & -0.000 & 0.067 & 0.067 & 0.944 & 0.255 \\
\bottomrule\bottomrule
\end{tabular}}
    \caption{First-Stage Effects (Simulations, Quadratic Homoskedastic, Unknown-Kink Bandwidth, ITT Bandwidth Selector).}
    \label{tab:sa-simuls-fs-quadratic-homoskedastic-kinkon-bwitt}
\end{table}

\begin{table}[p]
    \centering
    \resizebox{.8\textwidth}{!}{\begin{tabular}{lccccrrrrr}
\toprule\toprule
 & $h_0$ & $h_1$ & $N_{\mathrm{Co}}$ & $N_{\mathrm{Tr}}$ & Bias & SD & RMSE & EC & IL \\
\midrule
$\bb_{1}$ & 17.303 & 17.303 & 1816 & 3409 & -0.005 & 0.031 & 0.032 & 0.946 & 0.235 \\
$\bb_{2}$ & 16.910 & 16.910 & 1799 & 3295 & -0.005 & 0.032 & 0.032 & 0.952 & 0.240 \\
$\bb_{3}$ & 16.825 & 16.825 & 1837 & 3247 & -0.004 & 0.032 & 0.032 & 0.950 & 0.242 \\
$\bb_{4}$ & 17.058 & 17.058 & 1950 & 3248 & -0.003 & 0.032 & 0.033 & 0.948 & 0.240 \\
$\bb_{5}$ & 17.578 & 17.578 & 2135 & 3278 & -0.000 & 0.032 & 0.032 & 0.949 & 0.236 \\
$\bb_{6}$ & 18.313 & 18.313 & 2368 & 3312 & 0.003 & 0.032 & 0.032 & 0.945 & 0.230 \\
$\bb_{7}$ & 19.002 & 19.002 & 2581 & 3280 & 0.005 & 0.032 & 0.032 & 0.944 & 0.226 \\
\bottomrule\bottomrule
\end{tabular}}
    \caption{First-Stage Effects (Simulations, Quadratic Homoskedastic, \texttt{rdrobust} Bandwidth, ITT Bandwidth Selector).}
    \label{tab:sa-simuls-fs-quadratic-homoskedastic-rdrobustadj-bwitt}
\end{table}

\begin{table}[p]
    \centering
    \resizebox{.8\textwidth}{!}{\begin{tabular}{lccccrrrrr}
\toprule\toprule
 & $h_0$ & $h_1$ & $N_{\mathrm{Co}}$ & $N_{\mathrm{Tr}}$ & Bias & SD & RMSE & EC & IL \\
\midrule
$\bb_{1}$ & 18.213 & 18.213 & 1939 & 3752 & -0.005 & 0.029 & 0.030 & 0.951 & 0.223 \\
$\bb_{2}$ & 17.557 & 17.557 & 1894 & 3525 & -0.005 & 0.030 & 0.030 & 0.947 & 0.230 \\
$\bb_{3}$ & 17.203 & 17.203 & 1893 & 3370 & -0.004 & 0.030 & 0.030 & 0.950 & 0.234 \\
$\bb_{4}$ & 17.159 & 17.159 & 1958 & 3268 & -0.002 & 0.031 & 0.031 & 0.949 & 0.235 \\
$\bb_{5}$ & 17.317 & 17.317 & 2078 & 3171 & 0.000 & 0.031 & 0.031 & 0.945 & 0.234 \\
$\bb_{6}$ & 17.528 & 17.528 & 2213 & 3041 & 0.003 & 0.031 & 0.031 & 0.943 & 0.234 \\
$\bb_{7}$ & 17.616 & 17.616 & 2312 & 2838 & 0.006 & 0.031 & 0.031 & 0.948 & 0.236 \\
\bottomrule\bottomrule
\end{tabular}}
    \caption{First-Stage Effects (Simulations, Quadratic Heteroskedastic, Smooth-Boundary Bandwidth, ITT Bandwidth Selector).}
    \label{tab:sa-simuls-fs-quadratic-heteroskedastic-kinkoff-bwitt}
\end{table}

\begin{table}[p]
    \centering
    \resizebox{.8\textwidth}{!}{\begin{tabular}{lccccrrrrr}
\toprule\toprule
 & $h_0$ & $h_1$ & $N_{\mathrm{Co}}$ & $N_{\mathrm{Tr}}$ & Bias & SD & RMSE & EC & IL \\
\midrule
$\bb_{1}$ & 17.887 & 17.887 & 1892 & 3622 & -0.005 & 0.030 & 0.030 & 0.952 & 0.226 \\
$\bb_{2}$ & 16.808 & 16.808 & 1774 & 3244 & -0.004 & 0.031 & 0.031 & 0.948 & 0.238 \\
$\bb_{3}$ & 15.591 & 15.591 & 1615 & 2807 & -0.003 & 0.033 & 0.033 & 0.952 & 0.254 \\
$\bb_{4}$ & 13.963 & 13.963 & 1377 & 2248 & -0.002 & 0.037 & 0.037 & 0.946 & 0.283 \\
$\bb_{5}$ & 12.000 & 12.000 & 1082 & 1642 & -0.000 & 0.043 & 0.043 & 0.943 & 0.331 \\
$\bb_{6}$ & 10.000 & 10.000 & 788 & 1113 & 0.001 & 0.053 & 0.053 & 0.939 & 0.400 \\
$\bb_{7}$ & 8.168 & 8.168 & 543 & 715 & 0.001 & 0.066 & 0.066 & 0.935 & 0.495 \\
\bottomrule\bottomrule
\end{tabular}}
    \caption{First-Stage Effects (Simulations, Quadratic Heteroskedastic, Kink-Adaptive Bandwidth, ITT Bandwidth Selector).}
    \label{tab:sa-simuls-fs-quadratic-heteroskedastic-adaptive-bwitt}
\end{table}

\begin{table}[p]
    \centering
    \resizebox{.8\textwidth}{!}{\begin{tabular}{lccccrrrrr}
\toprule\toprule
 & $h_0$ & $h_1$ & $N_{\mathrm{Co}}$ & $N_{\mathrm{Tr}}$ & Bias & SD & RMSE & EC & IL \\
\midrule
$\bb_{1}$ & 7.979 & 7.979 & 540 & 716 & -0.003 & 0.069 & 0.069 & 0.942 & 0.264 \\
$\bb_{2}$ & 7.692 & 7.692 & 514 & 675 & -0.002 & 0.070 & 0.070 & 0.945 & 0.272 \\
$\bb_{3}$ & 7.537 & 7.537 & 499 & 652 & -0.002 & 0.070 & 0.070 & 0.952 & 0.276 \\
$\bb_{4}$ & 7.517 & 7.517 & 497 & 648 & -0.000 & 0.071 & 0.071 & 0.945 & 0.276 \\
$\bb_{5}$ & 7.587 & 7.587 & 500 & 653 & 0.002 & 0.071 & 0.071 & 0.944 & 0.274 \\
$\bb_{6}$ & 7.679 & 7.679 & 501 & 656 & 0.002 & 0.070 & 0.070 & 0.945 & 0.272 \\
$\bb_{7}$ & 7.718 & 7.718 & 493 & 643 & 0.001 & 0.071 & 0.071 & 0.944 & 0.273 \\
\bottomrule\bottomrule
\end{tabular}}
    \caption{First-Stage Effects (Simulations, Quadratic Heteroskedastic, Unknown-Kink Bandwidth, ITT Bandwidth Selector).}
    \label{tab:sa-simuls-fs-quadratic-heteroskedastic-kinkon-bwitt}
\end{table}

\begin{table}[p]
    \centering
    \resizebox{.8\textwidth}{!}{\begin{tabular}{lccccrrrrr}
\toprule\toprule
 & $h_0$ & $h_1$ & $N_{\mathrm{Co}}$ & $N_{\mathrm{Tr}}$ & Bias & SD & RMSE & EC & IL \\
\midrule
$\bb_{1}$ & 17.349 & 17.349 & 1823 & 3432 & -0.006 & 0.031 & 0.032 & 0.952 & 0.236 \\
$\bb_{2}$ & 16.879 & 16.879 & 1796 & 3288 & -0.005 & 0.032 & 0.032 & 0.949 & 0.241 \\
$\bb_{3}$ & 16.600 & 16.600 & 1802 & 3173 & -0.004 & 0.032 & 0.032 & 0.956 & 0.246 \\
$\bb_{4}$ & 16.606 & 16.606 & 1872 & 3104 & -0.003 & 0.033 & 0.033 & 0.949 & 0.248 \\
$\bb_{5}$ & 16.893 & 16.893 & 2007 & 3068 & -0.001 & 0.034 & 0.034 & 0.946 & 0.246 \\
$\bb_{6}$ & 17.267 & 17.267 & 2163 & 3001 & 0.002 & 0.034 & 0.034 & 0.946 & 0.244 \\
$\bb_{7}$ & 17.265 & 17.265 & 2238 & 2782 & 0.005 & 0.034 & 0.034 & 0.951 & 0.248 \\
\bottomrule\bottomrule
\end{tabular}}
    \caption{First-Stage Effects (Simulations, Quadratic Heteroskedastic, \texttt{rdrobust} Bandwidth, ITT Bandwidth Selector).}
    \label{tab:sa-simuls-fs-quadratic-heteroskedastic-rdrobustadj-bwitt}
\end{table}

\clearpage

\begin{table}[p]
    \centering
    \resizebox{.8\textwidth}{!}{\begin{tabular}{lccccrrrrr}
\toprule\toprule
 & $h_0$ & $h_1$ & $N_{\mathrm{Co}}$ & $N_{\mathrm{Tr}}$ & Bias & SD & RMSE & EC & IL \\
\midrule
$\bb_{1}$ & 19.534 & 19.534 & 2120 & 4260 & 0.003 & 0.055 & 0.055 & 0.960 & 0.397 \\
$\bb_{2}$ & 19.152 & 19.152 & 2138 & 4114 & 0.003 & 0.056 & 0.056 & 0.956 & 0.400 \\
$\bb_{3}$ & 18.902 & 18.902 & 2185 & 3961 & 0.004 & 0.057 & 0.057 & 0.959 & 0.403 \\
$\bb_{4}$ & 18.805 & 18.805 & 2271 & 3805 & 0.007 & 0.058 & 0.058 & 0.957 & 0.404 \\
$\bb_{5}$ & 18.832 & 18.832 & 2383 & 3639 & 0.011 & 0.058 & 0.059 & 0.960 & 0.403 \\
$\bb_{6}$ & 19.003 & 19.003 & 2517 & 3478 & 0.015 & 0.057 & 0.059 & 0.958 & 0.402 \\
$\bb_{7}$ & 19.115 & 19.115 & 2621 & 3265 & 0.017 & 0.057 & 0.059 & 0.957 & 0.406 \\
\bottomrule\bottomrule
\end{tabular}}
    \caption{Fuzzy Effects (Simulations, Linear Homoskedastic, Smooth-Boundary Bandwidth, ITT Bandwidth Selector).}
    \label{tab:sa-simuls-fuzzy-linear-homoskedastic-kinkoff-bwitt}
\end{table}

\begin{table}[p]
    \centering
    \resizebox{.8\textwidth}{!}{\begin{tabular}{lccccrrrrr}
\toprule\toprule
 & $h_0$ & $h_1$ & $N_{\mathrm{Co}}$ & $N_{\mathrm{Tr}}$ & Bias & SD & RMSE & EC & IL \\
\midrule
$\bb_{1}$ & 18.862 & 18.862 & 2022 & 3994 & 0.004 & 0.056 & 0.056 & 0.957 & 0.408 \\
$\bb_{2}$ & 17.546 & 17.546 & 1877 & 3512 & 0.004 & 0.058 & 0.058 & 0.953 & 0.432 \\
$\bb_{3}$ & 15.909 & 15.909 & 1661 & 2914 & 0.004 & 0.063 & 0.063 & 0.953 & 0.471 \\
$\bb_{4}$ & 13.997 & 13.997 & 1382 & 2257 & 0.003 & 0.070 & 0.070 & 0.958 & 0.535 \\
$\bb_{5}$ & 12.000 & 12.000 & 1081 & 1641 & 0.001 & 0.081 & 0.081 & 0.961 & 0.629 \\
$\bb_{6}$ & 10.000 & 10.000 & 788 & 1113 & -0.000 & 0.097 & 0.097 & 0.957 & 0.774 \\
$\bb_{7}$ & 8.562 & 8.562 & 596 & 783 & 0.001 & 0.118 & 0.118 & 0.967 & 0.956 \\
\bottomrule\bottomrule
\end{tabular}}
    \caption{Fuzzy Effects (Simulations, Linear Homoskedastic, Kink-Adaptive Bandwidth, ITT Bandwidth Selector).}
    \label{tab:sa-simuls-fuzzy-linear-homoskedastic-adaptive-bwitt}
\end{table}

\begin{table}[p]
    \centering
    \resizebox{.8\textwidth}{!}{\begin{tabular}{lccccrrrrr}
\toprule\toprule
 & $h_0$ & $h_1$ & $N_{\mathrm{Co}}$ & $N_{\mathrm{Tr}}$ & Bias & SD & RMSE & EC & IL \\
\midrule
$\bb_{1}$ & 8.558 & 8.558 & 609 & 823 & 0.005 & 0.118 & 0.119 & 0.956 & 0.458 \\
$\bb_{2}$ & 8.391 & 8.391 & 598 & 803 & 0.005 & 0.120 & 0.120 & 0.952 & 0.463 \\
$\bb_{3}$ & 8.281 & 8.281 & 589 & 788 & 0.004 & 0.120 & 0.120 & 0.958 & 0.467 \\
$\bb_{4}$ & 8.239 & 8.239 & 583 & 779 & 0.003 & 0.122 & 0.122 & 0.959 & 0.468 \\
$\bb_{5}$ & 8.251 & 8.251 & 579 & 773 & 0.001 & 0.120 & 0.120 & 0.959 & 0.468 \\
$\bb_{6}$ & 8.325 & 8.325 & 577 & 772 & -0.001 & 0.121 & 0.121 & 0.957 & 0.467 \\
$\bb_{7}$ & 8.375 & 8.375 & 575 & 752 & 0.001 & 0.122 & 0.122 & 0.956 & 0.471 \\
\bottomrule\bottomrule
\end{tabular}}
    \caption{Fuzzy Effects (Simulations, Linear Homoskedastic, Unknown-Kink Bandwidth, ITT Bandwidth Selector).}
    \label{tab:sa-simuls-fuzzy-linear-homoskedastic-kinkon-bwitt}
\end{table}

\begin{table}[p]
    \centering
    \resizebox{.8\textwidth}{!}{\begin{tabular}{lccccrrrrr}
\toprule\toprule
 & $h_0$ & $h_1$ & $N_{\mathrm{Co}}$ & $N_{\mathrm{Tr}}$ & Bias & SD & RMSE & EC & IL \\
\midrule
$\bb_{1}$ & 19.325 & 19.325 & 2100 & 4182 & 0.002 & 0.058 & 0.058 & 0.961 & 0.406 \\
$\bb_{2}$ & 19.171 & 19.171 & 2152 & 4123 & 0.003 & 0.059 & 0.059 & 0.960 & 0.406 \\
$\bb_{3}$ & 19.127 & 19.127 & 2235 & 4046 & 0.004 & 0.059 & 0.060 & 0.960 & 0.404 \\
$\bb_{4}$ & 19.198 & 19.198 & 2351 & 3949 & 0.006 & 0.061 & 0.061 & 0.959 & 0.403 \\
$\bb_{5}$ & 19.404 & 19.404 & 2495 & 3845 & 0.010 & 0.061 & 0.061 & 0.963 & 0.400 \\
$\bb_{6}$ & 19.638 & 19.638 & 2635 & 3703 & 0.013 & 0.060 & 0.061 & 0.959 & 0.399 \\
$\bb_{7}$ & 19.901 & 19.901 & 2763 & 3534 & 0.015 & 0.059 & 0.061 & 0.957 & 0.400 \\
\bottomrule\bottomrule
\end{tabular}}
    \caption{Fuzzy Effects (Simulations, Linear Homoskedastic, \texttt{rdrobust} Bandwidth, ITT Bandwidth Selector).}
    \label{tab:sa-simuls-fuzzy-linear-homoskedastic-rdrobustadj-bwitt}
\end{table}

\begin{table}[p]
    \centering
    \resizebox{.8\textwidth}{!}{\begin{tabular}{lccccrrrrr}
\toprule\toprule
 & $h_0$ & $h_1$ & $N_{\mathrm{Co}}$ & $N_{\mathrm{Tr}}$ & Bias & SD & RMSE & EC & IL \\
\midrule
$\bb_{1}$ & 19.822 & 19.822 & 2159 & 4374 & 0.001 & 0.073 & 0.073 & 0.962 & 0.522 \\
$\bb_{2}$ & 19.210 & 19.210 & 2146 & 4137 & 0.001 & 0.074 & 0.074 & 0.962 & 0.525 \\
$\bb_{3}$ & 18.737 & 18.737 & 2155 & 3903 & 0.002 & 0.074 & 0.074 & 0.963 & 0.526 \\
$\bb_{4}$ & 18.364 & 18.364 & 2185 & 3658 & 0.005 & 0.074 & 0.074 & 0.957 & 0.528 \\
$\bb_{5}$ & 18.091 & 18.091 & 2234 & 3405 & 0.009 & 0.075 & 0.076 & 0.957 & 0.530 \\
$\bb_{6}$ & 17.917 & 17.917 & 2293 & 3150 & 0.013 & 0.076 & 0.077 & 0.961 & 0.531 \\
$\bb_{7}$ & 17.759 & 17.759 & 2342 & 2876 & 0.015 & 0.077 & 0.078 & 0.960 & 0.535 \\
\bottomrule\bottomrule
\end{tabular}}
    \caption{Fuzzy Effects (Simulations, Linear Heteroskedastic, Smooth-Boundary Bandwidth, ITT Bandwidth Selector).}
    \label{tab:sa-simuls-fuzzy-linear-heteroskedastic-kinkoff-bwitt}
\end{table}

\begin{table}[p]
    \centering
    \resizebox{.8\textwidth}{!}{\begin{tabular}{lccccrrrrr}
\toprule\toprule
 & $h_0$ & $h_1$ & $N_{\mathrm{Co}}$ & $N_{\mathrm{Tr}}$ & Bias & SD & RMSE & EC & IL \\
\midrule
$\bb_{1}$ & 19.027 & 19.027 & 2044 & 4060 & 0.001 & 0.074 & 0.074 & 0.962 & 0.540 \\
$\bb_{2}$ & 17.579 & 17.579 & 1882 & 3526 & 0.002 & 0.077 & 0.077 & 0.959 & 0.568 \\
$\bb_{3}$ & 15.902 & 15.902 & 1660 & 2912 & 0.003 & 0.081 & 0.081 & 0.959 & 0.613 \\
$\bb_{4}$ & 13.995 & 13.995 & 1382 & 2257 & 0.002 & 0.090 & 0.090 & 0.954 & 0.687 \\
$\bb_{5}$ & 12.000 & 12.000 & 1081 & 1641 & 0.000 & 0.103 & 0.103 & 0.957 & 0.798 \\
$\bb_{6}$ & 10.000 & 10.000 & 788 & 1112 & -0.002 & 0.123 & 0.123 & 0.961 & 0.966 \\
$\bb_{7}$ & 8.181 & 8.181 & 545 & 717 & -0.003 & 0.152 & 0.152 & 0.966 & 1.360 \\
\bottomrule\bottomrule
\end{tabular}}
    \caption{Fuzzy Effects (Simulations, Linear Heteroskedastic, Kink-Adaptive Bandwidth, ITT Bandwidth Selector).}
    \label{tab:sa-simuls-fuzzy-linear-heteroskedastic-adaptive-bwitt}
\end{table}

\begin{table}[p]
    \centering
    \resizebox{.8\textwidth}{!}{\begin{tabular}{lccccrrrrr}
\toprule\toprule
 & $h_0$ & $h_1$ & $N_{\mathrm{Co}}$ & $N_{\mathrm{Tr}}$ & Bias & SD & RMSE & EC & IL \\
\midrule
$\bb_{1}$ & 8.684 & 8.684 & 625 & 848 & 0.001 & 0.153 & 0.153 & 0.959 & 0.593 \\
$\bb_{2}$ & 8.416 & 8.416 & 601 & 807 & 0.002 & 0.154 & 0.154 & 0.958 & 0.599 \\
$\bb_{3}$ & 8.209 & 8.209 & 580 & 773 & 0.003 & 0.156 & 0.156 & 0.958 & 0.605 \\
$\bb_{4}$ & 8.045 & 8.045 & 559 & 740 & 0.001 & 0.160 & 0.160 & 0.955 & 0.610 \\
$\bb_{5}$ & 7.926 & 7.926 & 539 & 711 & -0.002 & 0.160 & 0.160 & 0.955 & 0.616 \\
$\bb_{6}$ & 7.850 & 7.850 & 521 & 684 & -0.004 & 0.161 & 0.161 & 0.962 & 0.619 \\
$\bb_{7}$ & 7.780 & 7.780 & 501 & 651 & -0.003 & 0.162 & 0.162 & 0.958 & 0.625 \\
\bottomrule\bottomrule
\end{tabular}}
    \caption{Fuzzy Effects (Simulations, Linear Heteroskedastic, Unknown-Kink Bandwidth, ITT Bandwidth Selector).}
    \label{tab:sa-simuls-fuzzy-linear-heteroskedastic-kinkon-bwitt}
\end{table}

\begin{table}[p]
    \centering
    \resizebox{.8\textwidth}{!}{\begin{tabular}{lccccrrrrr}
\toprule\toprule
 & $h_0$ & $h_1$ & $N_{\mathrm{Co}}$ & $N_{\mathrm{Tr}}$ & Bias & SD & RMSE & EC & IL \\
\midrule
$\bb_{1}$ & 18.920 & 18.920 & 2042 & 4036 & -0.000 & 0.084 & 0.084 & 0.962 & 0.554 \\
$\bb_{2}$ & 18.614 & 18.614 & 2065 & 3924 & -0.000 & 0.083 & 0.083 & 0.961 & 0.551 \\
$\bb_{3}$ & 18.397 & 18.397 & 2110 & 3794 & 0.000 & 0.083 & 0.083 & 0.963 & 0.548 \\
$\bb_{4}$ & 18.284 & 18.284 & 2182 & 3647 & 0.003 & 0.083 & 0.083 & 0.959 & 0.543 \\
$\bb_{5}$ & 18.172 & 18.172 & 2256 & 3454 & 0.006 & 0.084 & 0.084 & 0.960 & 0.541 \\
$\bb_{6}$ & 18.104 & 18.104 & 2332 & 3236 & 0.010 & 0.084 & 0.085 & 0.966 & 0.540 \\
$\bb_{7}$ & 17.715 & 17.715 & 2330 & 2898 & 0.011 & 0.086 & 0.086 & 0.962 & 0.551 \\
\bottomrule\bottomrule
\end{tabular}}
    \caption{Fuzzy Effects (Simulations, Linear Heteroskedastic, \texttt{rdrobust} Bandwidth, ITT Bandwidth Selector).}
    \label{tab:sa-simuls-fuzzy-linear-heteroskedastic-rdrobustadj-bwitt}
\end{table}

\begin{table}[p]
    \centering
    \resizebox{.8\textwidth}{!}{\begin{tabular}{lccccrrrrr}
\toprule\toprule
 & $h_0$ & $h_1$ & $N_{\mathrm{Co}}$ & $N_{\mathrm{Tr}}$ & Bias & SD & RMSE & EC & IL \\
\midrule
$\bb_{1}$ & 17.945 & 17.945 & 1903 & 3648 & 0.002 & 0.060 & 0.060 & 0.957 & 0.443 \\
$\bb_{2}$ & 17.380 & 17.380 & 1868 & 3459 & 0.001 & 0.061 & 0.061 & 0.953 & 0.450 \\
$\bb_{3}$ & 17.202 & 17.202 & 1891 & 3370 & 0.002 & 0.062 & 0.063 & 0.960 & 0.448 \\
$\bb_{4}$ & 17.380 & 17.380 & 1997 & 3341 & 0.007 & 0.063 & 0.064 & 0.960 & 0.439 \\
$\bb_{5}$ & 17.857 & 17.857 & 2186 & 3339 & 0.016 & 0.063 & 0.065 & 0.956 & 0.424 \\
$\bb_{6}$ & 18.430 & 18.430 & 2397 & 3310 & 0.026 & 0.060 & 0.065 & 0.950 & 0.407 \\
$\bb_{7}$ & 18.889 & 18.889 & 2574 & 3202 & 0.032 & 0.057 & 0.065 & 0.958 & 0.394 \\
\bottomrule\bottomrule
\end{tabular}}
    \caption{Fuzzy Effects (Simulations, Quadratic Homoskedastic, Smooth-Boundary Bandwidth, ITT Bandwidth Selector).}
    \label{tab:sa-simuls-fuzzy-quadratic-homoskedastic-kinkoff-bwitt}
\end{table}

\begin{table}[p]
    \centering
    \resizebox{.8\textwidth}{!}{\begin{tabular}{lccccrrrrr}
\toprule\toprule
 & $h_0$ & $h_1$ & $N_{\mathrm{Co}}$ & $N_{\mathrm{Tr}}$ & Bias & SD & RMSE & EC & IL \\
\midrule
$\bb_{1}$ & 17.655 & 17.655 & 1862 & 3533 & 0.003 & 0.061 & 0.061 & 0.957 & 0.448 \\
$\bb_{2}$ & 16.710 & 16.710 & 1760 & 3207 & 0.002 & 0.063 & 0.063 & 0.952 & 0.463 \\
$\bb_{3}$ & 15.625 & 15.625 & 1619 & 2818 & 0.002 & 0.065 & 0.065 & 0.958 & 0.485 \\
$\bb_{4}$ & 13.979 & 13.979 & 1379 & 2252 & 0.002 & 0.070 & 0.070 & 0.956 & 0.536 \\
$\bb_{5}$ & 12.000 & 12.000 & 1080 & 1642 & 0.002 & 0.081 & 0.081 & 0.958 & 0.622 \\
$\bb_{6}$ & 10.000 & 10.000 & 787 & 1114 & 0.001 & 0.096 & 0.096 & 0.962 & 0.752 \\
$\bb_{7}$ & 8.509 & 8.509 & 589 & 775 & 0.000 & 0.114 & 0.114 & 0.967 & 0.910 \\
\bottomrule\bottomrule
\end{tabular}}
    \caption{Fuzzy Effects (Simulations, Quadratic Homoskedastic, Kink-Adaptive Bandwidth, ITT Bandwidth Selector).}
    \label{tab:sa-simuls-fuzzy-quadratic-homoskedastic-adaptive-bwitt}
\end{table}

\begin{table}[p]
    \centering
    \resizebox{.8\textwidth}{!}{\input{tables/simuls_fuzzy_quadratic_homoskedastic_kinkon_bwitt.tex}}
    \caption{Fuzzy Effects (Simulations, Quadratic Homoskedastic, Unknown-Kink Bandwidth, ITT Bandwidth Selector).}
    \label{tab:sa-simuls-fuzzy-quadratic-homoskedastic-kinkon-bwitt}
\end{table}

\begin{table}[p]
    \centering
    \resizebox{.8\textwidth}{!}{\input{tables/simuls_fuzzy_quadratic_homoskedastic_rdrobustadj_bwitt.tex}}
    \caption{Fuzzy Effects (Simulations, Quadratic Homoskedastic, \texttt{rdrobust} Bandwidth, ITT Bandwidth Selector).}
    \label{tab:sa-simuls-fuzzy-quadratic-homoskedastic-rdrobustadj-bwitt}
\end{table}

\begin{table}[p]
    \centering
    \resizebox{.8\textwidth}{!}{\input{tables/simuls_fuzzy_quadratic_heteroskedastic_kinkoff_bwitt.tex}}
    \caption{Fuzzy Effects (Simulations, Quadratic Heteroskedastic, Smooth-Boundary Bandwidth, ITT Bandwidth Selector).}
    \label{tab:sa-simuls-fuzzy-quadratic-heteroskedastic-kinkoff-bwitt}
\end{table}

\begin{table}[p]
    \centering
    \resizebox{.8\textwidth}{!}{\input{tables/simuls_fuzzy_quadratic_heteroskedastic_adaptive_bwitt.tex}}
    \caption{Fuzzy Effects (Simulations, Quadratic Heteroskedastic, Kink-Adaptive Bandwidth, ITT Bandwidth Selector).}
    \label{tab:sa-simuls-fuzzy-quadratic-heteroskedastic-adaptive-bwitt}
\end{table}

\begin{table}[p]
    \centering
    \resizebox{.8\textwidth}{!}{\input{tables/simuls_fuzzy_quadratic_heteroskedastic_kinkon_bwitt.tex}}
    \caption{Fuzzy Effects (Simulations, Quadratic Heteroskedastic, Unknown-Kink Bandwidth, ITT Bandwidth Selector).}
    \label{tab:sa-simuls-fuzzy-quadratic-heteroskedastic-kinkon-bwitt}
\end{table}

\begin{table}[p]
    \centering
    \resizebox{.8\textwidth}{!}{\input{tables/simuls_fuzzy_quadratic_heteroskedastic_rdrobustadj_bwitt.tex}}
    \caption{Fuzzy Effects (Simulations, Quadratic Heteroskedastic, \texttt{rdrobust} Bandwidth, ITT Bandwidth Selector).}
    \label{tab:sa-simuls-fuzzy-quadratic-heteroskedastic-rdrobustadj-bwitt}
\end{table}

\clearpage

\begin{table}[p]
    \centering
    \resizebox{.8\textwidth}{!}{\input{tables/simuls_itt_linear_homoskedastic_kinkoff_bwmain.tex}}
    \caption{Intention-to-Treat Effects (Simulations, Linear Homoskedastic, Smooth-Boundary Bandwidth, Fuzzy Bandwidth Selector).}
    \label{tab:sa-simuls-itt-linear-homoskedastic-kinkoff-bwmain}
\end{table}

\begin{table}[p]
    \centering
    \resizebox{.8\textwidth}{!}{\input{tables/simuls_itt_linear_homoskedastic_adaptive_bwmain.tex}}
    \caption{Intention-to-Treat Effects (Simulations, Linear Homoskedastic, Kink-Adaptive Bandwidth, Fuzzy Bandwidth Selector).}
    \label{tab:sa-simuls-itt-linear-homoskedastic-adaptive-bwmain}
\end{table}

\begin{table}[p]
    \centering
    \resizebox{.8\textwidth}{!}{\input{tables/simuls_itt_linear_homoskedastic_kinkon_bwmain.tex}}
    \caption{Intention-to-Treat Effects (Simulations, Linear Homoskedastic, Unknown-Kink Bandwidth, Fuzzy Bandwidth Selector).}
    \label{tab:sa-simuls-itt-linear-homoskedastic-kinkon-bwmain}
\end{table}

\begin{table}[p]
    \centering
    \resizebox{.8\textwidth}{!}{\input{tables/simuls_itt_linear_homoskedastic_rdrobustadj_bwmain.tex}}
    \caption{Intention-to-Treat Effects (Simulations, Linear Homoskedastic, \texttt{rdrobust} Bandwidth, Fuzzy Bandwidth Selector).}
    \label{tab:sa-simuls-itt-linear-homoskedastic-rdrobustadj-bwmain}
\end{table}

\begin{table}[p]
    \centering
    \resizebox{.8\textwidth}{!}{\input{tables/simuls_itt_linear_heteroskedastic_kinkoff_bwmain.tex}}
    \caption{Intention-to-Treat Effects (Simulations, Linear Heteroskedastic, Smooth-Boundary Bandwidth, Fuzzy Bandwidth Selector).}
    \label{tab:sa-simuls-itt-linear-heteroskedastic-kinkoff-bwmain}
\end{table}

\begin{table}[p]
    \centering
    \resizebox{.8\textwidth}{!}{\input{tables/simuls_itt_linear_heteroskedastic_adaptive_bwmain.tex}}
    \caption{Intention-to-Treat Effects (Simulations, Linear Heteroskedastic, Kink-Adaptive Bandwidth, Fuzzy Bandwidth Selector).}
    \label{tab:sa-simuls-itt-linear-heteroskedastic-adaptive-bwmain}
\end{table}

\begin{table}[p]
    \centering
    \resizebox{.8\textwidth}{!}{\input{tables/simuls_itt_linear_heteroskedastic_kinkon_bwmain.tex}}
    \caption{Intention-to-Treat Effects (Simulations, Linear Heteroskedastic, Unknown-Kink Bandwidth, Fuzzy Bandwidth Selector).}
    \label{tab:sa-simuls-itt-linear-heteroskedastic-kinkon-bwmain}
\end{table}

\begin{table}[p]
    \centering
    \resizebox{.8\textwidth}{!}{\input{tables/simuls_itt_linear_heteroskedastic_rdrobustadj_bwmain.tex}}
    \caption{Intention-to-Treat Effects (Simulations, Linear Heteroskedastic, \texttt{rdrobust} Bandwidth, Fuzzy Bandwidth Selector).}
    \label{tab:sa-simuls-itt-linear-heteroskedastic-rdrobustadj-bwmain}
\end{table}

\begin{table}[p]
    \centering
    \resizebox{.8\textwidth}{!}{\input{tables/simuls_itt_quadratic_homoskedastic_kinkoff_bwmain.tex}}
    \caption{Intention-to-Treat Effects (Simulations, Quadratic Homoskedastic, Smooth-Boundary Bandwidth, Fuzzy Bandwidth Selector).}
    \label{tab:sa-simuls-itt-quadratic-homoskedastic-kinkoff-bwmain}
\end{table}

\begin{table}[p]
    \centering
    \resizebox{.8\textwidth}{!}{\input{tables/simuls_itt_quadratic_homoskedastic_adaptive_bwmain.tex}}
    \caption{Intention-to-Treat Effects (Simulations, Quadratic Homoskedastic, Kink-Adaptive Bandwidth, Fuzzy Bandwidth Selector).}
    \label{tab:sa-simuls-itt-quadratic-homoskedastic-adaptive-bwmain}
\end{table}

\begin{table}[p]
    \centering
    \resizebox{.8\textwidth}{!}{\input{tables/simuls_itt_quadratic_homoskedastic_kinkon_bwmain.tex}}
    \caption{Intention-to-Treat Effects (Simulations, Quadratic Homoskedastic, Unknown-Kink Bandwidth, Fuzzy Bandwidth Selector).}
    \label{tab:sa-simuls-itt-quadratic-homoskedastic-kinkon-bwmain}
\end{table}

\begin{table}[p]
    \centering
    \resizebox{.8\textwidth}{!}{\input{tables/simuls_itt_quadratic_homoskedastic_rdrobustadj_bwmain.tex}}
    \caption{Intention-to-Treat Effects (Simulations, Quadratic Homoskedastic, \texttt{rdrobust} Bandwidth, Fuzzy Bandwidth Selector).}
    \label{tab:sa-simuls-itt-quadratic-homoskedastic-rdrobustadj-bwmain}
\end{table}

\begin{table}[p]
    \centering
    \resizebox{.8\textwidth}{!}{\input{tables/simuls_itt_quadratic_heteroskedastic_kinkoff_bwmain.tex}}
    \caption{Intention-to-Treat Effects (Simulations, Quadratic Heteroskedastic, Smooth-Boundary Bandwidth, Fuzzy Bandwidth Selector).}
    \label{tab:sa-simuls-itt-quadratic-heteroskedastic-kinkoff-bwmain}
\end{table}

\begin{table}[p]
    \centering
    \resizebox{.8\textwidth}{!}{\input{tables/simuls_itt_quadratic_heteroskedastic_adaptive_bwmain.tex}}
    \caption{Intention-to-Treat Effects (Simulations, Quadratic Heteroskedastic, Kink-Adaptive Bandwidth, Fuzzy Bandwidth Selector).}
    \label{tab:sa-simuls-itt-quadratic-heteroskedastic-adaptive-bwmain}
\end{table}

\begin{table}[p]
    \centering
    \resizebox{.8\textwidth}{!}{\input{tables/simuls_itt_quadratic_heteroskedastic_kinkon_bwmain.tex}}
    \caption{Intention-to-Treat Effects (Simulations, Quadratic Heteroskedastic, Unknown-Kink Bandwidth, Fuzzy Bandwidth Selector).}
    \label{tab:sa-simuls-itt-quadratic-heteroskedastic-kinkon-bwmain}
\end{table}

\begin{table}[p]
    \centering
    \resizebox{.8\textwidth}{!}{\input{tables/simuls_itt_quadratic_heteroskedastic_rdrobustadj_bwmain.tex}}
    \caption{Intention-to-Treat Effects (Simulations, Quadratic Heteroskedastic, \texttt{rdrobust} Bandwidth, Fuzzy Bandwidth Selector).}
    \label{tab:sa-simuls-itt-quadratic-heteroskedastic-rdrobustadj-bwmain}
\end{table}

\clearpage

\begin{table}[p]
    \centering
    \resizebox{.8\textwidth}{!}{\input{tables/simuls_fs_linear_homoskedastic_kinkoff_bwmain.tex}}
    \caption{First-Stage Effects (Simulations, Linear Homoskedastic, Smooth-Boundary Bandwidth, Fuzzy Bandwidth Selector).}
    \label{tab:sa-simuls-fs-linear-homoskedastic-kinkoff-bwmain}
\end{table}

\begin{table}[p]
    \centering
    \resizebox{.8\textwidth}{!}{\input{tables/simuls_fs_linear_homoskedastic_adaptive_bwmain.tex}}
    \caption{First-Stage Effects (Simulations, Linear Homoskedastic, Kink-Adaptive Bandwidth, Fuzzy Bandwidth Selector).}
    \label{tab:sa-simuls-fs-linear-homoskedastic-adaptive-bwmain}
\end{table}

\begin{table}[p]
    \centering
    \resizebox{.8\textwidth}{!}{\input{tables/simuls_fs_linear_homoskedastic_kinkon_bwmain.tex}}
    \caption{First-Stage Effects (Simulations, Linear Homoskedastic, Unknown-Kink Bandwidth, Fuzzy Bandwidth Selector).}
    \label{tab:sa-simuls-fs-linear-homoskedastic-kinkon-bwmain}
\end{table}

\begin{table}[p]
    \centering
    \resizebox{.8\textwidth}{!}{\input{tables/simuls_fs_linear_homoskedastic_rdrobustadj_bwmain.tex}}
    \caption{First-Stage Effects (Simulations, Linear Homoskedastic, \texttt{rdrobust} Bandwidth, Fuzzy Bandwidth Selector).}
    \label{tab:sa-simuls-fs-linear-homoskedastic-rdrobustadj-bwmain}
\end{table}

\begin{table}[p]
    \centering
    \resizebox{.8\textwidth}{!}{\input{tables/simuls_fs_linear_heteroskedastic_kinkoff_bwmain.tex}}
    \caption{First-Stage Effects (Simulations, Linear Heteroskedastic, Smooth-Boundary Bandwidth, Fuzzy Bandwidth Selector).}
    \label{tab:sa-simuls-fs-linear-heteroskedastic-kinkoff-bwmain}
\end{table}

\begin{table}[p]
    \centering
    \resizebox{.8\textwidth}{!}{\input{tables/simuls_fs_linear_heteroskedastic_adaptive_bwmain.tex}}
    \caption{First-Stage Effects (Simulations, Linear Heteroskedastic, Kink-Adaptive Bandwidth, Fuzzy Bandwidth Selector).}
    \label{tab:sa-simuls-fs-linear-heteroskedastic-adaptive-bwmain}
\end{table}

\begin{table}[p]
    \centering
    \resizebox{.8\textwidth}{!}{\input{tables/simuls_fs_linear_heteroskedastic_kinkon_bwmain.tex}}
    \caption{First-Stage Effects (Simulations, Linear Heteroskedastic, Unknown-Kink Bandwidth, Fuzzy Bandwidth Selector).}
    \label{tab:sa-simuls-fs-linear-heteroskedastic-kinkon-bwmain}
\end{table}

\begin{table}[p]
    \centering
    \resizebox{.8\textwidth}{!}{\input{tables/simuls_fs_linear_heteroskedastic_rdrobustadj_bwmain.tex}}
    \caption{First-Stage Effects (Simulations, Linear Heteroskedastic, \texttt{rdrobust} Bandwidth, Fuzzy Bandwidth Selector).}
    \label{tab:sa-simuls-fs-linear-heteroskedastic-rdrobustadj-bwmain}
\end{table}

\begin{table}[p]
    \centering
    \resizebox{.8\textwidth}{!}{\input{tables/simuls_fs_quadratic_homoskedastic_kinkoff_bwmain.tex}}
    \caption{First-Stage Effects (Simulations, Quadratic Homoskedastic, Smooth-Boundary Bandwidth, Fuzzy Bandwidth Selector).}
    \label{tab:sa-simuls-fs-quadratic-homoskedastic-kinkoff-bwmain}
\end{table}

\begin{table}[p]
    \centering
    \resizebox{.8\textwidth}{!}{\input{tables/simuls_fs_quadratic_homoskedastic_adaptive_bwmain.tex}}
    \caption{First-Stage Effects (Simulations, Quadratic Homoskedastic, Kink-Adaptive Bandwidth, Fuzzy Bandwidth Selector).}
    \label{tab:sa-simuls-fs-quadratic-homoskedastic-adaptive-bwmain}
\end{table}

\begin{table}[p]
    \centering
    \resizebox{.8\textwidth}{!}{\input{tables/simuls_fs_quadratic_homoskedastic_kinkon_bwmain.tex}}
    \caption{First-Stage Effects (Simulations, Quadratic Homoskedastic, Unknown-Kink Bandwidth, Fuzzy Bandwidth Selector).}
    \label{tab:sa-simuls-fs-quadratic-homoskedastic-kinkon-bwmain}
\end{table}

\begin{table}[p]
    \centering
    \resizebox{.8\textwidth}{!}{\input{tables/simuls_fs_quadratic_homoskedastic_rdrobustadj_bwmain.tex}}
    \caption{First-Stage Effects (Simulations, Quadratic Homoskedastic, \texttt{rdrobust} Bandwidth, Fuzzy Bandwidth Selector).}
    \label{tab:sa-simuls-fs-quadratic-homoskedastic-rdrobustadj-bwmain}
\end{table}

\begin{table}[p]
    \centering
    \resizebox{.8\textwidth}{!}{\input{tables/simuls_fs_quadratic_heteroskedastic_kinkoff_bwmain.tex}}
    \caption{First-Stage Effects (Simulations, Quadratic Heteroskedastic, Smooth-Boundary Bandwidth, Fuzzy Bandwidth Selector).}
    \label{tab:sa-simuls-fs-quadratic-heteroskedastic-kinkoff-bwmain}
\end{table}

\begin{table}[p]
    \centering
    \resizebox{.8\textwidth}{!}{\input{tables/simuls_fs_quadratic_heteroskedastic_adaptive_bwmain.tex}}
    \caption{First-Stage Effects (Simulations, Quadratic Heteroskedastic, Kink-Adaptive Bandwidth, Fuzzy Bandwidth Selector).}
    \label{tab:sa-simuls-fs-quadratic-heteroskedastic-adaptive-bwmain}
\end{table}

\begin{table}[p]
    \centering
    \resizebox{.8\textwidth}{!}{\input{tables/simuls_fs_quadratic_heteroskedastic_kinkon_bwmain.tex}}
    \caption{First-Stage Effects (Simulations, Quadratic Heteroskedastic, Unknown-Kink Bandwidth, Fuzzy Bandwidth Selector).}
    \label{tab:sa-simuls-fs-quadratic-heteroskedastic-kinkon-bwmain}
\end{table}

\begin{table}[p]
    \centering
    \resizebox{.8\textwidth}{!}{\input{tables/simuls_fs_quadratic_heteroskedastic_rdrobustadj_bwmain.tex}}
    \caption{First-Stage Effects (Simulations, Quadratic Heteroskedastic, \texttt{rdrobust} Bandwidth, Fuzzy Bandwidth Selector).}
    \label{tab:sa-simuls-fs-quadratic-heteroskedastic-rdrobustadj-bwmain}
\end{table}

\clearpage

\begin{table}[p]
    \centering
    \resizebox{.8\textwidth}{!}{\input{tables/simuls_fuzzy_linear_homoskedastic_kinkoff_bwmain.tex}}
    \caption{Fuzzy Effects (Simulations, Linear Homoskedastic, Smooth-Boundary Bandwidth, Fuzzy Bandwidth Selector).}
    \label{tab:sa-simuls-fuzzy-linear-homoskedastic-kinkoff-bwmain}
\end{table}

\begin{table}[p]
    \centering
    \resizebox{.8\textwidth}{!}{\input{tables/simuls_fuzzy_linear_homoskedastic_adaptive_bwmain.tex}}
    \caption{Fuzzy Effects (Simulations, Linear Homoskedastic, Kink-Adaptive Bandwidth, Fuzzy Bandwidth Selector).}
    \label{tab:sa-simuls-fuzzy-linear-homoskedastic-adaptive-bwmain}
\end{table}

\begin{table}[p]
    \centering
    \resizebox{.8\textwidth}{!}{\input{tables/simuls_fuzzy_linear_homoskedastic_kinkon_bwmain.tex}}
    \caption{Fuzzy Effects (Simulations, Linear Homoskedastic, Unknown-Kink Bandwidth, Fuzzy Bandwidth Selector).}
    \label{tab:sa-simuls-fuzzy-linear-homoskedastic-kinkon-bwmain}
\end{table}

\begin{table}[p]
    \centering
    \resizebox{.8\textwidth}{!}{\input{tables/simuls_fuzzy_linear_homoskedastic_rdrobustadj_bwmain.tex}}
    \caption{Fuzzy Effects (Simulations, Linear Homoskedastic, \texttt{rdrobust} Bandwidth, Fuzzy Bandwidth Selector).}
    \label{tab:sa-simuls-fuzzy-linear-homoskedastic-rdrobustadj-bwmain}
\end{table}

\begin{table}[p]
    \centering
    \resizebox{.8\textwidth}{!}{\input{tables/simuls_fuzzy_linear_heteroskedastic_kinkoff_bwmain.tex}}
    \caption{Fuzzy Effects (Simulations, Linear Heteroskedastic, Smooth-Boundary Bandwidth, Fuzzy Bandwidth Selector).}
    \label{tab:sa-simuls-fuzzy-linear-heteroskedastic-kinkoff-bwmain}
\end{table}

\begin{table}[p]
    \centering
    \resizebox{.8\textwidth}{!}{\input{tables/simuls_fuzzy_linear_heteroskedastic_adaptive_bwmain.tex}}
    \caption{Fuzzy Effects (Simulations, Linear Heteroskedastic, Kink-Adaptive Bandwidth, Fuzzy Bandwidth Selector).}
    \label{tab:sa-simuls-fuzzy-linear-heteroskedastic-adaptive-bwmain}
\end{table}

\begin{table}[p]
    \centering
    \resizebox{.8\textwidth}{!}{\input{tables/simuls_fuzzy_linear_heteroskedastic_kinkon_bwmain.tex}}
    \caption{Fuzzy Effects (Simulations, Linear Heteroskedastic, Unknown-Kink Bandwidth, Fuzzy Bandwidth Selector).}
    \label{tab:sa-simuls-fuzzy-linear-heteroskedastic-kinkon-bwmain}
\end{table}

\begin{table}[p]
    \centering
    \resizebox{.8\textwidth}{!}{\input{tables/simuls_fuzzy_linear_heteroskedastic_rdrobustadj_bwmain.tex}}
    \caption{Fuzzy Effects (Simulations, Linear Heteroskedastic, \texttt{rdrobust} Bandwidth, Fuzzy Bandwidth Selector).}
    \label{tab:sa-simuls-fuzzy-linear-heteroskedastic-rdrobustadj-bwmain}
\end{table}

\begin{table}[p]
    \centering
    \resizebox{.8\textwidth}{!}{\input{tables/simuls_fuzzy_quadratic_homoskedastic_kinkoff_bwmain.tex}}
    \caption{Fuzzy Effects (Simulations, Quadratic Homoskedastic, Smooth-Boundary Bandwidth, Fuzzy Bandwidth Selector).}
    \label{tab:sa-simuls-fuzzy-quadratic-homoskedastic-kinkoff-bwmain}
\end{table}

\begin{table}[p]
    \centering
    \resizebox{.8\textwidth}{!}{\input{tables/simuls_fuzzy_quadratic_homoskedastic_adaptive_bwmain.tex}}
    \caption{Fuzzy Effects (Simulations, Quadratic Homoskedastic, Kink-Adaptive Bandwidth, Fuzzy Bandwidth Selector).}
    \label{tab:sa-simuls-fuzzy-quadratic-homoskedastic-adaptive-bwmain}
\end{table}

\begin{table}[p]
    \centering
    \resizebox{.8\textwidth}{!}{\input{tables/simuls_fuzzy_quadratic_homoskedastic_kinkon_bwmain.tex}}
    \caption{Fuzzy Effects (Simulations, Quadratic Homoskedastic, Unknown-Kink Bandwidth, Fuzzy Bandwidth Selector).}
    \label{tab:sa-simuls-fuzzy-quadratic-homoskedastic-kinkon-bwmain}
\end{table}

\begin{table}[p]
    \centering
    \resizebox{.8\textwidth}{!}{\input{tables/simuls_fuzzy_quadratic_homoskedastic_rdrobustadj_bwmain.tex}}
    \caption{Fuzzy Effects (Simulations, Quadratic Homoskedastic, \texttt{rdrobust} Bandwidth, Fuzzy Bandwidth Selector).}
    \label{tab:sa-simuls-fuzzy-quadratic-homoskedastic-rdrobustadj-bwmain}
\end{table}

\begin{table}[p]
    \centering
    \resizebox{.8\textwidth}{!}{\input{tables/simuls_fuzzy_quadratic_heteroskedastic_kinkoff_bwmain.tex}}
    \caption{Fuzzy Effects (Simulations, Quadratic Heteroskedastic, Smooth-Boundary Bandwidth, Fuzzy Bandwidth Selector).}
    \label{tab:sa-simuls-fuzzy-quadratic-heteroskedastic-kinkoff-bwmain}
\end{table}

\begin{table}[p]
    \centering
    \resizebox{.8\textwidth}{!}{\input{tables/simuls_fuzzy_quadratic_heteroskedastic_adaptive_bwmain.tex}}
    \caption{Fuzzy Effects (Simulations, Quadratic Heteroskedastic, Kink-Adaptive Bandwidth, Fuzzy Bandwidth Selector).}
    \label{tab:sa-simuls-fuzzy-quadratic-heteroskedastic-adaptive-bwmain}
\end{table}

\begin{table}[p]
    \centering
    \resizebox{.8\textwidth}{!}{\input{tables/simuls_fuzzy_quadratic_heteroskedastic_kinkon_bwmain.tex}}
    \caption{Fuzzy Effects (Simulations, Quadratic Heteroskedastic, Unknown-Kink Bandwidth, Fuzzy Bandwidth Selector).}
    \label{tab:sa-simuls-fuzzy-quadratic-heteroskedastic-kinkon-bwmain}
\end{table}

\begin{table}[p]
    \centering
    \resizebox{.8\textwidth}{!}{\input{tables/simuls_fuzzy_quadratic_heteroskedastic_rdrobustadj_bwmain.tex}}
    \caption{Fuzzy Effects (Simulations, Quadratic Heteroskedastic, \texttt{rdrobust} Bandwidth, Fuzzy Bandwidth Selector).}
    \label{tab:sa-simuls-fuzzy-quadratic-heteroskedastic-rdrobustadj-bwmain}
\end{table}

\clearpage
\bibliography{CTY_2026_JOE--bib}
\bibliographystyle{plainnat}
